\documentclass[preprint]{aastex}

\usepackage{mathptmx}
\usepackage{amsmath}
\usepackage{graphicx}
\usepackage{latexsym}

\newcommand{\npo}{{n+1}}

\newcommand{\half}{\frac{1}{2}}
\newcommand{\lhalf}{(1/2)}

\newcommand{\lthreehalf}{(3/2)}
\newcommand{\kn}{{R_\epsilon}}
\newcommand\gtsim{\,\lower0.7ex\hbox{$\stackrel{>}{\sim}$}\,}
\newcommand\ltsim{\,\lower0.7ex\hbox{$\stackrel{<}{\sim}$}\,}
\newcommand{\varv}{{\upsilon}}
\newcommand{\smpmb}{\pmb}
\newcommand{\bfnabla}{{ \makebox{$\nabla$} \makebox[-7.80pt]{ }\raisebox{-0.1pt}[0pt][0pt]{$\nabla$} \makebox[-7.90pt]{ } \raisebox{-0.2pt}[0pt][0pt]{$\nabla$}\makebox[-7.90pt]{ } \raisebox{-0.3pt}[0pt][0pt]{$\nabla$} \makebox[-7.95pt]{ } \raisebox{-0.4pt}[0pt][0pt]{$\nabla$}}}

\def\ldbrack{\lbrack\!\!\!\left\lbrack}
\def\rdbrack{\rbrack\!\!\!\right\rbrack}

\shorttitle{Numerical Methods for 2-D Neutrino-RHD}
\shortauthors{Swesty and Myra}

\begin{document}

\title{A Numerical Algorithm for Modeling
Multigroup Neutrino-Radiation Hydrodynamics in Two Spatial Dimensions
\footnote{Submitted to {\em The Astrophysical Journal}.
This preprint, complete with full resolution figures, is a\-vail\-a\-ble
from http://nuclear.astro.sunysb.edu/emyra/pubs/swesty-myra\_methods.pdf.}
}

\author{F.\ Douglas Swesty and Eric S.\ Myra}
\affil{Department of Physics and Astronomy, \\
       State University of New York at Stony Brook, \\
       Stony Brook, NY 11794--3800}
\email{dswesty@mail.astro.sunysb.edu \\ emyra@mail.astro.sunysb.edu}

\begin{abstract} {\small It is now generally agreed that
multidimensional, multigroup, radiation
hydrodynamics is an indispensable element of any
realistic model of stellar-core collapse,
core-collapse supernovae, and proto-neutron star
instabilities.  We have developed a new,
two-dimensional, multigroup algorithm that can
model neutrino-radiation-hydrodynamic flows in
core-collapse supernovae.  Our algorithm uses an
approach that is similar to the ZEUS family of
algorithms, originally developed by Stone and
Norman. However, we extend that previous work in
three significant ways: First, we incorporate
multispecies, multigroup, radiation hydrodynamics
in a flux-limited-diffusion approximation.  Our
approach is capable of modeling pair-coupled
neutrino-radiation hydrodynamics, and includes
effects of Pauli blocking in the collision
integrals. Blocking gives rise to nonlinearities
in the discretized radiation-transport equations,
which we evolve implicitly in time. We employ
parallelized Newton-Krylov methods to obtain a
solution of these nonlinear, implicit equations.
Our second major extension to the ZEUS algorithm
is inclusion of an electron conservation equation,
which describes evolution of electron-number
density in the hydrodynamic flow.  This permits
following the effects of deleptonization in a
stellar core.  In our third extension, we have
modified the hydrodynamics algorithm to
accommodate realistic, complex equations of state,
including those having non-convex behavior.  In
this paper, we present a description of our
complete algorithm, giving detail sufficient to
allow others to implement, reproduce, and extend
our work.  Finite-differencing details are
presented in appendices.  We also discuss
implementation of this algorithm on
state-of-the-art, parallel-computing
architectures.  Finally, we present results of
verification tests that demonstrate the numerical
accuracy of this algorithm on diverse
hydrodynamic, gravitational, radiation-transport,
and radiation-hydrodynamic sample problems.  We
believe our methods to be of general use in a
variety of model settings where radiation
transport or radiation hydrodynamics is important.
Extension of this work to three spatial dimensions
is straightforward. }
\end{abstract}


\section{Introduction}

Development of a numerical description of neutrino-radiation-hydrodynamic
phenomena presents numerous modeling challenges.  These challenges
result from the complexity of the system---complexity stemming both
from the sheer number of important physical components, and from the
diversity and variability of microphysical interactions among these
components.  In the core-collapse supernova problem, this 
complexity presents itself in a
number of ways.  There is complexity associated with the interacting
flows of matter and neutrino radiation.  These flows exhibit coupling
strengths that vary spatially, and change as a region evolves. There
is typically strong coupling between matter and neutrino-radiation
in dense regions near the center, weak
coupling in the more diffuse outer regions of the core, and some kind
of intermediate-strength coupling elsewhere.  Where coupling is
weaker, neutrino radiation is not in local thermodynamic equilibrium
(LTE) with the matter.  Adding to this is the presence of multiple
species of neutrinos that are coupled through pair-production
processes and through the exchange of energy and lepton number with
matter.  Hence, neutrino distributions span a large range of classical
and quantum-mechanical behavior.  There is also a complex nuclear
chemistry, mediated by both the strong and weak interactions, present
in the problem.  Finally, the matter is described by a non-ideal-gas
equation of state (EOS), which includes phase transitions and other
complex behavior.

Modeling such a phenomenon requires a set of 
neutrino-radiation-hydrodynamic
equations that accounts for all the aforementioned complexities, which
must then be integrated forward in time from an initial model.  In
this paper, we present an algorithm for accomplishing this in two
spatial dimensions (2-D).  The extension of this work to three spatial
dimensions (3-D) is straightforward, and 
long-timescale 3-D simulations of core-collapse
supernovae using this approach should be computationally tractable in
the next few years.  In the sections that follow, we provide a
description of our algorithm, supplying detail sufficient for others
to implement and replicate this work.  Although other algorithms for
solving 1-D neutrino-radiation-hydrodynamics equations, or portions of
these equations, have been published
\citep[see, for example,][]{yb77,ss82,bruenn85,mbhlsv,schinder88,sbp88,sb89,mb93b,fds95},
the only, albeit partial, published descriptions of an algorithm for
solving multidimensional Eulerian neutrino-radiation equations 
are those 
of \cite{mwm93} and \cite{buras06}.

In our development of a radiation-hydrodynamic
algorithm relevant to supernovae, we have drawn
heavily from work originally performed for the
development of the ZEUS family of multidimensional
Eulerian radiation and radiation-hydrodynamic
algorithms
\citep{sn92a,sn92b,smn92,ts01,hn03,hayes05}.  This approach 
relies on staggered-mesh schemes to treat both the
hydrodynamic and radiation components of the flow.
In the original paper, \cite{sn92a} lay out a
hydrodynamics algorithm, which we extend to treat
dense-matter hydrodynamics.  More recently,
\cite{ts01} have extended this work to treat
radiation-hydrodynamics in a comoving-frame, gray,
flux-limited diffusion (FLD) approximation.  The
algorithm we describe in this paper extends this
work to a multispecies, nonlinear, multigroup
approach that is capable of treating non-LTE
continuum problems.  Ideally, we would like to
solve the comoving-frame Boltzmann equation in
multiple dimensions.  However, the computational
cost of such an effort would be so great that
multidimensional supernovae simulations could be
carried out only for short times, even when using
the most advanced, present generation of
parallel-computing architectures.  The multispecies, multigroup
flux-limited diffusion (MGFLD) approach we detail
here is an important stepping-stone to this goal.

Finite-difference hydrodynamic algorithms possess
several advantages that make them especially
suited to core-collapse supernova simulations.
First, these algorithms are formulated in a
generalized, orthogonal coordinate system that
allows for the easy interchange of Cartesian-,
cylindrical-, or spherical-polar-coordinate
systems.  Second, these algorithms avoids use of
Riemann solvers, either exact or approximate.
This allows these algorithms to be extended to
incorporate an arbitrary complex, and possibly
non-convex, EOS.  Additionally, such
finite-difference methods can be easily extended
to add new physics.  Finally, these methods are
relatively straightforward to implement on
massively parallel, distributed-memory computing
architectures.

Our scheme similarly builds on this staggered-mesh approach, but makes
major extensions to the treatment of radiation.  This permits
incorporation of numerous important aspects of neutrino flow.  The
first of these extensions accommodates multiple species of
radiation ($\nu_e, \bar{\nu}_e, \nu_\mu, \bar{\nu}_\mu, \nu_\tau,
\bar{\nu}_\tau$), in which particle-antiparticle types are coupled
through pair-production processes.  The particle-antiparticle coupling
mandates the simultaneous solution of the transport equations for both
particles and antiparticles.  The second extension is our multigroup
treatment of the radiation spectrum, which includes ability to treat
full coupling between energy groups that occurs by processes such as
neutrino-electron scattering.  Opacities and emissivities are included
in an energy-dependent form, obviating need for any mean-opacity
approximations.  The third extension of capabilities is the treatment
of Pauli-blocking effects in the radiation microphysics.  This
extension adds nonlinearities to the neutrino transport equations and
leads to additional steps in the numerical solution of those
equations.  Our algorithm also allows modeling of matter
deleptonization, through the addition of an equation describing
evolution of electrons.  Finally, our treatment of the hydrodynamics
equations permits a complex EOS by incorporating a set of nonlinear
solution algorithms for the Lagrangean portion of the gas energy
equation.

Much recent attention has been focused on the need for systematic
processes of verification and validation of computer simulation codes
\citep{roache98,knupp02, calder02,calder04,phystoday}.  These
procedures are required to achieve a reasonable degree of quality
assurance of simulation results.  Verification has been loosely
defined \citep{knupp02} as testing to ensure that equations are
being correctly solved, while validation has been defined
\citep{roache98} as testing to ensure that microphysical models are
adequate descriptions of nature.  In this paper, we present results of
a number of verification problems that stress important components of
our algorithm.  Here, we do not concern ourselves with problem-specific
microphysics. Validation tests for core-collapse supernova 
problems will be addressed in
a forthcoming publication (Swesty \& Myra, in preparation).

An important consideration in the design of
neutrino-radiation-hydrodynamics algorithms is that they be
implementable on state-of-the-art massively-parallel computing
architectures.  We have designed our algorithm
with this goal in mind, and have realized a parallel implementation in
the form of a code (V2D), which we currently employ to simulate
supernova convection (Myra \& Swesty, in preparation) 
and proto-neutron star instabilities (Swesty \& Myra, in preparation).
Although the focus of this paper is on the algorithm,
not the implementation, we have aimed to provide all detail 
necessary to allow
other developers to implement the algorithm and reproduce our
results.

This remainder of this paper is organized as follows: In \S2, we
introduce the coupled equations of neutrino-radiation hydrodynamics.
Section 3 contains a description, in schematic form, of our algorithm
for solving these equations numerically. (In appendices, we provide a
detailed description of the finite differencing, numerical solution, 
and implementation of
boundary conditions.)  We present in \S4 the results of verification
tests that we have used to benchmark this algorithm.  Finally, in \S5,
we present our conclusions about this algorithm.
\section{Equations of Neutrino Radiation Hydrodynamics\label{sec-equations}}

The equations of neutrino radiation hydrodynamics
must describe the time evolution of two primary
components: matter and neutrino radiation.  The
matter is assumed at all times to be in local
thermodynamic equilibrium (LTE).  The neutrinos
are never assumed to be in LTE, although such a
situation may obtain in certain situations. For
the moment, we assume that the radiation can be of
an arbitrary form ({\em e.g.}, photons, neutrinos,
{\em etc.}), and we will make the distinction
specific as needed.  This allows our algorithm to
be used for a variety of radiation-hydrodynamic
situations that may, or may not, involve
neutrinos.  However, we will assume that multiple
species of radiation are present.  In the case
where the radiation component consists of
neutrinos, it is necessary to describe six different
species of neutrino: $\nu_e$,
$\bar{\nu}_e$,$\nu_\mu$, $\bar{\nu}_\mu$,
$\nu_\tau$ and $\bar{\nu}_\tau$.

For the purposes of this paper, we assume the spatial domain
to be free of macroscopic electric and 
magnetic fields. In principal, there is no 
reason that the algorithm we present here could not
be extended to encompass magneto-hydrodynamic phenomena,
but such extensions are beyond the scope of this work.

In the subsections that follow, we first consider 
the hydrodynamic equations that describe the flow of 
dense matter, the equations that describe the evolution
of the comoving multigroup radiation energy density in 
the flux-limited diffusion approximation, and the 
microphysical coupling between matter and radiation.


\subsection{Hydrodynamics with Neutrino-Radiation Coupling} 
\label{sec-hydro-eqns}

The starting point for a description of the
material evolution is the set of Euler equations, which
describe the dynamics of the matter.
The corresponding starting
point for a description of the radiation is the set of
multigroup flux-limited diffusion equations, which we
address in the next section.  Since matter and
radiation do not evolve independently, there are
coupling terms that appear in both sets of
equations to describe the transfer of energy, lepton number,
and momentum between matter and radiation.
The Euler equations, for the system under 
consideration, are:
\begin{equation} \label{eq:cont}
\frac{\partial \rho}{\partial t} +
{\bfnabla} \cdot \left( \rho {\bf v} \right) = 0,
\end{equation}
\begin{equation} \label{eq:ne}
\frac{\partial n_e}{\partial t} +
{\bfnabla} \cdot \left( n_e {\bf v} \right) = {\mathbb N}, 
\end{equation}
\begin{equation} \label{eq:energy}
\frac{ \partial E}{\partial t} +
{\bfnabla} \cdot \left( E {\bf v} \right) +
P {\bfnabla} \cdot {\bf v} 
+{\mathsf Q}:{\bfnabla}{\bf v}
= {\mathbb S},
\end{equation}
\begin{equation} \label{eq:mom}
\frac{ \partial \left( \rho {\bf v} \right) }{\partial t} +
{\bfnabla} \cdot \left( \rho {\bf v}{\bf v} \right) +
{\bfnabla} P + 
{\bfnabla}\cdot {\mathsf Q} +
\rho {\bfnabla} \Phi +
{\bfnabla} \cdot  {\mathsf P}_{{\rm rad}}
= {\mathbb P}.
\end{equation}
Equation~(\ref{eq:cont}) is the continuity equation for mass, where
$\rho$ is the mass density, and ${\bf v}$ is the matter velocity.
These quantities, and those in the following equations, are understood
to be functions of position {\bf x} and time
$t$. Equation~(\ref{eq:ne}) expresses the evolution of electronic
number density, where $n_e$ is the net number density of electrons over
positrons.  It is only relevant to include this equation if there is a
variation in the ratio of the number density of electrons to the number 
density of baryons within the spatial domain, or if there are
processes that can change the net number of electrons in the system.
Thus, if the radiation being considered is
electromagnetic, equation~(\ref{eq:ne}) is a redundant linear multiple of
equation~(\ref{eq:cont}). However, in the presence of weak
interactions in dense matter, equation~(\ref{eq:ne}) is usually 
independent of equation~(\ref{eq:cont}), and its
right-hand side is non-zero.  Here, we express that right-hand-side
term---the net number production rate of electrons, having dimensions
of number per unit volume per unit time---by ${\mathbb N}$.  To
conserve lepton number, such reactions also imply a net number
production of radiation from neutrinos or some other
lepton. Therefore, evaluation of ${\mathbb N}$ involves integration
of production rates over all neutrino energies and a summation over
all neutrino flavors for any electron-number
changing weak reactions (see
\S\S\ref{sec-coll} and \ref{sec-coup-eq}).  The detailed microphysics
of such reactions have been explored elsewhere, including 
\citet{ffn2,fuller82,ffn85,bruenn85,lang2003,hix2003,hix2005} and are 
beyond the scope of this paper. 

Evolution of the internal energy of the matter is
given by the gas-energy equation
(\ref{eq:energy}), where $E$ is the matter
internal energy density, $P$ is the matter pressure,
and ${\sf Q}$ is the viscous stress tensor.  
Again, the right-hand side of this
equation is non-zero whenever energy, of any sort,
is transferred between matter and radiation.  This
can occur with neutrinos as a result of
weak interactions or with photons as a result of
electromagnetic interactions.  For the moment we
lump all such exchanges into the quantity
${\mathbb S}$, which has dimensions of energy per
unit volume per unit time, and represents the net
transfer rate of energy from radiation to matter.
The reactions that comprise ${\mathbb S}$ depend
on the physical phenomena being modeled.  However a
general form for these reactions that encompasses
most situations will be delineated in a later
section of this paper.  In the case of photons a
detailed description of such reactions can be
found in \cite{mih84,castor04,pomraning05}.  For the case of neutrinos,
in addition to the references for neutrino
number-changing reactions listed above, additional
reactions have been studied by
\citet{bps,yb76,yb77,ss82,bruenn85,schinder87,mb93c,rat2003,dut2004},
among others.

Finally, equation~(\ref{eq:mom}) is the gas-momentum equation, where
$\Phi$ is the gravitational potential, ${\mathsf P}_{{\rm rad}}$ is
the radiation-pressure tensor, and ${\mathbb P}$ is the net transfer
rate of momentum due to microphysical interactions between radiation 
and matter.

In equations~(\ref{eq:energy}) and (\ref{eq:mom}), we have
followed \cite{sn92a} in our addition of a viscous dissipation tensor
to the Euler equations in order to 
account for dissipation that occurs in shocks.  The details of
of this viscous dissipation tensor are discussed
in Appendix~\ref{app:artvisc}.

We note that it is also possible to substitute for
equation~(\ref{eq:energy}) linear combination of the 
gas energy and gas momentum equations to get an evolution equation
for the total matter energy \citep[cf.][eq.~24.5]{mih84}.
\begin{equation} \label{eq:totenergy}
\frac{\partial}{\partial t} \left( \rho E + \frac{1}{2} \rho \varv^2 \right) 
+ {\bfnabla} \cdot \left\{ \left( \rho E + P + \frac{1}{2} \rho \varv^2
\right) {\bf v} \right\} = - \rho {\bf v}\cdot{\bfnabla}\Phi + \rho {\mathbb S}.
\end{equation}
For core-collapse supernova
simulations, however, an internal energy formulation, as given in
equation~(\ref{eq:energy}), has advantages.  This is because there is a
vast amount of internal energy in matter relative to kinetic energy.  This
follows from the thermodynamic domination of degenerate electrons, which
contribute a large amount of zero-temperature energy and pressure. 
Given this
situation, our choice of solving the gas-energy equation helps insure
an accurate calculation of entropy, which is critical in degenerate
regimes where a small change in energy can lead to a large change in
temperature. In other problems, such as high mach-number flows,
where kinetic energy dominates, a
system may be better solved by using equation~(\ref{eq:totenergy}).

Closure of this system of equations requires additional relations.
First, is an equation of state (EOS), which is a parametric
description of gas pressure and internal energy in terms of
temperature, density, and composition. We discuss this further in
\S\ref{sec-eos}. Second, is an expression for the gravitational
potential $\Phi$, which is discussed in Appendix~\ref{app:gmomsrc}.
Finally, microphysical expressions are needed to evaluate
${\mathbb N}$, ${\mathbb S}$, and
${\mathbb P}$. A general form for these terms is discussed in 
\S\ref{sec-coup-eq} and Appendix~\ref{app:collision}.


\subsection{Radiation Transport}

For simulations of neutrino radiation-hydrodynamic
phenomena, solutions of the full discrete
ordinates Boltzmann equation, including
comoving-frame and group-to-group coupling terms,
have been attempted only in one spatial dimension
\citep{mb93a,mb93b,mb93c,lieb2001,lieb2004}.  This
is because of the computational cost associated
with the high dimensionality of the Boltzmann
equation.  For simulations in more than one
spatial dimension, the computational burden of
solving the Boltzmann equation currently
necessitates resort to an approximate solution.
The recent 2-D work of \cite{livne04} ignored
coupling terms to achieve computational simplicity
and the 2-D work of \cite{buras06} used a variable
Eddington factor method at low angular resolution
to render the calculation tractable.

In contrast, we implement a fully two-dimensional, Eulerian, multigroup,
flux-limited diffusion scheme that keeps all order $\varv/c$ coupling
terms and which is practical for high spatial resolution simulations 
on current parallel architectures.  This scheme extends our earlier work
\citep{mbhlsv,sss04} as well as that of \cite{ts01}
and involves the solution of the zeroth angular
moment of the Boltzmann equation.  These equations
take the form of a pair of angle-integrated,
monochromatic, radiation energy equations in the co-moving
frame that describe radiation of a particle and,
where applicable, its antiparticle:
\begin{equation}\label{eq:bte0}
\frac{\partial E_{\epsilon}}{\partial t} +
{\bfnabla} \cdot \left( E_{\epsilon} {\bf v}
\right) + {\bfnabla} \cdot {\bf F}_{\epsilon} -
\epsilon \frac{\partial}{\partial \epsilon} 
\left( {\mathsf P}_{\epsilon}:
{\bfnabla} {\bf v} \right) = {\mathbb S}_{\epsilon},
\end{equation}
\begin{equation}\label{eq:bte0bar}
\frac{\partial \bar{E}_{\epsilon}}{\partial t} +
{\bfnabla} \cdot \left( \bar{E}_{\epsilon} {\bf v} \right) +
{\bfnabla} \cdot \bar{{\bf F}}_{\epsilon} -
\epsilon \frac{\partial}{\partial \epsilon} 
\left( \bar{{\mathsf P}}_{\epsilon}:
{\bfnabla} {\bf v} \right) = \bar{{\mathbb S}}_{\epsilon}.
\end{equation}
These expressions are equivalent to equation~(6.49) in
\citet{castor04}, as derived by \citet{buch83}.  The scalar quantities
$E_{\epsilon}$ and $\bar{E}_{\epsilon}$ are the particle and
antiparticle monochromatic radiation-energy densities at
position ${\bf x}$ and time $t$.  The particle and antiparticle
monochromatic radiation-energy flux densities are given by
vectors ${\bf F}_{\epsilon}$ and $\bar{{\bf F}}_{\epsilon}$. The
particle and antiparticle monochromatic radiation pressure
are given by ${\mathsf P}_{\epsilon}$ and $\bar{{\mathsf
P}}_{\epsilon}$, which take the form of second-rank tensors.  
For the definition of all of these quantities we refer the reader
to the comprehensive work  of \cite{mih84}.
The right-hand side quantities, ${\mathbb S}_{\epsilon}$ and
$\bar{{\mathbb S}}_{\epsilon}$, account for coupling between matter
and radiation.  They contribute to the quantities ${\mathbb N}$,
${\mathbb S}$ and ${\mathbb P}$ of equations
(\ref{eq:ne})--(\ref{eq:mom}).  The form of this contribution is
described in \S\ref{sec-coup-eq} and
Appendix~\ref{app:collision}. Expressions of the form ${\mathsf  
P}_{\epsilon}:{\bfnabla} {\bf v}$, indicate contraction in both
indices of the second-rank tensors ${\mathsf P}_{\epsilon}$ and
${\bfnabla} {\bf v}$.  For photons, and other particles that are their
own  antiparticles, the barred expressions have no meaning and equation
(\ref{eq:bte0bar}) can be ignored.

Equations~(\ref{eq:bte0}) and (\ref{eq:bte0bar}) actually represent a
large set of equations. There is a pair of such equations for each
wavelength or frequency in the spectrum of radiation.
Additionally, if one is transporting more than one species of radiation
particle
({\em e.g.}, neutrinos of different flavors or some other collection
of diverse particles), there will be additional sets of equations 
of this form to
account for these additional species.

Although the set of moment equations represented by
equations~(\ref{eq:bte0}) and (\ref{eq:bte0bar}) is exact, it does not
possess a unique solution because of the multiple unknowns (radiation
energy density, flux density, and pressure) in each equation.
(However, note that in the hydrostatic limit, the terms involving the
pressure tensor vanish.)  The solution of the monochromatic radiation
energy equation requires the specification of a closure relationship
relating $E_\epsilon$, ${\bf F}_\epsilon$ and $\mathsf{P}_\epsilon$.
Unless one has already solved the full Boltzmann equation (obviating
the present discussion), the true relationships among these quantities
are only known in the asymptotic limits of transport
behavior---diffusion, where the optical depth is large; and
free-streaming, where the optical depth is small.  Therefore, solution
of the monochromatic energy equation in general situations requires an
approximate closure relationship.  

One of the most common approximations invokes the assumption that
radiation is diffusive and obeys Fick's Law
\begin{equation}\label{eq:fick}
{\bf F}_\epsilon \equiv -D_\epsilon {\bfnabla} E_\epsilon.
\end{equation}
In the diffusive limit, this Fick's law approximation becomes exact
and the diffusion coefficient is given by
\begin{equation} \label{eq:diff_simple}
D_\epsilon = \frac{c}{3\kappa^T_\epsilon},
\end{equation}
where $\kappa^T_\epsilon$ is the total opacity, given by
\begin{equation} \label{eq:ktot}
\kappa^T_\epsilon = 
\kappa^{a}_\epsilon + 
\kappa^{c}_\epsilon + 
\int d\epsilon^\prime \kappa^s(\epsilon,\epsilon^\prime).
\end{equation}
This total opacity consists of contributions from the absorption
opacity $\kappa_\epsilon^a$, the total conservative scattering opacity
$\kappa_\epsilon^c$, and the total non-conservative scattering
opacity, which is expressed by the integral in
equation~(\ref{eq:ktot}). 
We use the subscript $\epsilon$ to indicate that the quantity is a 
function of radiation wavelength, frequency, or energy.
We discuss the absorption and
non-conservative scattering opacities further in \S\ref{sec-coll}.

However, the
flux can become acausal if both Fick's Law and the diffusion coefficient
of equation~(\ref{eq:diff_simple}) are employed unconditionally.  
This
is because ${\bf F}_\epsilon$ is proportional to the ${\bfnabla}
E_\epsilon$, which is unbounded.   
This problem usually manifests itself in optically translucent 
or optically thin (free-streaming) situations.
Maintaining causality demands that
$|{\bf F}_\epsilon| \leq c E_\epsilon$ always.
One standard technique developed to maintain causality
is flux-limiting, which has an extensive history associated with it
\citep{minerbo78,pomraning81,lp81,bw82,lund83,levermore84,cvc89,cv90,
janka91,janka92,jdvh92,cb94,svhb00}.  In this paper, 
we will follow \cite{mbhlsv} and \cite{ts01} and
make use of the flux-limiting scheme
derived by \cite{lp81}. 
However, our algorithm can easily be used,
with virtually no modification, with other flux-limiting schemes 
that are based on the Knudsen number. 

In the Levermore-Pomraning closure, the flux is written in the form of
Fick's law, but modifications are made to the diffusion coefficient to
insure that causality is maintained and correct physical behavior
occurs in the free streaming limit.  A general form of a flux-limited
diffusion coefficient is given as
\begin{equation} \label{eq:lpd}
D_\epsilon \equiv \frac{c \lambda_\epsilon(\kn)}
{\kappa^T_\epsilon},
\end{equation}
which becomes the Levermore-Pomraning specification by defining
the flux-limiter $\lambda_\epsilon(\kn)$ as 
\begin{equation}  \label{eq:lpfl}
\lambda_\epsilon(\kn)
\equiv
\frac{2 + \kn}
{6 + 3\kn + \kn^2}.
\end{equation}
The quantity, $\kn$, is the radiation Knudsen number, which is the
dimensionless ratio of the radiation mean free path to a
representative length scale.  Thus, the Knudsen number is given by
\begin{equation}
\kn
\equiv
\frac{\left|{\bfnabla} E_\epsilon \right|}{\kappa^T_\epsilon E_\epsilon},
\label{eq:knudsen}
\end{equation}
where the $\epsilon$ subscripts emphasize that all these quantities
are dependent on the energy of the radiation under consideration.
This definition of the Knudsen number ensures correct limiting
behavior, both when $\kn\rightarrow 0$ in the diffusive limit and
$\kn\rightarrow \infty$ in the free streaming limit.

Irrespective of any approximation, the tensorial Eddington factor,
${\mathsf X}_\epsilon$,
which relates radiation pressure and energy, is defined by
\begin{equation} \label{eq:edddef}
{\mathsf{P}}_\epsilon \equiv {\mathsf X}_\epsilon E_\epsilon.
\end{equation}
The quantity ${\mathsf X}_\epsilon$ is often written in the form of
another general expression,
\begin{equation} \label{eq:chidef}
{\mathsf X}_\epsilon \equiv \onehalf \left(
1-\chi_\epsilon \right) 
{\mathsf{I}}
+
\onehalf \left(
  3 \chi_\epsilon - 1
\right)
{\bf n}{\bf n},
\end{equation}
where $\mathsf{I}$ is the identity tensor and where
${\bf n}{\bf n}$ is a dyad constructed from ${\bf n}$, the unit
vector parallel to the radiative flux. The quantity $\chi_{\epsilon}$
is referred to as the scalar Eddington factor.  Upon applying the
Levermore-Pomraning prescription, we obtain the useful expression,
\begin{equation} \label{eq:chismdef}
\chi_\epsilon = \lambda_\epsilon (\kn) + 
\left\{\lambda_\epsilon(\kn)\right\}^2 \: \kn^2,
\end{equation}
which gives us the full Eddington tensor,
\begin{equation}
{\mathsf X}_\epsilon \equiv \frac{1}{2}\left(
1-\lambda_\epsilon (\kn) -
\left\{\lambda_\epsilon (\kn)\right\}^2 \:
\kn^2 \right)
{\mathsf{I}}
+
\frac{1}{2}
\left(
  3 \lambda_\epsilon (\kn)
+ 3 \left\{\lambda_\epsilon (\kn)\right\}^2 \:
          \kn^2 
- 1
\right)
{\bf n}{\bf n},
\label{eq:chi_tensor}
\end{equation}
in the Levermore-Pomraning scheme.  

With the application of flux-limited diffusion, closure relations are
now entirely determined, and all moments of radiation are expressed in
terms of $E_\epsilon$.  
In similar fashion, a closure scheme for equation~(\ref{eq:bte0bar})
follows by direct analogy.
Thus, equations~(\ref{eq:bte0}) and (\ref{eq:bte0bar}) can be cast in
a form in which they possesses, at least formally, a unique solution,
\begin{equation}\label{eq:bte0f}
\frac{\partial E_{\epsilon}}{\partial t} +
{\bfnabla} \cdot \left( E_{\epsilon} {\bf v} \right) -
{\bfnabla} \cdot (D_\epsilon {\bfnabla} E_{\epsilon}) -
\epsilon \frac{\partial}{\partial \epsilon} 
\left\{
({\mathsf X}_{\epsilon} E_\epsilon):
{\bfnabla} {\bf v} \right\} = {\mathbb S}_{\epsilon},
\end{equation}
\begin{equation}\label{eq:bte0barf}
\frac{\partial \bar{E}_{\epsilon}}{\partial t} +
{\bfnabla} \cdot \left( \bar{E}_{\epsilon} {\bf v} \right) -
{\bfnabla} \cdot (\bar{D}_\epsilon {\bfnabla} \bar{E}_{\epsilon}) -
\epsilon \frac{\partial}{\partial \epsilon} 
\left\{
(\bar{{\mathsf X}}_{\epsilon} \bar{E}_\epsilon):
{\bfnabla} {\bf v} \right\} = \bar{{\mathbb S}}_{\epsilon}.
\end{equation}
It is this form of the transport equation for which we describe a solution method
in \S3.


\subsection{Collision Integral} \label{sec-coll}

The right-hand side of equation~(\ref{eq:bte0}), the collision
integral, can be expressed in a general particle- and
species-independent way as
\begin{equation}\label{eq:coll}
{\mathbb S}_{\epsilon} \equiv 
\left\ldbrack {\mathbb S}_{\epsilon} \right\rdbrack_{\rm emis-abs} +
\left\ldbrack {\mathbb S}_{\epsilon} \right\rdbrack_{\rm pairs} +
\left\ldbrack {\mathbb S}_{\epsilon} \right\rdbrack_{\rm scat}.
\end{equation}
These terms account for various mechanisms by which energy may
transfer between matter and radiation.  
Most radiation processes fall into one of these three forms
and can be included in our algorithm.

The first term on the right-hand side of equation~(\ref{eq:coll}),
$\left\ldbrack {\mathbb S}_{\epsilon} \right\rdbrack_{\rm emis-abs}$,
represents emission-absorption of radiation by processes that change
the monochromatic radiation energy or number densities.
In photon transport, an
example of such a process is the transition of an atom between
different energy states that results in emission of a photon.  In
neutrino transport, an example is the capture of an electron by a
nucleus that results in emission of a neutrino.  In general, these
processes can be expressed as
\begin{equation}\label{eq:emis_abs}
\left\ldbrack {\mathbb S}_{\epsilon} \right\rdbrack_{\rm emis-abs} = 
S_\epsilon 
\left(1 + \eta \frac{\alpha}{\epsilon^3}E_\epsilon \right)
- c \kappa^a_\epsilon E_\epsilon,
\end{equation}
where $S_\epsilon$ is the emissivity of the radiation field (with
dimensions of energy per unit volume per unit time per
radiation-energy interval).  It is the rate at which energy is 
added to the radiation, while $\kappa^a_\epsilon$ is the absorption
opacity for the reverse process (in units of inverse length).  
By making the flux-limited diffusion approximation
we have made the assumption that the distribution function
in the collision integral is isotropic and thus the
expression $\alpha E_\epsilon /
\epsilon^3$ is the quantum-mechanical phase-space occupation number
for the radiation field at position ${\bf x}$, time $t$, and energy
$\epsilon$.  The quantity $\alpha$ is given by $(hc)^3/4\pi g = $ 9.4523 ${\rm
MeV}^4$ ${\rm cm}^3$ ${\rm erg}^{-1}$ for both photons and neutrinos.
This follows from the statistical weight factor, $g$, being unity for both
particles.

The factor $\eta$ takes on different values,
depending on the quantum statistics of the
radiation field under consideration.  It is unity
for photons and all other bosons, leading to the
well-known stimulated emission of photons.  For
neutrinos and all other fermions, $\eta = -1$,
leading to {\em inhibited emission}---a term we
find more physically intuitive than {\em
stimulated absorption} \citep{bludman77}, which is
frequently used in the literature. This form
follows naturally from the Pauli exclusion
principle, which allows only a single fermion per
quantum state.  In the case of neutrinos, once the
Fermi sea is fully occupied the emissivity drops to
zero.  \cite{bruenn85} gives a more complete
description of the quantum mechanical origin of
this factor in the case of neutrinos.
Finally, for a classical radiation field, $\eta =
0$, reflecting the Maxwell-Boltzmann character of
classical particles.

The quantum mechanical principle of detailed
balance requires that $S_\epsilon$ and
$\kappa^a_\epsilon$ be related. When radiation and
matter are in chemical equilibrium, we have must
have a relationship between emission and
absorption (the forward and inverse reactions)
such that the right-hand side of equation
(\ref{eq:emis_abs}) vanishes.  This balance
relationship is expressed in Kirchoff's Law,
\begin{equation} \label{eq:kirchoff}
S_{\epsilon} = c B_{\epsilon} \kappa_{\epsilon}^a
(1 - \eta e^{(\mu_{\epsilon} - \epsilon)/T}),
\end{equation}
where $B_{\epsilon}$ is the generalized Planck ``black-body'' function
given by 
\begin{equation} \label{planck}
B_{\epsilon} = g 
\frac{4 \pi \epsilon^{3}}{\left( hc \right)^{3}}
\left( \frac{1}{e^{(\epsilon - \mu_{\epsilon})/T} - \eta} \right).
\end{equation}
When radiation is in chemical equilibrium with matter, it makes sense
to assign the radiation a chemical potential, which we represent by
$\mu_{\epsilon}$.  A chemical potential obviously has little meaning
in non-equilibrium conditions; however, Kirchoff's Law always has
meaning for determining the microphysical relationship between
emission and absorption under {\em any} radiative conditions, whenever
the matter is in LTE (\cite{mih84}, p.\ 387).  To satisfy the detailed
balance requirement when matter and radiation are out of equilibrium,
one substitutes for $\mu_\epsilon$ in equation~({\ref{eq:kirchoff})
the value it would take if matter and radiation were already
equilibrated.  This allows Kirchoff's Law to set the relationship
between $S_\epsilon$ and $\kappa_\epsilon^a$.  The correctness of this
procedure is a consequence of detailed balance, which must be
satisfied microscopically, regardless of any macroscopic state of the
system.  As an example of its application, in the weak charged-current
reaction $e^- + p \rightarrow
\nu_e + n$, we substitute $\mu_\epsilon = \mu_e + \mu_p -
\mu_n$, where $\mu_{e,p,n}$ are the electron, proton, and neutron
chemical potentials for the matter in LTE.  Note that if we were to
apply this procedure for a process emitting photons,
$\mu_{\epsilon} = 0$ at all times.

The second term on the right-hand side of equation~(\ref{eq:coll}),
$\left\ldbrack {\mathbb S}_{\epsilon}\right\rdbrack_{\rm pairs}$,
represents processes that create a particle-antiparticle pair.  (Since
a photon is its own antiparticle, this process also represents
production of a pair of photons.)  In photon transport, an example of
such a process is the mutual annihilation of a positron-electron pair
to produce a pair of gamma rays.  A corresponding analogy in neutrino
processes is $e^- + e^+$ annihilation to produce a
neutrino-antineutrino pair.  In general, these processes may be
expressed in the form
\begin{equation}\label{eq:pairs}
\left\ldbrack {\mathbb S}_{\epsilon} \right\rdbrack_{\rm pairs} = 
+\left(1 + \eta \frac{\alpha}{{\epsilon}^3}
E_{\epsilon} \right)
\epsilon   \int d\epsilon^\prime
G(\epsilon,\epsilon^\prime)
\left(1 + \eta \frac{\alpha}{{\epsilon^\prime}^3}
\bar{E}_{\epsilon^\prime} \right),
\end{equation}
where $G(\epsilon,\epsilon^\prime)$ is the
pair-production kernel for production of a
particle with energy $\epsilon$ and antiparticle
energy $\epsilon^\prime$. The monochromatic energy
density of the produced antiparticle is given by
$\bar{E}_{\epsilon^\prime}$.  In fermion
transport, the $\eta = -1$ factor gives two
final-state Pauli blocking terms, reflecting the
inability of a pair-production process to produce
a fermion-antifermion pair in which either of the
presumed final states is already occupied.  In
contrast, for bosons, final-state degeneracy is
actually enhanced, since $\eta = 1$.

We note that equation~(\ref{eq:pairs}) makes no
provision for an inverse reaction.  Although an
inverse radiation pair-annihilation reaction is of
potential importance under some conditions, and
its addition to the algorithm is straightforward,
we do not consider it in this paper.  Radiation
pair-annihilation reaction rates are usually
strongly dependent on the angular distribution of
the radiation and this situation is fundamentally
incompatible with the assumption of isotropy that
was made in deriving the monochromatic radiation
energy equation from the Boltzmann equation.
Inclusion of pair-annihilation terms thus requires
some {\it ad hoc} assumption about the angular distribution of radiation
that is dependent on the particular phenomenon
being modeled.  Detailed discussion of the role of
pair annihilation will appear in future work on
core-collapse supernovae where such effects are
potentially relevant.

The final term on the right-hand side of equation~(\ref{eq:coll}),
$\left\ldbrack {\mathbb S}_{\epsilon} \right\rdbrack_{\rm scat}$,
represents general, non-conservative, scattering processes.  These
processes result in no net creation or destruction of
particles. Hence, their contribution to ${\mathbb N}$ is zero (see
\S\ref{sec-coup-eq}).  However, they will change the energy
distribution of the radiation field as a result of interactions with
matter.  

For non-conservative scattering processes, the contribution to
the collision integral is
\begin{equation}\label{eq:scat}
\left\ldbrack {\mathbb S}_{\epsilon} \right\rdbrack_{\rm scat} = 
\left(1 + \eta \frac{\alpha}{\epsilon^3}E_\epsilon \right)
c \int d\epsilon^\prime 
\kappa^s(\epsilon,\epsilon^\prime)
E_{\epsilon^\prime}
- E_{\epsilon}
c \int d\epsilon^\prime 
\kappa^s(\epsilon^\prime,\epsilon)
\left(1 + \eta \frac{\alpha}{{\epsilon^\prime}^3}
E_{\epsilon^\prime} \right),
\end{equation}
where $\kappa^s(\epsilon,\epsilon^\prime)$ is the scattering
opacity for particles in energy state $\epsilon$ scattering
into energy state $\epsilon^\prime$.  
In analogy to the other processes discussed in
this section, there is also final-state enhancement (or blocking) in
these expressions for bosons (or fermions) resulting from the $1 +
\eta \alpha E_\epsilon/\epsilon^3$ terms.

Finally, we note that for each interaction presented in this section,
there is possibly a conjugate reaction of importance that entails
interactions involving antiparticles.  These produce $\ldbrack
\bar{{\mathbb S}}_\epsilon\rdbrack$ versions of each of the terms we
have presented.  To calculate these interactions, one substitutes
``barred'' versions for each of the production and opacity terms,
and interchanges $E_\epsilon$ and $\bar{E}_\epsilon$.  In analogy
with equation~(\ref{eq:coll}) the sum of all such contributions
yields $\bar{{\mathbb S}}_\epsilon$.  The sole exception to this
is equation~(\ref{eq:pairs}) where the antiparticle analog is given by
\begin{equation}\label{eq:apairs}
\left\ldbrack \bar{\mathbb S}_{\epsilon} \right\rdbrack_{\rm pairs} = 
+\left(1 + \eta \frac{\alpha}{{\epsilon}^3}
\bar{E}_{\epsilon} \right)
\epsilon   \int d\epsilon^\prime
G(\epsilon^\prime,\epsilon)
\left(1 + \eta \frac{\alpha}{{\epsilon^\prime}^3}
{E}_{\epsilon^\prime} \right).
\end{equation}
Note that the same pair-production kernel $G$ appears in equations
(\ref{eq:pairs}) and (\ref{eq:apairs}) with the order of the 
arguments reversed between the two equations. 


\subsection{Radiation-Matter Coupling} \label{sec-coup-eq}

A key step in closing equations~(\ref{eq:cont})--(\ref{eq:mom})
is providing a method for evaluating the right-hand-side
collision terms that couple matter and radiation.  
The evaluation of
the sources for (${\mathbb N}$, ${\mathbb S}$, and ${\mathbb P}$) in
terms of the results of \S\ref{sec-coll} is straightforward.
The right-hand side of equation~(\ref{eq:ne}) gives the rate of lepton
exchange between the radiation an matter. It can be written as
\begin{equation}\label{eq:n-exch0}
{\mathbb N} = - \sum_l \int \frac{1}{\epsilon} 
\left( ^l{\mathbb S}_\epsilon - ^l\bar{{\mathbb S}}_\epsilon \right) d\epsilon,
\end{equation}
where the form of the emissivity, $^l{\mathbb
S}_\epsilon$, is given by equation~(\ref{eq:coll})
and its specifics by the subsequent equations in
\S\ref{sec-coll} for each flavor of radiation.  
The leading superscript $l$ is used to denote
the flavor of the radiation, {\em e.g.}, for neutrinos
$l = e$, $\mu$, or $\tau$.
The integral over $\epsilon$
accounts for contributions from the complete
spectrum of the radiation field. The minus sign in
equation~(\ref{eq:n-exch0}) accounts for the fact
that a gain in lepton number for the radiation field is a
loss for lepton number in matter. The sum over $l$
accounts for the possibility of multiple species
of radiation that can engage in number exchange
with the matter.  In practice, the number exchange
is always due to electron neutrino-antineutrino
emission-absorption.  The barred term is non-zero
if there are distinct antiparticles being evolved
separately from the particles ({\em e.g.}, both
electron neutrinos and antineutrinos). Note that
scattering is not a number-exchanging interaction
and, hence, the contribution of the scattering
terms, when integrated and summed, is zero.

Regardless of whether there is a number quantity that is exchanged
between matter and radiation, there is generally a non-zero energy
exchange between the two.  The net result of this exchange on the
matter side is given by, ${\mathbb S}$, the right-hand side of
equation~(\ref{eq:energy}), which can be written as
\begin{equation} \label{eq:e_xch0}
{\mathbb S} = - \sum_{\ell} \int \left( ^\ell{\mathbb S}_\epsilon +
^\ell\bar{{\mathbb S}}_\epsilon \right) d\epsilon.
\end{equation}
It is important to
note that unlike equation~(\ref{eq:n-exch0}), the sum is now over
$\ell$ (rather than $l$), which is a summation over {\em all} species of radiation,
not just those that exchange net number with the matter.  Once again,
the barred term is non-zero whenever antiparticles are being evolved
distinctly.

Finally, we will ignore momentum exchange, ${\mathbb P}$, between matter and
radiation in our present flux-limited diffusion scheme, 
{\em i.e.},
\begin{equation}
{\mathbb P} = 0.
\end{equation}
The calculation of microphysical momentum exchange
between matter and radiation requires a knowledge
of the angular distribution of the radiation or,
at least, the knowledge of the angular averaged
absorption opacity.  In the case of neutrino
transport in core-collapse supernovae, 1-D
Boltzmann simulations have shown this effect to be
negligible, and thus we ignore it for the remainder
of this paper.  However, it would be easy to
include this effect for photons or neutrinos if
the angular distribution of radiation were known.


\subsection{Enforcing the Pauli Exclusion Principal for Neutrinos} \label{sec-fd}

The equations of neutrino radiation hydrodynamics described in the
previous subsections are semi-classical in that the Pauli exclusion
principal for neutrinos is taken into account only in the collision
integral terms.  The left-hand-side of equations~(\ref{eq:bte0f}) and
(\ref{eq:bte0barf}) are purely classical and do not guarantee that 
the occupancy of a specific neutrino energy state is less than or equal
to unity.  This constraint can be stated mathematically as
\begin{equation}
0 \le f_\epsilon \equiv \frac{\alpha}{\epsilon^3}E_\epsilon \leq 1
\label{eq:pauli}
\end{equation}
where $f_\epsilon$ is the neutrino distribution function and where
where $\alpha = (hc)^3/4\pi g = $ 9.4523 ${\rm MeV}^4$ ${\rm cm}^3$
${\rm erg}^{-1}$ for both photons and neutrinos (for which $g=1$).
The numerical solution of equations~(\ref{eq:bte0f}) and
(\ref{eq:bte0barf}) for the neutrino and 
antineutrino spectral radiation energy densities can produce values
of $E_\epsilon$ and $\bar{E}_\epsilon$ for which the distribution 
functions have values greater than unity.  This is most likely 
to occur in situations where the neutrino distribution function 
becomes highly degenerate, such as in the core of a proto-neutron
star.  This problem has been known for some time \citep{bruenn85}.

Since distribution function values greater than
unity are obviously unphysical for neutrinos we
need to supplement the solution of the neutrino
radiation-hydrodynamic equations by adopting an
``enforcement'' algorithm that is ensures that the
constraint represented by equation~(\ref{eq:pauli})
is satisfied after a new value of the neutrino
spectral energy densities is computed.  We detail
this enforcement algorithm in Appendix~\ref{app:pauli}.

\subsection{Conservation} \label{sec-cons}

The hydrodynamic equations that we have presented in
this section are not in a conservative form that
would admit a finite-volume approach to their
discretization.  Therefore, the equations do not
guarantee exact numerical conservation of either
energy or momentum.  No discretization scheme can
simultaneously conserve all physically conserved
quantities.  An excellent example of such
quantities are linear and angular momentum,
which cannot, in general, both be conserved
by the same discretization.  The neutrino
radiation-diffusion equations we have presented
are also not in conservative form.  Some attempts
have been made to arrive at a discretization of
neutrino transport equations \citep{lieb2004}, in
the 1-D Boltzmann case, that conserves both
neutrino energy and number. However, in general no
such discretization has been discovered.  In order
to ensure that sufficient accuracy is being
achieved in a simulation, one must monitor the
conservation of various physical quantities that
are important.  The relative importance of
conservation is problem dependent and we will
address this for this algorithm in the context of
core-collapse supernovae in a future work.
\section{The Numerical Method}

The algorithm that we employ for the solution of
the 2-D monochromatic radiation-hydrodynamic (RHD)
equations, described in the previous
section, is an extension of the ZEUS family of 
algorithms of \cite{sn92a,sn92b};
\cite{smn92}; and \cite{ts01}.  The extensions include the
incorporation of a multigroup treatment of the
radiation spectrum with group-to-group coupling,
the incorporation of multiple radiation species
with particle-antiparticle coupling and Pauli blocking, 
the solution
of an additional advection equation describing the
evolution of the electron number density, and the
incorporation of a complex equation of state.  We
also employ a methodology for solving the
implicitly discretized radiation-diffusion
equations that is different from the approach set
forth by \cite{ts01}.

The algorithm we employ solves the hydrodynamic
equations~(\ref{eq:cont}), (\ref{eq:ne}),
(\ref{eq:energy}), and (\ref{eq:mom}), along with
flux-limited diffusive transport equations
(\ref{eq:bte0f}) and (\ref{eq:bte0barf}).  Our
algorithm is Eulerian and employs a staggered mesh
similar to other ZEUS-type algorithms
\citep{sn92a,sn92b,smn92,ts01,hn03,hayes05}.  Like
these other algorithms, our time evolution scheme
utilizes a combination explicit techniques to
evolve the hydrodynamic portions of the RHD
equations while employing implicit techniques to
solve the transport portions of these equations.
Our time evolution scheme is dissimilar to these
other schemes in that the order of solution of
substeps differs from these schemes.

In the subsections that follow, we describe the
generalized computational mesh that we employ, the
use of operator splitting, the order of
operator-split substeps, the discretization of the
equations, and the parallel implementation of the
method.


\subsection{Computational Mesh} \label{sec-mesh}

We employ a spatially staggered mesh on an
orthogonal coordinate system identical to that
employed by the ZEUS family of algorithms
\citep{sn92a,sn92b,smn92,ts01,hn03,hayes05}.  For
the sake of simplicity,  we have not allowed the
mesh to adapt dynamically in any way and we
consider the mesh as fixed in time.  It would
be straightforward to extend the algorithm we
describe in this paper to accommodate the moving
mesh that is described in \cite{sn92a}.

We number cell edges in each of our two coordinate
directions, $x_1$ and $x_2$, by integer indices.
The $i$th cell edge in the $x_1$ direction has an
$x_1$ coordinate given by $[x_1]_i$, while the
$j$th cell edge in the $x_2$ direction has an
$x_2$ coordinate given by $[x_2]_j$.  The cell
centers have coordinates in each direction given
by $[x_1]_{i+\lhalf}$ and $[x_2]_{j+\lhalf}$.  Thus, the
location of a cell-centers in this mesh is fully
specified by a pair of discretized coordinates in
the $x_1$ and $x_2$ directions
$([x_1]_{i+\lhalf},[x_2]_{j+\lhalf})$.  This staggered
mesh is illustrated in Figure~\ref{fig:mesh_2}.

Intensive quantities, such as pressure, mass
density, internal energy density, temperature,
electron number density (or electron fraction),
etc. are defined at cell centers.  Components of
vector quantities, such velocities, momenta,
gradients of intensive quantities, and fluxes are
defined on the corresponding cell faces.  We refer
to these latter quantities as face-centered
variables. The spatial location for each of these
types of variables is depicted in
Figure~\ref{fig:mesh_2}.  Note that we use
standard subscript notation to define the
discretized analogs of all quantities,
{\em e.g.}, $[T]_{i+\lhalf,j+\lthreehalf}$ is the discretized
temperature variable defined at coordinates
$([x_1]_{i+\lhalf},[x_2]_{j+\lthreehalf})$.
Our finite difference notation is described in 
Appendix~\ref{app:diff}.

It is clear from Figure~\ref{fig:mesh_2} that
vector components are not co-located at a single
spatial point on the mesh.  Occasionally, quantities derived
from these components are needed at alternate
locations, in which case a scheme that averages
values from nearby spatial locations is employed
to compute values at the needed location.  We
discuss such averaging on a case-by-case basis
when we detail the discretization of the
equations.

\begin{figure}[htbp]
\begin{center}
\includegraphics[scale=0.60]{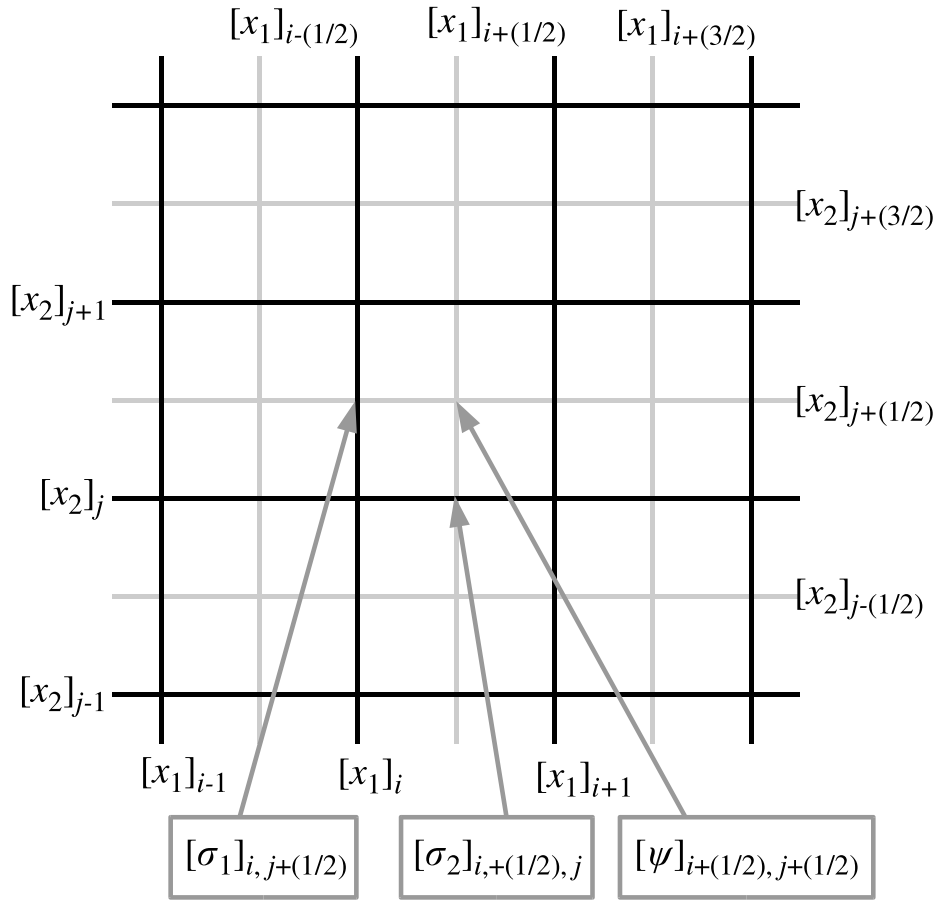}
\end{center}
\caption{\label{fig:mesh_2} A portion of the staggered 
spatial mesh, showing the location of spatial
coordinates.  The bold black lines define cell
edges while the gray lines show the
coordinates of cell-centers.  The location of
a typical cell-centered quantity, $\psi$, and of a
typical vector quantity, ${\smpmb \sigma}$, with
components $\sigma_1$ and $\sigma_2$, are also
shown.}
\end{figure}

In order to discretize the spectral variables, we
also define a mesh over the energy dimension,
{\em i.e.,} the spectrum of radiation energies.  The
range of the domain in the energy dimension is
discretized into groups, {\em i.e.,} cells in energy
space.  The $k$th group has a lower edge with an
energy coordinate $[\epsilon]_k$ and the center of
the $k$th group has an energy coordinate given by
$[\epsilon]_{k+\lhalf}$.  Discretized spectral
quantities, such as spectral radiation energy
densities and spectral flux densities, are usually
defined at group centers.  Since such quantities
usually have a dependence on spatial location as
well as energy, the discretized analogs of these
quantities carry an extra subscript indicating
their location in the energy dimension.  For
example, the radiation energy density
$E_\epsilon$ at spatial coordinates
$([x_1]_{i+\lhalf},[x_2]_{j+\lhalf})$ and energy
coordinate $[\epsilon]_{k+\lhalf}$ is denoted as
$[E_\epsilon]_{k+\lhalf,i+\lhalf,j+\lhalf}$.
\begin{figure}[htbp]
\begin{center}
\includegraphics[scale=0.75]{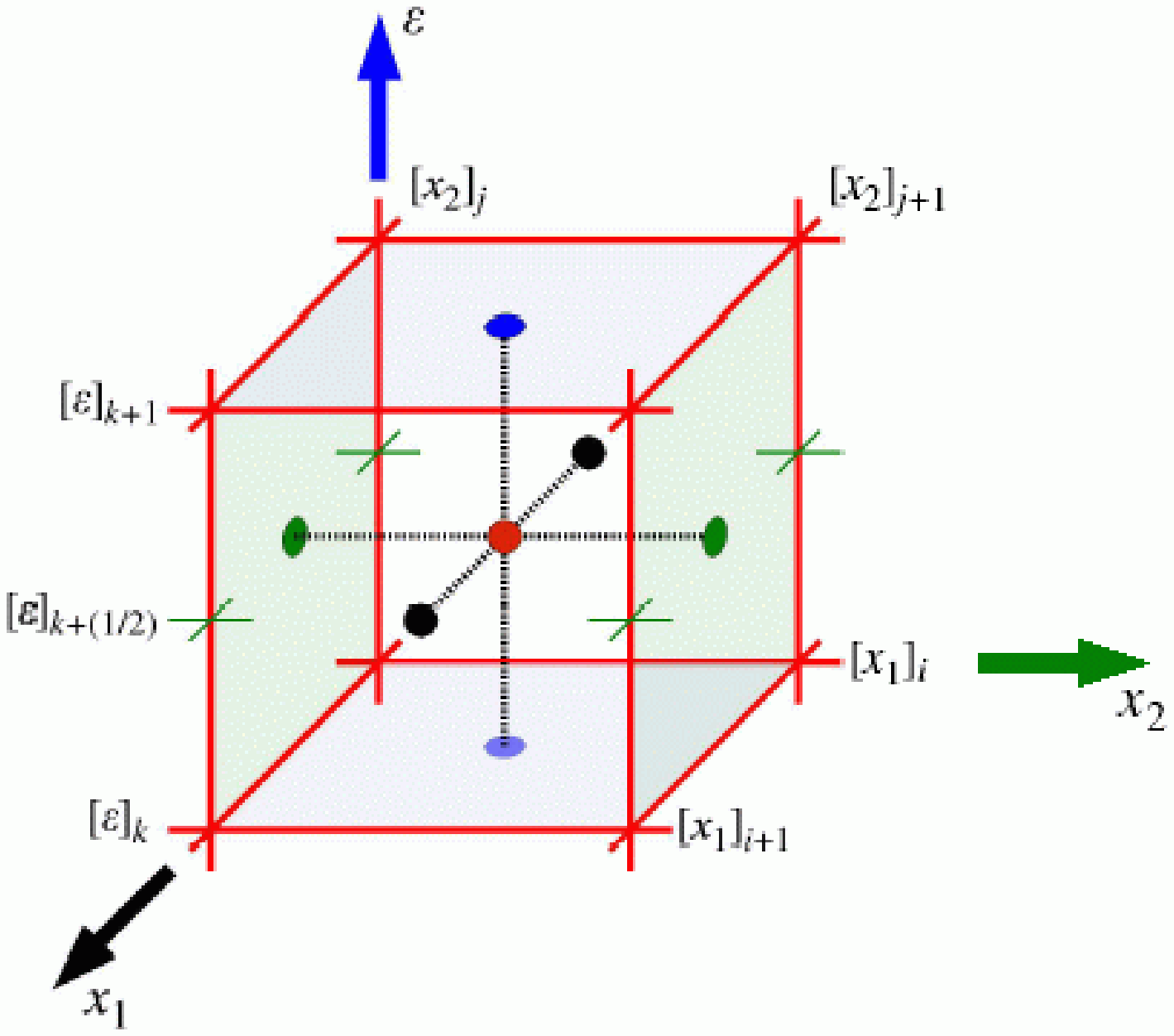}
\end{center}
\caption{\label{fig:full_grid_2} The full 
three-dimensional mesh in $x_1$--$x_2$--$\epsilon$, showing the
positions for evaluation of the spectral radiation quantities within a
cell.  These quantities are evaluated at points on the plane 
that passes through the
mid-point of the grid in the radiation-energy dimension.  The
face-centered radiation variables ({\em e.g.}, spectral radiation flux
density) are evaluated at the positions shown by the black or green dots
on the cell faces.  The cell-centered variables ({\em e.g.}, the
spectral radiation energy density), are evaluated at the cell center,
the position of which is shown by the red dot. The blue dots, which
lie on cell faces in the energy direction show the position of fluxes
representing transfer of energy between energy groups.  }
\end{figure}
The location of energy-dependent intensive
quantities such as the spectral radiation energy
density, and energy-dependent vector quantities,
such as the spectral flux density are illustrated
in Figure~\ref{fig:full_grid_2}.  A complete
listing, delineating where various physical
quantities are defined in the spatial-energy meshes,
is found in Appendix~\ref{app:list}.


\subsection{Covariant Formulation} \label{sec-covariant}

We choose to follow \cite{sn92a} by writing all
finite-difference expressions in terms of a
generalized orthogonal coordinate system that is
capable of describing Cartesian, cylindrical, and
spherical-polar coordinate systems.  Our goal is
to enable a single code that is easily adaptable
to any 2-D curvilinear coordinate system, avoiding
the labor that would otherwise be required to
implement a code in each individual coordinate
system desired.  This technique is well described
in \citep{sn92a} and we refer the reader there for
details.  The notation for coordinates and other
geometrical quantities that we employ in each
coordinate system are described in
Table~\ref{metric_scaling}, located in Appendix~\ref{app:cov}.
The detailed form of the metric coefficients, the gradient and divergence
operators, and tensor expressions that are needed to evaluate the
radiation-hydrodynamic equations are described in their entirety in 
Appendices~\ref{app:rad-trans} and ~\ref{app:gmom-rad}. 


\subsection{Operator Splitting}

Our algorithm employs operator splitting to decouple the 
overall time integration of the radiation-hydrodynamic 
equations into substeps.  The motivation for this procedure is discussed
in \cite{sn92a}, to which we refer the reader.
In general, we split the right-hand-sides of the 
time evolution equations into advective, source, viscous, and 
radiation-matter-coupling terms and solve these split equations
to update the hydrodynamic and radiation quantities accordingly.

The following describes the application of this operator-splitting
approach to the equations in our model. The time integration of
the continuity
equation~(\ref{eq:cont}) requires no operator splitting, since there
is only a single advective term, and no source or collision term, in the equation.
Thus, we can restate equation~(\ref{eq:cont}) as
\begin{equation} 
\left\ldbrack \frac{ \partial \rho}{\partial t}\right\rdbrack_{\rm total}  =
\left\ldbrack \frac{ \partial \rho}{\partial t}\right\rdbrack_{\rm advection},
\end{equation}
where
\begin{equation} 
\left\ldbrack \frac{ \partial \rho}{\partial t}\right\rdbrack_{\rm advection}
=-{\bfnabla} \cdot (\rho {\bf v}) .
\label{eq:cont_adv}
\end{equation}

The electron conservation equation~(\ref{eq:ne}) is split into two
terms,
\begin{equation} \label{eq:ye-split}
\left\ldbrack \frac{\partial n_e}{\partial t}\right\rdbrack_{\rm total}  =
\left\ldbrack \frac{\partial n_e}{\partial t}\right\rdbrack_{\rm advection}  +
\left\ldbrack \frac{\partial n_e}{\partial t}\right\rdbrack_{\rm collision},
\end{equation}
where 
\begin{equation} \label{eq:ye-adv}
\left\ldbrack \frac{\partial n_e}{\partial t}\right\rdbrack_{\rm advection}
=-{\bfnabla} \cdot (n_e {\bf v})
\end{equation}
is the advective term and
\begin{equation} \label{eq:ye-source}
\left\ldbrack \frac{\partial n_e}{\partial t}\right\rdbrack_{\rm collision}
= {\mathbb N}
\end{equation}
is the source or collision-integral term.
In a similar manner, the gas-energy
equation~(\ref{eq:energy}) is split into
four separate sets of terms: advection terms, 
the Lagrangean or source terms, viscous dissipation terms,
and the collision-integral terms. 
\begin{equation} \label{eq:e-split}
\left\ldbrack \frac{ \partial E }{\partial t}\right\rdbrack_{\rm total}  =
\left\ldbrack \frac{ \partial E }{\partial t}\right\rdbrack_{\rm advection}  +
\left\ldbrack \frac{ \partial E }{\partial t}\right\rdbrack_{\rm source}  +
\left\ldbrack \frac{ \partial E }{\partial t}\right\rdbrack_{\rm visc}  +
\left\ldbrack \frac{ \partial E }{\partial t}\right\rdbrack_{\rm collision}
\end{equation}
where
\begin{equation} \label{eq:e-advective}
\left\ldbrack \frac{ \partial E }{\partial t}\right\rdbrack_{\rm advection}  =
-{\bfnabla} \cdot (E {\bf v}),
\end{equation}
\begin{equation} \label{eq:e-source}
\left\ldbrack \frac{ \partial E }{\partial t}\right\rdbrack_{\rm source}  =
-P {\bfnabla}\cdot{\bf v},
\end{equation}
\begin{equation} \label{eq:e-visc}
\left\ldbrack \frac{ \partial E }{\partial t}\right\rdbrack_{\rm visc}  =
-{\mathsf Q}:{\bfnabla}\cdot{\bf v},
\end{equation}
and
\begin{equation} \label{eq:e-coll}
\left\ldbrack \frac{ \partial E }{\partial t}\right\rdbrack_{\rm collision}
= {\mathbb S}.
\end{equation}
The gas-momentum equation
(\ref{eq:mom}) is operator split into five sets of terms
\begin{eqnarray} \label{eq:s-split}
\left\ldbrack \frac{ \partial \left( \rho {\bf v} \right) }{\partial
t}\right\rdbrack_{\rm total}  & = &
\left\ldbrack \frac{ \partial \left( \rho {\bf v} \right) }{\partial
t}\right\rdbrack_{\rm advection}  +
\left\ldbrack \frac{ \partial \left( \rho {\bf v} \right) }{\partial
t}\right\rdbrack_{\rm source}  +
\left\ldbrack \frac{ \partial \left( \rho {\bf v} \right) }{\partial
t}\right\rdbrack_{\rm radiation} 
\nonumber \\ \nonumber \\ & &
+
\left\ldbrack \frac{ \partial \left( \rho {\bf v} \right) }{\partial
t}\right\rdbrack_{\rm visc} +
\left\ldbrack \frac{ \partial \left( \rho {\bf v} \right) }{\partial
t}\right\rdbrack_{\rm collision},
\end{eqnarray}
where the advection term is
\begin{equation} \label{eq:s-advect}
\left\ldbrack \frac{ \partial \left( \rho {\bf v} \right) }{\partial
t}\right\rdbrack_{\rm advection}  =
-{\bfnabla} \cdot (\rho {\bf v} {\bf v}),
\end{equation}
the source terms are
\begin{equation} \label{eq:s-source}
\left\ldbrack \frac{ \partial \left( \rho {\bf v} \right) }{\partial
t}\right\rdbrack_{\rm source}  =
-{\bfnabla} P 
-\rho{\bfnabla} \Phi,
\end{equation}
the viscous dissipation terms are
\begin{equation} \label{eq:s-visc}
\left\ldbrack \frac{ \partial \left( \rho {\bf v} \right) }{\partial
t}\right\rdbrack_{\rm visc}  =
-{\bfnabla} \cdot {\mathsf Q},
\end{equation}
the radiation pressure terms are
\begin{equation} \label{eq:s-rad}
\left\ldbrack \frac{ \partial \left( \rho {\bf v} \right) }{\partial
t}\right\rdbrack_{\rm radiation}  =
-{\bfnabla} \cdot {\mathsf P}_{{\rm rad}},
\end{equation}
and the collision integral terms are
\begin{equation} \label{eq:s-coll}
\left\ldbrack \frac{ \partial \left( \rho {\bf v} \right) }{\partial
t}\right\rdbrack_{\rm collision}  =
{\mathbb P}.
\end{equation}
Finally, the radiation-energy equation~({\ref{eq:bte0f}) is
operator split as
\begin{equation} 
\left\ldbrack \frac{ \partial E_\epsilon }{\partial t}\right\rdbrack_{\rm total}  =
\left\ldbrack \frac{ \partial E_\epsilon }{\partial t}\right\rdbrack_{\rm advection}  +
\left\ldbrack \frac{ \partial E_\epsilon }{\partial t}\right\rdbrack_\text{diff-coll},
\end{equation}
where
\begin{equation} 
\left\ldbrack \frac{ \partial E_\epsilon }{\partial t}\right\rdbrack_{\rm advection} =
-{\bfnabla} \cdot (E_\epsilon {\bf v}),
\label{eq:nu-advect}
\end{equation}
while the combination of the diffusive and collision integral terms are
defined by
\begin{equation} 
\left\ldbrack \frac{ \partial E_\epsilon }{\partial t}\right\rdbrack_\text{diff-coll} =
{\bfnabla} \cdot (D_\epsilon {\bfnabla} E_{\epsilon}) +
\epsilon \frac{\partial}{\partial \epsilon}
\left\{
({\mathsf X}_{\epsilon} E_\epsilon):
{\bfnabla} {\bf v} \right\} + {\mathbb S}_{\epsilon}.
\label{eq:nu-diff}
\end{equation}
The antiparticle monochromatic diffusion equation~(\ref{eq:bte0barf}) is 
operator split in the analogous fashion to equation~(\ref{eq:bte0f}).

We also note that each of the aforementioned
advection terms is itself directionally operator
split into two pieces corresponding to advection
in each of the two coordinate directions which we
generically denote as $x_1$ and $x_2$.  Thus for
each set of advection terms we can write
\begin{equation} \label{eq:dir_split}
\left\ldbrack \frac{ \partial }{\partial t}\right\rdbrack_{\rm advection} =
\left\ldbrack \frac{ \partial }{\partial t}\right\rdbrack_\text{adv-1} +
\left\ldbrack \frac{ \partial }{\partial t}\right\rdbrack_\text{adv-2}.
\end{equation}

Because of the complexity of the operator split
equations, we restrict the discussion of the
numerical methods used to solve the individual
pieces of the operator equations to Appendices
\ref{app:advect}--\ref{app:pauli}.  In the next
section we concentrate on the order of updates,
based on these operator split pieces, employed to
evolve the equations from time $[t]^n$ to time
$[t]^\npo$.


\subsection{Order of Solution of Operator Split Equations
\label{sec:oos}}

Our algorithm employs the following order for
solution of the operator-split equations detailed
in the previous subsection.  The complete sequence
of solving each of these operator split pieces
constitutes the algorithm for evolving the
equations from time $[t]^n$ to time
$[t]^\npo \equiv [t]^n+\Delta t$. A schematic
illustration of our algorithm, for a single
timestep, is provided in Figure~\ref{fig:timestep}.
The details of each substep are provided in
Appendices~\ref{app:advect}--\ref{app:pauli}.  The
hydrodynamic portions of the operator-split
equations are solved explicitly, while the
radiative portions of the equations are solved
implicitly.  The motivation for this choice of a
hybrid explicit-plus-implicit approach is well-described in
\cite{smn92}, and we refer the reader  
there for a detailed description of the issues
involved.

Following \cite{sn92a}, we denote partial updates
of variables of quantities by means of
superscripts.  Thus, at the beginning of a timestep,
the matter internal energy density is denoted by
$[E]^n$.  The partially updated
internal energy resulting from, for example, substep $f$
(updated via eq.~[\ref{eq:e-coll}]) is denoted by $[E]^{n+f}$.  
The superscript $n+f$ serves
to indicate that the quantity includes all partial
updates prior to and including substep $f$.  The
final update of each quantity, within a given timestep, 
is labeled by superscript $n+1$.  When we denote a discretized
quantity without spatial- or energy-index
subscripts, we are referring to the entire spatial
or energy range of that discretized quantity.  In
our description of each substep, we will describe
what quantities are updated as a result of that substep.

\begin{figure}[htbp]
\begin{center}
\includegraphics[scale=0.75]{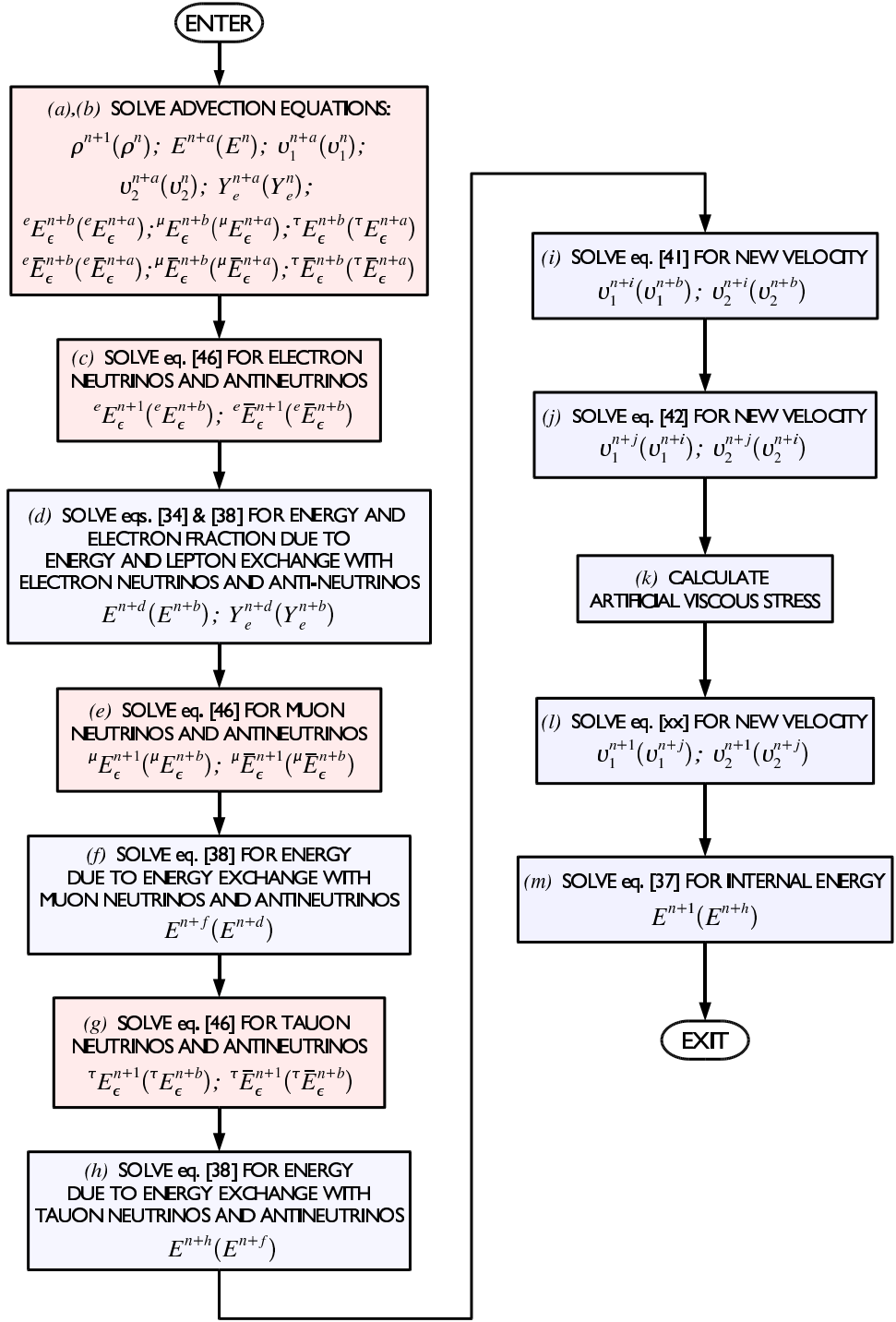}
\end{center}
\caption{\label{fig:timestep} The algorithm for 
advancing the model by a single timestep from
$[t]^n$ to $[t]^{n+1}$. The red boxes indicate
steps where the Pauli exclusion principal
constraint is enforced after new values of the
neutrino energy densities are calculated. 
The variables listed in each box 
are those that result from the update
(those in parenthesis are inputs to the update
step).  ``Enter'' signifies the beginning of the
timestep, while ``exit'' signifies the end of the 
timestep.
}
\end{figure}

The substep $a$-$b$ in the algorithm consists of
explicit numerical solution of the advection
portions of the operator-split equations.  This
substep is actually a combination of two
directionally split substeps corresponding to
advection in the $x_1$ and $x_2$ directions.  In
this substep, all advective portions of the
operator split equations are solved, namely,
equations~(\ref{eq:cont_adv}), (\ref{eq:ye-adv}),
(\ref{eq:e-advective}), (\ref{eq:s-advect}), and
(\ref{eq:nu-advect}).  Note that, since the
radiation energy density is a function of spectral
energy, equation~(\ref{eq:nu-advect}) must be
solved for every energy group and for each type of
radiation present, {\em i.e.,} for each of the six types
of neutrinos.  Thus, equation~(\ref{eq:nu-advect})
represents a set of $6N_g$ equations that must be
solved, while equations~(\ref{eq:cont_adv}),
(\ref{eq:ye-adv}), (\ref{eq:e-advective}), and
(\ref{eq:s-advect}) represent five equations (the
momentum equation is actually two
equations, one for each of the the two components of
the velocity).  Thus, when the number of energy
groups $N_g$ is large, the computational cost of
the advection substep is dominated by the cost of
solving the radiation-advection equations
represented by equation~(\ref{eq:nu-advect}).  Our
algorithm for explicit solution of the
advection portion of the operator-split equations
is exactly the same as that of \cite{sn92a} and is
detailed in Appendix~\ref{app:advect}.  We note
that after the calculation of the updated values
of the radiation energy densities, the Pauli
exclusion principal constraint enforcement is
applied.  This is indicated in
Figure~\ref{fig:timestep} by red shading of the
advective update box.  We also note that the
advective update substep is itself composed of
numerous substeps in which the advection equations
are solved by directionally split substeps.  The
directionally split advection algorithm utilizes
Norman's consistent advection scheme \citep{nwb80}
in which the advection of all quantities is tied
to the mass-flux.  The details of the advection
substeps are illustrated in
Figure~\ref{fig:advection} (see Appendix~\ref{app:advect}), 
to which the reader is referred for
more detail.  The net result of this substep
are the partially updated quantities
$[ \rho ]^{n+b}$, $[ E ]^{n+b}$
$[ T ]^{n+b}$, $[ P ]^{n+b}$,
$[ n_e ]^{n+b}$, $[ Y_e ]^{n+b}$,
$[ {\bf v} ]^{n+b}$,$[ {\bf s} ]^{n+b}$,
$[ ^e E_\epsilon ]^{n+b}$,
$[ ^e \bar{E}_\epsilon ]^{n+b}$,
$[ ^\mu E_{\epsilon} ]^{n+b}$,
$[ ^\mu \bar{E}_{\epsilon} ]^{n+b}$,
$[ ^\tau E_{\epsilon} ]^{n+b}$
and
$[ ^\tau \bar{E}_{\epsilon} ]^{n+b}$.

The second and third substeps (substeps $c$ and $d$) 
in the solution of the
operator-split equations involve the evolution of
equations describing the
radiative evolution of electron neutrinos and
antineutrinos and the exchange of energy
and lepton number between matter and these 
neutrinos.  This substep involves the implicit solution
of the set of radiation diffusion equations for
electron neutrinos and antineutrinos represented by
equation~(\ref{eq:nu-diff}) and the collision-integral
equations represented by equations~(\ref{eq:ye-source}) and
(\ref{eq:e-coll}).  In 
the second step, represented by the second box ($c$) in
Figure~\ref{fig:timestep}, the complete set of
implicitly differenced diffusion equations for
electron neutrinos and antineutrinos represented
by equation~(\ref{eq:nu-diff}) is solved
simultaneously via Newton-Krylov iteration.  The
details of the finite-differencing and numerical
solution of these equations is detailed in
Appendix~\ref{app:rad-trans}. Once the implicit
solution of this set of equations has been
accomplished, the amount of lepton and energy
exchange between matter and electron
neutrinos and antineutrinos is fixed.  After the new
values of the electron-neutrino-antineutrino
energy densities are calculated, the Pauli
exclusion principal constraint enforcement
algorithm is applied to the electron-neutrino-antineutrino energy
densities.  The 
application of this constraint-enforcement
algorithm is indicated by the red shading of the
second box ($c$) in Figure~\ref{fig:timestep}.  The
second substep results in updated
quantities 
$[ E ]^{n+c}$, 
$[ T ]^{n+c}$, 
$[ P ]^{n+c}$, and fully updated radiation quantities
$[ ^e E_{\epsilon} ]^{n+1}$ and 
$[ ^e \bar{E}_{\epsilon}]^{n+1}$.

In the substep $d$ (represented by the third box [$d$]
of Figure~\ref{fig:timestep}), since the amount of
energy and lepton exchange with matter has been
fixed by the previous substep, equation~(\ref{eq:ye-source})
is solved for 
the new value of electron number density and,
thus, the new value of electron fraction $Y_e$.
Subsequently equation~(\ref{eq:e-coll}) is
solved implicitly for the new value of
internal energy density. Once the new internal
energy density is determined, the equation of state
determines the new value of matter temperature and
pressure corresponding to the new
internal energy density.  The details of this
substep are described in Appendix~\ref{app:rad-trans}.
The third substep results in the fully updated quantities 
$[ E ]^{n+d}$,
$[ T ]^{n+d}$,
$[ P ]^{n+d}$,
$[ n_e ]^{n+d}$,
and
$[ Y_e ]^{n+d}$.

The second and third substeps (substeps $c$ and $d$)
are subsequently repeated for the muon
neutrinos and antineutrinos in substeps $e$ and $f$
(shown as boxes $e$ and $f$ the flowchart) and tauon
neutrinos and antineutrinos in 
substeps $g$ and $h$ (shown as boxes $g$ and $h$
in the flowchart).  In substeps $e$ and $g$, the
Pauli exclusion principal constraint algorithm is
applied to the muon neutrino and antineutrino energy
densities and the tauon neutrino and antineutrino
energy densities, respectively.  This is indicated
by the red shading of the boxes corresponding to
substeps $e$ and $g$ in Figure~\ref{fig:timestep}.  In
substeps $f$ and $h$, the solution of equation
(\ref{eq:ye-source}) is not required since the
production of muon neutrinos and antineutrinos and
tauon neutrinos and antineutrinos results in no
change in lepton number---these
neutrinos are always produced in
particle-antiparticle pairs.  Equation~(\ref{eq:e-coll}) is solved for
the new matter 
internal energy density, temperature, and pressure,
as described in the case of the second substep.
Substep $e$ results in the updated quantities
$[ E ]^{n+e}$, 
$[ T ]^{n+e}$,
$[ P ]^{n+e}$, 
$[ ^\mu E_{\epsilon} ]^{n+1}$, and 
$[ ^\mu \bar{E}_{\epsilon}]^{n+1}$.  
Substep $f$ results in the updated quantities 
$[ E ]^{n+f}$, 
$[ T ]^{n+f}$, and 
$[ P ]^{n+f}$.
Substep $g$ results in the updated quantities
$[ E ]^{n+g}$, 
$[ T ]^{n+g}$,
$[ P ]^{n+g}$, 
$[ ^\tau E_{\epsilon} ]^{n+1}$, and 
$[ ^\tau \bar{E}_{\epsilon} ]^{n+1}$.  
Substep $h$ results in the updated quantities 
$[ E ]^{n+h}$, 
$[ T ]^{n+h}$, and 
$[ P ]^{n+h}$.

In substep $i$, the momentum and velocities are updated 
via the solution of equation~(\ref{eq:s-source}) to account for
gravitational- and pressure-induced accelerations.  This substep 
is almost identical in detail to that of \cite{sn92a}, but we
describe this in detail in Appendix~\ref{app:gmomsrc}.
In this paper, we consider the gravitational force to be spherically
symmetric based on the mass constained interior to a given radius.
The description of the calculation of the gravitational mass
is also detailed in Appendix~\ref{app:gmomsrc}.  This substep
results in the updated quantities 
$[{\bf v} ]^{n+i}$ and $[{\bf s}]^{n+i}$.

In substep $j$, the momentum and velocities are updated 
via the solution of equation~(\ref{eq:s-rad}) to account for
radiation-pressure-induced accelerations.  This substep 
relies on the Eddington factor differencing
of the gray transport algorithms of \cite{smn92} and
\cite{ts01} which, for our multigroup transport, is applied on a
group-by-group basis.  We described this 
approach in detail in Appendix~\ref{app:gmom-rad}.
This substep results in the updated quantities
$[{\bf v}]^{n+j}$ and 
$[{\bf s}]^{n+j}$.

In substep $k$, the components of the artificial viscous stress are 
calculated according to the prescription of \cite{sn92a}.
This calculation is described in Appendix~\ref{app:artvisc}.

In substep $\ell$ the momentum and velocities are
updated via the solution of equation~(\ref{eq:s-visc})
to account for accelerations induced by the
gradients of the viscous stress.  This substep is
identical in detail to that of \cite{sn92a}, and we
describe this in detail in Appendix~\ref{app:artvisc}.  This substep
results in the 
updated quantities $[{\bf v}]^{n+1}$
and $[{\bf s}]^{n+1}$.

In substep $m$, the internal energy density is
updated via the solution of equation~(\ref{eq:e-visc})
to account for viscous dissipation.
Like substep $\ell$, this substep
identical in detail to that of \cite{sn92a}, and we
describe the update equation in Appendix~\ref{app:artvisc}.  This
substep results in the 
updated quantity $[E]^{n+m}$.  We point
out that the temperature and pressure are not updated in this
step, as they will be updated after the following step.

In substep $n$, the Lagrangean portion of the gas energy
equation, described by equation~(\ref{eq:e-source}) is solved 
to account for compression or expansion of the gas and the
effects of viscous stresses.  The time differencing of this equation
is implicit.  However, the since the divergence of
the velocity in equation~(\ref{eq:e-source}) is evaluated 
based on the partially updated velocities 
$[{\bf v}]^{n+1}$, there is no spatial coupling between
the unknowns in equation~(\ref{eq:e-source}).  This equation can thus
be solved by a local, nonlinear iterative solution 
algorithm in each spatial zone.
The finite differencing of equation~(\ref{eq:e-source}) and our
solution algorithm are described in Appendix~\ref{app:gas-lagrange}.
This substep results in the updated quantities
$[ E ]^{n+1}$,
$[ T ]^{n+1}$,
and
$[ P ]^{n+1}$.

This sequence of partial updates 
represented by substeps $a,b$ to $m$ constitutes the 
algorithm for evolving the equations of neutrino 
radiation hydrodynamics from time $[ t]^n$ to
time $[t]^{n+1} = [t]^n +\Delta t $.


\subsection{Boundary Conditions} \label{sec-bound}

Up to this point, we have neglected any discussion
of boundary conditions and how they are applied
within the algorithm.  In general, boundary
conditions for a specific quantity are applied immediately
after any update of that quantity.  Thus, any given
quantity may have boundary-condition updates several
times during the course of a single timestep.  The
details of how specific boundary conditions are
applied are delineated in Appendix~\ref{app:bconds}, and we refer the
reader there for more information.

In a parallel implementation of our algorithm,
where parallelism is achieved via spatial domain
decomposition, there are also internal ``processor
boundaries.''  Values of variables at these
boundaries must be kept consistent among
processors.  Thus, boundary updates are a frequent occurrence, since
internal processor boundaries are in the middle of
actively computed regions.  Such consistency
requires update of values of a quantity at the
edge of processor boundaries after each update of
that quantity---before it is needed for another
calculation requiring spatial derivatives of that
quantity.  We discussion this
issue in \S\ref{sec-parallel}.


\subsection{Timestep Selection} \label{sec-time}

The stability of our algorithm is restricted by the stability
of the solution of explicitly solved operator-split equations.  
In calculating the
timestep we follow the algorithm laid out in \cite{sn92a}.
This algorithm for selecting the timestep depends on several 
different types of stability criteria which are then combined
as an RMS average to yield a stable timestep.

The calculation of the timestep is based on four key criteria.
The first is the Courant timescale
(calculated in both the $x_1$ and $x_2$ directions), which represents
the minimum sound-crossing time for a particular zone in each
dimension. More formally,
\begin{equation}
\left[\Delta t_{\rm Courant}\right]^{n+1}_{i+\lhalf, j+\lhalf} = 
\frac{1}{c_s} \min \left(
\left[ \Delta x_1 \right]_{i+\lhalf}, 
\left[ g_2 \right]_{i} \left[ \Delta x_2 \right]_{j+\lhalf}
\right),
\end{equation}
where $c_s$ is the local speed of sound at
coordinate ($[x_1]_{i+(1/2)},[x_2]_{j+(1/2)}$).
An accurate calculation of $c_s$ requires the
equation of state to supply an adiabatic index.
For the purposes of timestep selection, however, a
conservative overestimate of $c_s$ is obtained by
making a conservative approximation ({\em i.e.},
overestimate) of the sound speed by using a
polytropic EOS having an overestimate of $\gamma$.

The second and third metrics are the flow timescales
in the $x_1$ and $x_2$ directions, which are the timescales over which a
particle in the fluid, located at one cell face, can travel to the
opposite face, in each respective direction.  These timescales are
expressed as
\begin{equation}
\left[\Delta t_\text{$x_1$ flow}\right]^{n+1}_{i+\lhalf, j+\lhalf} = 
\min \left(
\frac{\left[ \Delta x_1 \right]_{i+\lhalf}}
{\left[ \varv_1 \right]_{i+1, j+\lhalf}},
\frac{\left[ \Delta x_1 \right]_{i+\lhalf}}
{\left[ \varv_1 \right]_{i, j+\lhalf}}
\right)
\end{equation}
and
\begin{equation}
\left[\Delta t_\text{$x_2$ flow}\right]^{n+1}_{i+\lhalf, j+\lhalf} = 
\min \left(
\frac{\left[ g_2 \right]_{i+\lhalf} \left[ \Delta x_2 \right]_{j+\lhalf}}
{\left[ \varv_2 \right]_{i+\lhalf, j+1}},
\frac{\left[ g_2 \right]_{i+\lhalf} \left[ \Delta x_2 \right]_{j+\lhalf}}
{\left[ \varv_2 \right]_{i+\lhalf, j}}
\right).
\end{equation}
Finally, the fourth timescale is a viscous dissipation timescale
set by the magnitude of the viscous stress.
This timescale is defined as
\begin{equation}
\left[\Delta t_{\rm conv}\right]^{n+1}_{i+\lhalf, j+\lhalf} = 
2 \sqrt{l_q} \min \left(
\frac{ \left[ \Delta x_1 \right]_{i+\lhalf}}
{\left[ \varv_1 \right]_{i+1,j+\lhalf} - \left[ \varv_1 \right]_{i,j+\lhalf}},
\frac{ \left[ g_2 \right]_{i} \left[ \Delta x_2 \right]_{j+\lhalf}}
{\left[ \varv_2 \right]_{i+\lhalf,j+1} - \left[ \varv_2 \right]_{i+\lhalf,j}}
\right),
\label{eq:t1t4}
\end{equation}
where we compare the timescales in each direction and take the
minimum.  The quantity $l_q$ is a number of order unity and is
defined in Appendix~\ref{app:artvisc}.

The inverse squares of each of these timescales
are added at each mesh point.  The maximum value
of the quantity in parenthesis in equation~(\ref{eq:t1t4})
is found for the entire spatial domain. 
The inverse square
root of this quantity represents the minimum
timescale in the entire domain.  A fraction, which
we refer to as the CFL fraction and designate by
$f_{\rm CFL}$, of this time is used as the new
timestep size.  It is our practice to fix $f_{\rm
CFL}$ throughout the course of a given
simulation. Typically, we set $f_{\rm CFL} = 0.5$.

Thus, the timestep value eventually used is
\begin{eqnarray}
\left[ \Delta t \right]^{n+1} & = &
f_{\rm CFL} 
\left\{
\max_{\text{all } i,j}
\left( 
\frac{1}{\left[ \Delta t_{\rm Courant} \right]_{i+\lhalf,j+\lhalf}^2} +
\frac{1}{\left[ \Delta t_\text{$x_1$ flow}\right]_{i+\lhalf,j+\lhalf}^2} 
\right.
\right.
\nonumber \\ & &
\left.
\left.
\quad \quad \quad \quad \; \;
+
\frac{1}{\left[ \Delta t_\text{$x_2$ flow}\right]_{i+\lhalf,j+\lhalf}^2} +
\frac{1}{\left[ \Delta t_{\rm conv} \right]_{i+\lhalf,j+\lhalf}^2}
\right)
\right\}^{-1}.
\end{eqnarray}


\subsection{Equation of State and Opacity Interface}
\label{sec-eos}

This algorithm makes no assumption about the equation of state 
other than assuming that the EOS is of the form
\begin{equation}
E \equiv E\left(T,\rho,Y_e\right)
\label{eq:eos1}
\end{equation}
and
\begin{equation}
P \equiv P\left(T,\rho,Y_e\right).
\label{eq:eos2}
\end{equation}
Our numerical algorithm can accommodate an arbitrary
EOS of this form.  The algorithm does not rely on solution of
a Riemann problem and makes no assumptions about convexity in the
EOS.  Thus, the EOS can be supplied as a simple formula that can 
be evaluated in a small subroutine or, alternatively, in the 
more general form of a thermodynamically 
consistent tabular interpolation \citep{fds96}.  
The algorithm does require that 
it be possible for the relationship described by equation 
(\ref{eq:eos1}) be inverted, either analytically or numerically
to yield
\begin{equation}
T \equiv T\left(E,\rho,Y_e\right).
\label{eq:eos3}
\end{equation}
In particular, whenever a new value of the matter internal energy
$E$ is calculated, it must be possible to compute a new value of
the temperature $T$ corresponding to that internal energy.  A
polytropic EOS can be easily accommodated within the relationships
described by equations~(\ref{eq:eos1})--(\ref{eq:eos3}).

We also assume that the absorption opacity, the conservative
scattering opacity, and the emissivity be of the form
\begin{equation}
\kappa^a_\epsilon \equiv \kappa^a_\epsilon
\left(T,\rho,Y_e,\epsilon \right),
\label{eq:op1}
\end{equation}
\begin{equation}
\kappa^c_\epsilon \equiv \kappa^c_\epsilon
\left(T,\rho,Y_e,\epsilon \right),
\label{eq:op2}
\end{equation}
and
\begin{equation}
S_\epsilon \equiv S_\epsilon
\left(T,\rho,Y_e,\epsilon \right).
\label{eq:op3}
\end{equation}
The absorption opacity and the emissivity should 
be related in such a manner as to preserve 
the quantum mechanical principle of detailed balance
(see \S\ref{sec-coll}). 

The scattering opacity is assumed to be of the form
\begin{equation}
\kappa^s_{\epsilon,\epsilon^\prime} \equiv 
\kappa^s_{\epsilon,\epsilon^\prime}
\left(T,\rho,Y_e,\epsilon,\epsilon^\prime \right),
\label{eq:op4}
\end{equation}
where, in the scattering reaction, $\epsilon$ is the energy of the
incoming particle  and $\epsilon^\prime$ is the
energy of the outgoing particle.
This opacity should also preserve detailed balance.

Finally, the pair-production rate is assumed to be of the
form
\begin{equation}
G_{\epsilon,\epsilon^\prime} \equiv 
G_{\epsilon,\epsilon^\prime}
\left(T,\rho,Y_e,\epsilon,\epsilon^\prime \right),
\label{eq:op5}
\end{equation}
where $\epsilon$ is the energy of the neutrino that 
is produced and $\epsilon^\prime$ is the energy of the
antineutrino that is produced.


\subsection{Parallel Implementation} \label{sec-parallel}

The size of problem encountered in
multidimensional radiation-hydrodynamic models,
particularly in stellar collapse, necessitates our
use of massively parallel computing resources in order 
to solve the discretized equations.
Since we solve a long-timescale problem, it is
necessary that we achieve strong-scaling, {\em
i.e.,} we wish a fixed-size problem to scale well
to a large number of processors and, therefore,
reduce our wall-clock time to solution.  In
addition, the number of variables in the problem
requires a large amount of memory, further
necessitating a parallel solution strategy.

A parallel implementation of our algorithm can be
achieved via spatial domain decomposition of the
$\text{2-D}$ spatial domain into a logically Cartesian
topology of 2-D subdomains.  Such a decomposition
is illustrated in Figure~\ref{fig:bc_3}.
\begin{figure}[htbp]
\begin{center}
\includegraphics[scale=1.0]{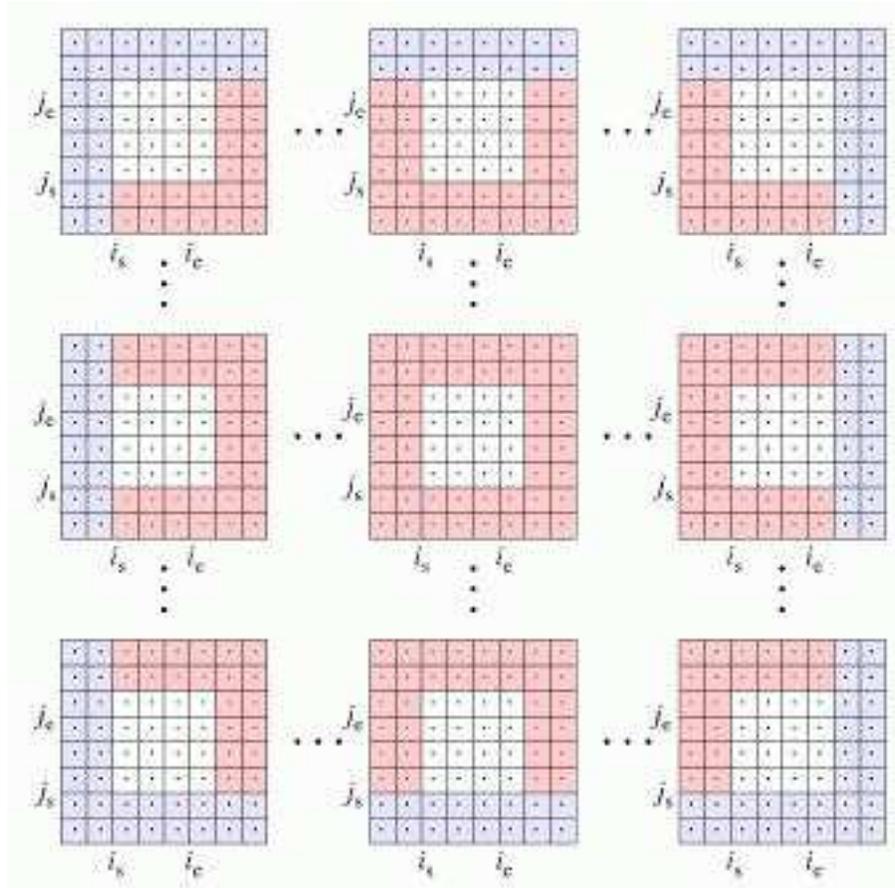}
\end{center}
\caption{\label{fig:bc_3} A schematic of the outer edges and center
of the entire computational
mesh, showing spatial domain decomposition into a
logical Cartesian mesh for parallel computing.
Blue shaded zones represent genuine boundary zones
of the computational mesh.  Zones that are 
shaded red represent internal ghost zones that
are duplicate copies of zones that are resident
on other processors. The actual size of the subdomains
is problem dependent.
In the present
implementation of the algorithm the width of the
boundary and the ghost zones is always two.  This
is a result of the present differencing scheme, in
which coupling between cells within a given equation
never extends beyond next-nearest neighbor.
The $x_1$ and $x_2$  starting and ending indices on each subdomain
are labeled as $i_s$,~$i_e$ and $j_s$,~$j_e$ respectively.
}
\end{figure}
This is a well-established scheme for the
parallelization of numerical methods for partial
differential equations \citep{mpi1}.  The
logically Cartesian process topology is
straightforward to create using MPI (Message
Passing Interface) subroutine calls
\citep{mpi2} and can be configured to allow for 
periodic boundaries if so desired.  This logically
Cartesian spatial decomposition is independent of
the choice of coordinate system and is carried out
with an orthogonal spatial mesh defined by
generalized coordinates, which we have described
previously.  The partitioning of this mesh into
subdomains is accomplished by specifying the size
of the process topology in each of its two
dimensions.  Once the size of the process topology
has been specified, the mesh is divided in an
approximately even fashion over the process
topology to achieve a balancing of
computational work.  By specifying the process
topology to be as square as possible, the ratio of
the number of ghost zones to non-ghost zones can
be minimized, thus reducing the
communication-to-computation ratio and improving
the scalability of the algorithm.

In our parallelization scheme, we do not decompose
in the spectral energy dimension of our mesh.
Thus the 2-D quadrilateral subdomains illustrated
in Figure~\ref{fig:bc_3} are actually 3-D
hexahedra, where the third dimension is the
spectral energy dimension.  By not decomposing the
problem in the energy dimension we avoid costly
communication that would be associated with the
evaluation of the integral terms in equations~(\ref{eq:pairs}),
(\ref{eq:scat}), 
(\ref{eq:n-exch0}), and (\ref{eq:e_xch0}), as well
as the application of the Pauli exclusion
principal constraint enforcement algorithm.

Under a logically Cartesian spatial domain
decomposition, our discretization algorithms for
the hydrodynamic and radiation-transport equations
require a limited set of communication patterns.
Evaluation of fluxes or spatial derivatives gives
rise to a local discretization stencil that 
requires information from ``ghost zones''
surrounding each subdomain.  The values of
variables in these ghost zones must be obtained
from adjacent subdomains by means of
message-passing to and from nearest-neighbor
processes.  This process of ghost-zone exchange is
illustrated in Figure~\ref{fig:ghost_zone_exchange}.
\begin{figure}[htbp]
\vspace{0.5in}
\begin{center}
\includegraphics[scale=0.5]{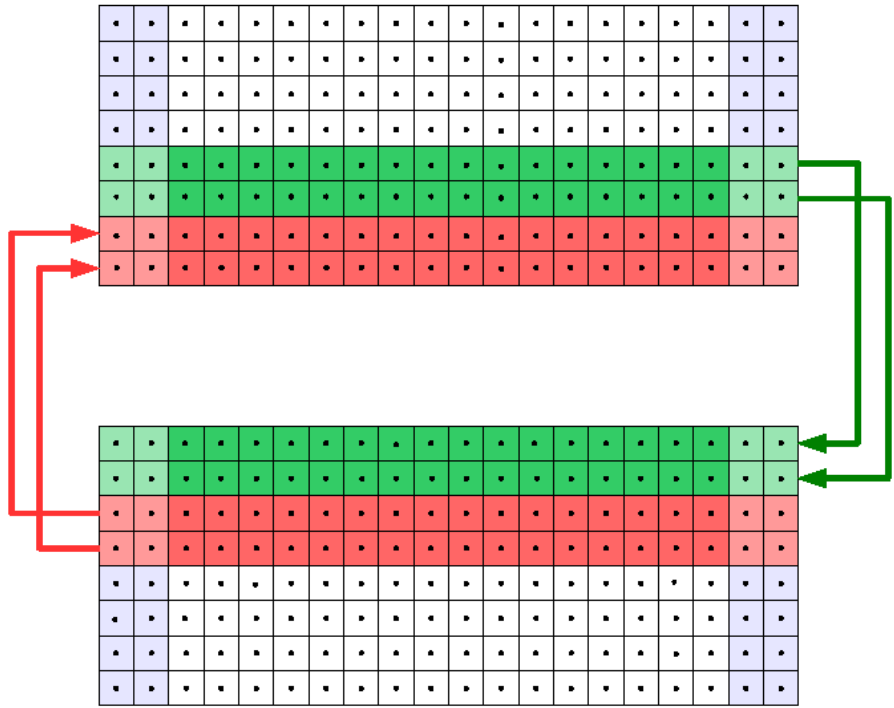}
\end{center}
\caption{\label{fig:ghost_zone_exchange} A schematic showing how
logically adjacent processes can exchange internal boundary
information through the exchange of ghost zones.  Ghost zones contain
data from the computational domain that is resident on a processor
other than the one being considered.  The mutual exchange of this
internal ``boundary'' information ensures that each process in the
calculation is working with up-to-date information. The arrows show
the flow of information.  Active computational zones in one
processor's domain ({\it e.g.,} the red-shaded zones in the bottom
subdomain) are transferred to the ghost zones in the other processor's
subdomain. The green-shaded zones show the corresponding flow in the
opposite direction. In the figure, the blue-shaded zones represent
ghost or genuine boundary zones that are not involved in the exchange.  
The flow of information to and from
these zones is governed by exchange rules for the transverse dimension.
}
\vspace{0.5in}
\end{figure}
Ghost zone values of a specific quantity, such as
density, must be exchanged before those values are
needed in the evaluation of any expression in
which those variables appear.  These exchanges can
be accomplished asynchronously, but we avoid
discussion of the complexities of doing this, since that
topic is beyond the scope of this paper.

The number of the ghost zones required is a function of the
discretization scheme.  An examination of the finite difference
equations, presented in the appendices, indicates that the maximum
number of zones that couple within a single equation is five in each
spatial dimension. For example, equation~(\ref{eq:cont}), for
van Leer advection of scalars, couples five zones $i-\frac{3}{2}$,
$i-\onehalf$, $i+\onehalf$, $i+\frac{3}{2}$, and $i+\frac{5}{2}$
centered about the $i+\onehalf$ cell center in the $x_1$
direction. If a spatially higher-order differencing scheme were
implemented, there would be longer-range coupling among zones.  Therefore,
the width of the ghost-zone region would be correspondingly
larger. Readers who desire a more complete description of ghost-zone
exchange are referred to Chapter 4 of
\citet{mpi1}.  For the implementation of the algorithm 
described in this paper, two ghost zones in each 
spatial dimension are required.

Unfortunately, nearest-neighbor message passing is
not the only communication pattern required by the
algorithm.  Global reduction operations are
required in two instances.  First, calculation of
the timestep size, $\Delta t$ (see \S\ref{sec-time}),
requires a global reduction to determine the
minimum CFL time for the entire domain.  Second,
the Newton-Krylov-based iterative solution of the
implicitly differenced radiation-transport
equations (Appendix~\ref{app:rad-trans}), requires
global reduction operations to evaluate vector
inner products. In the BiCGSTAB algorithm, which
we use to solve the linear systems in the
Newton-Krylov scheme, this can mean multiple
global reductions per iteration.  This can impose
a bottleneck to scalability for
simulations running on large numbers of processors.  To
reduce this bottleneck, we have developed an
algebraically equivalent variant of the BiCGSTAB
algorithm, which requires only one global
reduction per iteration (Swesty 2006, in preparation).  
This
improvement can be seen in Figure~\ref{fig:f1},
where we plot the parallel speedup of the algorithm when
calculating a supernova model on {\em seaborg},
the IBM-SP at NERSC.
\begin{figure}[htbp]
\includegraphics[width=35pc]{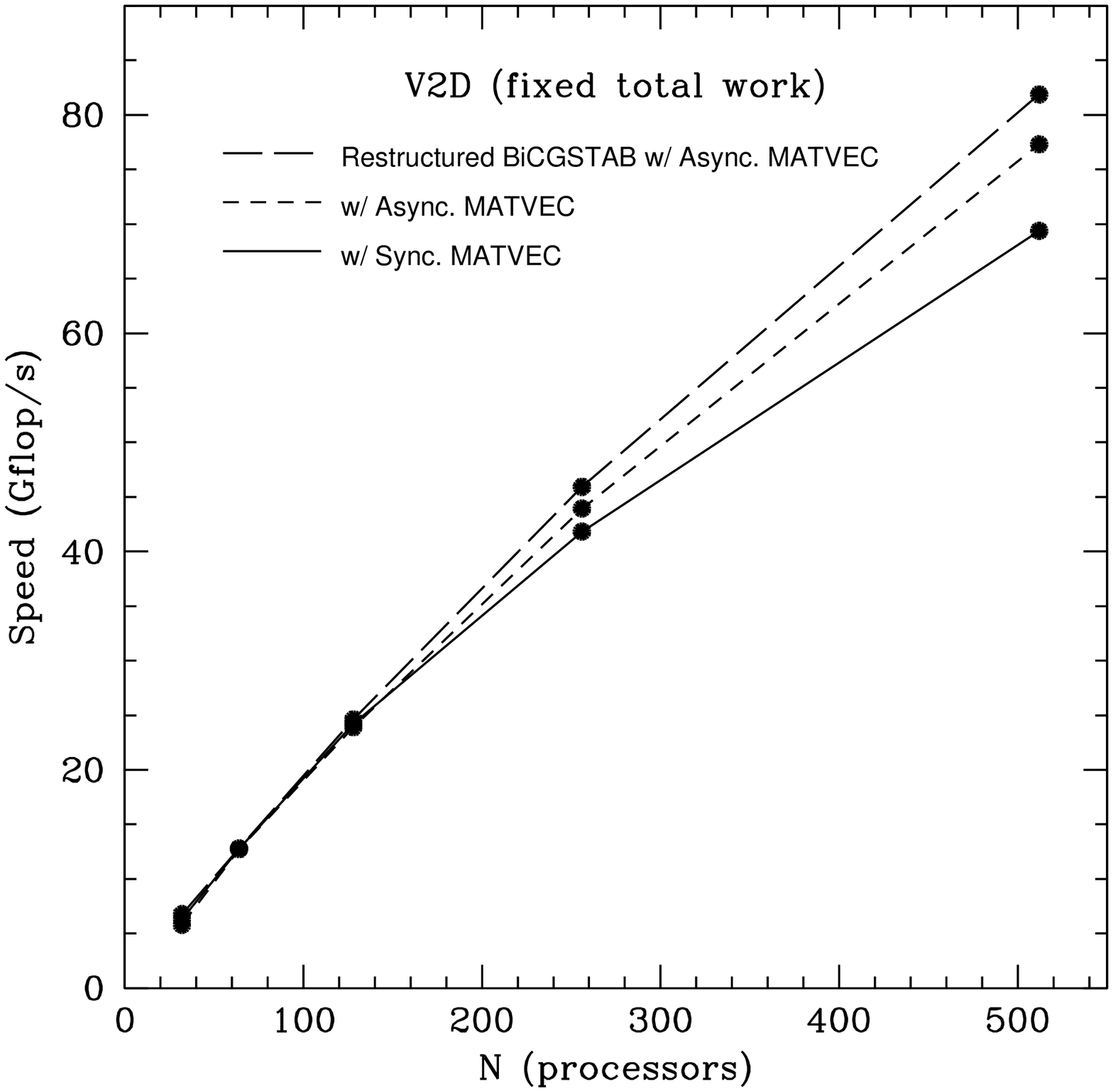}
\caption{\label{fig:f1}Parallel speedup of the algorithm,
with fixed problem size, on the 
NERSC IBM-SP ({\tt seaborg}).  Note that the roughly 82~Gflop/s, which
is obtained running with 512 processors, represents at speedup of
about 14 over a 32-processor run---a parallel efficiency of about
85\%.  }
\end{figure}
The major floating point cost of the Newton-Krylov algorithm is
expended in the evaluation of the finite-differenced
nonlinear diffusion equation. This operation requires only 
nearest-spatial-neighbor communication to
evaluate the finite-difference stencil of the divergence operator.
Whenever the nearest neighbor is a zone whose data is resident on
another processor, we amortize the communication cost by performing
the nearest neighbor-communication asynchronously.  This allows us to
carry out, simultaneously, the portions of the matrix-vector multiply
operation corresponding to interior (local) zones of each subdomain.
This yields a further improvement in scalability, as seen in
Figure~\ref{fig:f1}.


\subsection{Code Structure}
\label{sec-code}

Although time advancement is the most important part of the algorithm,
it is only the central portion of a sequence of many operations that
manage an overall simulation.  For completeness, we present a diagram
of the remaining portion of the algorithm in
Figure~\ref{fig:v2d_full}.  
\begin{figure}[htbp]
\begin{center}
\includegraphics[scale=0.75]{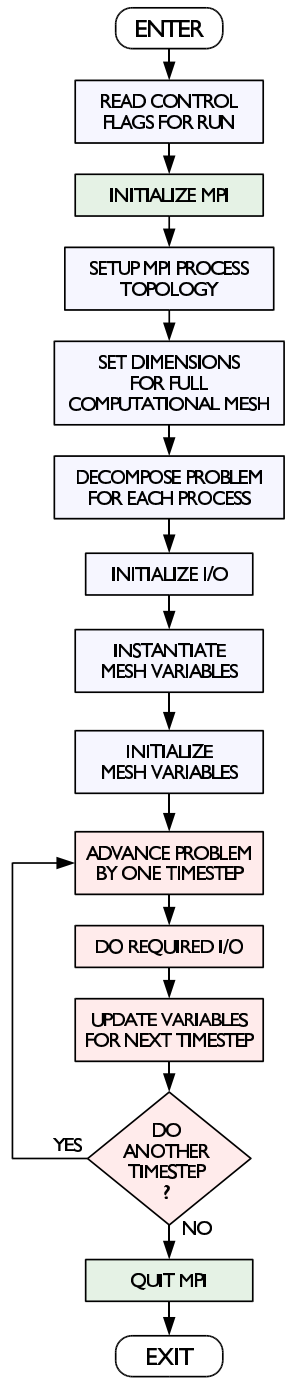}
\end{center}
\caption{\label{fig:v2d_full} An overview of the complete
algorithm from program initiation to finish, focusing on the steps
required to set up the problem.  The timestep advancement algorithm,
which is the red-colored loop, is detailed in Figure~\ref{fig:timestep}.  
}
\end{figure}
Several of the initial operations, depicted by blue boxes in
Figure~\ref{fig:v2d_full},
reflect a parallel implementation of the algorithm.  The main 
computational effort described by the flowchart is the time
evolution loop depicted by the red boxes in Figure~\ref{fig:v2d_full}.
The core of this loop is the timestep evolution previously 
described in Figure~\ref{fig:timestep}.  The purpose of the 
loop in  Figure~\ref{fig:v2d_full} is simply to evolve
the equations forward in time.

Of note in Figure~\ref{fig:v2d_full} is
that we periodically write out data from a run in the form of checkpoint
files.  These files capture the state of the simulation and serve two
functions.  They act as restart files, so that a simulation can be
resumed from any checkpointed timestep.  They also serve as a resource
for post-processing analysis.  Use of parallel file systems,
MPI-I/O \citep{mpi2}, and HDF5\footnote{http://hdf.ncsa.uiuc.edu/HDF5/}
in our simulations allows us to create a checkpoint file, using a
particular number of processors and processor topology, and restart it
at a later time, using a different processor count or topology.  In
fact, it is routine for us to use checkpoint files in this manner.
The file portability built into both MPI and HDF5 also allows our use
of these files across diverse computer architectures.

\section{Code Verification Tests\label{sec:validation}}

To test our algorithm, we have subjected the code
that implements it, V2D, to a suite of
verification test problems that stress individual
components of the code in a variety of different
contexts.  These tests are broken out into five
main classes: (1) hydrodynamic tests, which test
only hydrodynamic portions of the code, {\em i.e.,}
without radiation transport; (2) gravity tests,
which also involve hydrodynamics, but
which stress self-gravity of the system; (3)
transport tests, which test the radiation
transport portion of the code, but in a static
medium ({\em i.e.,} without hydrodynamics); (4)
radiation hydrodynamics tests, which test the
coupled hydrodynamic and radiation portions of
the code; and (5) correctness and scalability
tests, which are a diverse set of tests used to
test such things as the nonlinear solvers
implemented within the code and the correctness
and scalability of the parallel implementation of
the code.

Because of the length of this
manuscript, we have chosen only a few problems for
each of these categories.  These problems were
chosen to encompass the most important aspects of
algorithm correctness.  In some categories, such
as hydrodynamics, there are many classic
verification test problems that we have not
included for reasons of brevity.  
All of the test problems we have included in this section, along
with many others, are run as automated regression
tests that are executed daily against our source
code repository.   

In the following subsections we provide 
verification test descriptions and results 
for each of the aforementioned categories.
\subsection{Hydrodynamic Tests}

The hydrodynamic verification tests we have implemented employ solely the
hydrodynamics aspects of the algorithm.  These problems only involve
equations~(\ref{eq:cont})--(\ref{eq:mom}),
as presented in
\S\ref{sec-hydro-eqns}, with all radiation-coupling terms omitted.

We perform three standard tests of hydrodynamics.
The first is a passive advection test, which
tests the ability of an advection scheme to preserve
features in a the profile of a physical quantity when the
quantity is advected by a moving medium.
The second and third problems, the shock tube and
the Sedov-Taylor blast wave, test a code's ability
to reproduce the solutions to two standard
problems whose solutions are known analytically.

Two of these tests, the passive advection problem
and the shock tube problem were performed by
\cite{sn92a}, but since our hydrodynamic algorithm
differs slightly in the order of solution
steps, we repeat these tests to establish the
performance of our algorithm on the classic
problems.  The Sedov-Taylor blastwave problem was
not addressed by \cite{sn92a}, and so we include
the problem here.


\subsubsection{Passive Advection}

The passive advection test exclusively exercises
advection equations described in Appendix~\ref{app:advect}.  The
goal of this test is to delineate how faithfully
a waveform is preserved as it propagates
through a moving medium.  Hence, it serves as an
excellent measure of the diffusivity of the
advection algorithm.
For comparison purposes, our setup closely follows the passive
advection test results reported in \citet{sn92a}.  Source terms in the
hydrodynamics are ignored and only the advection equations are evolved
in time.  There are no gravitational forces, no
pressure gradients, and no radiation.  In Cartesian coordinates, the
test uses a uniform, time-independent velocity field, which is set to
$5\times 10^4$~cm~s$^{-1}$, parallel to one of the axes.  As an initial
waveform, we set up a square pulse 50 zones wide near the left-hand
boundary of a domain 400 zones wide, where each zone is 1~cm wide.
This pulse is applied to the density, and since we assume the material
is an ideal gas, it also appears in the pressure and internal energy.
Specifically, for passive advection in the $x_1$ direction,
we set 
$\rho = 1$ g cm$^{-3}$
and $P = 1$ erg cm$^{-3}$ in the region $5$~cm~$ < x_1 < 55 $~cm, and
$\rho = 10^{-10}$~g~cm$^{-3}$
and $P = 10^{-10}$ erg cm$^{-3}$ elsewhere.
The velocity in the $x_1$ direction is set to $5\times 10^4$~cm~s$^{-1}$
and the velocity in the $x_2$ direction is set to zero.
The matter internal energy density is set to $E = P/(\gamma-1)$
everywhere with $\gamma=5/3$.
These initial conditions are evolved 
until the pulse propagates a distance of five times its
initial width.  With the background velocity, and using a timestep of
5~$\mu$s, this is achieved in 1000 timesteps.

\begin{figure}[htbp]
\includegraphics[width=20pc]{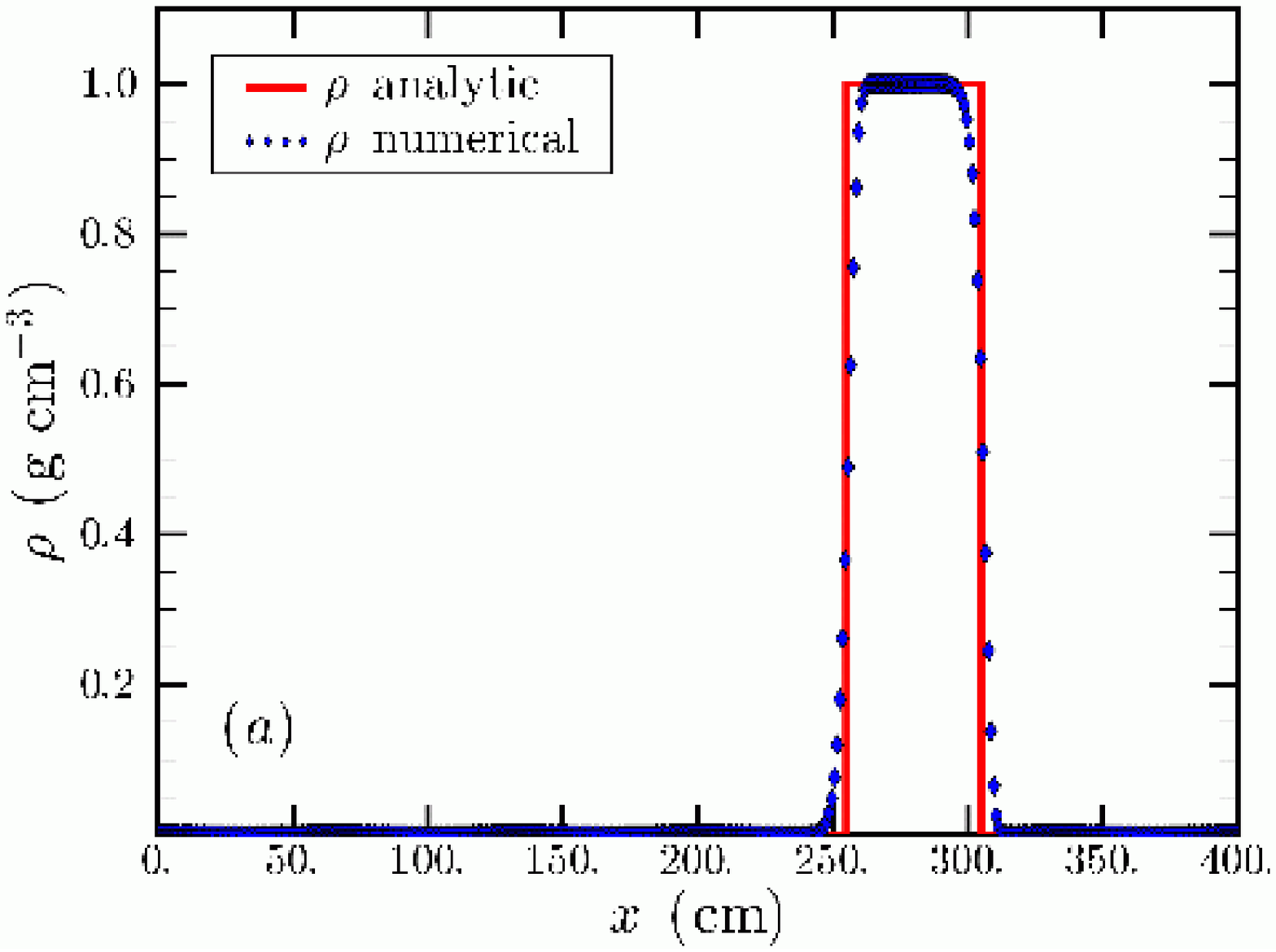}
\includegraphics[width=20pc]{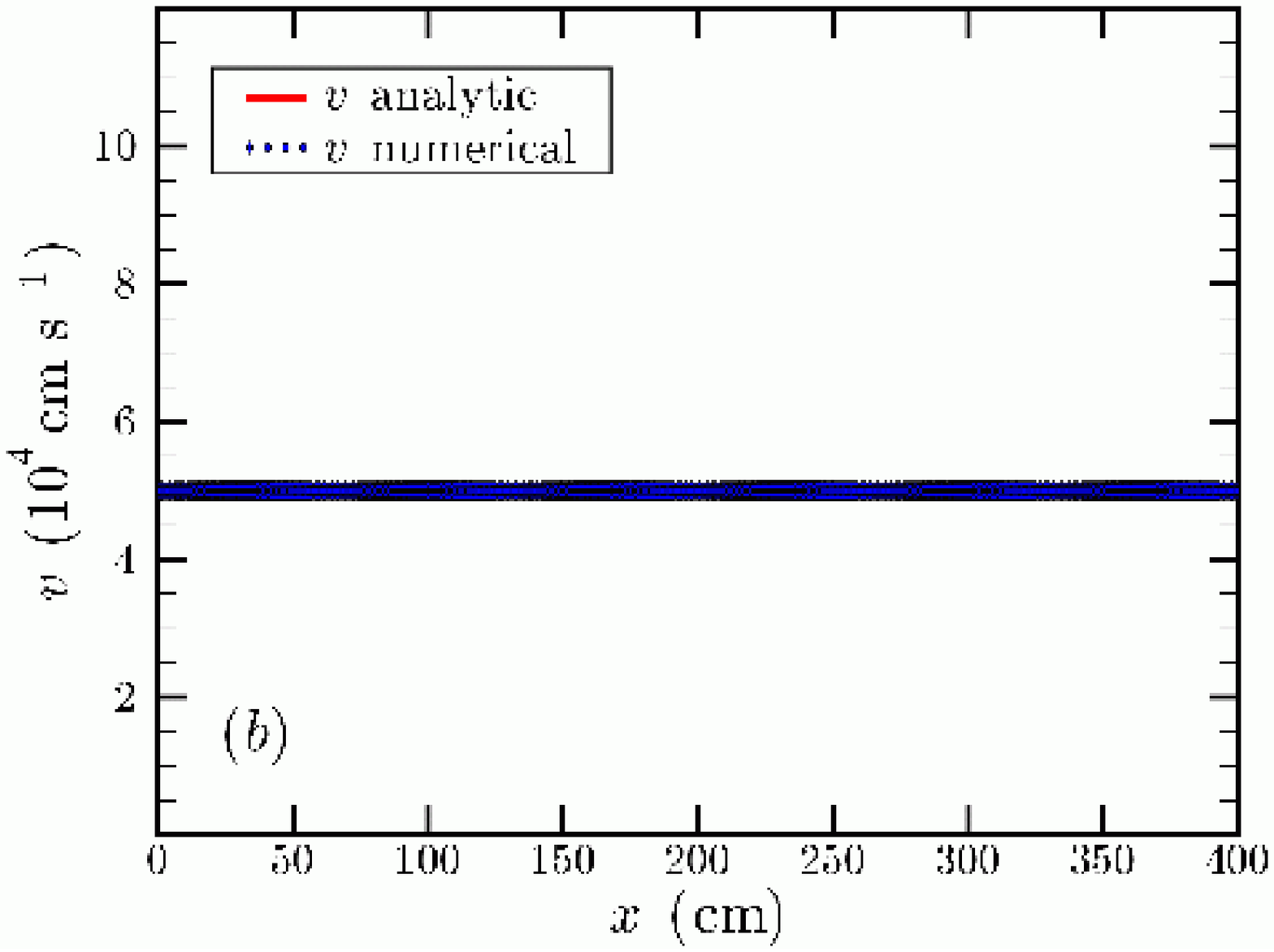}
\includegraphics[width=20pc]{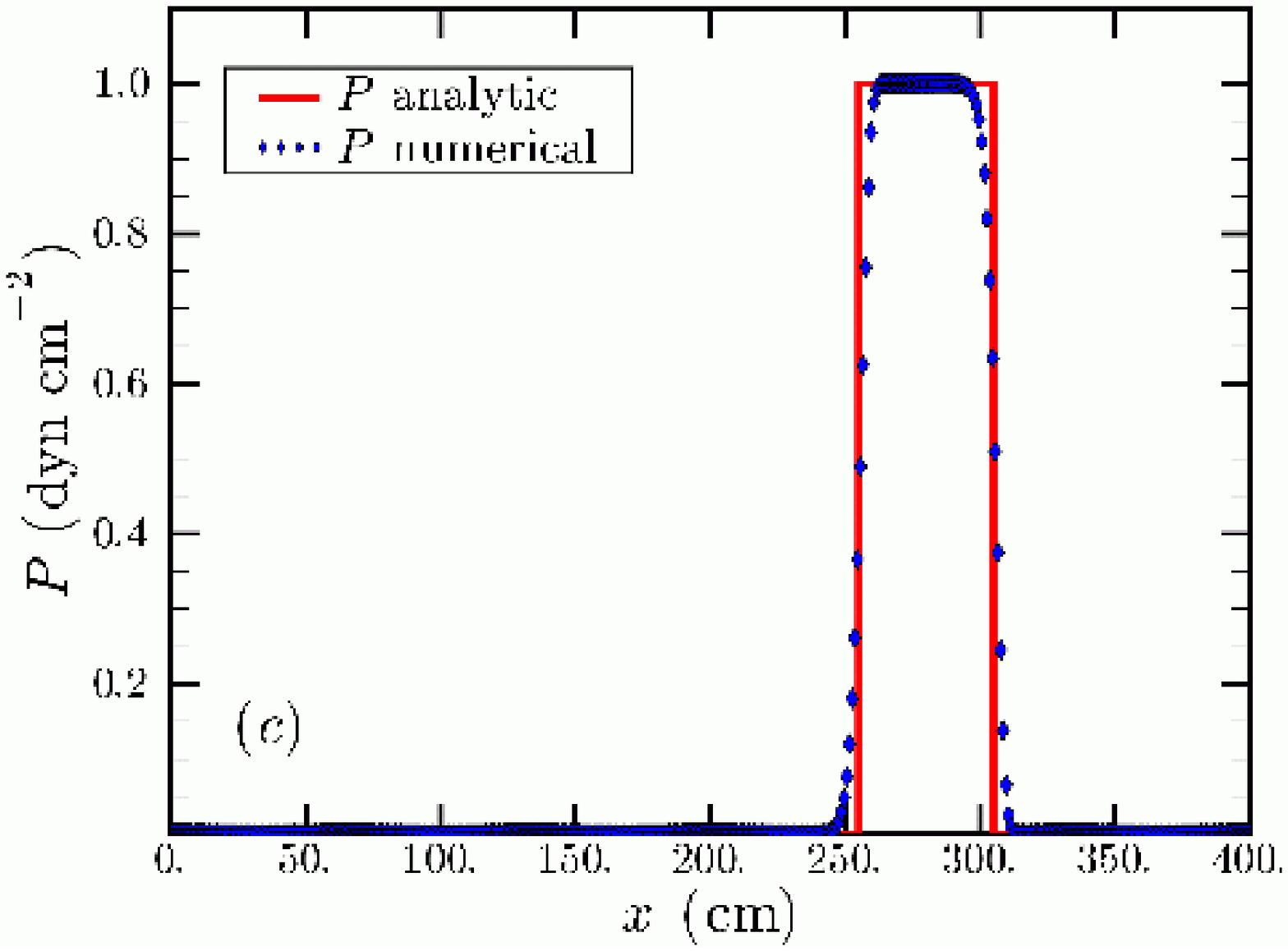}
\includegraphics[width=20pc]{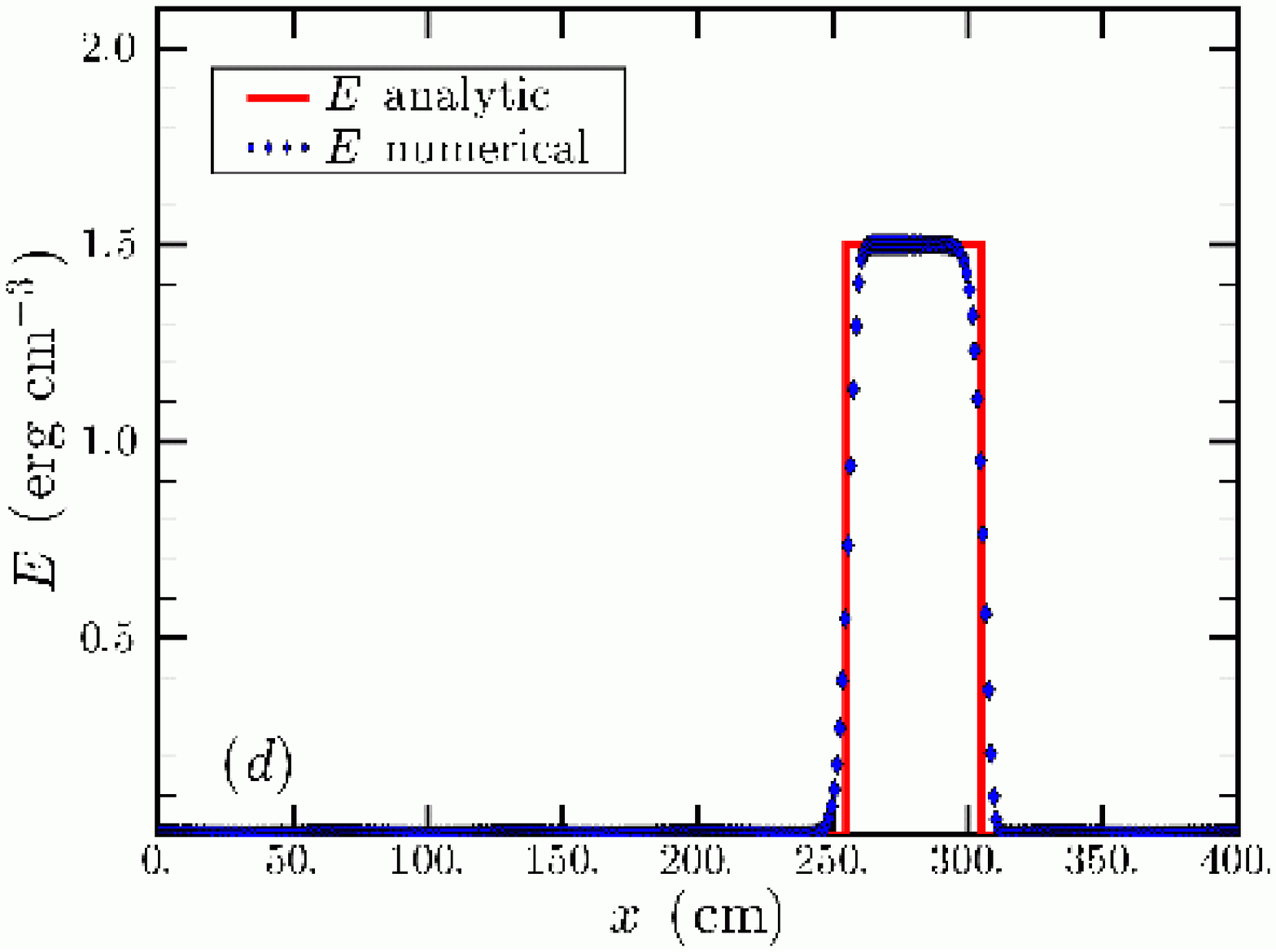}
\caption{\label{fig:pass_adv_x} Results in Cartesian coordinates for
passive advection of a 50-zone wide pulse carried along in the
$x$ direction for 1000 timesteps---a distance corresponding to 5 pulse
widths.  The advection scheme used is that of van Leer. We show
comparisons of the following quantities: $(a)$ mass density, $(b)$
velocity, $(c)$~pressure, and $(d)$ internal energy density.  Exact
results, which would be expected from a ``perfect'' advection scheme,
are shown in solid red, whose vertical lines are located at 255 and
305~cm.  Numerical results are plotted with diamonds.  In the
velocity plot $(b)$, the numerical data has been set to agree precisely
with the exact velocity and, hence, completely obscures the analytic
line. }
\end{figure}

Results for our test are shown in Figure~\ref{fig:pass_adv_x}.
Although there is some diffusion evident, given that the square pulse
shape is not preserved exactly, the performance of V2D's
implementation of advection is in good agreement with the results of
\citet{sn92a}, who have also implemented van Leer's scheme.  
We perform numerous variations on this test.  In
one such Cartesian variation, we run the mirror
image of the test in the negative $x_1$
direction. After making the necessary changes to
adjust for the symmetry, results of the tests in
the $+x_1$ and $-x_1$ directions are found to be
bitwise identical.  We also perform the same pair
of tests in the $+x_2$ and $-x_2$ directions.
Once again, results of this pair are bitwise
identical, {\em mutatis mutandis}.  This
$X_2$-direction pair also has bitwise identical
results to the pair of tests in the $x_1$
direction.


\subsubsection{Shock Tube \label{sec-sod}}

\begin{figure}[htbp]
\includegraphics[width=20pc]{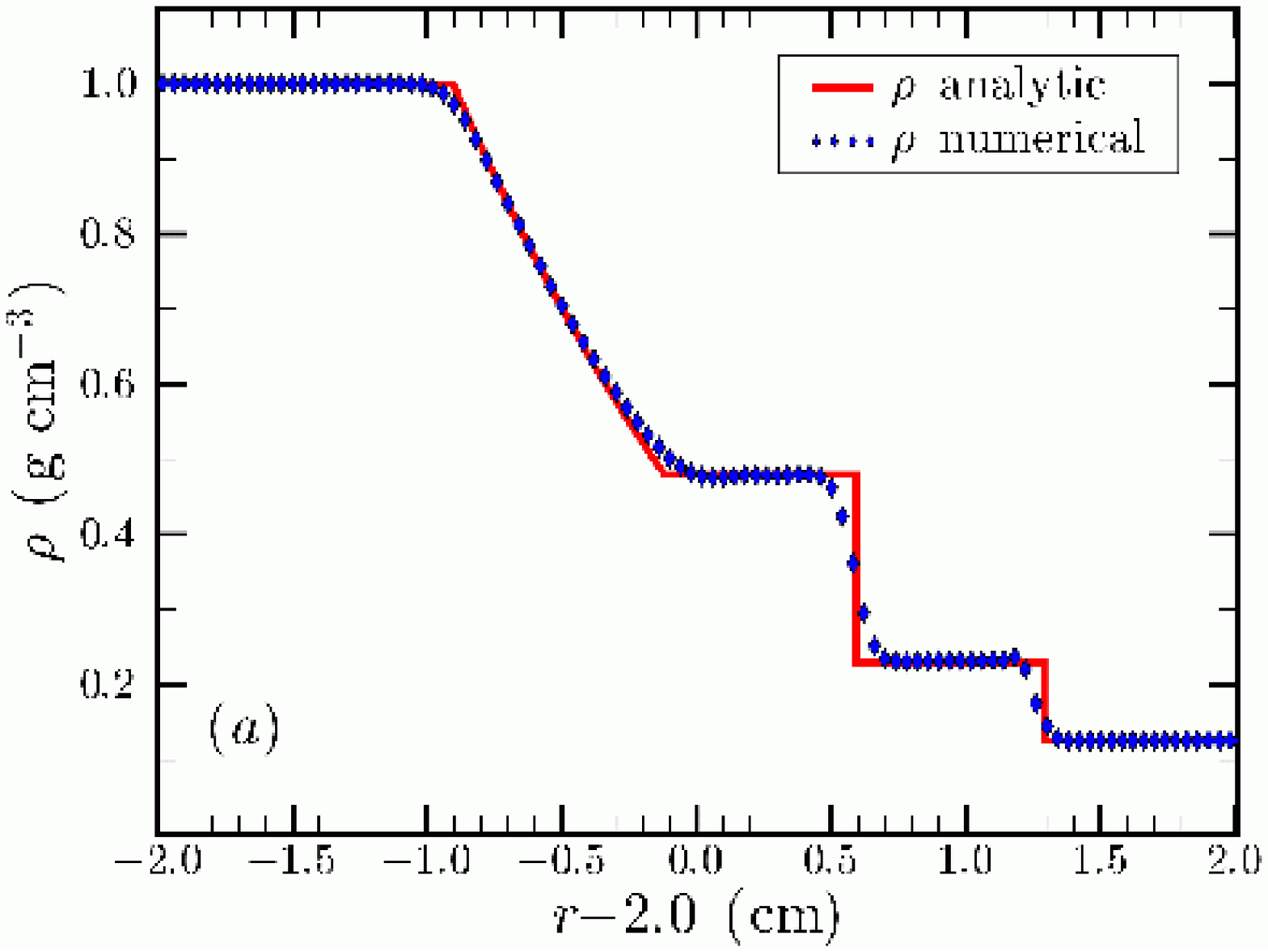}
\includegraphics[width=20pc]{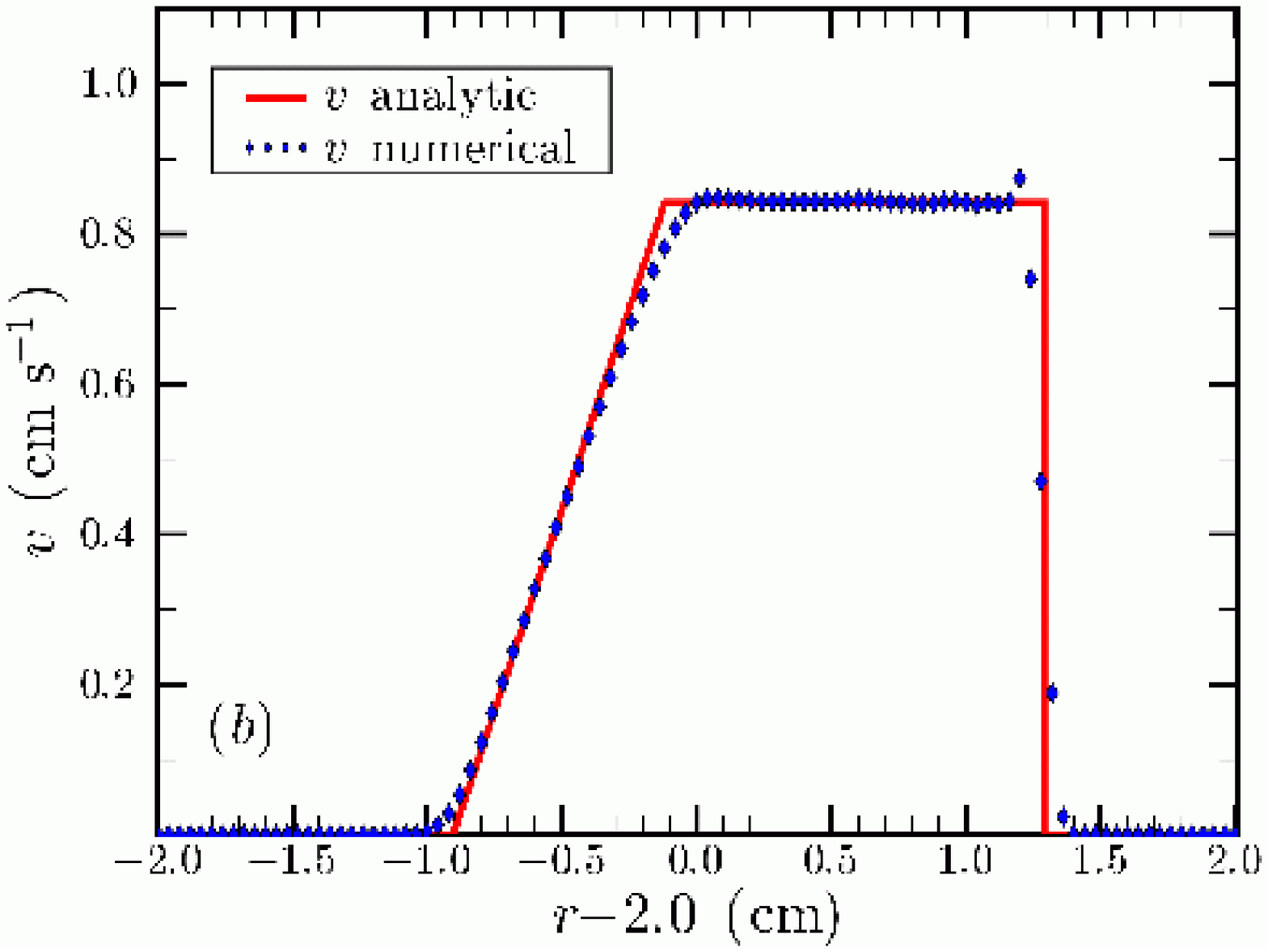}
\includegraphics[width=20pc]{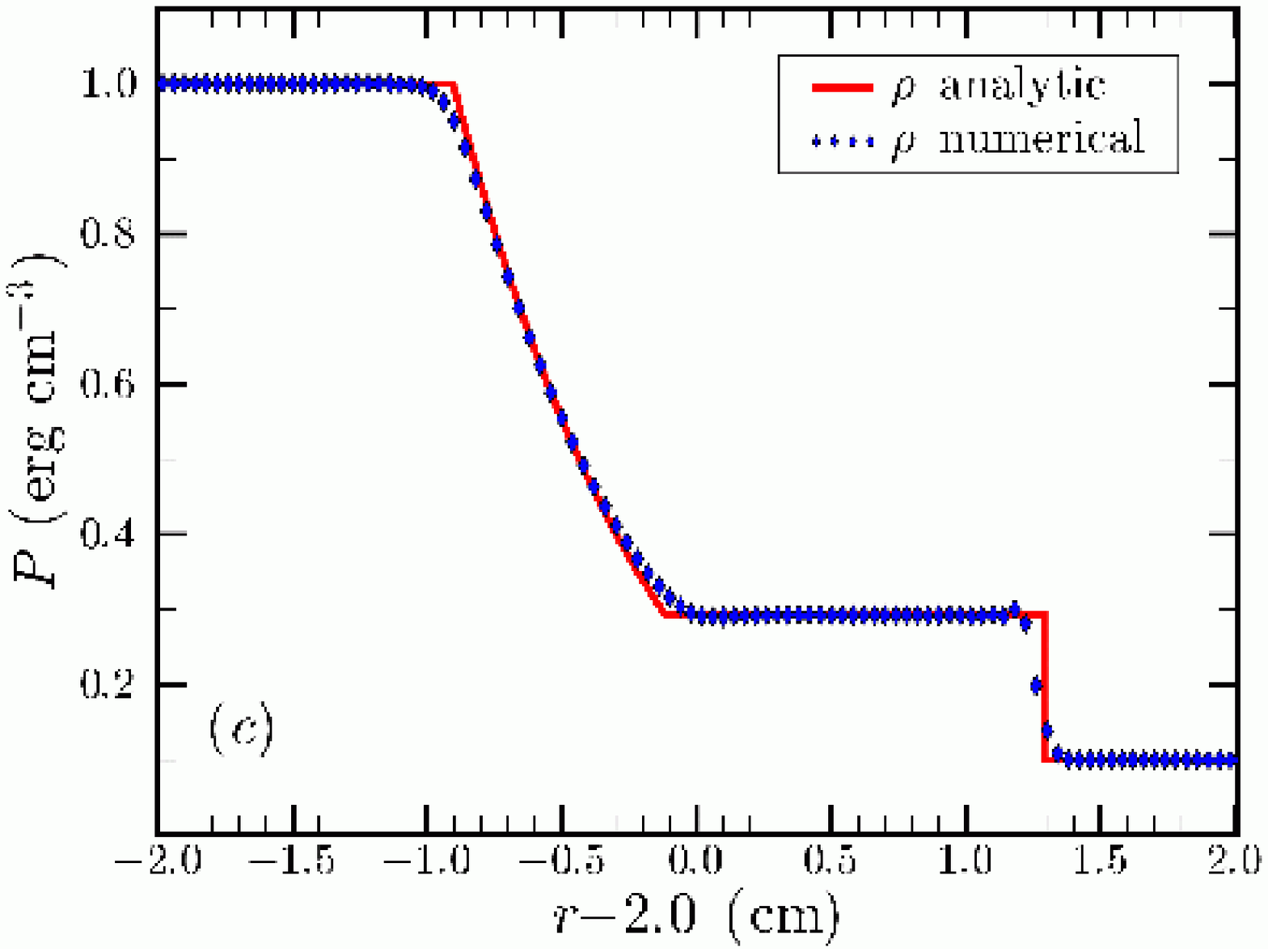}
\includegraphics[width=20pc]{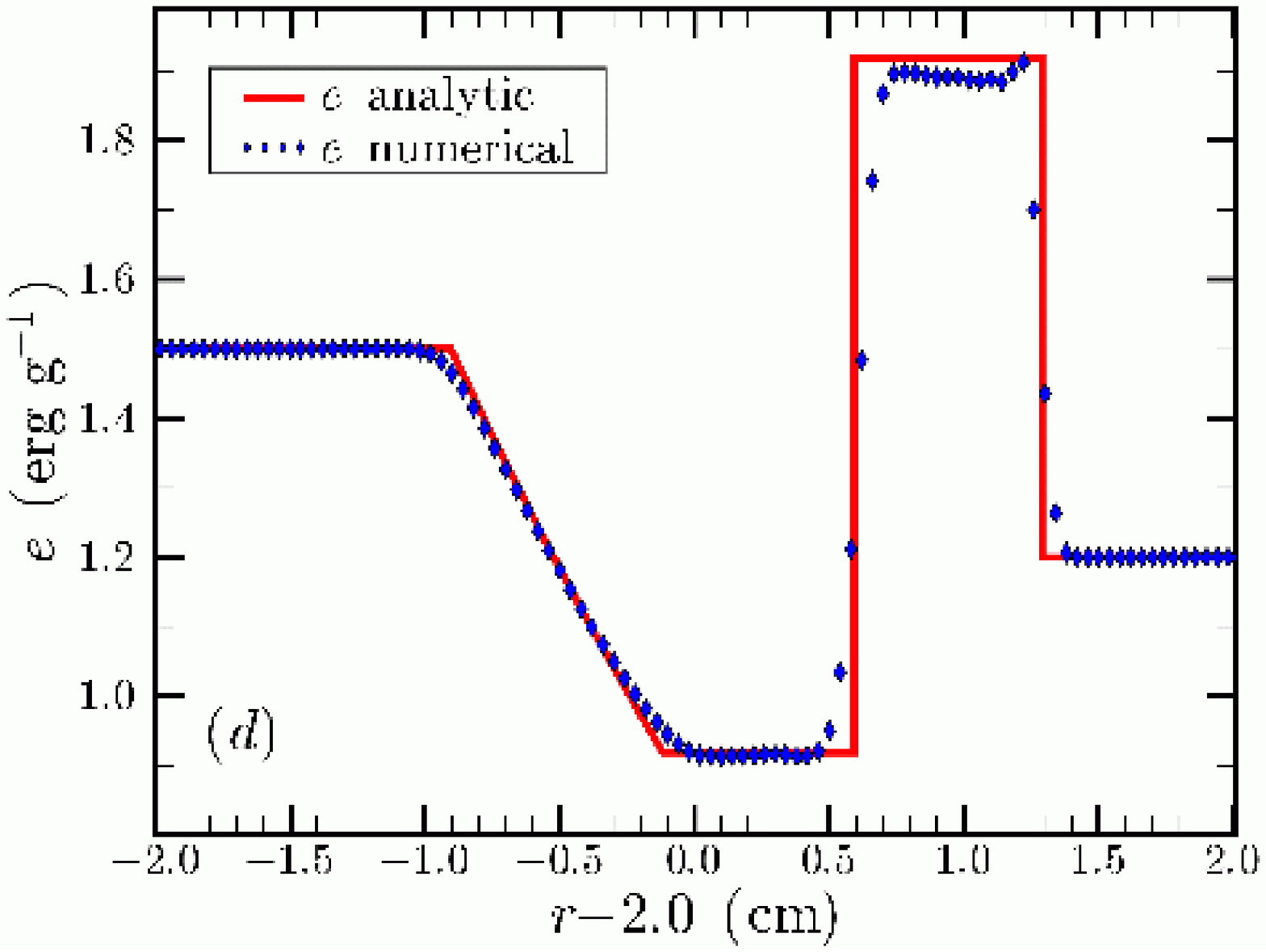}
\caption{\label{fig:sodx} Results for the 100-zone, non-relativistic,
shock-tube problem in spherical coordinates, with the shock
propagating outward in the radial direction. To assist in comparison
of these results with those of Cartesian shock tubes, we define a
radial coordinate $x_1$ such that $x_1 = 0$ is the initial position of
the shock wave.  With this definition, $x_1 = r - 2$~cm, where $r$ is
the distance from the center of the spherical ``tube''.  We show
comparisons for the following: $(a)$ mass density, $(b)$ velocity,
$(c)$ pressure, and $(d)$ specific internal energy. These results
compare favorably with those of \citet{sn92a} and the best schemes in
\citet{haw84b}. }
\end{figure}

The shock tube problem serves as an important test
of both the advection scheme employed by a code
and of the overall performance of a hydrodynamic
algorithm.  First introduced by \citet{sod78}, it
is now widely used as a standard verification
test.  Although essentially a one-dimensional
problem, it turns out to be a useful test for a
2-D code.  This is because, in addition to
checking for the correct behavior in the principal
direction of interest, it can also check that this
correct behavior is exactly replicated at all
points in the second dimension.

We set up and run the shock tube problem in each of the
three coordinate systems (Cartesian, cylindrical,
and spherical) and in each direction.  For the
purposes of this discussion, we choose $\xi =
x_1$, where $x_1$ can be related either to the
$x$-direction in Cartesian coordinates, the radius
$\Re$ in cylindrical coordinates, or the radius
$r$ in spherical coordinates.  We construct the
numerical solution with 100 zones in the $x_1$
direction and 12 zones in the $x_2$ direction.  As
just discussed, the $x_1$ coordinate is centered
about zero. It extends 4 cm in total such that
$-2~{\rm cm}~\leq~ x_1~\leq~+2~{\rm cm}$. Zones in
each direction are uniform in spatial extent.  
The initial conditions are those set forth by Sod:
in the left half of the domain we have 
$\rho = 1$ g cm$^{-3}$
and $P = 1$~erg~cm$^{-3}$, 
while in the right half we 
have
$\rho = 0.125$ g cm$^{-3}$
and $P = 0.1$ erg cm$^{-3}$.  
All velocities are initially zero.
The internal energy density
is initialized to $E = P/(\gamma-1)$ everywhere.
In this, and all the shock tube problems in our test suite, we have
set the polytropic index, $\gamma = 5/3$.

In our tests, we follow the subsequent evolution until $t =
0.70~\text{s}$.  The decision of how long to run the problem differs
among authors.  For comparison purposes, the particular choice is
important only in that the problem needs to run long enough so that
the resulting shock wave can propagate a significant distance into the
low-density gas.  However, the calculation must not run so long that
the shock reaches the right-hand boundary of the computational domain.
Our choice of timestep size is governed by the CFL condition (see
\S\ref{sec-time}), and our fraction of the CFL time is set to 0.5.
With these choices, each test run takes 59 timesteps to reach 0.70~s.
Results for a test run using spherical coordinates are shown in
Figure~\ref{fig:sodx}.  In all our tests, we use our implementation of
the van Leer advection scheme combined with Norman's consistent
advection, as described in Appendix~\ref{app:advect}.  Our result
compares favorably to \citet{sn92a} ({\em cf.} their Figure~11).  The
similarity of these results is not surprising since the hydrodynamic
algorithms of ZEUS and V2D have a similar approach, and both
calculations use van Leer advection.  Our results also compare well to
the best results in
\citet{haw84b} ({\em cf}. their Figures~6--15). In particular, the
results using their consistent advection scheme are comparable to
those shown here.

In addition to running the shock tube problem in
spherical coordinates, as shown in
Figure~\ref{fig:sodx}, we also perform the shock
tube problem in cylindrical and Cartesian
coordinates.  In the case of the latter, we
perform the tests in the $+x$ and $-x$ directions
and in the $+y$ and $-y$ directions.  After
adjustment for the different coordinate systems
and directions, the results of these tests are
compared and found to be bitwise identical to each
other.  As a check of the parallelization of the
hydrodynamics algorithm, we also run the shock tube
problem in spherical coordinates in the $+r$
direction on several different processor counts
and topologies: a serial (1$\times$1)
single-processor run and parallel 4-, 16-, 64-,
and 256-processor runs in 2$\times$2, 4$\times$4,
8$\times$8, and 16$\times$16 topologies,
respectively. When the results of these runs are
compared, they are found to be bitwise identical
to each other.


\subsubsection{Sedov-Taylor Blast Wave}

The Sedov-Taylor blast wave is another example of
an idealized problem for which a self-similar
analytic solution exists.  The solution was
originally found by \cite{taylor50} and
\cite{sed59}.  The problem is as follows: At time
$t = 0$, a large amount of energy is deposited at
a single point in an otherwise uniform cold,
diffuse, ideal gas.  The result is an expanding
spherical shock front, behind which is an
expanding, spherically symmetric, distribution of
matter.  Since the solution to the problem is
self-similar, there exist similarity variables
such that radial profiles of the physical
quantities are invariant in time with respect to
the similarity variables.
\begin{figure}[htbp]
\begin{center}
\includegraphics[scale=0.6]{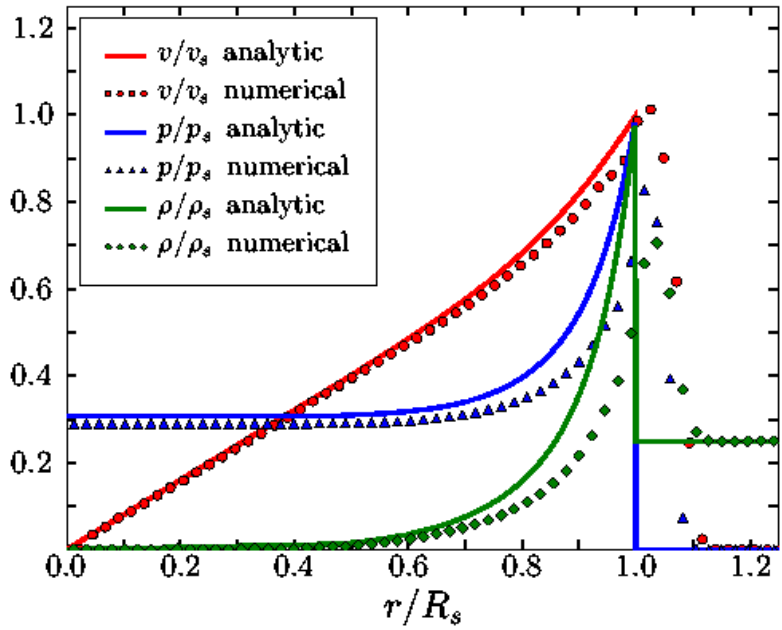}
\end{center}
\caption{\label{fig:sedov2} 
Results for the three-zone bomb 
Sedov-Taylor blast wave in spherical coordinates
plotted versus the self-similar analytic solution.  All
quantities are plotted in terms of the dimensionless similarity
variables for the problem, as indicated in the legend. (The ``$s$''
subscript on the quantities in the ratios indicates that they are
analytic values at the shock front.)  The solid lines represent the
analytic solution; the data points are the numerical solution.  }
\end{figure}
\begin{figure}[htbp]
\vspace{0.25in}
\begin{center}
\includegraphics[scale=0.6]{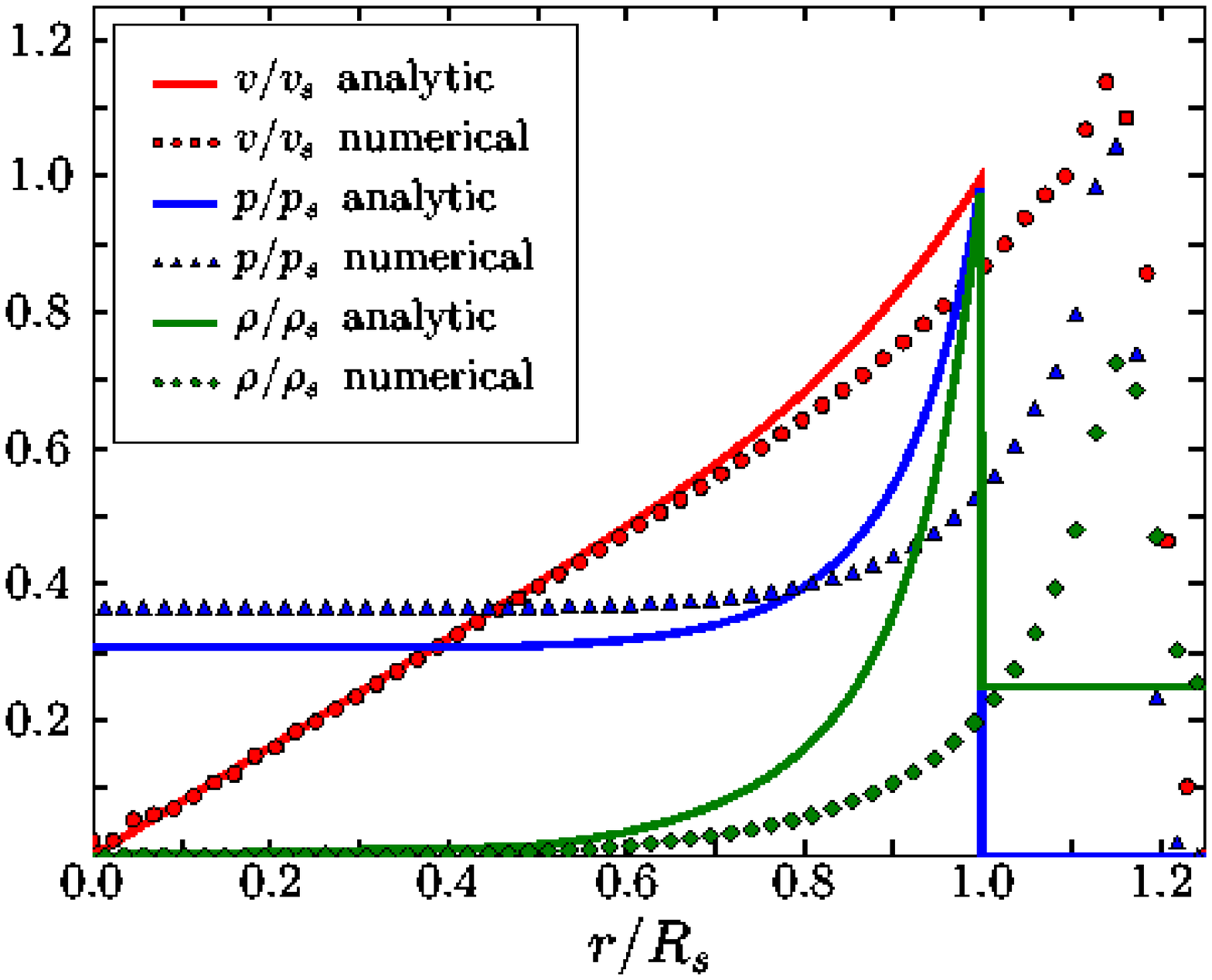}
\end{center}
\caption{\label{fig:sedov1} 
\vspace{0.25in}
Same as the previous figure except
the bomb is confined to one zone.}
\end{figure}

For the numerical test, we set up the problem in
spherical coordinates.  We choose a spatial domain
that is 1-cm in radius and is spanned by 100
uniform width zones in the $x_1$ (radial)
direction.  In the $x_2$ (angular) direction, 
the domain size $\pi$ radians with 6 equally spaced
angular zones (the number of angular zones should
be truly irrelevant to the outcome of the
problem).  A solution with multiple angular zones
is performed purely as a check that the
hydrodynamics code is capable of maintaining
perfect spherical symmetry.  The initial
conditions are $\rho=0.1$ g cm$^{-3}$ and
$T=10^{-8}$ MeV.  We initialize the problem in two
different ways.  In the first case
we place a high energy density
``bomb'' of $T=1$ MeV, $\rho=0.1$ g cm$^{-3}$
matter in the the innermost
three zones.  We refer to this case as the
the three-zone bomb case.
In the second case, the bomb is located only in the
centermost zone and the temperature is set to $T=27$ MeV
so as to keep the total bomb energy constant.
We refer to this as the one-zone bomb case.
The precise values of the initial
conditions are somewhat arbitrary as long as the
vast majority of the energy in the problem is
initially confined to a small region near the
center.  We assume an ideal gas with $\gamma =
5/3$.  We evolve the solution until a time of
$7\times 10^{-9}$ seconds using a CFL factor of
$0.5$ which takes 1496 steps for the one-zone bomb
case and 458 steps for the three-zone bomb
case. The results for both cases are shown in
Figures \ref{fig:sedov1} and \ref{fig:sedov2}.
Our results for the three-zone bomb case agree
well with other codes where the bomb is spread out
over multiple zones.  However, the one-zone bomb
case is perhaps a more authentic way of
initializing the problem which is rarely seen in
the published literature. As we can see from
Figure \ref{fig:sedov1}, the agreement of the
numerical and analytic solutions is substantially
worse in the one-zone bomb case.  The one-zone
bomb result can be improved somewhat by using
non-uniform zoning.

Defects caused by zoning in
numerical solutions of the Sedov-Taylor problem
are common.  The problem is especially difficult
because most of the material piles up behind the
shock, and it is difficult to maintain the spatial
resolution at all the points necessary to
track accurately its movement.  Even codes that
can draw on adaptive mesh refinement (AMR) still
have issues.  For example, \citet{flash}, using
the FLASH code, manage to track the position of
the shock very well, but are less successful in
obtaining good post-shock profiles for the
quantities of interest.  \citet{pen98}, also
employing an AMR-based code, is forced to choose
between optimizing either resolution and
positioning of the shock or resolution of the
post-shock material.

Finally, we note that when we check our results for spherical
symmetry, we observe that the solutions of all radial rays are bitwise
identical to each other, and that all velocities in the angular
($x_2$) direction are identically zero at all times.

\subsection{Gravity Tests} \label{sec-gravity-tests}

Although V2D is a code intended for calculations in two spatial
dimensions, self gravity is currently approximated by assuming that
the mass distribution interior to a point is spherically symmetric
(see Appendix~\ref{app:gmomsrc}).  This treatment naturally leads to
gravity verification tests that have spherical geometry.  
The two gravity verification tests that we consider in this paper 
(a unit density sphere and a polytrope of index 1), both have spherical 
geometry and are solved in spherical coordinates.


\subsubsection{Unit-Density Sphere}   \label{sec-unit-dens-sph}

In this test problem we compute the gravitational
potential for a sphere of unit density.  

Since the gravitational potential given by
equation~(\ref{eq:potential_sph}) is a trivial
function of the gravitational mass interior to a
point, this problem reduces to a test of our
algorithm for computing the mass interior to a
point.  In a serial implementation, this summation
could be accomplished trivially.  However, in a
parallel implementation, this summation requires a
series of reduction operations across the process
topology.  In practice, we accomplish this by
first summing the mass in a shell at a given
radius by a series of MPI global reduction
operations in the $x_2$ direction.  Once the
amount of mass at a given radius is known, the
total mass interior to each radial zone is summed
in an outward fashion starting from the inner edge
of the grid.  These summation steps involve
numerous message-passing operations and the
purpose of this test also serves to verify the
correctness of our implementation.

The problem is set up by constructing a 10-km sphere of unit density,
which consists of 100 radial zones of equal spatial width and 40
angular zones of equal angular width, extending from 0--$\pi$ radians.
The gravitational mass is computed using the method given in
Appendix~\ref{app:gmomsrc}.  The result is compared against the simple
exact computation that is possible for this problem.

The value of this test is the comparison that is possible between the
exact method and the various modes of calculation that may take place
when V2D is executed.  
One of the pitfalls in parallel computing is the potential 
effect in global summation operations
caused by floating point arithmetic
being neither associative nor distributive.  
Thus, the answer to a
summation operation that is distributed over multiple
processors will, in general, produce a result that is dependent on the
number of processors used.  Furthermore, results can vary slightly from
run to run, even on identical numbers of processors, since MPI
(which is commonly used for message passing) offers no
guarantees about the order of arrival of inter-processor messages.
Because of this, the order of individual operations may vary from run to
run, possibly producing variable results.

To estimate the magnitude of these effects, we calculate the
gravitational mass in a number of different ways and compare each
result against the exact result.  The deviations of these various
results from the exact result are shown in Table~\ref{gravmass}.  In
what is labeled ``MPI with deterministic reduction'' in
Table~\ref{gravmass}, we have replaced the MPI\_ALLREDUCE operation in
the standard MPI library with one of our own construction.  Our
version is designed to guarantee an identical order of all critical
arithmetic operations, regardless of processor count.  As the table
shows, the deterministic results agree well with the exact result.  As
indicated in the table, these deterministic results also agree with
each other bitwise, {\em i.e.}, to the full numerical precision of the
system on which they were run.  This is to be expected from the
deterministic design of these calculations.  Although determinism is a
desirable feature, it is obtained at the cost of serializing certain
key time-intensive portions of the computation.  Therefore, it is not
a practical method to employ in long-running production calculations.

\begin{figure}[htbp]
\vspace{0.25in}
\begin{center}
\includegraphics[scale=0.6]{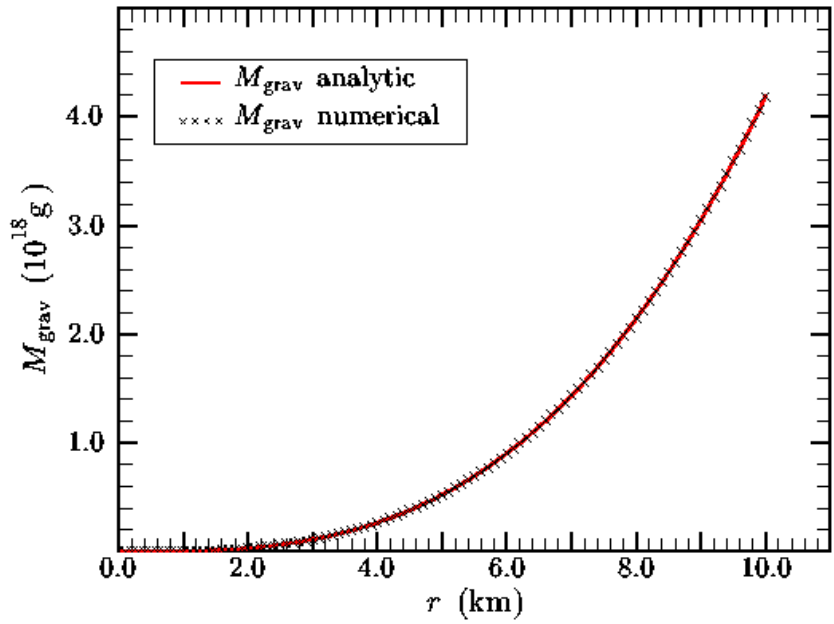}
\end{center}
\caption{\label{fig:grav_mass_1x1_deter}  A test of V2D's spherical
gravity solver.  A 10-km sphere of unit density is constructed and the
gravitational mass at each of 100 grid points is compared against an
exact calculation.  The results shown here are for a single-processor
calculation corresponding to the first row of
Table~\ref{gravmass}. The maximum discrepancy of the total mass
calculated by the two methods produces relative differences of $2.4
\times 10^{-14}$.  Plots of similar data, for the other versions of
this test shown in Table~\ref{gravmass}, are visually and (in the case
of the deterministic runs) bitwise indistinguishable from this figure.
 }
\vspace{0.25in}
\end{figure}
\begin{deluxetable}{cccc} 
\tabletypesize{\normalsize}
\tablecaption{\sc Gravitational Mass Calculation Comparisons: Maximum
Deviations of Various Test Runs vs. the Exact Result  
\label{gravmass}}
\tablewidth{0pt}
\tablehead{
\colhead{method used} & \colhead{processor} & \colhead{processor} &  
\colhead{relative deviation} \\
& \colhead{count} & \colhead{topology\tablenotemark{a}} & \colhead{from exact method\tablenotemark{ b}} 
}
\startdata
MPI with deterministic reduction    &  1   & 1$\times$1   & $2.4 \times 10^{-14}$
\\
                                    &  4   & 2$\times$2   & $2.4 \times 10^{-14}$
\\
                                    &  16  & 4$\times$4   & $2.4 \times 10^{-14}$
\\
                                    &  64  & 8$\times$8   & $2.4 \times 10^{-14}$
\\
                                    &  256 & 16$\times$16 & $2.4 \times 10^{-14}$
\\
standard MPI &  1   & 1$\times$1   & $5.2 \times 10^{-16}$
\\
             &  4   & 2$\times$2   & $4.1 \times 10^{-16}$
\\
             &  16  & 4$\times$4   & $3.6 \times 10^{-16}$
\\
             &  64  & 8$\times$8   & $2.6 \times 10^{-16}$
\\
             & 256  & 16$\times$16 & $4.1 \times 10^{-16}$
\\
\enddata
\tablenotetext{a}{Processor topology dimensions refer to the logical
structure by which the two-dimensional computational grid was
decomposed among processors, not to any physical
processor-interconnect topology that the computer hardware employed.}
\tablenotetext{b}{The deviations shown here for each of the ``MPI with
deterministic reduction'' results, in fact, agree with each other
bitwise---they agree to the full numerical precision of the system on
which they were run.}
\end{deluxetable}

The results in Table~\ref{gravmass}, labeled ``standard MPI,'' use the
ordinary, standard-compliant MPI libraries for all message passing.
Such calculations proceed with no regard to determinism of results.
(This is also the mode in which we do production computations using
the gravity solver.)  The results from this method show, curiously, a
smaller deviation from the exact result than shown by the
deterministic calculations.  More significantly, this deviation
differs with processor count, as expected.  However, comparisons with
the exact result are so good, that it is inconceivable that the
overall results of a large-scale, long-running simulation could ever
vary in an important way with processor count, solely because of the
magnitude of variations exhibited here.  Machine epsilon is of the
same order ($\sim 10^{-16}$) as the deviations---for the architectures
on which both these tests and our production runs are performed.

On the basis of these tests, and of others that exercise other
parallelized portions of the code ({\it e.g.}, parallel solvers, the
SN tests---see \S\S4.5 and 4.6), we can proceed with confidence
that parallelization alone cannot produce incorrect results (due to
mistakes in parallelization) or deceptive results (due to answers that
depend on processor count).

Figure~\ref{fig:grav_mass_1x1_deter} shows a comparison of the the
exact result and that calculated by the deterministic method using a
single processor.  As already noted, the agreement is excellent.


\subsubsection{Stability of a Polytrope}

The polytrope stability verification test involves both 
gravity and hydrodynamics.
This test is initialized by numerically
constructing a static polytrope.  We then measure how well the static
solution is maintained when the polytrope is subjected to hydrodynamic
evolution.  This test provides a check on both the spherical-gravity
solver and on the correctness and stability of the hydrodynamic
portion of the V2D algorithm.

We choose a polytrope with index $n=1$, which corresponds to an
polytropic equation of state that has an adiabatic index of $\gamma =
2$.  With this choice, we can construct a hydrostatic ``star'' by
solving the Lane-Emden equation, which has an analytic solution for
$n=1$ (\citet{chandra67}, p.~84ff),
\begin{equation}
\rho(\xi) = \rho_c \frac{\sin \xi}{\xi},
\end{equation}
where $\rho_c$ is the central density and $\xi$ is a radial coordinate
defined such that the radius, 
\begin{equation} 
r = \beta \xi, 
\end{equation}
where 
\begin{equation} \label{eq:alpha_poly}
\beta = \left( \frac{(n+1) K}{4 \pi G} \rho_c^{(1/n)-1} \right)^{1/2}.
\end{equation}
The quantity $K$ is the polytropic constant and $G$ is the universal
gravitational constant.  As indicated by
equation~(\ref{eq:alpha_poly}), for $n=1$, $\beta$ is
independent of $\rho_c$.  For convenience, we set $\rho_c = 1~{\rm
g~cm}^{-3}$ and choose $K = 2 \pi G$, such that $\beta = 1$. Hence,
for this problem, the density range throughout this configuration
ranges between 0 and $1~{\rm g~cm}^{-3}$, and the surface is located
at $\pi$~centimeters from the center.

This analytic solution yields initial conditions for the test problem.
We
construct a model having 100 radial zones (equally spaced in radius)
and 32 angular zones (equally spaced angularly).  
The zone center coordinates are determined by
\begin{equation}
[r]_{i+\lhalf} = \frac{1}{3}
\sqrt{
\left([r]_{i+1}\right)^2+
[r]_{i+1} [r]_{i}+
\left([r]_{i}\right)^2
}\label{eq:rmid}
\end{equation}
in accordance with the coordinate regularization scheme of \cite{mm89}.
To initialize the
density, we assign values for $\rho$ at cell centers by averaging
the analytic solution evaluated at the adjacent cell faces, {\em i.e.},
\begin{equation}
\left[ \rho \right]_{i+\lhalf, j+\lhalf} = 
\frac{1}{2} 
\left( 
\frac{ \sin (\left[ x_1 \right]_{i}) }{\left[ x_1 \right]_{i}} +
\frac{ \sin (\left[ x_1 \right]_{i+1}) }{\left[ x_1 \right]_{i+1}}
\right).
\end{equation}
This initialization is carried out in an angularly symmetric
fashion. Apart from the angular symmetry, we find the exact procedure
by which the mesh is initialized to the analytic solution is unimportant
to the subsequent evolution.

\begin{figure}[htbp]
\vspace{0.25in}
\begin{center}
\includegraphics[scale=0.6]{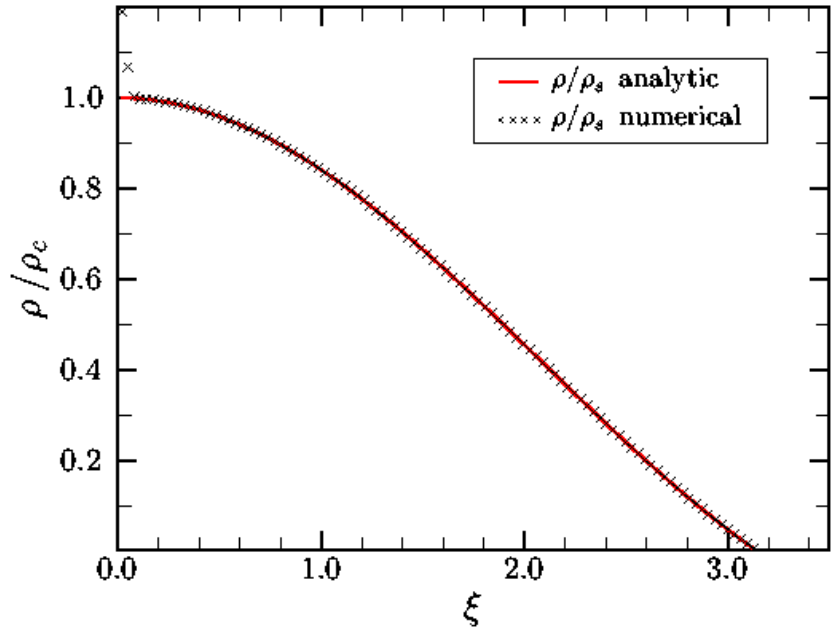}
\end{center}
\caption{\label{fig:poly_n1_1x1_mpi}  The normalized density 
for the polytropic stability test after 1000 timesteps.
Except near the center, where grid
resolution is an issue, the numerically evolved solution typically
agrees with the analytic solution to a few parts in $10^{4}$ or
better. }
\vspace{0.25in}
\end{figure}

Once initialized, the system is
evolved through 1000 timesteps.  The timestep size is chosen to
have a CFL factor of $0.5$, as determined by the procedure
described in \S\ref{sec-time}.  After 1000 timesteps, this corresponds
to a model time of about 14\:500~s.  
(To prevent the necessity of tiny
timestep sizes, given the relatively small angular zone size when
using 32 angular zones, we do not include contributions to the Courant
time in the angular direction from the first 10 radial zones.  To
ensure that unphysical motions do not occur because of this, we also
suppress all angular motion, that is motion in the $x_2$ direction, 
inside these 10 zones.)

Figure~\ref{fig:poly_n1_1x1_mpi} shows a
comparison between the evolved configuration and
the analytic solution.  Except near the center,
where grid resolution is an issue, the numerically
evolved solution typically agrees with the
analytic solution to a few parts in $10^{4}$ or
better.  The larger deviation near the center is a
result of gridding effects where the mesh provides
insufficient spatial resolution.  The radial
velocities are also checked.  After 1000
timesteps, the absolute value of the radial
component of velocity nowhere exceeds $9 \times
10^{-7}~{\rm cm~s}^{-1}$.  The transverse
component of velocity remains identically zero at
all timesteps throughout the spatial domain.  The
previous statement is the equivalent of saying
that the code allows all radial rays to evolve in
identical fashion---as expected, and as required.
Similar results are obtained for the case where
we choose $K=3.874 \times 10^4$~cm$^5$~s$^{-2}$~g$^{-1}$
and the mass of the configuration is $M=1.4 M_\odot$.
\subsection{Transport Tests}

The transport verification tests are intended to establish the 
numerical performance of the radiation diffusion aspects of the
algorithm.
Each of the tests described in this section assume
that radiation is propagating through a hydrostatic
medium.  Thus, the hydrodynamic variables are not evolved.
Problems that test coupling of radiation diffusion to 
hydrodynamics are incorporated in \S \ref{app:radhyd}.
In all transport and radiation-hydrodynamic tests, the Pauli 
blocking factors that appear in the collision integral
are retained with $\eta = -1$.  However, the problems
are such that the particle occupancies are much, much less than unity
and therefore the blocking factors obtain a 
classical value of unity.


\subsubsection{Diffusion of a 1-D Gaussian Pulse}
\label{sec-gauss_1d}

This problem makes use of an analytic solution of the time-dependent
diffusion equation in 1-D Cartesian coordinates for the case of a 
constant diffusion coefficient \citep{Crank,but68,rhb98}.  For 
initial conditions in the form a Gaussian profile centered at
point $r_0$, the analytic 
solution to the diffusion equation is given by
\begin{equation}
E_\epsilon = E_{\rm o} \sqrt{\frac{t_{\rm o}}{t_{\rm o}+\delta t}}
\exp \left\{-
\frac{\left(  r - r_{\rm o} \right)^2}{4D\:(t_{\rm o}+\delta t)}  \right\},
\end{equation}
where $t_{\rm o}$ is the initial time from which the problem is 
evolved, and $\delta t$ is the evolution time.
The quantity $E_{\rm o}$ the initial amplitude of the pulse and
$D$ is the diffusion coefficient.
This solution can be used to test the diffusion 
algorithm in both Cartesian and spherical coordinates.
In the case of spherical coordinates the pulse is centered 
at some large radius such that the plane-parallel limit of the 
spherical coordinate system is reached.

The initial conditions for the numerical solution are constructed as
follows. We use a single energy group of radiation, bounded by 0 and
1~MeV.  We use the arithmetic average of the group boundaries,
0.5~MeV, to define the energy at the group center.  Since the problem
is hydrostatic and there is no emission-absorption, there will be no
change in spectral shape.
Hence, the details concerning energy
grouping are irrelevant.  Spatially, we divide the domain into 100
radial zones, covering 1~cm radially, with each zone spaced uniformly.
In order to achieve the plane-parallel limit the inner
edge of the radial grid is placed at $x_1 = 10^4$ cm.
Flat (zero-flux) boundary conditions are applied at the inner and $\pm x_2$
edges of the grid with free streaming boundary conditions at the 
outer edge.
The number of
angular zones is arbitrary to the outcome of the problem (each radial ray
should evolve virtually identically).  
We
perform our test with 16 angular zones, equally spaced in angle,
covering a range of $10^{-4}$ radians and check to make sure
the solution in each radial ray is the same.  This choice 
of angular coordinates makes the
computational domain in the transverse direction cover about 1~cm and,
therefore, gives the domain the shape of a small curvilinear
``square.''
\begin{figure}[htbp]
\begin{center}
\includegraphics[scale=0.75]{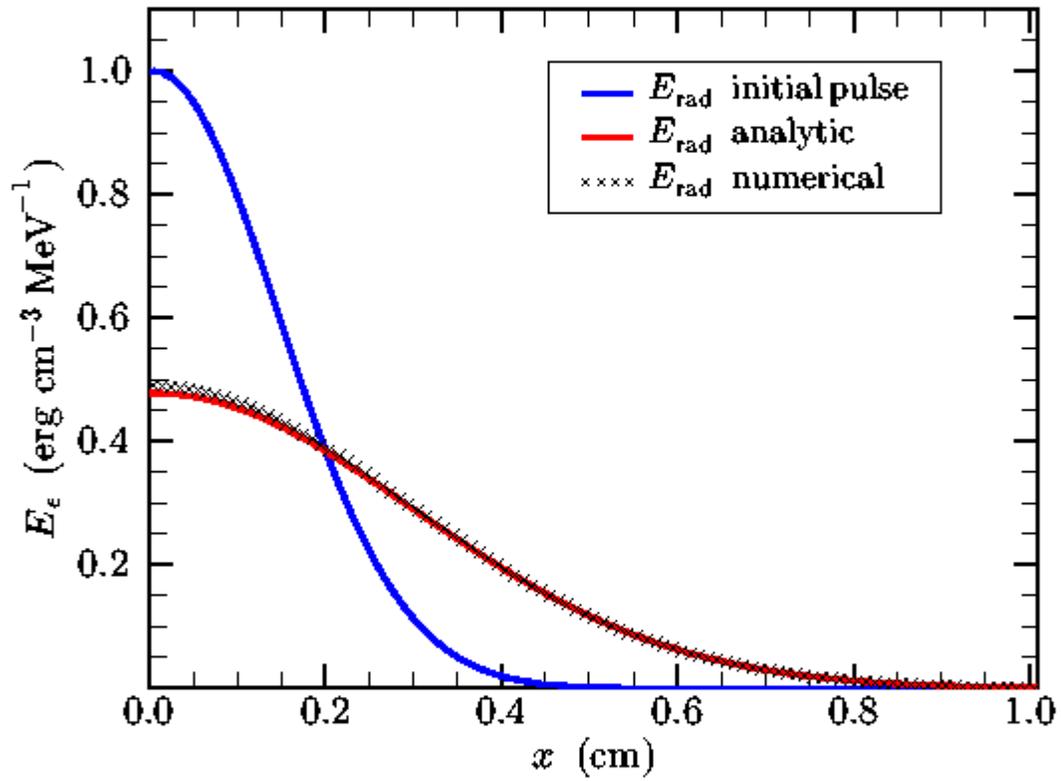}
\end{center}
\caption{\label{fig:rad_gauss_x1} Results after 
3.3~ns (100 timesteps) for a spreading Gaussian
pulse of radiation, propagating in the radial direction in spherical
polar coordinates.}
\end{figure}
To allow a more complete testing of the code, we
perform our calculation by setting up three
identical parallel transport problems, thus
testing all three species of radiation that V2D is
currently set up to track.  This is done for both
particles and antiparticles of each species.  In
the language of neutrino transport, this is the
equivalent of evolving electron, muon, and tauon
neutrinos and their antiparticles. This makes $\ell$
in the equations below have values from 1 to 6,
inclusive.

The discretized initial profile is given by
\begin{eqnarray}
^\ell\left[ E_\epsilon \right]_{k+\lhalf,i+\lhalf,j+\lhalf} = 
E_{\rm o}
\exp\left\{ - \frac{1}{4D\:t_{\rm o}} \left( \left[ r \right]_{i+\lhalf} - 
r_{\rm o} \right)^2 \right\}
& \text{for all $k,\ell,i,j$},
\end{eqnarray}
where, as indicated, the radial profile is replicated
over all angular zones, and applied to all species of radiation.
Here, $k=1$ for the single energy group of radiation being
evolved. We have set the diffusion coefficient $D =
10^{7}$~cm$^2$~s$^{-1}$, which in the diffusive limit corresponds to 
a total opacity, $\kappa_T = 10^3$~cm$^{-1}$ via
equation~(\ref{eq:diff_simple}). The constant $E_{\rm o}
= 1$~erg~cm$^{-3}$~MeV$^{-1}$.  After evolving for a time $t$, the
analytic solution to this problem leads us to expect a numerical
solution of the form
\begin{eqnarray} \label{eq:gauss_1d}
^\ell\left[ E_\epsilon \right]_{k+\lhalf,i+\lhalf,j+\lhalf} = 
E_{\rm o}
\sqrt {\frac{t_{\rm o}}{t_{\rm o}+\delta t}}
\exp\left\{ - \frac{1}{4D\:(t_{\rm o}+\delta t)}
 \left( \left[ r \right]_{i+\lhalf} - 
r_{\rm o} \right)^2 \right\} 
\nonumber \\
\text{\hspace{3.0in} for all $k,\ell,i,j$},
\end{eqnarray}
where we have set $t_{\rm o} = 1.0$~ns for this test.  We run
the simulation to where $t_{\rm o}+\delta t = 4.3$~ns, a total of 100
timesteps.  The timestep size was chosen to be 
$\Delta t = 100 ~\Delta r/c$.
Figure~\ref{fig:rad_gauss_x1} shows a comparison between the analytic
solution and that achieved by the numerical test.  Except near the
center, where the numerical solution slightly lags the analytic
solution, the agreement is excellent.  With the setup described, and
using the familiar Euclidean norm, the test exhibits a small relative
residual at the final timestep,
\begin{equation}
^\ell{\cal R}_2 \equiv
\left(
\frac{
\sum_i \sum_j 
\left( ^\ell\left[ E_\epsilon 
\right]^{\rm analytic}_{k+\lhalf,i+\lhalf,j+\lhalf}
- ^\ell\left[ E_\epsilon 
\right]^{\rm numerical}_{k+\lhalf,i+\lhalf,j+\lhalf}
\right)^2}
{\sum_i \sum_j 
\left( ^\ell\left[ E_\epsilon 
\right]^{\rm analytic}_{k+\lhalf,i+\lhalf,j+\lhalf}
\right)^2}
\right)^{1/2}
\simeq 0.019,
\end{equation}
for each radiation species individually.  As a check, we confirm that
the results for each species are bitwise identical.

We also check how well the code maintains spherical symmetry.  Because
the metric tensor, when using spherical coordinates, contains
trigonometric functions of the angular coordinates, it cannot be
expected that spherical symmetry will be identically preserved.  In
our 16-angular-zone model, we find that the maximum relative deviation
in the radiation energy densities between corresponding points along
radial rays is 4~$\times$~$10^{-15}$.  Thus, spherical symmetry is well
maintained, and the departure from it is of the degree expected for a
system whose machine epsilon is of order $10^{-16}$.


\subsubsection{Diffusion of a Gaussian Pulse in Two Dimensions}
\label{sec-gauss_2d}

This problem extends the 1-D Gaussian pulse
problem to two dimensions and tests the
performance of the diffusion algorithm in multiple
dimensions.  The computational domain is similar
to the spherically symmetric problem: The same
single radiation-energy group extending from 0--1
MeV is used. The spatial domain is a curvilinear
square similar to that used in the previous
problem, but this time measuring 2~cm on a side.
The center of this domain is located at a radius
of $10^4$~cm from the origin in a spherical-polar
coordinate system.  The radial domain is divided
into 101 zones; the angular domain of $2
\times 10^{-4}$~rad is also divided into 101 zones.  
The use of an odd number of zones ensures that the pulse can be 
placed at the exact center of the domain.
Since the
computational domain is small relative to its distance from the
origin, it resembles a Cartesian grid, which is the way we view it in
Figure~\ref{fig:rad_gauss_2d}, which shows both initial conditions and
results.  Nevertheless, the numerical problem is actually solved in
spherical polar coordinates and the solution 
is compared against the analytic
solution in that same coordinate system.  
\begin{figure}[htbp]
\begin{center}
\includegraphics[width=20pc]{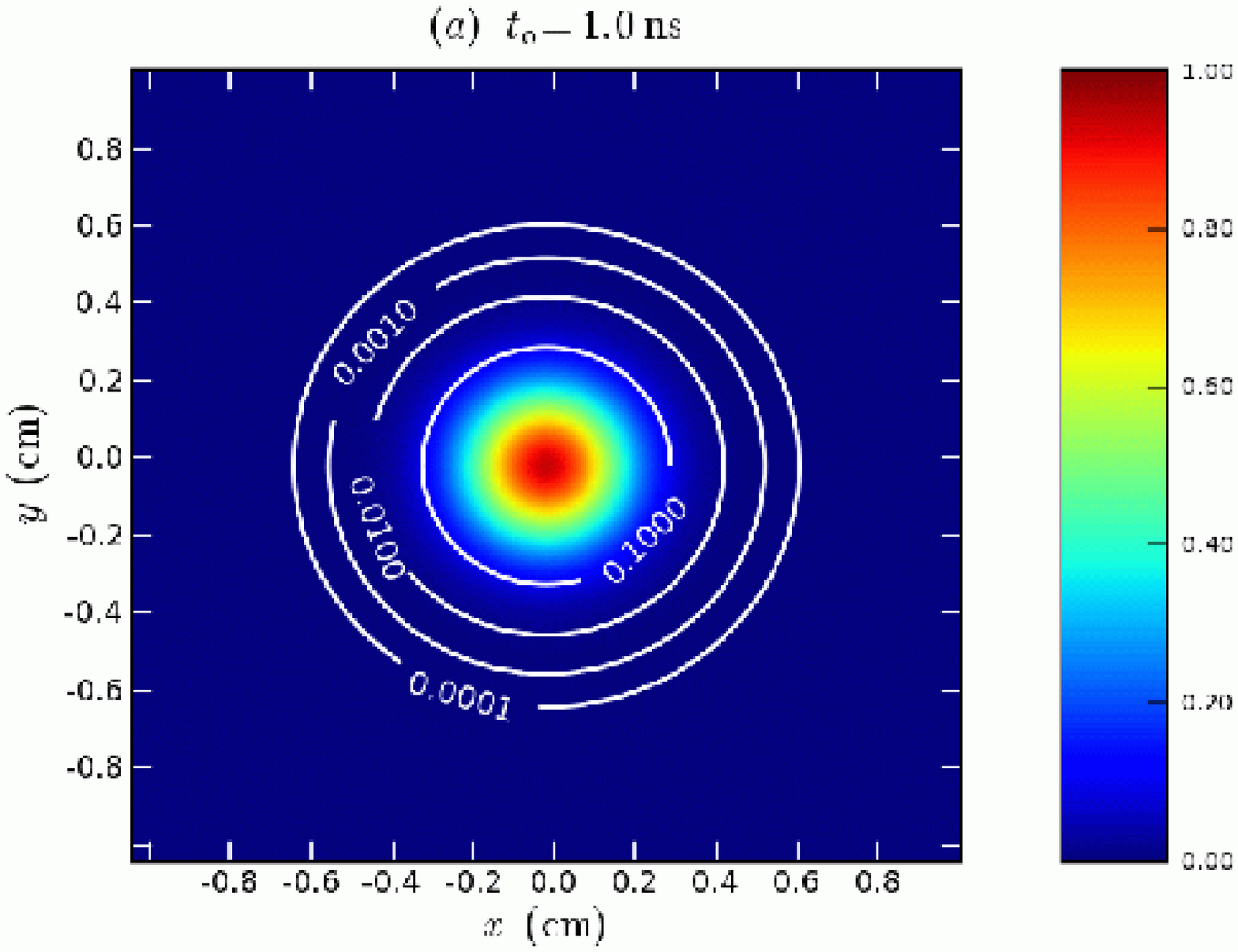}
\includegraphics[width=20pc]{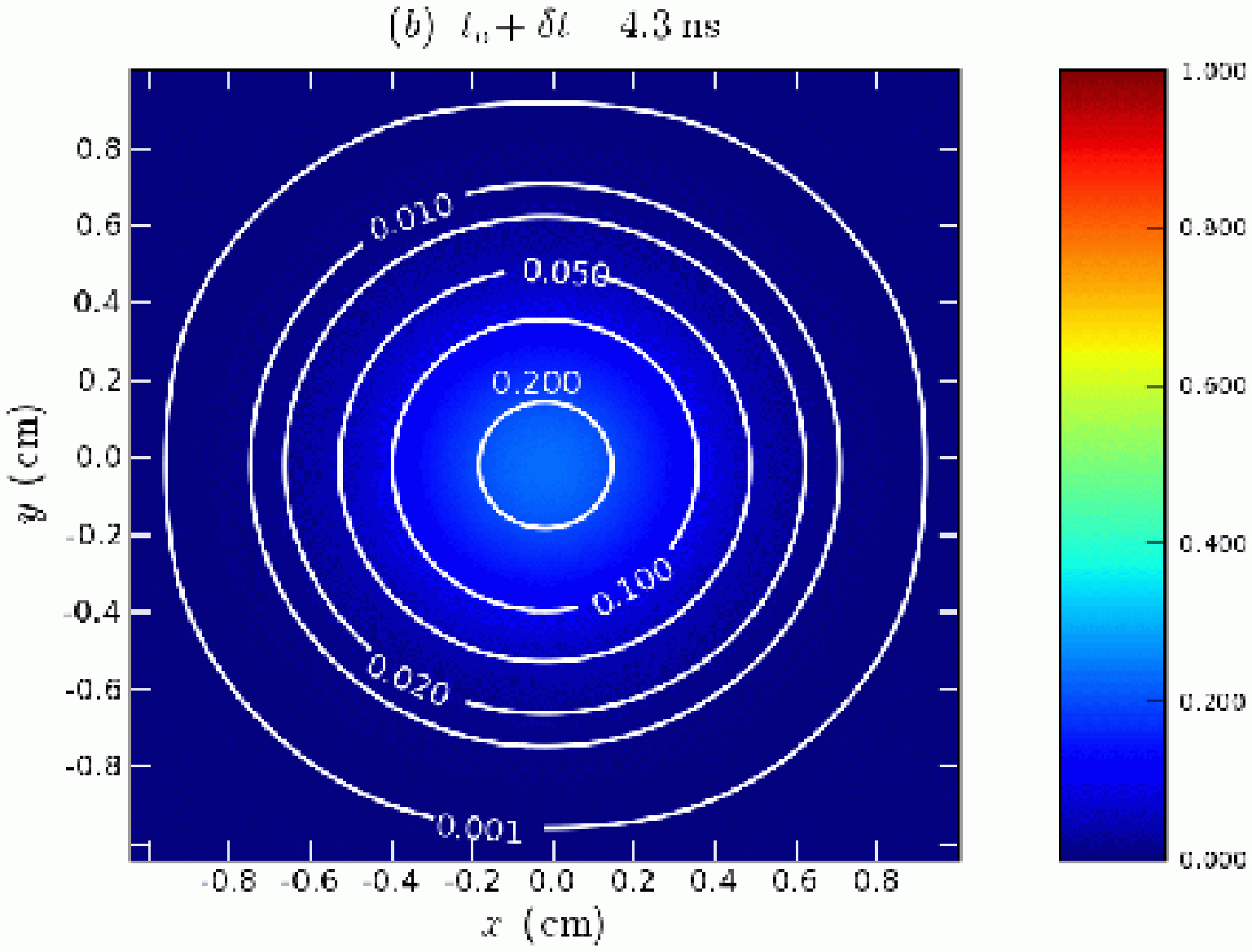}
\end{center}
\caption{\label{fig:rad_gauss_2d} 
Initial conditions and solution for the diffusive propagation of a
two-dimensional Gaussian pulse.  For display purposes, we show
everything in an approximate Cartesian coordinate system, though the
actual calculations are carried out in spherical polar coordinates. The
``Cartesian'' coordinate $x$ is defined such that $x = r - r_{\rm o}$,
where $r_{\rm o} = 10^4$~cm. Similarly, the ``Cartesian'' coordinate
$y$ is defined as $r_{\rm o} \theta$. The initial Gaussian pulse is
shown in ({\em a}), where both the colormap and the contours display
the energy density over the spatial domain. In ({\em b}), the solution
is shown after $\simeq 3.3$~ns of evolution, using the same colormap
as ({\em a}).  Except for areas close to the boundary, where edge
effects are present, the pulse precisely maintains its axisymmetric
form about $(x,y)=(0,0)$ throughout the evolution. }
\end{figure}

The initial Gaussian pulse is given by
\begin{equation}
E_\epsilon = E_{\rm o} \exp \left\{
\frac{-\left| {\bf r} - {\bf r_{\rm o}} \right|^2}
{4D\:t_{\rm o}} 
\right\}
\end{equation}
which gives an azimuthally symmetric distribution about the point
${\bf r_{\rm o}}$, which is the displacement of the center of the
pulse from the origin.  For our test problem, we have selected $r_{\rm
o} = 10^4$~cm and the direction of the displacement to be located at
the angular center of our chosen domain.  Once again, we have chosen a
uniform, constant diffusion coefficient $D = 10^{7}$~cm$^2$~s$^{-1}$,
$E_{\rm o} = 1$~erg~cm$^{-3}$~MeV$^{-1}$, and $t_{\rm o} = 1$~ns.

This pulse is initialized numerically at time $t_{\rm o}$ to the
values given by
\begin{eqnarray} \label{eq:g2d_init}
\lefteqn{
^\ell\left[ E_\epsilon \right]_{k+\lhalf,i+\lhalf,j+\lhalf}  = } 
\nonumber \\ & &
E_{\rm o}
\exp\left\{ - \frac{1}{4D\:t_{\rm o}}
\left( \left[ r \right]_{i+\lhalf} 
\cos(\Psi - \left[\theta\right]_{j_{\rm s}+\lhalf}-
\left[\theta\right]_{j+\lhalf}) - 
r_{\rm o}
\cos(\Psi - \left[\theta\right]_{j_{\rm s}+\lhalf}-\theta_{\rm o})
\right)^2 \right\}
\nonumber \\ & &
\;
\times
\exp\left\{ - \frac{1}{4D\:t_{\rm o}}
\left( \left[ r \right]_{i+\lhalf} 
\sin(\Psi - \left[\theta\right]_{j_{\rm s}+\lhalf}-\left[\theta\right]_{j+\lhalf}) - 
r_{\rm o}
\sin(\Psi - \left[\theta\right]_{j_{\rm s}+\lhalf}-\theta_{\rm o})
\right)^2 \right\}
\nonumber \\ & &
\text{\hspace{3.0in} for all $k,\ell,i,j$},
\end{eqnarray}
where $\Psi$ is {\em half} of the angle subtended by the computational
domain at the origin, $\left[\theta\right]_{j_{\rm s}+\lhalf}$ is
the angular coordinate at the first half-integer mesh point, and
$\left[\theta\right]_{j+\lhalf}$ is the angular coordinate of a
general cell center where the distribution is being initialized.  The
quantity $\theta_{\rm o}$ is the angular coordinate of the center of
the initial pulse.
Flat (zero-flux) boundary conditions are applied at the edges of the 
grid.
We allow the initial pulse to evolve for 3.3~ns in time. 
As in the 1-D problem, the timestep was chosen to be
$\Delta t = 100~\Delta r/c$.
By direct substitution, it is
straightforward to show \citep{rhb98} that at time $t_{\rm o} + t$,
the analytic solution for this problem is
\begin{equation}
E_\epsilon = E_{\rm o} \left( \frac{t_{\rm o}}{t_{\rm o}+t} \right)
\exp \left\{- \frac{1}{4D\:(t_{\rm o}+t)} 
\left| {\bf r} - {\bf r_{\rm o}} \right|^2 \right\}.
\end{equation}
We note that for this two-dimensional problem, the fraction 
containing $t_{\rm o}$ and $t$, which multiplies the Gaussian function,
appears as a linear power.  This contrasts with the 1-D
case, where this factor appears under a square root
(eq.~[\ref{eq:gauss_1d}]).

The analytic solution corresponds to a predicted numerical solution of
\begin{eqnarray}
\lefteqn{
^\ell\left[ E_\epsilon \right]_{k+\lhalf,i+\lhalf,j+\lhalf}  = 
E_{\rm o} \left( \frac{t_{\rm o}}{t_{\rm o}+t} \right) } 
\nonumber \\ & &
\times 
\exp\left\{ - \frac{1}{4D\:(t_{\rm o}+t)}
\left( \left[ r \right]_{i+\lhalf} 
\cos(\Psi - \left[\theta\right]_{j_{\rm s}+\lhalf}-\left[\theta\right]_{j+\lhalf}) - 
r_{\rm o}
\cos(\Psi - \left[\theta\right]_{j_{\rm s}+\lhalf}-\theta_{\rm o})
\right)^2 \right\}
\nonumber \\ & &
\times
\exp\left\{ - \frac{1}{4D\:(t_{\rm o}+t)}
\left( \left[ r \right]_{i+\lhalf} 
\sin(\Psi - \left[\theta\right]_{j_{\rm s}+\lhalf}-\left[\theta\right]_{j+\lhalf}) - 
r_{\rm o}
\sin(\Psi - \left[\theta\right]_{j_{\rm s}+\lhalf}-\theta_{\rm o})
\right)^2 \right\}
\nonumber \\ & &
\text{\hspace{3.0in} for all $k,\ell,i,j$}.
\end{eqnarray}

Figure~\ref{fig:rad_gauss_2d} shows the results of solving the test
problem numerically with V2D.  In Figure~\ref{fig:rad_gauss_2d}({\em
a}), we show the initial conditions as given by
equation~(\ref{eq:g2d_init}). Both the colormap and the plotted
contours show the initial radiation energy density.

Figure~\ref{fig:rad_gauss_2d}({\em b}) shows
results after $\simeq 3.3$~ns of evolution using
the same scaled colormap for $E_\epsilon$ as in
({\em a}).  The initial pulse has spread.  Except
for areas close to the boundary, where edge
effects are present, the pulse precisely maintains
its axisymmetric form about $(x,y)=(0,0)$
throughout the evolution.  The boundary
discrepancies result from our imposition of flat
boundary conditions (see Appendix~\ref{app:bconds}) and our use (by
necessity!) of a 
finite domain.  This results in a numerical
solution that is too small at the boundary.  The
behavior of the solution near the boundaries could
likely be improved through the imposition of
time-dependent boundary conditions.  Nevertheless,
at 0.1~cm inside a boundary, the deviation of the
numerical from the analytic solution is already
reduced to $<$~1.4\%, which is typical of the
deviation over the rest of the domain.  The
character of the deviation of the numerical from
the analytic solution is similar to the 1-D
problem of \S\ref{sec-gauss_1d}.  The numerical
pulse diffuses too slowly, so that it is slightly
too large near the center and slightly too small
further out.

Running with the setup described and using the results displayed in
Figure~\ref{fig:rad_gauss_2d}, we calculate the residual as a
quantitative measure of the solution quality.  With the Euclidean
norm, the test exhibits a relative residual at the final timestep of
\begin{equation}
^\ell{\cal R}_2 = 
\left(
\frac{
\sum_i \sum_j 
\left( ^\ell\left[ E_\epsilon 
\right]^{\rm analytic}_{k+\lhalf,i+\lhalf,j+\lhalf}
- ^\ell\left[ E_\epsilon 
\right]^{\rm numerical}_{k+\lhalf,i+\lhalf,j+\lhalf}
\right)^2}
{\sum_i \sum_j 
\left( ^\ell\left[ E_\epsilon 
\right]^{\rm analytic}_{k+\lhalf,i+\lhalf,j+\lhalf}
\right)^2}
\right)^{1/2}
\simeq 0.008.
\end{equation}


\subsubsection{Flux Divergence Test}	\label{sec-flux-div}

The flux divergence test was posed by \citet{ts01} as a means of testing
the divergence term of the diffusion equation.
Our problem setup is modeled after that test described in \S5.2
of \citet{ts01}.  
An analytic solution is known for the case of the diffusion equation
on a unit square in Cartesian coordinates with periodic boundary
conditions and a constant diffusion coefficient.
For this problem, the diffusion equation can be solved via separation
of variables.  \cite{lamb95} outlines an analytic solution:
\begin{equation}\label{eq:fluxdivan}
E_\epsilon(x,y,t) = 2 + e^{-8 \pi^2 t} \{\sin (2\pi x)\}\{\sin (2 \pi y)\}.
\end{equation}

For the numerical test, we discretize a unit square
in curvilinear coordinates into 100 uniform zones
in each spatial dimension in a fashion similar to
the previous 2-D Gaussian verification problem.
The inner edge of the grid is located at $r_0 = 10^4$ cm.
For the radiation, we use a single energy group,
bounded by 0 and 1~MeV, with the arithmetic
average of the group boundaries, 0.5~MeV, assigned
as the group-center value.  Since this is a pure
transport test and the radiation microphysical
cross-sections are set such that opacities are
independent of material properties ($D=1)$, the
values of the material variables are irrelevant
(other than the specification that the medium is
hydrostatic).
\begin{figure}[htbp]
\vspace{0.5in}
\begin{center}
\includegraphics[scale=0.75]{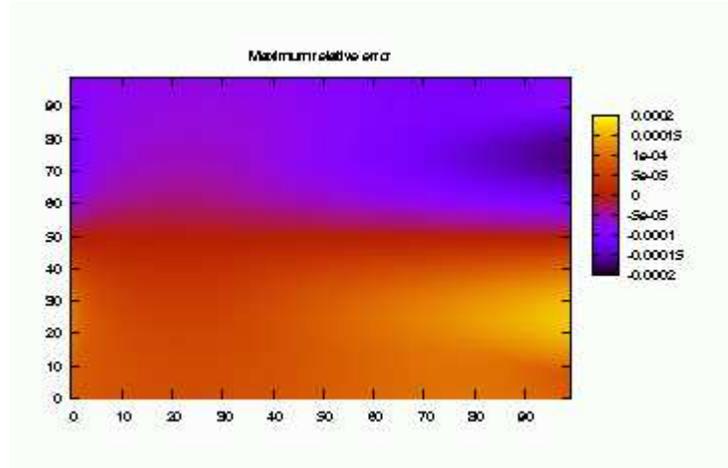}
\end{center}
\caption{\label{fig:rad_flux_diverg} The results of the radiative
flux-divergence test showing the maximum relative error in each of 
the spatial zones.  In our test calculation, on the final timestep, the
maximum relative error that any point reaches over the entire domain
is about $-1.6 \times 10^{-8}$. The maximum relative error that any
point reaches over the entire domain at any time in the course of the
simulation is about $1.7 \times 10^{-4}$.  }
\vspace{0.5in}
\end{figure}
The system is initialized to the differenced form of the analytic
solution at a small time, $t_{\rm o} = 10^{-14}$~s,
\begin{equation}\label{eq:fluxdivnu}
\left[ E_\epsilon \right]^{\rm o}_{i+\lhalf, j+\lhalf} = 
2 + e^{-8 \pi^2 t_{\rm o}} 
\sin \left\{2\pi (\left[ x \right]_{i+\lhalf}-
\left[ x \right]_{i_{\rm s}+\lhalf})\right\}
\sin \left\{2\pi (\left[ y \right]_{j+\lhalf}-
\left[ y \right]_{j_{\rm s}+\lhalf})\right\},
\end{equation}
and is left to evolve for 1 second (approximately 100 timesteps).  
Timesteps are chosen such that
$\Delta t = 100 \Delta r/c$, where $\Delta r$ is the spacing of
the radial zones.  

The comparison between the numerical and analytical results is very
good.  Figure~\ref{fig:rad_flux_diverg} shows the maximum relative
error of the numerical solution across the spatial domain.  This is
the maximum relative amount by which each spatial point deviates, at
any time, during the course of the calculation.  Therefore, at a
typical timestep, the solution is actually better than that plotted.
In our benchmark calculation, on the final timestep, the maximum
relative error that any point reaches over the entire domain is about
$-1.6 \times 10^{-8}$. The maximum relative error that any point
reaches over the entire domain at any time in the course of the
simulation is about $1.7 \times 10^{-4}$.

At the final timestep, using the Euclidean norm, we find our results
exhibit a very small relative residual,
\begin{equation} \label{eq:l2norm_flux_div}
^\ell{\cal R}_2 = 
\left(
\frac{
\sum_i \sum_j 
\left( ^\ell\left[ E_\epsilon 
\right]^{\rm analytic}_{k+\lhalf,i+\lhalf,j+\lhalf}
- ^\ell\left[E_\epsilon 
\right]^{\rm numerical}_{k+\lhalf,i+\lhalf,j+\lhalf}
\right)^2}
{\sum_i \sum_j 
\left( ^\ell\left[E_\epsilon 
\right]^{\rm analytic}_{k+\lhalf,i+\lhalf,j+\lhalf}
\right)^2}
\right)^{1/2}
\simeq 9.1 \times 10^{-9}.
\end{equation}
This number is small at all timesteps throughout the calculation.  It
sees its maximum value ($\simeq 7 \times 10^{-5}$) right at the
beginning of the calculation. It then declines almost
monotonically to the final value shown in
equation~(\ref{eq:l2norm_flux_div}). 



\subsubsection{Light-Front Propagation}		\label{sec-light-front}

The light-front propagation problem tests the ability of the 
algorithm to propagate a front in the optically thin (free-streaming)
limit where the behavior of the Boltzmann equation is hyperbolic.
This presents a challenging test for a flux-limited diffusion algorithm.
This test was used for the flux-limited diffusion algorithm of 
\citet{ts01} (see their \S5.5).

The problem consists of modeling the propagation of a light front,
following it from its start, at the inner edge of the domain,
until it reaches the halfway point of the domain.  The time for this
to occur is just the straightforward expectation, based on a front
traveling at speed $c$.

The numerical version of this problem is set up in spherical polar
coordinates, at a large radius $r_{\rm o}$, which we choose as
$10^4$~cm. We divide a 1-cm spatial domain in the radial direction
into 100 zones, equally spaced.  A single radiative energy group is
also assigned.  It is arbitrarily bounded by 0 and 1~MeV and given the
arithmetic average value, 0.5~MeV, at the group center.  We set the
total radiative opacity, $\kappa_T = 10^{-5}
\text{ cm$^{-1}$}$, which is constant in space and time.  
Like the other transport verification problems in
this subsection, there is no emission or
absorption.  The initial value for the radiation
energy density within the computational domain is
set to a suitably small value ($\lesssim
10^{-5}$~erg~cm$^{-3}$~MeV$^{-1}$) to avoid
encountering possible problems with zero or
negative energy densities.

Energy is supplied to the domain by a Dirichlet boundary 
condition at the inner edge of the domain.  This is
accomplished by fixing all left-hand-boundary zones at all times to
1~erg~cm$^{-3}$~MeV$^{-1}$.  This positions the radiation front at $x=0$
when $t=0$.  Since the problem is of limited duration, the boundary
conditions chosen for $x=1$~cm, the right-hand boundary, are largely
irrelevant. We run this problem with flat boundary conditions (see
Appendix~\ref{app:rad-trans}). 

We also run this test as a nominal two-dimensional
problem, with 2 zones in the $\theta$ direction
that span $10^{-4}$~rad.  Periodic boundary
conditions are applied to quantities in this
second dimension.

These initial conditions are slightly different that those used by
\citet{ts01}, who fix $\kappa_T$ to achieve an optical depth of 0.01
for the domain.  This places our version of the problem in an
optically thinner medium, since our smaller value of $\kappa_T$ gives
our domain an optical depth of $10^{-5}$.  In addition, Turner \&
Stone position the radiation front initially at $x = 0.1$~cm.

Our results are shown in Figure~\ref{fig:rad_front_x1}.
This plot corresponds to
a time of about $1.7 \times 10^{-11}$~s.  
The curve shown in
Figure~\ref{fig:rad_front_x1}({\em a}) shows results from our standard
version of the test.  The timestep is selected as 0.5 times the minimum
transport CFL time, $\Delta r/c$, in the problem.  
Since the zones are equally
spaced, this corresponds to a timestep of $1.7 \times 10^{-13}$~s.
The problem runs 100 timesteps to reach the outcome shown.  Clearly,
the numerical solution is somewhat diffusive.

\begin{figure}[htbp]
\includegraphics[width=20pc]{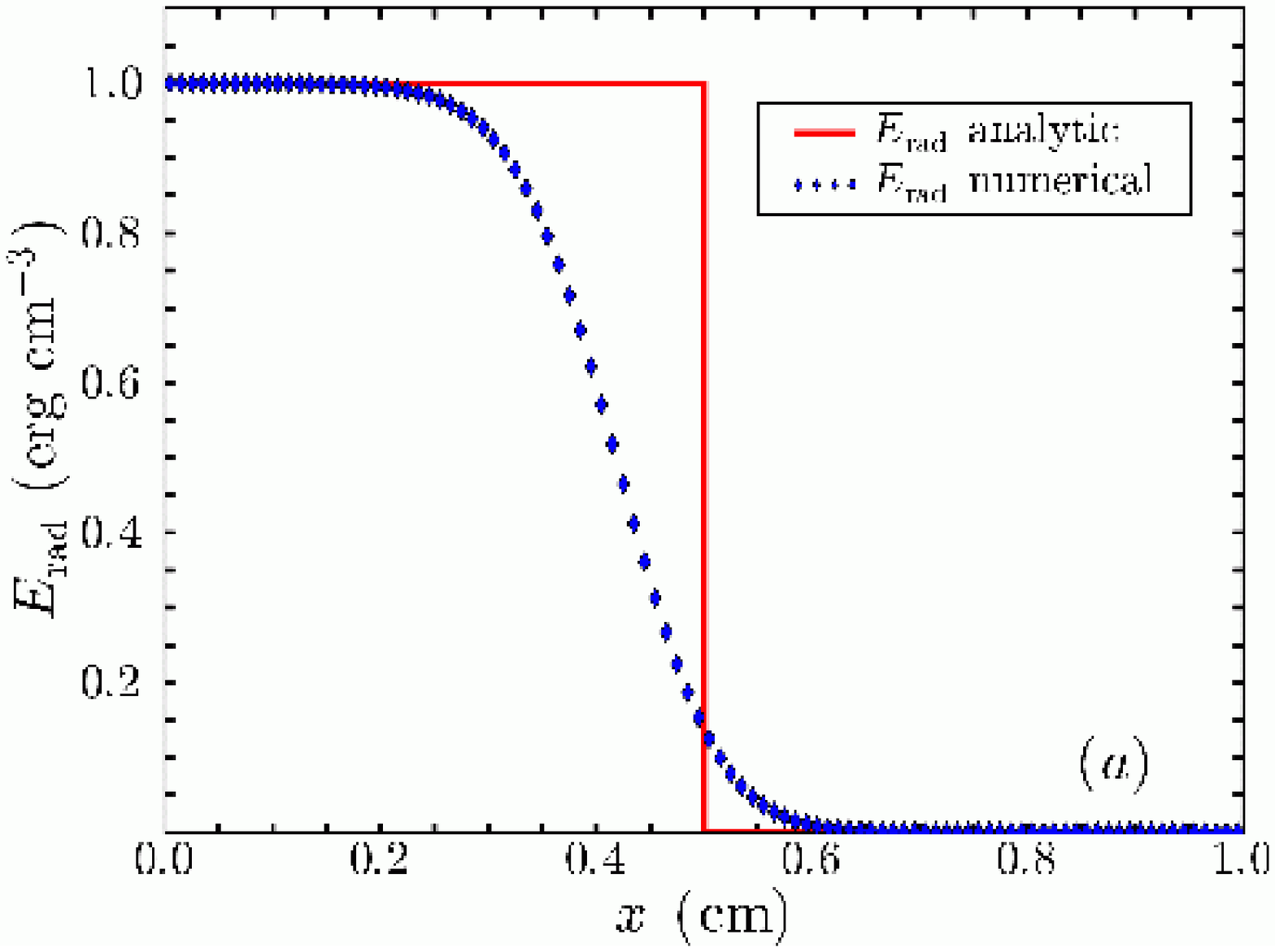}
\includegraphics[width=20pc]{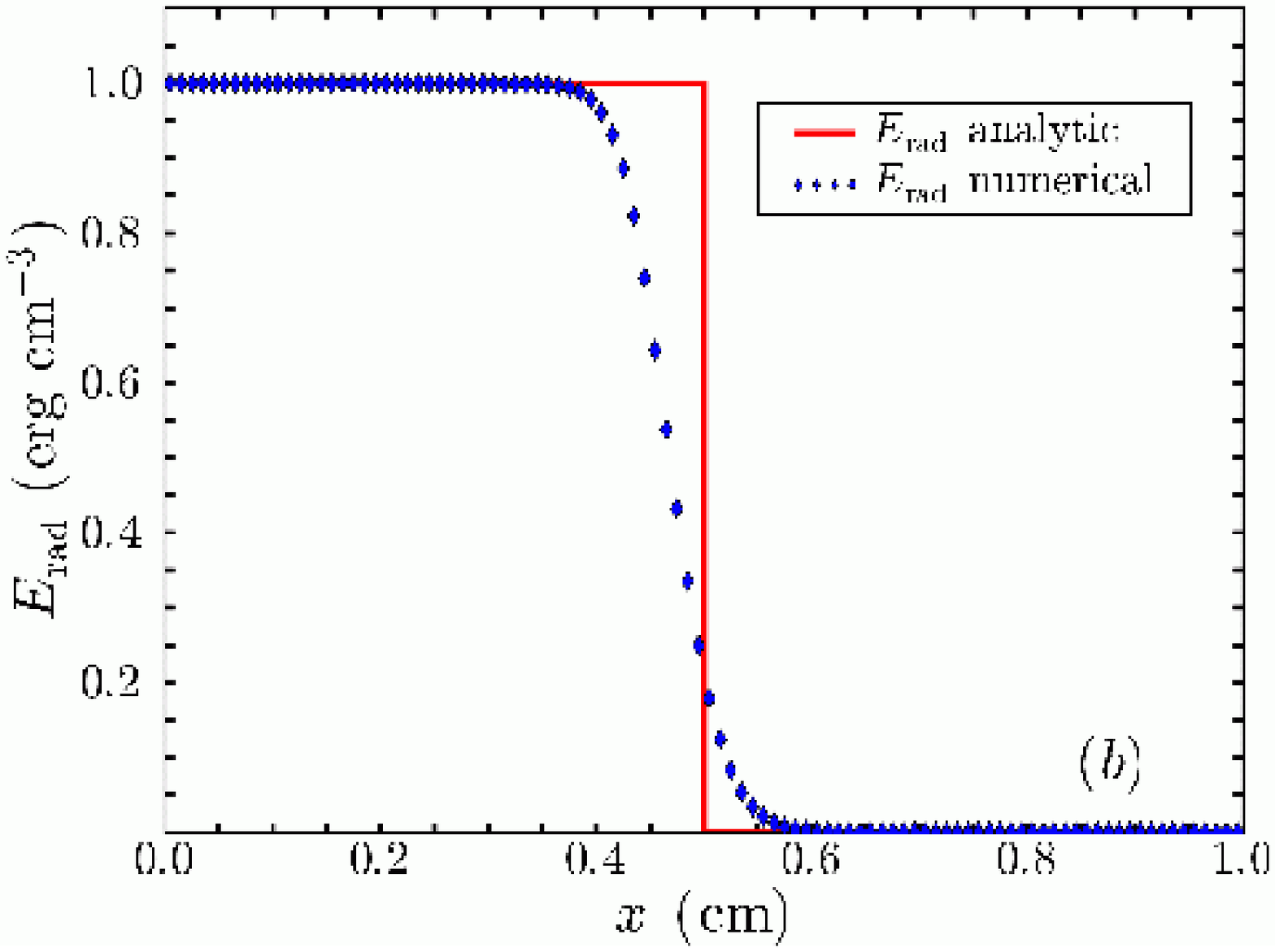}
\includegraphics[width=20pc]{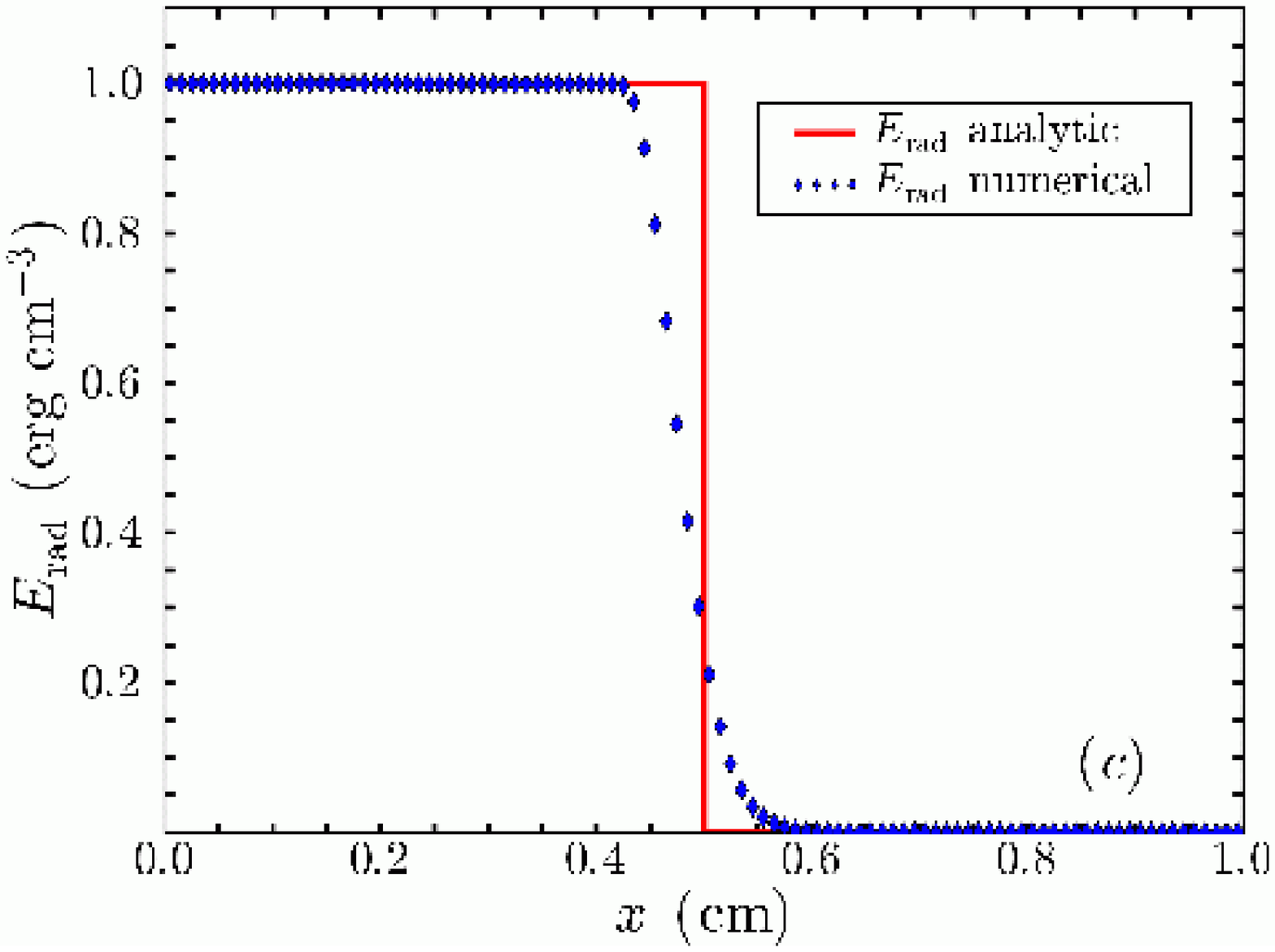}
\includegraphics[width=20pc]{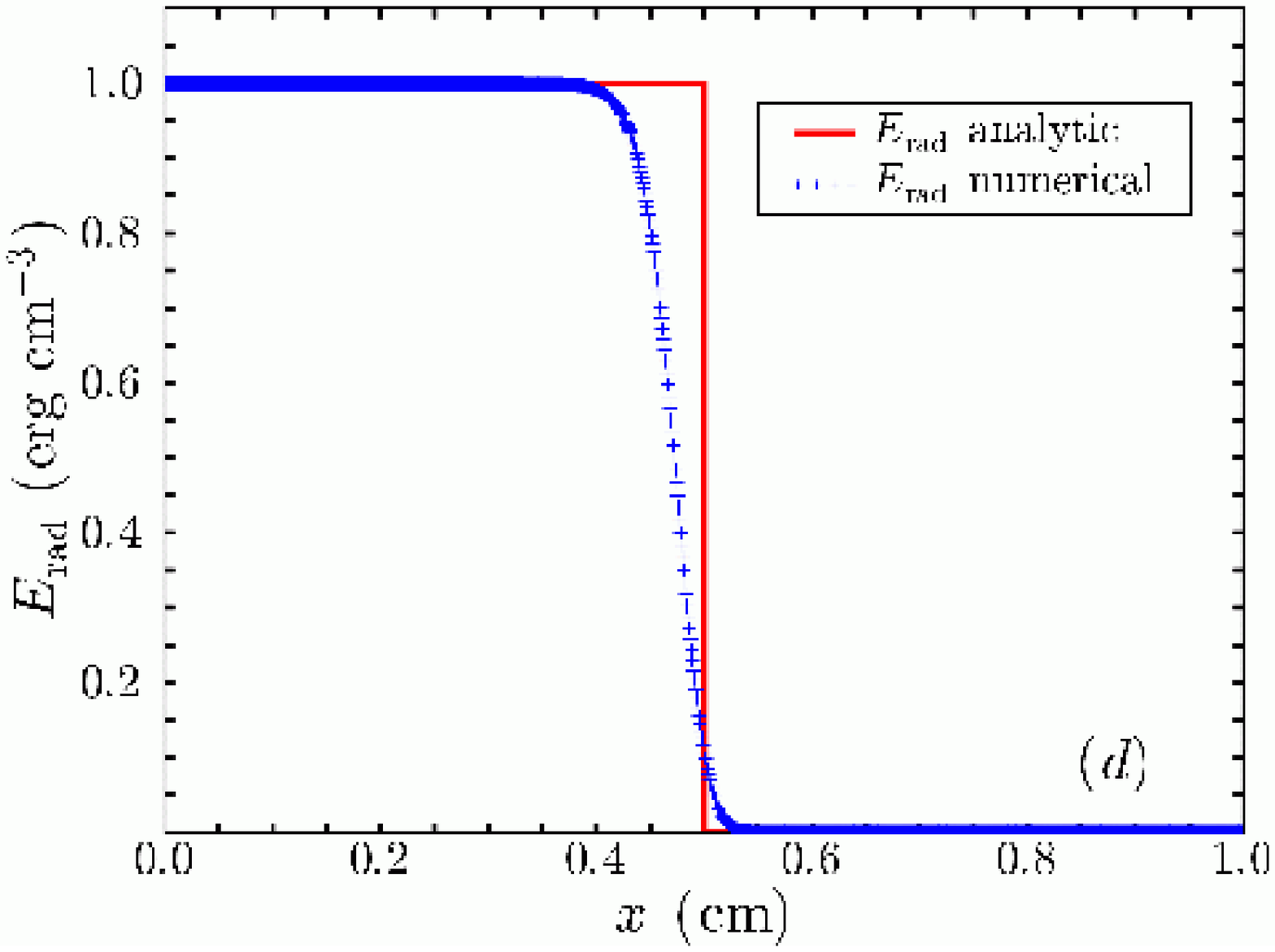}
\caption{\label{fig:rad_front_x1} Results of the light-front test
in spherical polar coordinates.
The coordinate value $x=0$
is located at radius, $r_{\rm o}=10^4$~cm.  
The plots show the calculated solutions using
various fractions of the minimum transport CFL time: ({\em a}) 0.5,
({\em b}) 0.05, ({\em c}) 0.005, and in the case of ({\em d}), 0.5,
but calculated using 1000 radial zones.}  
\end{figure}

Improvement can be obtained when timesteps are reduced.  However, even
then, the advantage realized is limited.  This is illustrated in
Figures~\ref{fig:rad_front_x1}({\em b}) and ({\em c}), where
calculations were made using CFL fractions of 0.05 and 0.005,
respectively. Although both the radiation fronts are much better
resolved than for Figure~\ref{fig:rad_front_x1}({\em a}), 10 and 100
times the amount of computational effort was expended in producing
results ({\em b}) and ({\em c}) relative to ({\em a}).  Furthermore,
the resolution improvement in ({\em c}) relative to ({\em b}) is
probably not significant enough to warrant 10 times the computational
effort.

In Figure~\ref{fig:rad_front_x1}({\em d}), we set the CFL fraction
back to 0.5, but divide the domain into 1000 zones in the radial
direction. The combination of more zones, which are thinner and thus
decrease the timestep size, results in a calculation that requires
roughly the same computational work as performed for ({\em c}).  The
resolution of the radiation front is also comparable, though the
increased spatial resolution noticeably sharpens the front and
reduces the presence of its numerical precursor.

An analogous set of tests are performed for light fronts 
propagating in the $x_2$ direction.  These tests 
achieves similar results to those of the $x_1$ direction tests.

All the results we obtain compare favorably with those of
\citet{ts01}, who also use the Levermore-Pomraning flux limiter in a
diffusion scheme to model their radiation.  However, in both cases,
the diffusivity evident in the solutions demonstrates some ultimate
limitations in using flux-limited diffusion in a regime that is
maximally far away from the diffusion limit.

\subsection{Radiation Hydrodynamics Tests
\label{app:radhyd}}

We have conducted a set of verification tests 
designed to elucidate the performance of our algorithm on
radiation-hydrodynamic phenomena.  These tests focus on the 
exchange of energy between matter and radiation under a
variety of conditions.  The first of these verification
problems tests the exchange of energy in cases where the 
matter is static.  The second tests the exchange of energy
under conditions where the dynamics of the matter drives
the exchange.

\subsubsection{Relaxation to Thermal Equilibrium Tests}

This test, suggested by \cite{ts01}, is designed
to test the ability of a code to achieve thermal
equilibrium between matter and radiation by means
of emission and absorption.  This problem is a
gray problem where the emission is blackbody in
nature.  Our multigroup algorithm can accommodate
this problem by simply setting the number of
groups to one and setting the group width to
$\Delta\epsilon =1$ MeV.  The gray opacities and
emissivity can then be used for this single group
to model a gray problem.  We consider two cases:
one where the initial matter temperature is above
the equilibrium value and the other where the
initial matter temperature is below the
equilibrium value.  In both cases, the initial
radiation energy density is set to $E =
10^{12}$~erg~cm$^{-3}$.  The equation of state is
described by a $\gamma=5/3$ ideal gas with a mean
molecular weight of $\mu=0.6$.  The density of the
material is taken to be $\rho = 10^{-7}$~g~cm$^{-3}$
and the opacity is taken to be 
$\kappa^a = 4\times 10^{-8}$~cm$^{-1}$.  In the
first case, the initial value of the matter energy
density is chosen as $e=10^{2}$~erg~cm$^{-3}$; in
the second case the initial value is 
$e=10^{9}$~erg~cm$^{-3}$.  The timestep was chosen to
be $\Delta t = 10^{-11}$~s with the initial
time set to $t_0 = 10^{-16}$~s.  The
numerical results of these tests are displayed in
Figure~\ref{fig:rhd_heating}.
\begin{figure}[htbp]
\includegraphics[width=20pc]{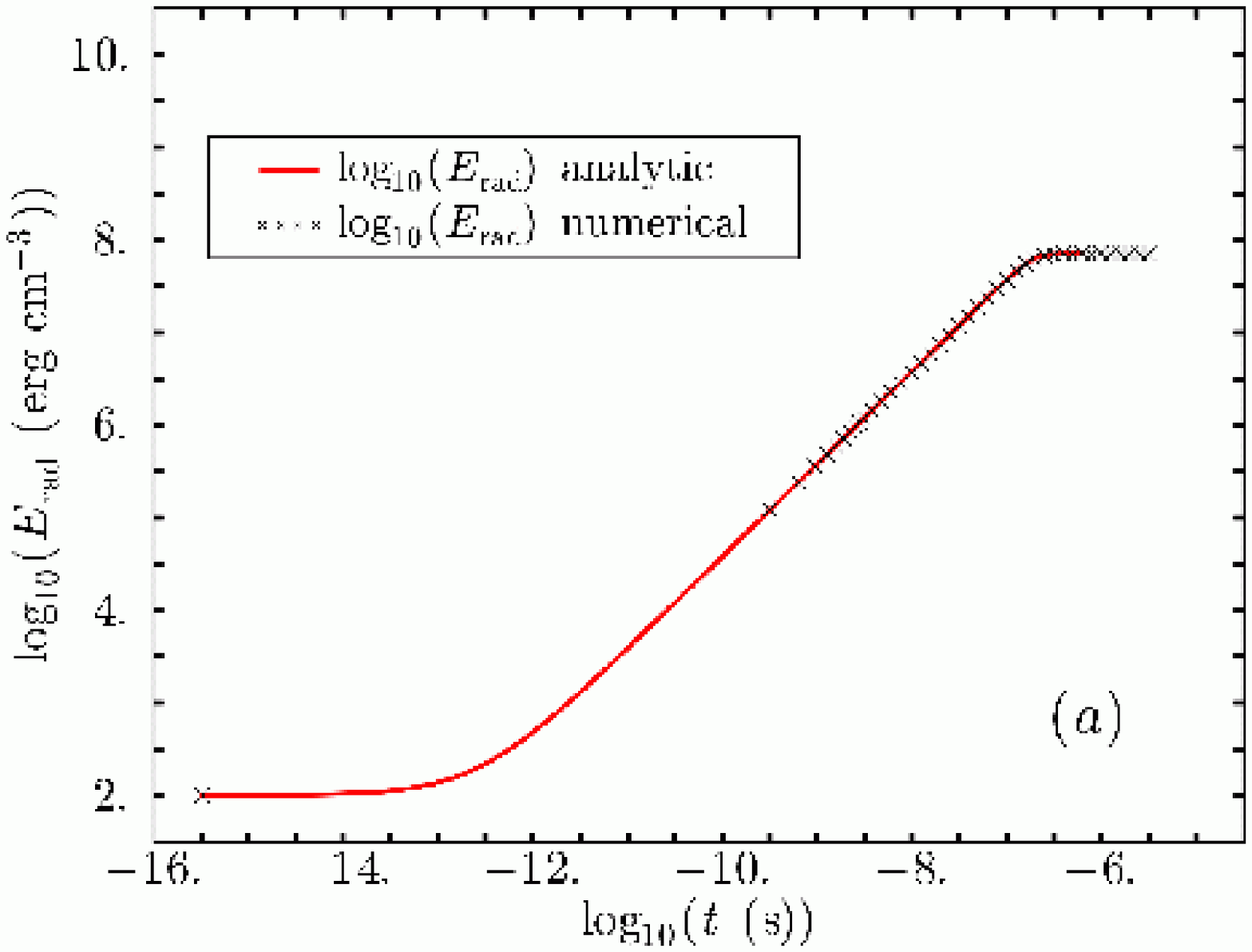}
\includegraphics[width=20pc]{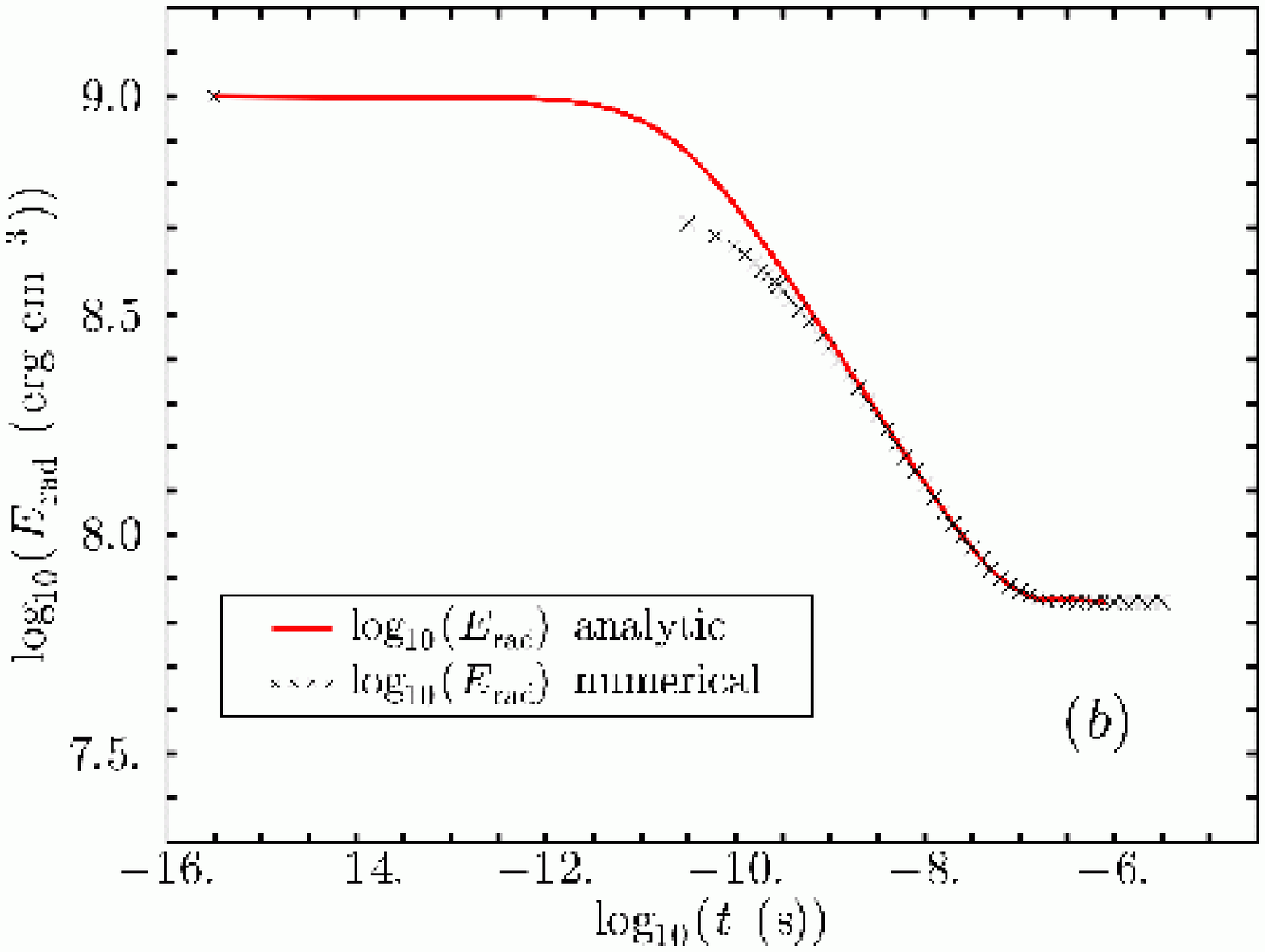}
\caption{\label{fig:rhd_heating}  Two trajectories towards
equilibrium: $(a)$ via radiative heating and $(b)$ via radiative cooling.
In both plots, the analytic solution is plotted in red and
representative samplings from the numerical solutions are indicated by
crosses.}
\end{figure}

The analytic solution to this problem can be easily obtained by integrating
the ordinary differential equation defined by the emission-absorption terms:
\begin{equation}
\frac{de}{dt} = c\kappa^a E - 4\pi B(T),
\label{eq:eatest}
\end{equation}
where 
\begin{equation}
B(T) \equiv \frac{a_R c}{4\pi}T^4,
\label{eq:bb}
\end{equation}
\begin{equation}
T = \frac{\mu m_b (\gamma-1) e}{\rho},
\label{eq:eosid}
\end{equation}
and $a_R$ is the radiation constant and $m_b$ is
the baryon mass.  If we make the assumption that
the radiation energy density is approximately
constant, then by substituting equation~(\ref{eq:eosid}) into
equation~(\ref{eq:bb}), we  
can rewrite (\ref{eq:eatest}) as
\begin{equation}
\frac{de}{dt} = \beta^4 - \eta^4 e^{4},
\label{eq:eatest2}
\end{equation}
where $\eta$ and $\beta$ are constants.
Clearly, the initial value of the matter energy density $e_o$, at time
$t_0$, reaches a final equilibrium point $e_f = \beta/\eta$ at some
time $t_f$.  Thus, we can integrate equation~(\ref{eq:eatest2}) to find
\begin{equation}
\int^{t}_{t_0} dt^\prime  = 
\int^{e}_{e_0} \frac{de^\prime}{\beta^4 - \eta^4 e^{\prime 4}}.
\label{eq:eatest3}
\end{equation}
The indefinite form of the integral on the right-hand-side of
equation~(\ref{eq:eatest3}) can be expressed in two forms 
\citep{selby72,grad94}
\begin{equation}
f(x) \equiv 
\int \frac{dx}{\beta^4 - \eta^4 x^4} =
\left\{
\begin{array}{ll}
\displaystyle{
\frac{1}{2\eta\beta^3}
\left\{
\frac{1}{2} \ln\left(\frac{\beta/\eta-x}{\beta/\eta+x}\right) +
 \tan^{-1}\left(\frac{\eta x}{\beta}\right) 
\right\}} & \mbox{if $ x < \beta/\eta$} 
\vspace{0.2in} \\
\displaystyle{
\frac{1}{2\eta\beta^3}
\left\{
\frac{1}{2} \ln\left(\frac{x+\beta/\eta}{x-\beta/\eta}\right) +
 \tan^{-1}\left(\frac{\eta x}{\beta}\right) 
\right\}}  & \mbox{if $ x > \beta/\eta$},
\end{array}
\right.
\label{eq:eatest4}
\end{equation}
which depend on the relationship of $x$ to $\beta/\eta$. 
Thus, we can rewrite equation~(\ref{eq:eatest3}) to yield
an analytic relationship between $e$ and $t$:
\begin{equation}
t(e) = t_0+f(e)-f(e_0).
\end{equation}
In this form, it is easy to compute the time $t(e)$ at which a value of the
energy density $e$ obtains.

In Figure~\ref{fig:rhd_heating}, we compare the numerical and analytic
solutions that we obtain for both the heating and cooling cases.  In
both cases, the solution achieves the equilibrium value, as expected.
In the radiative cooling case, we note that the numerical results
lead the analytic solution slightly for the first few steps.
This is expected, since the cooling timescale is initially somewhat
shorter than the numerical timestep---resulting in an slightly inaccurate
solution.  As time evolution proceeds and the matter radiatively
cools towards equilibrium, the cooling timescale exceeds the time step, and the numerical
solution becomes nearly identical to the analytic solution for
subsequent times.  The cooling-case result makes an interesting
contrast to the results reported in Figure~3 of
\cite{ts01}, who indicate that their numerical solution for cooling
{\em lags} the analytic result at early times.  
The difference between are solution and theirs arises
because our algorithm does not solve the Lagrangean portion of the gas
energy equation simultaneously with the radiation-diffusion equation.

\subsubsection{Radiation-damping of Acoustic Hydrodynamic Waves}

This verification test, first carried out by
\citet{ts01}, is based on the
linearized analysis of the radiation-hydrodynamic
equations carried out by
\citet{mih83} (hereafter MM83).  This analysis predicts the
radiation-damping of acoustically driven
hydrodynamic disturbances.  The dynamics of the
acoustically driven wave depend on the ratio of the
optical depth of the problem to the damping
length.

Our test utilizes the same parameters as those
employed by Turner and Stone.  The gas is an
ideal gas with adiabatic index of $\gamma=5/3$.
The unperturbed matter density is specified as
$\rho_0 = 3.216 \times 10^{-9}$~g~cm$^{-3}$.  The
unperturbed matter internal energy density is
given by $E_0 = 26\;020$~erg~cm$^{-3}$ which
corresponds to an unperturbed temperature of $T =
5.59043\times 10^{-6}$~MeV and an adiabatic sound
speed of $c_s = 2\;998\;295$~cm~s$^{-1}$ ($10^{-4} c$). 
The unperturbed
radiation energy density is specified as
$E_{\epsilon 0} = 17\;340$~erg~cm$^{-3}$~MeV$^{-1}$.  We
utilize a single energy group centered on an
energy of $1$ MeV with a width of $1$ MeV.  This
choice of grouping makes the MGFLD equations
behave as though they were gray flux-limited
diffusion equations.  This choice of problem
parameters yields a Boltzmann number for the
system of
\begin{equation}
{\mathbb B} \equiv \frac{4\gamma c_s E_0}{c E_{\epsilon 0}} = 10^{-3},
\end{equation}
where $c_s$ is the adiabatic sound speed of the gas given by
\begin{equation}
c_s = \sqrt{\frac{\gamma P}{\rho}}.   
\end{equation}
In the linearized analysis 
of MM83, the system is also characterized by the 
quantity $r$ given by
\begin{equation}
r \equiv \frac{c_s}{c} {\mathbb B}
\end{equation}
which, for the parameters described above, yields
the value of $r = 0.1$.  Finally, the matter is
assumed to have an absorption opacity given by
$\kappa^a = 0.4 \rho$ cm$^{-1}$.

MM83 posit plane-wave solutions to the linearized 
radiation-hydrodynamic equations which allows us to write the 
density, matter internal 
energy density, and velocity in the form
\begin{equation}
\rho_1 = \rho_0 \alpha_\rho e^{i(\omega t-kx)},
\label{eq:waved}
\end{equation}
\begin{equation}
E_1 = E_0 \alpha_E e^{i(\omega t-kx)},
\label{eq:wavee}
\end{equation}
\begin{equation}
\varv_1 = \varv_0  e^{i(\omega t-kx)},
\label{eq:wavev}
\end{equation}
where $\alpha_\rho$, $\alpha_E$, and $\varv_0$ are constants
specifying the amplitude of the acoustic perturbation.

Based on these perturbations MM83 derive 
a dispersion relationship for the time-dependent 
linear radiation-hydrodynamics equations described by the
equation
\begin{equation}
A z^4 + Bz^2 + C = 0
\label{eq:disp1}
\end{equation}
where $A$, $B$, and $C$ are complex coefficients given by
\begin{equation}
A \equiv 1 - \frac{16\tau_{c_s}}{\mathbb B} i,
\end{equation}
\begin{equation}
B \equiv 3\tau_{c_s}\left(1+\frac{i}{\tau_{c}} \right)^2
-1 +\left(\frac{16\gamma \tau_{c_s}}{\mathbb B} \right)i+
\frac{16c_s \tau_{c_s}^2}{c \mathbb B}
\left(1+\frac{i}{\tau_{c}}\right)
\left(
5+\frac{i}{3\left(\gamma-1\right)\tau_{c}}+
\frac{16c_s\gamma}{3 c {\mathbb B} \left(\gamma-1\right)}
\right),
\end{equation}
\begin{equation}
C \equiv -3\tau_{c_s}^2
\left\{
\left(1+\frac{i}{\tau_{c}} \right)^2 +
\left(\frac{16\gamma c_s}{c {\mathbb B}} \right)
\left(1+\frac{i}{\tau_{c}}\right)
\right\},
\end{equation}
and where we define
\begin{equation}
\tau_{c_s} \equiv \frac{c_s \kappa^a}{\omega}
\end{equation}
and
\begin{equation}
\tau_{c} \equiv \frac{c \kappa^a}{\omega}.
\end{equation}
The quantity $\tau_{c_s}$ is related to the optical depth
per acoustic perturbation wavelength.
MM83 suggest that equation~(\ref{eq:disp1}) must be solved numerically,
but an simple analytic solution of this particular form of quartic
equation exists.  By the substitution $u = z^2$ equation 
(\ref{eq:disp1}) is reduced to a complex quadratic equation, for which 
the standard quadratic formula applies.  The roots are then given
by 
\begin{equation}
z \equiv z_r + iz_i = \pm \sqrt{u}.
\end{equation} 
The wave number $k$ can then be written as
\begin{equation}
k \equiv \frac{\omega}{c_s} z = \frac{\omega}{c_s} z_r -
\frac{\omega}{c_s} z_i i,
\end{equation} 
and, thus, the perturbation takes the form
\begin{equation}
e^{i(\omega t-kx)} = e^{-x/L} 
e^{i\left(\omega t - 2\pi x/\Lambda\right)},
\label{eq:pert2}
\end{equation}
where the damping length $L$ and the perturbation wavelength $\Lambda$
are defined by
\begin{equation}
L = \frac{c_s}{\omega z_i}
\end{equation} 
and
\begin{equation}
\Lambda = \frac{2\pi c_s}{\omega z_r}.
\end{equation} 
The complex relationship between the damping length and the 
perturbation wavelength is illustrated in Figure~\ref{fig:disp1}
for the aforementioned values of ${\mathbb B}$ and $r$. 
\begin{figure}[htbp]
\vspace{0.25in}
\begin{center}
\includegraphics[scale=0.50]{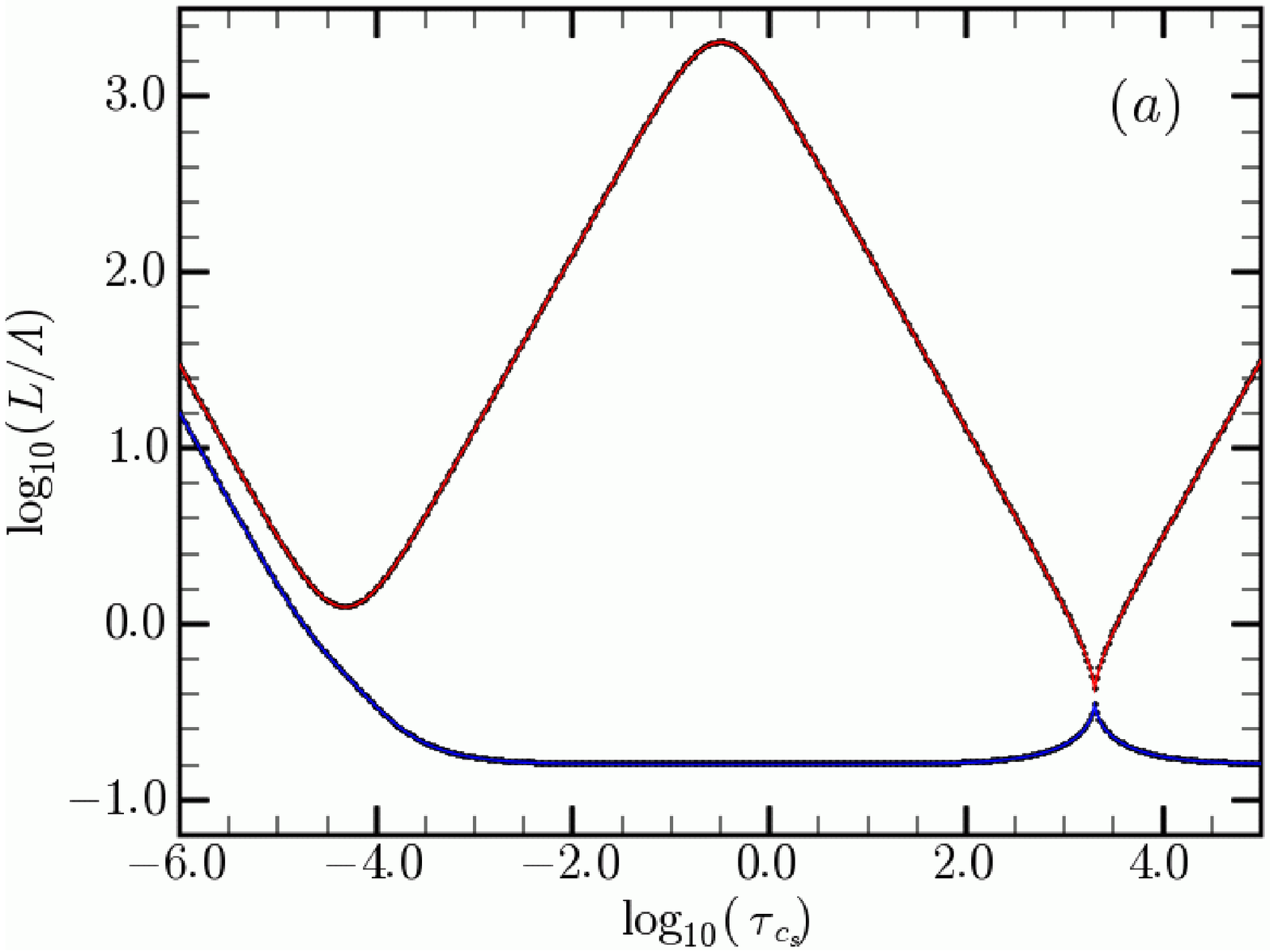}
\includegraphics[scale=0.50]{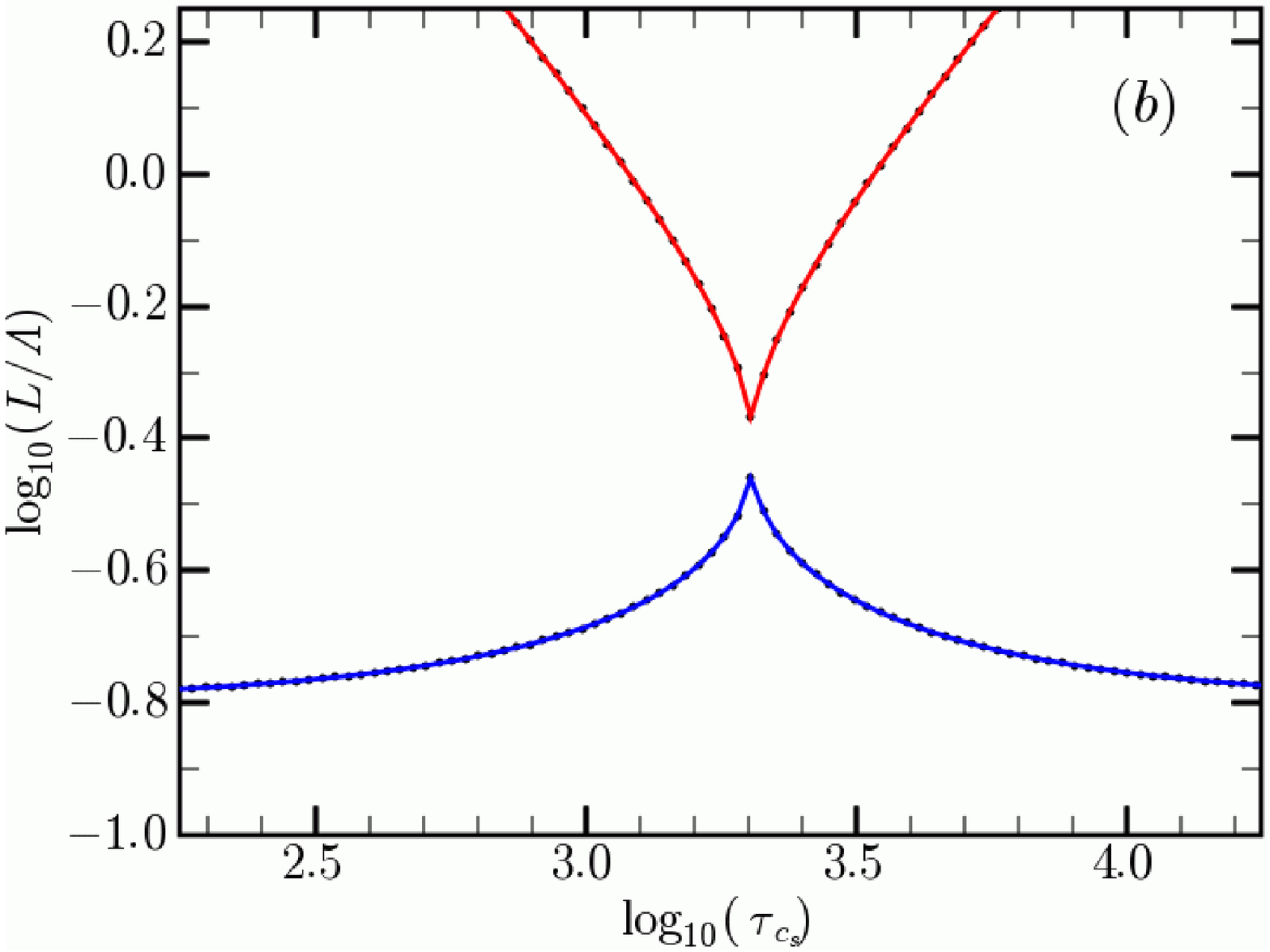}
\end{center}
\caption{\label{fig:disp1}  The damping length to perturbation
length ratio versus optical depth per perturbation wavelength.  In
({\em a}), the upper (red) curve shows the root corresponding to
matter acoustic waves, while the lower (blue) solution corresponds to
radiation dominated waves.  These two curves do not cross near
$\log_{10} \tau_{c_s} \approx 3.4$. However, they actually approach
much more closely than the artist-rendered curves in Figure~4 of 
MM83 leads one to
believe.  An expanded view of the close approach of these curves is shown
in ({\em b}).}
\vspace{0.25in}
\end{figure}
\begin{deluxetable}{llllll}
\tabletypesize{\normalsize}
\tablecaption{{\sc Parameters for Acoustic Wave Tests}
\label{tab:acoust}}
\tablewidth{0pt}
\tablehead{
\colhead{$\tau_\Lambda$} & \colhead{$\omega$} &\colhead{ $L$ } &
\colhead{$\Lambda$} & \colhead{ $L/\Lambda$} & 
\colhead{Behavior\tablenotemark{a}}
\\
\colhead{} & \colhead{(rad s$^{-1}$)} &\colhead{(cm)} &
\colhead{(cm)} & \colhead{} & \colhead{}
}
\startdata
$10^{3}$   &   $2.423 \times 10^{-5}$   &   $4.912 \times 10^{12}$   &
$7.774\times 10^{11}$        &    $\quad \;\; \: 6.316$  &  damped   
\\
$1$        &    $2.423 \times 10^{-2}$   &  $9.775 \times 10^{11}$   &   
$7.774 \times 10^8$  &  $1257 $ & undamped 
\\
$10^{-3}$  &   $2.423 \times 10^1$   &   $1.374 \times 10^6$   &   
$7.774\times 10^5$  &  $\quad \;\;\: 1.766$ & damped 
\\
\enddata
\tablenotetext{a}{The behavior of the material acoustic wave is listed
for each set of parameters.} 
\end{deluxetable}

This figure reveals that acoustic waves with a values of 
$\tau_{c_s} \sim 10^{-3}$ or $\tau_{c_s} \sim 10^{3}$
should be only slightly undamped while those with
$\tau_{c_s} \sim 1 $ should be heavily attenuated.
We point out that, despite appearances, 
the two curves do not intersect near 
$\tau_{c_s} = 10^{3.4}$.  Our curves, however, show a much
closer approach at these points than do the artist-rendered
curves of MM83.

To formulate test problems based on this analytic solution
of the radiation-hydrodynamic equations,
it is more convenient to work with the
variable
\begin{equation}
\tau_\Lambda \equiv 2\pi \tau_{c_s} = \kappa^a \Lambda,
\end{equation}
which is the optical depth per wavelength.

  We choose three test cases corresponding to values of $\tau_\Lambda$
of $10^{-3}$, $1$, and $10^3$.  The corresponding values of 
$\omega$ and $L$ are also given in Table~\ref{tab:acoust}.  
By combining equation~(\ref{eq:pert2}) with 
equations~(\ref{eq:waved})--(\ref{eq:wavev}), 
and retaining only the real portion, we obtain the solutions
\begin{equation}
\rho = \rho_0 \left\{1 + \alpha_\rho e^{-x/L} 
\cos \left( \omega t-\frac{2\pi x}{\Lambda} \right)
\right\},
\label{eq:waved3}
\end{equation}
\begin{equation}
E = E_0 \left\{ 1 + \alpha_E e^{-x/L} 
\cos \left( \omega t-\frac{2\pi x}{\Lambda} \right)
\right\},
\label{eq:wavee3}
\end{equation}
and
\begin{equation}
\varv = \varv_0 e^{-x/L} 
\cos \left( \omega t-\frac{2\pi x}{\Lambda} \right).
\label{eq:wavev3}
\end{equation}

For our test problems we choose the amplitude for the perturbations 
to be one percent in all cases giving $\alpha_\rho = \alpha_E = 
\alpha_{ER} = 0.01$ and $\varv_0 = 0.1 c_s$.
The problem is spatially discretized by choosing
the zoning such that the domain is divided
into 150 zones, with the zone width of 10 zones per
perturbation wavelength ($\Delta r =
\Lambda/10$). The left-end of the domain is
located at a radius large enough to
put the spherical coordinate system into the
plane-parallel limit.  We have retained the same
equation of state and energy group structure as
used in the radiative heating/cooling problem.
The timestep is chosen to be approximately a factor
of 10 greater than the heating/cooling timescale
in each problem.

The analytic solution describes a wave moving from left to right
across the domain.  The density, velocity, matter internal energy density,
and radiation energy density are initialized to their
analytic values at $t=0$.  The perturbations are 
driven by reseting these quantities to their analytic 
values  in the leftmost ten zones, after the 
completion of every timestep.

Our numerical results for late-times are displayed in
Figures~\ref{fig:rhd_thick}--\ref{fig:rhd_thin}.  The behavior in all
cases agrees well with the analytic solution.  In the optically thick
and thin cases, the numerical solution is damped just as predicted
while the optically translucent case produces a solution that is
almost unattenuated---again, as predicted.  Since the timesteps are much
longer than the radiative diffusion timescales, the radiation energy
density is essentially flat in all cases.  If we choose timesteps
shorter than the heating/cooling timescales, we can also reproduce
oscillatory behavior in the radiation-energy density.
\begin{figure}[htbp]
\begin{center}
\includegraphics[width=40pc]{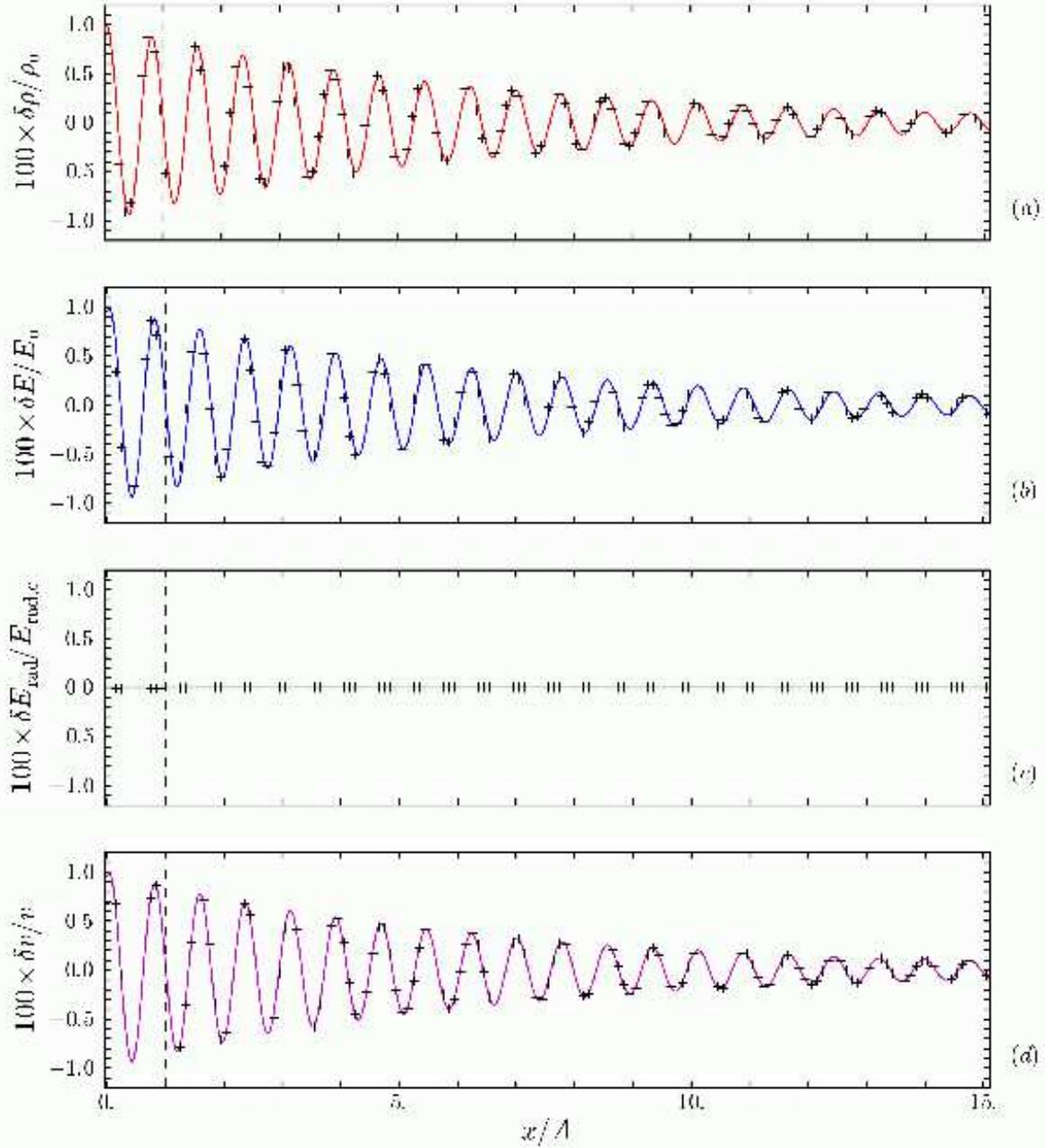}
\end{center}
\caption{\label{fig:rhd_thick}  Acoustic damping for the 
$\tau_\Lambda=10^3$ case. The y-axes represent $100\delta X/X_0$
where $\delta X \equiv X-X_0$ where $X$ is $\rho$, $e$, $v$, or $E$.
The region where the solution is reset after every timestep is to the
left of the vertical dashed line.}
\end{figure}
\begin{figure}[htbp]
\begin{center}
\includegraphics[width=40pc]{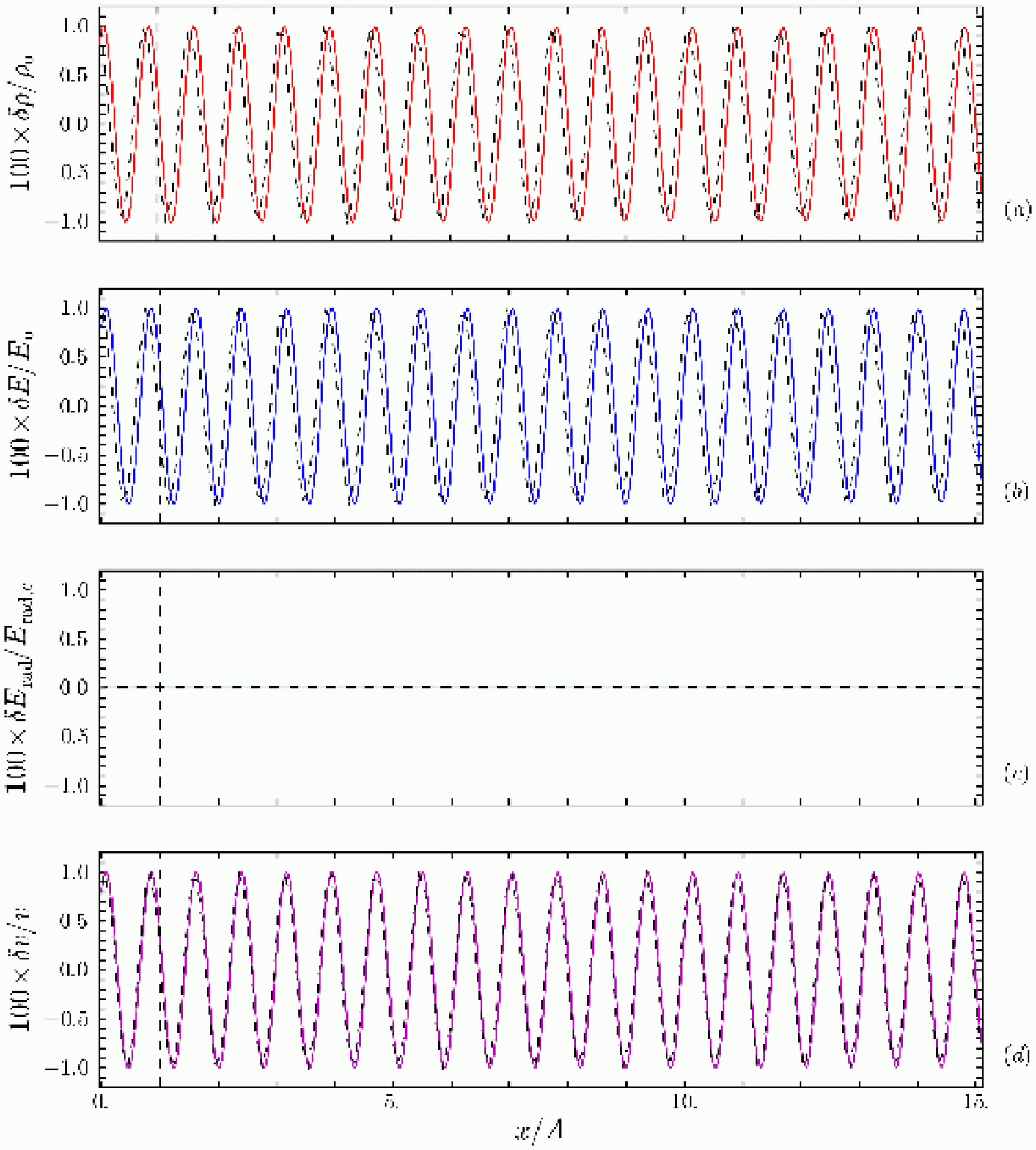}
\end{center}
\caption{\label{fig:rhd_trans}  Same as figure \ref{fig:rhd_thick} but
for the $\tau_\Lambda=1$ case.}
\end{figure}
\begin{figure}[htbp]
\begin{center}
\includegraphics[width=40pc]{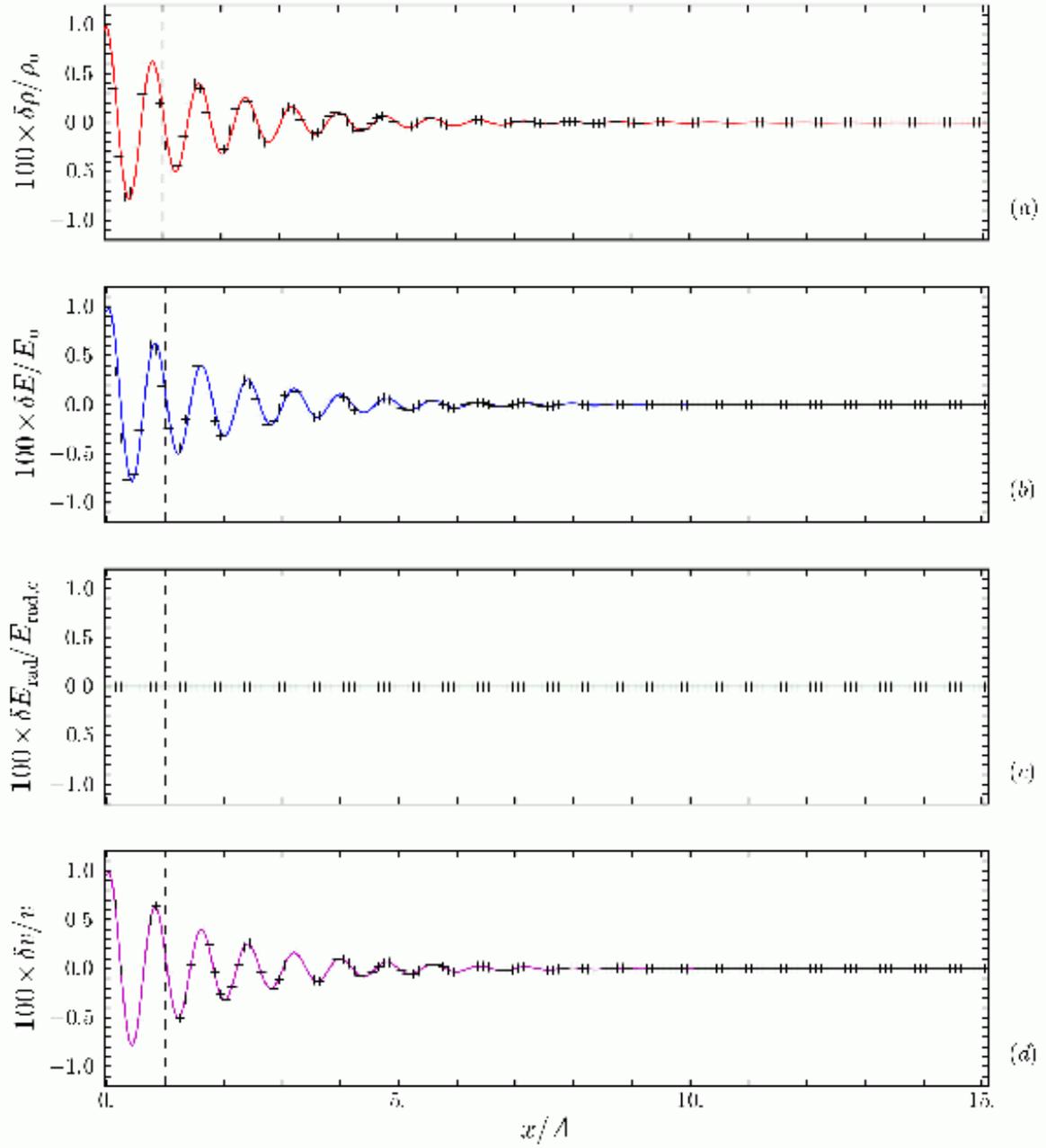}
\end{center}
\caption{\label{fig:rhd_thin}  Same as figure \ref{fig:rhd_thick} but
for the $\tau_\Lambda=10^{-3}$ case.}
\end{figure}
\subsection{Parallel-Integrity Tests}

The parallel-integrity tests are a set of miscellaneous tests that are
intended to confirm that the parallel implementation of V2D is
correct.  As such, they are more in the nature of computer-code
implementation tests instead of physics tests.
Nevertheless, since a portion of this paper is devoted to the 
parallelization of this algorithm, we include a 
brief description of these
tests.  The regular execution of these tests is critical to ensure
that the parallel implementation is bug-free.

We describe three tests in this section.  The
first two check for correct parallelization of I/O in the
code.  The third tests for the correctness of parallelization of the
entire V2D code on a test neutrino-radiation-hydrodynamic 
problem.  There are numerous
other parallel integrity tests we run that are not 
described here for reasons of brevity.  
Although they are described in other sections, we note that the
shock tube and unit-density sphere tests described in
\S\S\ref{sec-sod} and \ref{sec-unit-dens-sph} are performed in such a
way that makes them parallel-integrity tests as
well. These tests are run on different process
topology configurations and the results are
compared to ensure that answers are bitwise
identical in all variables regardless of the
process topology.  In fact, many of our verification tests
are likewise run on a variety of process
topology configurations and bitwise comparisons
are performed among the results.


\subsubsection{I/O Process Topology Comparison Test}

This test serves as a test of V2D's implementation
of parallel I/O, which uses files formatted in
HDF5.  An initial HDF5-formatted file is created
via a serial program.  V2D is rerun multiple times
on diverse numbers of processors.  Each time, an
initial data file is read in via parallel I/O and
written back out in parallel.  Following this, the
output files and input files are then compared to
verify that they are identical.

These V2D parallel I/O runs are carried out with several
different processor topologies: 1$\times$1,
2$\times$2, 4$\times$4, 8$\times$8, and
16$\times$16, which correspond to 1-, 2-, 4-, 16-,
64-, and 256-processor runs.  The processor
topology refers to the manner in which the
contents of each array have been assigned to a
processor.  In each case, the data has been
decomposed according the spatial dimensions of the
arrays. As an example, the 16$\times$16-processor
run sees a 32$\times$32 array divided into 256
portions---each of dimension
2$\times$2, with one portion
resident on each processor.
The six distinct output files produced by the multiple runs are
compared to each other.  We find that these files are bitwise
identical, thus providing confidence in the correctness of our
parallel implementation of I/O in V2D.


\subsubsection{Parallel Memory-File Comparison Test}

This test is a variation on the test in the previous section.  Whereas
the purpose of the ``I/O Process Topology Comparison Test'' is the
integrity of data when different processor topologies
are used to do reading and writing, the focus of this test is to
confirm integrity of data resident in a file relative to data resident
in the distributed memory of an executing parallel program.  The two
tests also stress two different functional processes by which data is
used.  The first test stresses a characteristic that is important in
the post-processing of production data, which occurs across diverse
platforms and processor counts.  This second test stresses a
characteristic that is important to the checkpointing, suspension,
restarting and reading of checkpointed data during the process of
performing a single production run.

A parallel program instantiates three four-dimensional arrays of
dimension 40$\times$2$\times$132$\times$132, which are initialized to
specific floating-point values and written to an HDF5-formatted file.
The file is subsequently read back by the same program, and the data
read from the file is compared to data resident in memory.
This test is performed with the same set of processor topologies
that are used in the previous test.  The decomposition of the data is
also handled in the same way.  In all cases, we find that the
data in memory and data written to files agree bitwise.


\subsubsection{Neutrino Radiation-Hydrodynamic Process Topology Comparison Test}

The final parallel-integrity test we perform is a test with the complete
neutrino radiation-hydrodynamics algorithm.  
This test stresses all parallel aspects of
the code---message passing and global operations in the hydrodynamic
and transport sections of the code, and parallel I/O of a realistic
data set.

The test is simply constructed.  For initial
conditions, we take a checkpoint file from a
supernova model.  The test is conducted by
evolving the model forward in time by a single
timestep on a number of different processor
topologies using a deterministic global reduction
operation for global summations.  As we discussed
in the gravitational unit-sphere-verification-test
section, the standard MPI\_ALLREDUCE global
reduction summation operation is non-deterministic
and can yield slightly different answers with each
run.  Therefore, for testing purposes, we use a
deterministic global summation operation.  We wish
to emphasize that the use of MPI\_ALLREDUCE for
calculating a global minimum CFL time {\em is}
deterministic and there are no run-to-run
variations in this quantity. For the test, a final
output file is produced by each run with varying
process topologies.  The test is deemed successful
if all variables in the checkpoint files are in
exact bitwise agreement.

The deterministic global summation operation that we employ uses a brute 
force approach in which all of the values involved in the summation 
are systematically communicated to process 0 where they are summed
in the same order that they would be summed if the code were purely 
sequential.  Although this approach is slow, it guarantees that the sum
is identical to all bits no matter what process topology is employed.

Although impractically slow for production work,
deterministic global summations are needed for
integrity tests to eliminate the possibility of
message passing errors.  With the deterministic
global sum in place, the output files produced by
each processor topology of this test should be
bitwise identical.  Any deviation indicates a
programming error in the parallel implementation
of the algorithm.  Hence, deterministic
programming is an invaluable development tool in
achieving and maintaining consistency and
correctness of code.


In performing this test, we set up the problem in spherical
coordinates with 256 radial zones, 32 angular zones, 20 neutrino
groups for each flavor of neutrino and corresponding antineutrino.
V2D is then run for a single timestep using a number of processor
topologies:  1$\times$1, 2$\times$2, 4$\times$4, 16$\times$4,
32$\times$8, where the first number 
indicates the number of processors across which the radial portion of
the domain is divided, and the second number is the same for the
angular portion.  A standard V2D checkpoint file is written at the
end of each of these runs and compared.  With the deterministic global
reduction in use, we find that all result files are bitwise
identical.

\subsection{Solver Tests}

The Newton-Krylov solver algorithms used to solve
the implicitly discretized radiation-diffusion
equations are also subjected to regular
verification and unit testing.  We construct
nonlinear and linear test problems that are
mathematically simple and for which
solutions are known.  The solvers are then tested
on these problems and the numerical solution is
compared to the known solution to ensure that they
agree within acceptable tolerances.  These
tests are carried out on a single processor runs
and with multiple processors.  The solutions are
compared with the known solutions in both cases in
order to ensure that no parallel implementation
errors are present.  We also have carried out
numerous tests using both MPI\_ALLREDUCE calls and
our own deterministic global summation to ensure
that the slight process-to-process variability
present in the summation of the MPI\_ALLREDUCE
poses no problem for the solution of the implicit
systems arising from the implicitly discretized
radiation-diffusion equations.  We have never 
found any physically 
significant difference in result between
the two methods for global summations.
\section{Conclusions}

We have developed a numerical algorithm for 2-D
multigroup neutrino-radiation-hydrodynamics that
is especially suited to modeling stellar core
collapse, core collapse supernovae, and
proto-neutron star convection.  However, this
algorithm is sufficiently general as to be useful
for a variety of 2-D radiation-hydrodynamic
applications involving radiation other than
neutrinos. Our algorithm solves the combined set
of comoving-frame neutrino-radiation hydrodynamics
equations, where radiation is treated in a
multigroup flux-limited diffusion approximation.
The combined set of equations is evolved in time
using an operator-splitting approach.  The
nonlinear diffusion equations are evolved
implicitly in pair-coupled fashion while
hydrodynamic portions of the operator-split
equations are evolved explicitly.  The explicit
portion of this approach draws on a numerical
scheme developed by \cite{sn92a} and \cite{smn92}.
The implicit differencing of this scheme is
motivated by the work of \cite{ts01}.  However,
the multigroup treatment of the radiation
distribution, the pair-coupling between particles
and antiparticles, the Pauli blocking effects,
the solution of implicitly differenced diffusion
equations via Newton-Krylov iteration, and the
means of including an arbitrary equation of state
have not been heretofore addressed in such an
algorithm.

The numerical scheme has been described in
sufficient detail to allow a complete
implementation of the algorithm by the reader.  We
have discussed it's parallel implementation
on distributed memory architectures via
spatial domain decomposition.  This includes
message-passing strategies to enhance the
scalability of the algorithm.

In order to benchmark the numerical performance of the algorithm, we
have presented the results of a suite of important verification tests.
These tests stress diverse portions of the algorithm and include
hydrodynamic, radiation-transport, radiation-hydrodynamic,
gravitational, and parallel integrity tests.  Our objective has been
to establish the numerical accuracy of our algorithm both as a whole
and by individual aspects.  This verification process is vital 
to establish the integrity of future simulations.
This process also helps insure that simulations carried out with 
this algorithm can be properly evaluated by other researchers.

\vspace{0.3in}

\acknowledgments

The authors thank the following for valuable advice which has us
in the development of this algorithm and
paper: John Blondin, Steve Bruenn, 
Christian Cardall, Ken DeNisco,
Frank Graziani, Raph Hix, Jim Lattimer, Dimitri Mihalas,
Tony Mezzacappa, Dan Reynolds, Dennis Smolarski,
and Carol Woodward. We gratefully acknowledge
support by the Office of Science of the
U.S. Dept. of Energy, through SciDAC Award
DE-FC02-01ER41185, under which this work was
funded. This research used resources of the
National Energy Research Scientific Computing
Center, which is supported by the Office of
Science of the U.S. Department of Energy under
Contract No. DE-AC03-76SF00098 at Lawrence
Berkeley National Laboratory.  This research also
used resources of the National Center for
Computational Sciences at Oak Ridge National
Laboratory, which is supported by the Office of
Science of the U.S. Department of Energy under
Contract No. DE-AC05-00OR22725.

\vspace{0.3in}
\appendix

\section{Finite Differencing Notation\label{app:diff}}

All discretized quantities are denoted by the use
of square brackets around the symbol for the
quantity.  
The location in time at which quantities are defined
are denoted through the use of standard superscript 
notation and is described in \S \ref{sec:oos}.
The location at which a discretized
quantity is located in space or energy is denoted
through the use of standard subscript notation as
follows.  Any subscripts inside the square bracket
denote a component of a vector, tensor, metric coefficient, or a
spectral (radiation-energy-dependent) quantity. 
energies.Any leading superscripts inside a square bracket are used to 
label the specific flavor of neutrino-related quantities, {\em e.g.,}
electron, muon, or tauon flavors.

A coordinate, or a quantity derived from a coordinate, which we
generalize by $\xi$, is naturally defined either at cell centers
or at cell edges in each dimension.  We 
enumerate cell edges with integer values
and cell centers with half-integer values in each
dimension.  With this convention, cell center $i+\onehalf$
is surrounded by cell edges $i$ and $i+1$.
Figure~\ref{fig:mesh_2} illustrates
the 2-D spatial mesh, including the enumeration of
cell edges and cell centers in each dimension.
The spectral or radiation dimension uses an identical
enumeration (see Figure~\ref{fig:full_grid_2}).

Discretized coordinates are denoted by a single subscript.  Therefore,
our notation for the discrete analog of coordinate $\xi$ at a cell
edge is
\begin{equation}
\xi \text{  at the $i$th cell edge } 
\rightarrow \left[ \xi \right]_{i}.
\end{equation}
The value of $\xi$ at a cell center is given by
\begin{equation}
\xi \text{  at cell center $i+{\scriptsize \onehalf}$ } 
\rightarrow \left[ \xi \right]_{i+\lhalf}.
\end{equation}
In these cases, $\xi$ might be spatial
coordinates, which we generically refer to as $x_1$
and $x_2$; an energy-space coordinate, $\epsilon$;
differences of coordinates, metric coefficients,
$g_2$, $g_{31}$, and $g_{32}$ or their spatial
derivatives; or the area and volume elements,
$\Delta A$ and $\Delta V$, respectively.  We use
the terms ``group'' or ``energy group''
synonymously to refer to cells in the spectral
dimension.  Usually when we use the terms ``cell'' 
or ``zone,'' we will be referring to the spatial mesh.

In general, scalar quantities are
defined at cell centers.   A generic scalar quantity $\psi(x_1,x_2)$,
that is a function of spatial coordinates $x_1$ and $x_2$, has a discretized
analog that is notated as
\begin{equation}
\psi \text{  at the cell center  $(i+{\scriptsize \onehalf}, j+{\scriptsize \onehalf})$  } 
\rightarrow 
\left[ \psi \right]_{i+\lhalf, j+\lhalf}.
\end{equation}
Figure~\ref{fig:mesh_2}
illustrates the locations for such cell-centered discretized variables.  
If the scalar quantity is spectral ({\em i.e.,} a function
of the spectral energy $\epsilon$), such
as the spectral radiation energy density,
we denote it by $\psi_\epsilon$.  The discretized analog of
this quantity, when 
cell centered in the spatial and energy dimensions, is denoted by
\begin{equation}
\psi_\epsilon \text{  at energy group $k+{\scriptsize \onehalf}$  
and cell-center $(i+{\scriptsize \onehalf}, j+{\scriptsize \onehalf})$  } 
\rightarrow 
\left[ \psi_\epsilon \right]_{k+\lhalf, i+\lhalf, j+\lhalf}.
\end{equation}
Our convention is that spatial indices follow any
spectral indices on all discretized quantities.
Figure~\ref{fig:full_grid_2} shows the location at
which such quantities are defined in the 3-D (two
spatial dimensions and one spectral dimension) mesh.

A general vector quantity, ${\smpmb \sigma}$, is
decomposed into its two components, $\sigma_1$ and
$\sigma_2$.  These quantities are usually defined on the
respective cell faces in the direction of the
component. Thus, we adopt the following notation:
\begin{equation}
\sigma_{1} \text{  at cell face $(i, j+{\scriptsize \onehalf})$} 
\rightarrow 
\left[ \sigma_{1} \right]_{i,j+\lhalf},
\end{equation}
\begin{equation}
\sigma_{2} \text{  at cell face $(i+{\scriptsize \onehalf}, j)$} \rightarrow
\left[ \sigma_{2} \right]_{i+\lhalf,j}.
\end{equation}
Note that indices inside the square brackets refer
to components of the vector.
Figure~\ref{fig:mesh_2} depicts the locations at
which the discrete analogs of the vector components
are defined.

If the vector quantity is a function of spectral energy 
({\em e.g.,} the spectral radiation flux
density), which we denote in general by ${\smpmb \sigma}_\epsilon$, and
its components by $\sigma_{\epsilon,1}$ and $\sigma_{\epsilon,2}$ , then it is
face-centered in the spatial dimensions, but group-centered
({\em i.e.,} cell-centered) in the energy dimension. Thus,
\begin{equation}
\sigma_{\epsilon,1} \text{  at group center $k+{\scriptsize \onehalf}$
and cell face $(i, j+{\scriptsize \onehalf})$  } \rightarrow  
\left[ \sigma_{\epsilon,1} \right]_{k+\lhalf,i,j+\lhalf},
\end{equation}
\begin{equation}
\sigma_{\epsilon,2} \text{  at group center $k+{\scriptsize \onehalf}$ 
and cell face $(i+{\scriptsize \onehalf}, j)$ } \rightarrow  
\left[ \sigma_{\epsilon,2} \right]_{k+\lhalf,i+\lhalf,j}.
\end{equation}
Figure~\ref{fig:full_grid_2} illustrates the location at which such quantities
are defined.

Most derivatives are calculated via finite differences, and these
places will be made obvious throughout the text that follows.  For
some quantities, however, we calculate derivatives analytically and
use these analytic expressions in the code.  In the present paper,
these analytic expressions are often used for derivatives of
the metric scale factors (see \S\ref{sec-covariant}). In
differenced expressions, such derivatives are notated as shown in the
following example
\begin{equation}
\frac{\partial g_2}{\partial x_1} \text{  when evaluated at grid point $i$  }
\rightarrow \left[ \frac{\partial g_2}{\partial x_1} \right]_{i}.
\end{equation}

\section{Discretized Variables\label{app:list}}

In this appendix, we present a partial list of the more important
physical quantities and the discretized notation for these
quantities. The conventions for the finite-difference notation we
utilize is discussed in detail in Appendix~\ref{app:diff}.  The
physical quantities and their discretized analogs are listed in
Table~\ref{discrete}.  Figures~\ref{fig:mesh_2} and
\ref{fig:full_grid_2} 
show the locations at which these variables are defined. 
{
%
%
\begin{deluxetable}{lclc}
\tabletypesize{\scriptsize}
\tabletypesize{\small}
\tablecaption{{\sc Sample Discretized Quantities} 
\label{discrete}}
\tablewidth{0pt}
\tablehead{
\colhead{Physical Quantity} & \colhead{Algebraic} & \colhead{Quantity Type} &  
\colhead{Notation} \\
& \colhead{Symbol} & \colhead{} &  
\colhead{}
}
\startdata
spatial coordinate: cell edge        &  $x_1$   & independent variable &  
$\left[ x_1 \right]_i$
\\
spatial coordinate: cell center   &  $x_1$   & independent variable &  
$\left[ x_1 \right]_{i+\lhalf}$
\\
spatial coordinate: cell edge        &  $x_2$   & independent variable &  
$\left[ x_2 \right]_j$
\\
spatial coordinate: cell center   &  $x_2$   & independent variable &  
$\left[ x_2 \right]_{j+\lhalf}$
\\
spectral energy coordinate:   &  $\epsilon$   & independent variable &  
$\left[ \epsilon \right]_{k}$
\\
\quad \quad \quad \quad group edge
\\
spectral energy coordinate:   &  $\epsilon$   & independent variable &  
$\left[ \epsilon \right]_{k+\lhalf}$
\\
\quad \quad \quad \quad group center
\\
$g_2$ metric coefficient:    &  $g_2$   & function of independent &  
$\left[ g_2 \right]_{i}$
\\
\quad \quad \quad \quad cell edge & & \quad \quad variable, $x_1$
\\
$g_2$ metric coefficient:    &  $g_2$   & function of independent &  
$\left[ g_2 \right]_{i+\lhalf}$
\\
\quad \quad \quad \quad cell center & & \quad \quad variable, $x_1$
\\
mass density                          &  $\rho$   & cell-centered, intensive &
$\left[ \rho \right]_{i+\lhalf,j+\lhalf}$
\\
electron fraction                       &  $Y_e$   & cell-centered, intensive &
$\left[ Y_e \right]_{i+\lhalf,j+\lhalf}$
\\
electron number density                       &  $n_e$   & cell-centered, intensive &
$\left[ n_e \right]_{i+\lhalf,j+\lhalf}$
\\
matter internal energy density         &  $E$     & cell-centered, intensive &
$\left[ E \right]_{i+\lhalf,j+\lhalf}$
\\
matter pressure                        &  $P$     & cell-centered, intensive &
$\left[ P \right]_{i+\lhalf,j+\lhalf}$
\\
matter temperature                     &  $T$     & cell-centered, intensive &
$\left[ T \right]_{i+\lhalf,j+\lhalf}$
\\
$x_1$ comp.: matter velocity &  $\varv_1$  & face-centered, vector &
$\left[ \varv_1 \right]_{i,j+\lhalf}$
\\
$x_2$ comp.: matter velocity &  $\varv_2$  & face-centered, vector &
$\left[ \varv_2 \right]_{i+\lhalf,j}$
\\
$x_1$ comp.: matter momentum &  $s_1$  & face-centered, vector &
$\left[ s_1 \right]_{i,j+\lhalf}$
\\
\quad \quad \quad \; \; density  
\\
$x_2$ comp.: matter momentum &  $s_2$  & face-centered, vector &
$\left[ s_2 \right]_{i+\lhalf,j}$
\\
\quad \quad \quad \; \; density  
\\
spectral radiation energy density        &  $E_\epsilon$  & cell-centered, intensive &
$\left[ E_\epsilon \right]_{k+\lhalf,i+\lhalf,j+\lhalf}$
\\
$x_1$ comp.: spectral radiation  &  $F_{\epsilon,1}$  & face-centered, vector &
$\left[ F_{\epsilon,1} \right]_{k+\lhalf,i,j+\lhalf}$
\\
\quad \quad \quad \; \; flux density  
\\
$x_2$ comp.: spectral radiation  &  $F_{\epsilon,2}$  & face-centered, vector &
$\left[ F_{\epsilon,2} \right]_{k+\lhalf,i+\lhalf,j}$
\\
\quad \quad \quad \; \; flux density  
\\
1,2 entry: spectral radiation &  $\{ {\mathsf P_\epsilon \}_{12}}$  & cell-centered, tensor &
$\left[ \{ {\mathsf P_\epsilon} \}_{12} \right]_{k+\lhalf,i+\lhalf,j+\lhalf}$
\\
\quad \quad \quad \; \; pressure
\\
\enddata
\end{deluxetable}
}

\section{Generalized Coordinates, Areas, and Volumes\label{app:cov}}

In this section, we describe how the covariant formulation introduced
in \S\ref{sec-covariant} is implemented in our algorithm. 
In adopting the formulation of \cite{sn92a} for generalized
coordinates, we also adopt a portion of their notation. 
Although there are other possible coordinate systems that
could be used with this method, we focus here on the three most common
systems: (i) Cartesian, (ii) cylindrical, and (iii) spherical.

Paralleling \cite{mih84} and \cite{sn92a}, we write the metric 
in a generalized orthogonal coordinate system as
\begin{equation}
ds^2 = 
(g_1)^2 dx_1^2 +
(g_2)^2 dx_2^2 +
(g_{31} g_{32})^2 dx_3^2.
\end{equation}
For Cartesian, cylindrical, and spherical coordinates,
Table~\ref{metric_scaling} gives the values of
these quantities and their derivatives.  In 
discretized form, these quantities are used both
on both the integer and staggered, half-integer
mesh.

  For Cartesian coordinates $(x,y)$, all metric coefficients are
unity:
\begin{eqnarray}
\left[ g_{1} \right]_i =
\left[ g_{2} \right]_i =
\left[ g_{31} \right]_i =
\left[ g_{32} \right]_j 
= 1, & 
\nonumber \\
\left[ g_{1} \right]_{i+\lhalf} =
\left[ g_{2} \right]_{i+\lhalf} =
\left[ g_{31} \right]_{i+\lhalf} =
\left[ g_{32} \right]_{j+\lhalf} 
= 1 & 
\end{eqnarray}
and all derivatives are obviously zero.  

For cylindrical coordinates $(z,\Re)$, the metric coefficients are
\begin{eqnarray}
\left[ g_{1} \right]_i =
\left[ g_{2} \right]_i =
\left[ g_{31} \right]_i 
= 1, &
\nonumber \\
\left[ g_{1} \right]_{i+\lhalf} =
\left[ g_{2} \right]_{i+\lhalf} =
\left[ g_{31} \right]_{i+\lhalf}
= 1, & 
\nonumber \\
\left[ g_{32} \right]_j = \left[ \Re \right]_j; &
\left[ g_{32} \right]_{j+\lhalf} = \left[ \Re \right]_{j+\lhalf},
\end{eqnarray}
with non-zero derivatives
\begin{equation}
\left[ \frac{\partial g_{32}}{\partial x_1} \right]_{i} = 
\left[ \frac{\partial g_{32}}{\partial x_1} \right]_{i+\lhalf} = 1.
\end{equation}

Similarly, for spherical
coordinates $(r, \theta)$, on the integer mesh, we have
\begin{eqnarray}
\left[ g_1 \right]_i = 1;~~
\left[ g_2 \right]_i = \left[ r \right]_i; & 
\left[ g_{31} \right]_i = \left[ r \right]_i; &
\left[ g_{32} \right]_j = \sin \left[ \theta \right]_j,
\end{eqnarray}
while on the half-integer mesh, we have
\begin{eqnarray}
\left[ g_{1} \right]_{i+\lhalf} = 1;~~
\left[ g_2 \right]_{i+\lhalf} = \left[ r \right]_{i+\lhalf}; & 
\left[ g_{31} \right]_{i+\lhalf} = \left[ r \right]_{i+\lhalf}; &
\left[ g_{32} \right]_{j+\lhalf} = \sin \left[ \theta \right]_{j+\lhalf}.
\end{eqnarray}
The non-zero derivatives, when differenced, are
\begin{eqnarray}
\left[ \frac{\partial g_2}{\partial x_1} \right]_i = 
\left[ \frac{\partial g_2}{\partial x_1} \right]_{i+\lhalf} = 
\left[ \frac{\partial g_{31}}{\partial x_1} \right]_i = 
\left[ \frac{\partial g_{31}}{\partial x_1} \right]_{i+\lhalf} = 1
\end{eqnarray}
and
\begin{eqnarray}
\left[ \frac{\partial g_{32}}{\partial x_2} \right]_j = 
\cos \left[ \theta \right]_j ; & & 
\left[ \frac{\partial g_{32}}{\partial x_2} \right]_{j+\lhalf} = 
\cos \left[ \theta \right]_{j+\lhalf}.
\end{eqnarray}
\begin{deluxetable}{lcccc}
\tabletypesize{\normalsize}
\tablecaption{{\sc Coordinates and Metric Scale Factors for Selected
Coordinate Systems}
\label{metric_scaling}}
\tablewidth{0pt}
\tablehead{
\colhead{Quantity} & \colhead{Symbol} &\colhead{Cartesian} &
\colhead{Cylindrical} &   
\colhead{Spherical}
}
\startdata
coordinates           &   $x_1$   &   $x$   &   $z$         &   $r$   \\
                      &   $x_2$   &   $y$   &   $\Re$       & $\theta$ \\
                      &   $x_3$   &   $z$   &   $\vartheta$ & $\phi$ \\
metric scale factors  &   $g_1$   &    1    &    1          & 1      \\	 
                      &   $g_2$   &    1    &    1       & $r$    \\	 
                      &   $g_{31}$ &   1    &    1          & $r$    \\
                      &   $g_{32}$ &   1    &   $\Re$       & $\sin \theta$ \\
scale factor derivatives     &   $\partial g_2/\partial x_1$ & 0 & 0 & 1 \\
                      &   $\partial g_{31}/\partial x_1$ & 0 & 0 & 1 \\
                      &   $\partial g_{32}/\partial x_1$ & 0 & 0 & 0 \\
                      &   $\partial g_2/\partial x_2$    & 0 & 0 & 0 \\
                      &   $\partial g_{31}/\partial x_2$ & 0 & 0 & 0 \\
                      &   $\partial g_{32}/\partial x_2$ & 0 & 1 & $\cos \theta$ \\
decomposed volume elements &   $d\Upsilon_1$ & $dx$ & $dz$ & $r^2 dr$ \\
                           &   $d\Upsilon_2$ & $dy$ & $\Re \; d\Re$ & $\sin
\theta \; d\theta$ \\
                           &   $\Upsilon_3$ & $1$ & $ 2\pi$ & $2\pi$ \\
area elements         &   $dA_1$ & $dy$ & $2 \pi \; \Re \; d\Re $ & $2 \pi r^2 \sin \theta \; d\theta$ \\
                      &   $dA_2$ & $dx$ & $2 \pi \; \Re \; dz$& $2 \pi r \sin \theta \; dr$ \\
\enddata
\end{deluxetable}

\begin{figure}[htbp]
\vspace{0.3in}
\begin{center}
\includegraphics[scale=0.35]{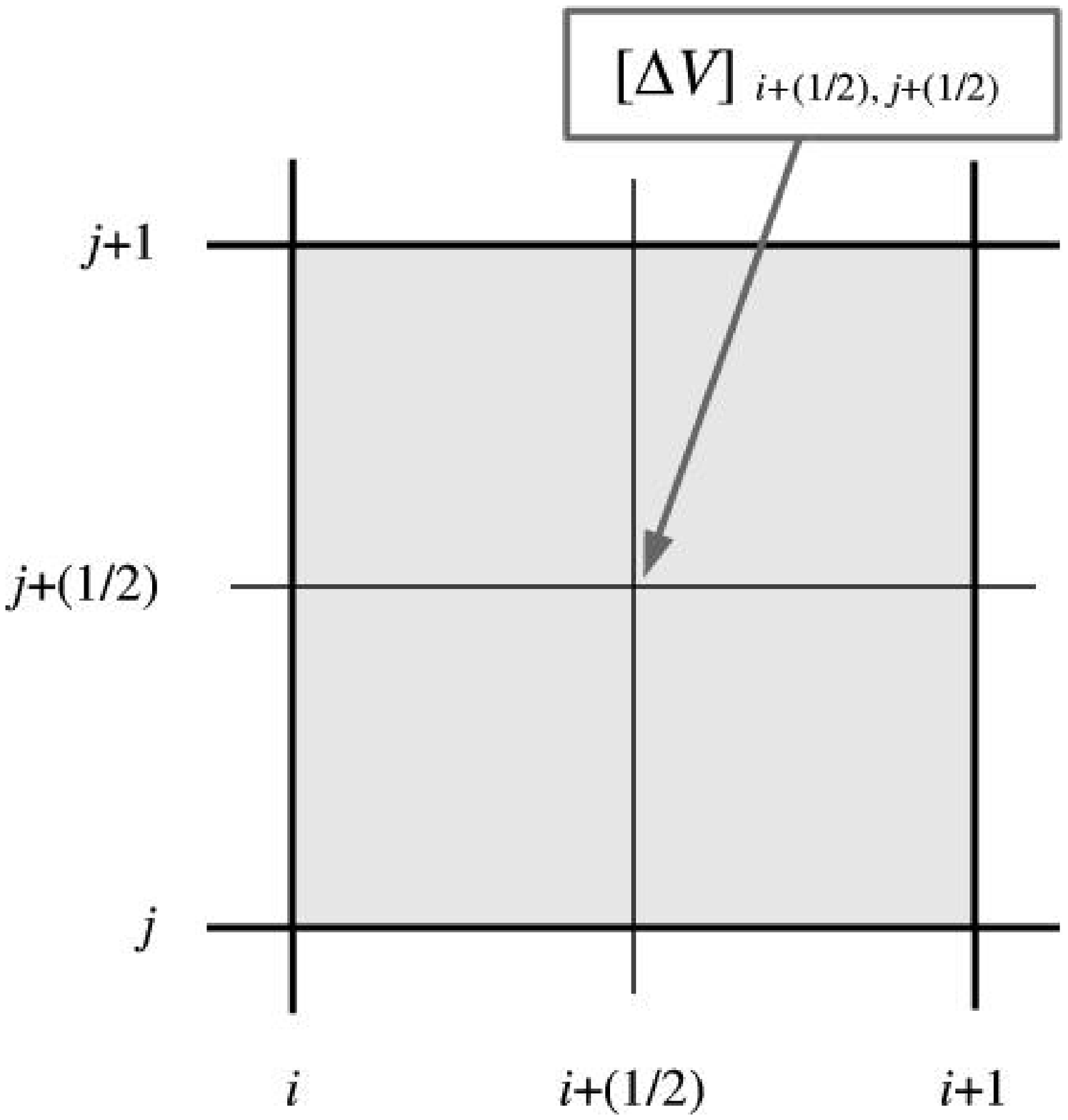}
\end{center}
\caption{\label{fig:volume} Location of the volume element that
corresponds to the volume of the computational cell $[\Delta
V]_{i+\lhalf,j+\lhalf}$.  Volume is defined as a cell-centered
quantity and is therefore defined at the junction of the half integer
mesh at the center of a cell. The third dimension, which makes this
enclosure a volume, is orthogonal to the page and is in the direction
of the assumed symmetry.}
\vspace{0.3in}
\end{figure}

We also make frequent use of the 2-D spatial volume element, 
$dV$, which in
general can be decomposed such that
\begin{equation}
dV =  d\Upsilon_1 d\Upsilon_2 \Upsilon_3.
\end{equation}
The decomposed volume elements, the $d\Upsilon$'s, are given by
\begin{eqnarray} \label{eq:dups}
d\Upsilon_1 = g_2 g_{31} \; dx_1; &
d\Upsilon_2 = g_{32} \; dx_2. &
\end{eqnarray}
The value of $\Upsilon_3$ depends on the specific coordinate
system.

   In Cartesian coordinates, the decomposed volume elements and their
discretization are trivial:
\begin{eqnarray}
d\Upsilon_x = dx; &
d\Upsilon_{y} = dy; &
\Upsilon_{z} = 1.
\end{eqnarray}
When differenced, they are defined in the code as
\begin{equation}
\left[ \Delta \Upsilon_x \right]_i = 
\left[x\right]_{i+\lhalf} - \left[x\right]_{i-\lhalf}
\end{equation}
and
\begin{equation}
\left[ \Delta \Upsilon_y \right]_i = 
\left[y\right]_{i+\lhalf} - \left[y\right]_{i-\lhalf}.
\end{equation}
Cell-centered and cell-edge coordinates in the $x$ direction
are usually related by
\begin{equation}
\left[x\right]_{i+\lhalf} =
\frac{1}{2} 
\left\{ 
   \left[ x \right]_{i+1}+
   \left[ x \right]_{i} 
 \right\},
\end{equation}
with the analogous relationship for the $y$ coordinates.
Cell-centered volume elements are also used:
\begin{equation}
\left[ \Delta \Upsilon_x \right]_{i+\lhalf} = 
\left[x\right]_{i+1} - \left[x\right]_{i}
\end{equation}
and
\begin{equation}
\left[ \Delta \Upsilon_y \right]_{i+\lhalf} = 
\left[y\right]_{i+1} - \left[y\right]_{i}.
\end{equation}

In cylindrical coordinates, with azimuthal symmetry, these elements become
\begin{eqnarray}
d\Upsilon_z =  dz; &
d\Upsilon_{\Re} = \Re \; d\Re; &
\Upsilon_{\vartheta} = 2 \pi,
\end{eqnarray}
which, upon discretization, become
\begin{equation}
\left[ \Delta \Upsilon_z \right]_i = 
\left[z\right]_{i+\lhalf} - \left[z\right]_{i-\lhalf},
\end{equation}
\begin{equation}
\left[ \Delta \Upsilon_{\Re} \right]_j = 
2 \pi \left[ \Re \right]_{j}
\left( \left[ \Re \right]_{j+\lhalf} - \left[ \Re \right]_{j-\lhalf} \right).
\end{equation}
Cell-centered and cell-edge coordinates in the $z$ direction
are usually related by
\begin{equation}
\left[z\right]_{i+\lhalf} =
\frac{1}{2} 
\left\{ 
   \left[ z \right]_{i+1}+
   \left[ z \right]_{i} 
 \right\},
\end{equation}
with the analogous relationship for the $\Re$ coordinates.

In spherical coordinates, with azimuthal symmetry imposed, 
these elements become
\begin{eqnarray}
d\Upsilon_r = r^2 dr; &
d\Upsilon_{\theta} = \sin \theta \; d\theta; &
\Upsilon_{\phi} = 2\pi.
\end{eqnarray}
When differenced, they appear in the code as
\begin{equation} \label{eq:ups_r}
\left[ \Delta \Upsilon_r \right]_i = 
\frac{1}{3} \left\{ (\left[ r \right]_{i+\lhalf})^2
+ \left[ r \right]_{i+\lhalf} \left[ r \right]_{i-\lhalf}
+ (\left[ r \right]_{i-\lhalf})^2 \right\}
( \left[r\right]_{i+\lhalf} - \left[r\right]_{i-\lhalf} ),
\end{equation}
\begin{equation}
\left[ \Delta \Upsilon_{\theta} \right]_j = 
- \cos \left[ \theta \right]_{j+\lhalf} + \cos \left[ \theta \right]_{j-\lhalf}.
\end{equation}
We note that writing the volume element in the form used in
equation~(\ref{eq:ups_r}) avoids the computation  
of the difference of two cubes, which can lead to a substantial loss of 
precision.
The choice of relationship between cell-centered and cell-edge coordinates
in spherical geometry is quite important.  Following \cite{mm89}
we usually choose
\begin{equation}
\left[r\right]_{i+\lhalf} = \sqrt{
\frac{1}{3} 
\left\{ 
   (\left[ r \right]_{i+1})^2
 + \left[ r \right]_{i+1} \left[ r \right]_{i} +
  (\left[ r \right]_{i})^2 
 \right\}
}
\end{equation}
for radial coordinates, while choosing
\begin{equation}
\left[\theta\right]_{j+\lhalf} =
\frac{1}{2} 
\left\{ 
   \left[ \theta \right]_{j+1}+
   \left[ \theta \right]_{j} 
 \right\}
\end{equation}
to relate cell-centered and cell-edge coordinates
for the angular dimension.
Cell-centered volume elements are also used:
\begin{equation}
\left[ \Delta \Upsilon_r \right]_{i+\lhalf} = 
\frac{1}{3} \left\{ (\left[ r \right]_{i+1})^2
+ \left[ r \right]_{i+1} \left[ r \right]_{i}
+ (\left[ r \right]_{i})^2 \right\}
( \left[r\right]_{i+1} - \left[r\right]_{i} ),
\end{equation}
\begin{equation}
\left[ \Delta \Upsilon_{\theta} \right]_{j+\lhalf} = 
- \cos \left[ \theta \right]_{j+1} + \cos \left[ \theta \right]_{j}.
\end{equation}

Also of importance are the components of the differential area vector,
$d{\bf A}$, in the $x_1$ and $x_2$ directions.  These are given, in
general, by
\begin{eqnarray}
dA_1 = \Upsilon_3 g_2 g_{31} d\Upsilon_2; &
dA_2 = \Upsilon_3 g_{31} g_{32} dx_1.
\end{eqnarray}
Since these elements are components of a vector, 
these quantities become face-centered when differenced
\begin{equation}
\left[ \Delta A_1 \right]_{i,j+\lhalf} = 
\Upsilon_3 \left[ g_2 \right]_i  \left[ g_{31} \right]_i 
\left[ \Delta \Upsilon_2 \right]_{j+\lhalf}
\end{equation} 
and
\begin{equation}
\left[ \Delta A_2 \right]_{i+\lhalf,j} = 
\Upsilon_3 \left[ g_{31} \right]_{i+\lhalf} \left[ g_{32} \right]_{j}
(\left[ x_1 \right]_{i+1} - \left[ x_1 \right]_i).
\end{equation}
Figure~\ref{fig:areas} shows the location and orientation of the
components of these area vectors with respect to the cell volume they
enclose.

\begin{figure}[htbp]
\vspace{0.3in}
\begin{center}
\includegraphics[scale=0.35]{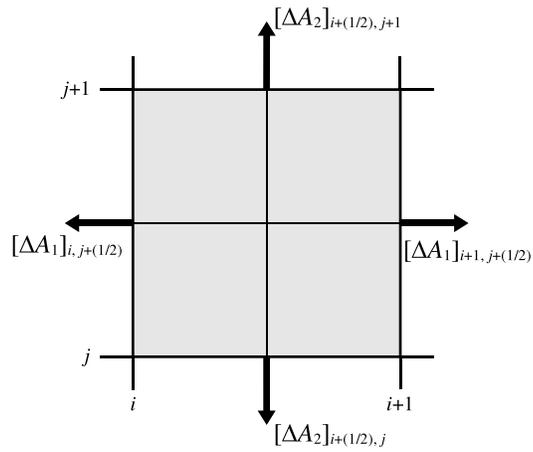}
\end{center}
\caption{\label{fig:areas} Location and orientation of the components
of the differential area vector with respect to the volume element
they enclose.  The third dimension, which makes these cell ``faces''
areas, is orthogonal to the page and is in the direction of an assumed
symmetry. Area, being a vector quantity, is defined at cell faces.  In
our staggered mesh algorithm, this means that vector components
originate from different locations.\vspace{0.3in}}
\end{figure}

For Cartesian coordinates, the area elements and their
discretization are again trivial.
\begin{eqnarray}
dA_x =  dy; &
dA_y =  dx,
\end{eqnarray}
which, when differenced, give
\begin{equation}
\left[ \Delta A_x \right]_{i,j+\lhalf} = 
\left[ y \right]_{j+1} - \left[ y \right]_j
\end{equation} 
and
\begin{equation}
\left[ \Delta A_y \right]_{i+\lhalf,j} = 
\left[ x \right]_{i+1} - \left[ x \right]_i.
\end{equation}

In cylindrical coordinates, imposing symmetry in the azimuthal direction,
the area elements are
\begin{eqnarray}
dA_z = 2 \pi \Re \; d\Re; &
dA_\Re = 2 \pi \Re \; dz,
\end{eqnarray}
which, when differenced, give
\begin{equation}
\left[ \Delta A_z \right]_{i,j+\lhalf} = 
2 \pi \left[ \Re \right]_{j+\lhalf}
(\left[ \Re \right]_{j+1} -
\left[ \Re \right]_j )
\end{equation} 
and
\begin{equation}
\left[ \Delta A_\Re \right]_{i+\lhalf,j} = 
2 \pi \left[ \Re \right]_{j} 
(\left[ z \right]_{i+1} - \left[ z \right]_i).
\end{equation}

In spherical coordinates, imposing azimuthal symmetry, the area
elements are
\begin{eqnarray}
dA_r = 2 \pi r^2 \sin \theta d\theta; &
dA_\theta = 2 \pi r \sin \theta dr,
\end{eqnarray}
which, when differenced, give
\begin{equation}
\left[ \Delta A_r \right]_{i,j+\lhalf} = 
2 \pi ( \left[ r \right]_i )^2  
(-\cos \left[ \theta \right]_{j+1} + 
\cos \left[ \theta \right]_j )
\end{equation} 
and
\begin{equation}
\left[ \Delta A_\theta \right]_{i+\lhalf,j} = 
2 \pi \left[ r \right]_{i+\lhalf} \sin \left[ \theta \right]_{j}
(\left[ r \right]_{i+1} - \left[ r \right]_i).
\end{equation}

\section{Discretization of Advection Equations\label{app:advect}}


In this appendix we discuss the solution of 
equations~(\ref{eq:cont_adv}), (\ref{eq:ye-adv}),
(\ref{eq:e-advective}), (\ref{eq:s-advect}), and
(\ref{eq:nu-advect}).
Our numerical method for these equations is that of \cite{sn92a} and 
the first subsection this appendix merely restates their approach 
for the convenience of the reader.  In later subsections we write
down the update equations in a convenient form for numerical
implementation.

\subsection{The ZEUS Advection Scheme}

Since equations~(\ref{eq:cont_adv}), (\ref{eq:ye-adv}),
(\ref{eq:e-advective}), (\ref{eq:s-advect}), and
(\ref{eq:nu-advect}).
have the similar form,
we can generalize these expressions. First, for advection of a scalar
quantity, we write
\begin{equation} \label{eq:diff_ad_sc}
\left\ldbrack \frac{ \partial \psi}{\partial t}
\right\rdbrack_{\rm advection} + 
{\bfnabla} \cdot \left( \psi {\bf v} \right) = 0,
\end{equation}
where $\psi$ represents the scalar field being advected.  Similarly,
for advection of a vector field ${\smpmb \sigma}$, we write
\begin{equation} \label{eq:diff_ad_vec}
\left\ldbrack \frac{ \partial {\smpmb \sigma}}{\partial t}
\right\rdbrack_{\rm advection} + 
{\bfnabla} \cdot \left( {\smpmb \sigma}{\bf v} \right) = 0.
\end{equation}
In solving the advection equation, it is common to solve
finite-difference approximations to integral forms of equations
(\ref{eq:diff_ad_sc}) and (\ref{eq:diff_ad_vec}).  For advection of a
scalar field, this takes the form
\begin{equation} \label{eq:int_ad_sc}
\frac{d}{dt} \int_V \psi dV = - \int_{{\bf A}} \psi {\bf v} \cdot d{\bf A},
\end{equation}
and for advection of a vector field,
\begin{equation}\label{eq:int_ad_vec}
\frac{d}{dt} \int_V {\smpmb \sigma}dV = 
- \int_{{\bf A}} {\smpmb \sigma} {\bf v} \cdot d{\bf A},
\end{equation}
where $V$ is an arbitrary volume whose bounding surface is ${\bf A}$.
%
%
\begin{figure}[htbp]
\begin{center}
\includegraphics[scale=0.35]{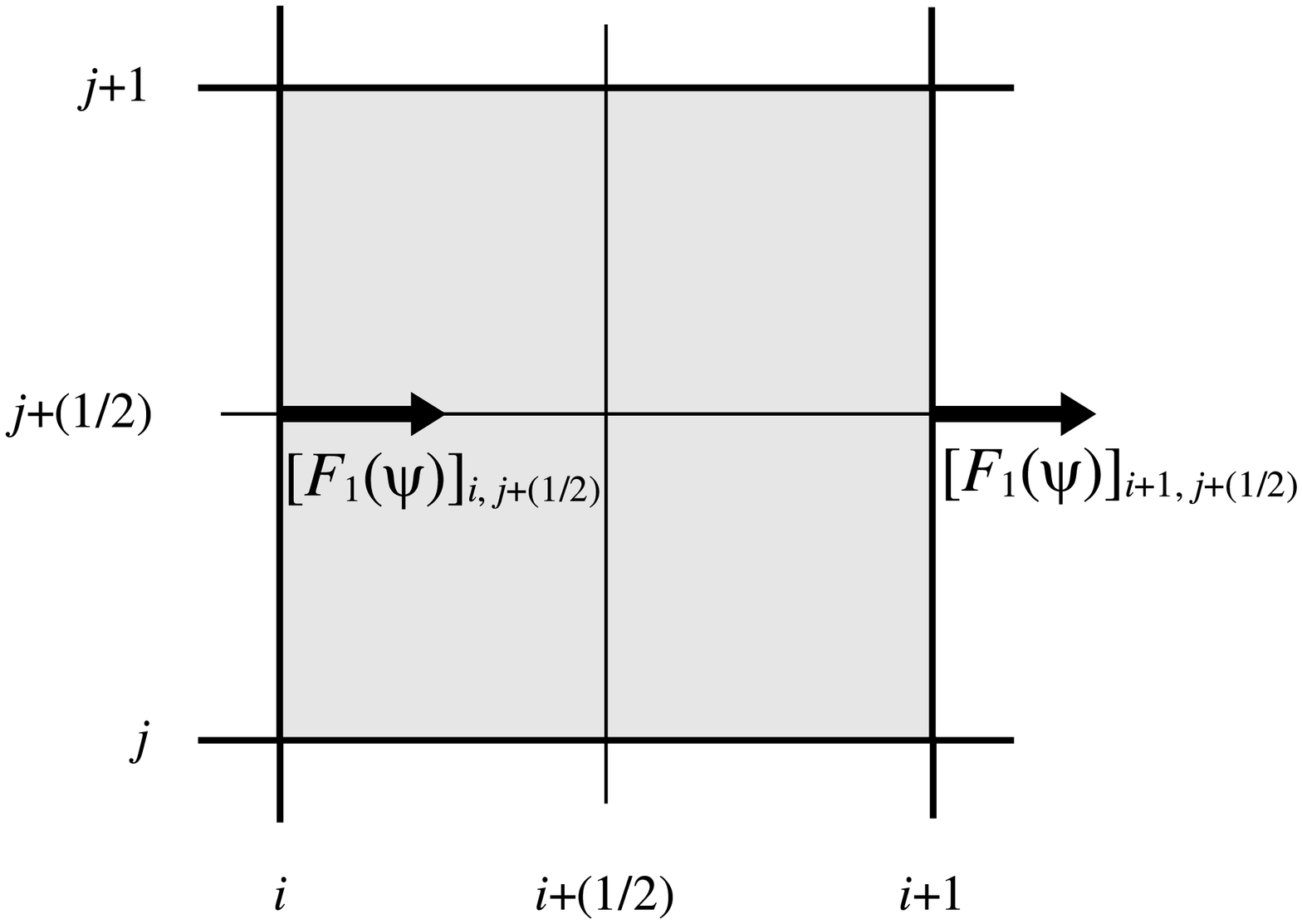} 
\hspace{0.2in}
\includegraphics[scale=0.35]{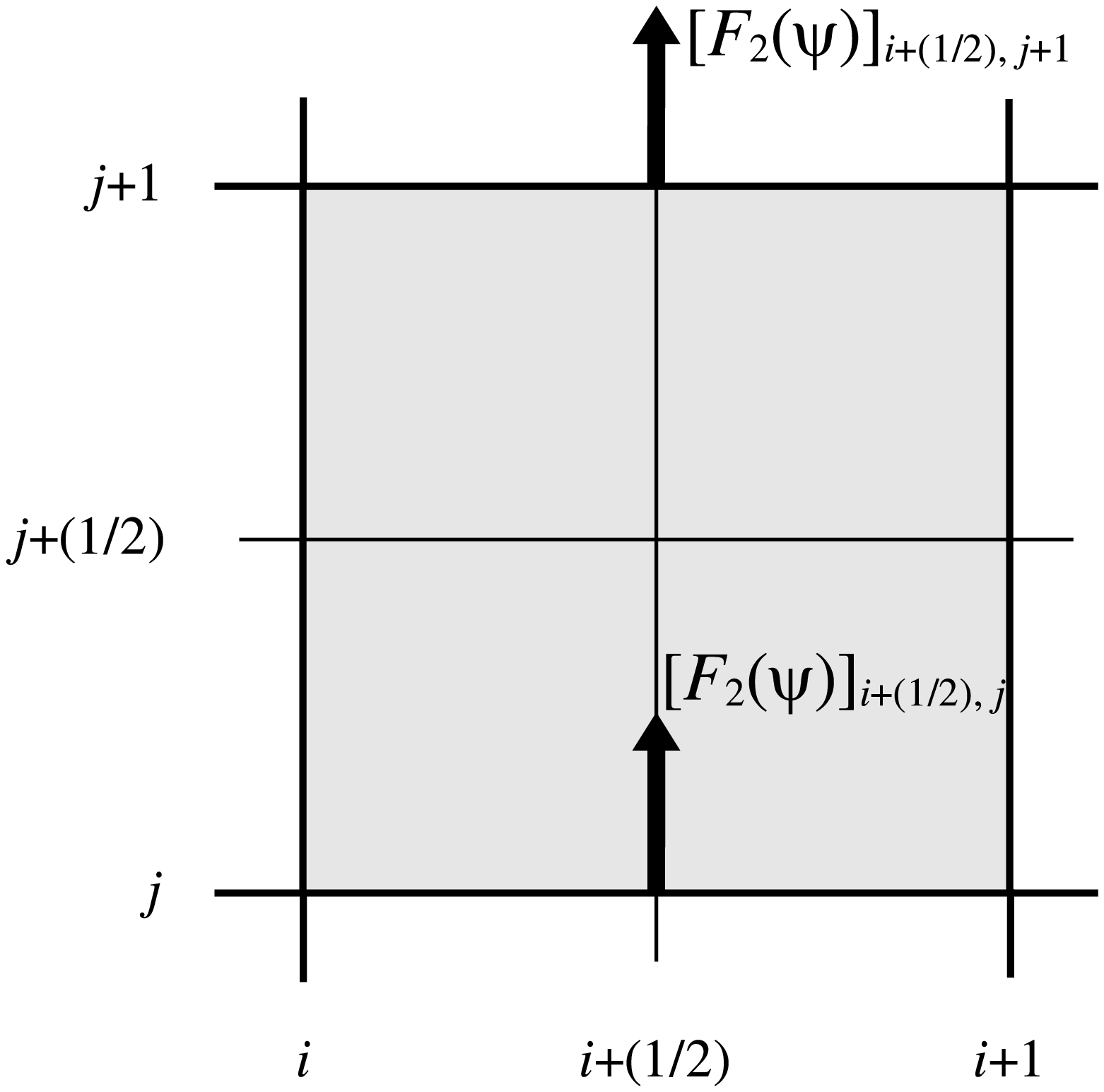}
\end{center}
\caption{\label{fig:fluxes} Definition of locations for the $x_1$
and $x_2$ components the flux densities used in the advection of a
general, cell-centered, scalar quantity
$[\psi]_{i+\lhalf,j+\lhalf}$. The shaded region indicates 
the volume element, $[\Delta V]_{i+\lhalf,j+\lhalf}$, used in the
differenced integral advection equation
(eq.~[\ref{eq:diff_ad_sc}]). Since flux density is a vector quantity,
its components are defined at the four face centers bounding the
volume element. \vspace{0.3in}}
\end{figure}
%
%

   Following \cite{sn92a} we directionally split the update of all 
equations of the form of (\ref{eq:int_ad_sc}) and 
(\ref{eq:int_ad_vec}) into separate sequential updates of the 
variables due to advection in the 
$x_1$ and $x_2$ directions.  We alternate the directional 
order of advection updates
($x_1$ updates followed by $x_2$ updates and {\em vice versa}) with each timestep.
A complete directional ``sweep'' is performed by solving all of the
advection equations corresponding to one direction, followed by 
a second sweep solving the equations corresponding to the orthogonal
direction.

The consistent advection scheme of \cite{sn92a} ties the advection of 
all variables to the mass flux.  Therefore, 
we first consider the update to the continuity equation due to advection in 
the $x_1$ direction.  We can write the discretized version 
of equation~(\ref{eq:cont_adv}) as
\begin{eqnarray} \label{eq:mflux_comb_x1}
\lefteqn{ \frac{ \left[ \Delta V \right]_{i+\lhalf,j+\lhalf}}{\Delta t} 
\left( \left[ \rho \right]^{n+\beta}_{i+\lhalf,j+\lhalf} - 
\left[ \rho \right]^{n+\alpha}_{i+\lhalf,j+\lhalf} \right) 
= }
\nonumber \\ & & 
- \left( \left[ F_1(\rho) \right]_{i+1,j+\lhalf}^{n+\alpha}
\left[ \Delta A_1 \right]_{i+1,j+\lhalf}  -
\left[ F_1(\rho) \right]_{i,j+\lhalf}^{n+\alpha}
\left[ \Delta A_1 \right]_{i,j+\lhalf} \right),
\end{eqnarray}
where the bounding volume is now $[\Delta
V]_{i+\lhalf,j+\lhalf}$, the volume of an
arbitrary cell in the computational mesh.  The
$\Delta A_i$'s are the areas of the cell faces
that, collectively, bound $\Delta V$.  (These
cell-face areas are defined in Appendix~\ref{app:cov}.)  The quantity 
$[ F_1(\rho)
]_{i,j+\lhalf}$ represents the mass flux density (or
mass flux per unit area) in the
$x_1$ direction across the cell face at
$(i,j+\lhalf)$.  
The superscripts indicate the
level of temporal update, with $n+\alpha$ being
the time-level at the beginning of the $x_1$
advection update, and $n+\beta$ being the
time-level upon completion of the update.  
If the
$x_1$ update is carried out before the $x_2$
update we have $\alpha = 0$ and $\beta=a$; if the
$x_2$ update is carried out first we have
$\alpha=a$ and $\beta = b$.
The update equation for mass advection in the $x_2$ direction is given
by
\begin{eqnarray} \label{eq:mflux_comb_x2}
\lefteqn{ \frac{ \left[ \Delta V \right]_{i+\lhalf,j+\lhalf}}{\Delta t} 
\left( \left[ \rho \right]^{n+\delta}_{i+\lhalf,j+\lhalf} - 
\left[ \rho \right]^{n+\gamma}_{i+\lhalf,j+\lhalf} \right) 
= }
\nonumber \\ & & 
- \left( \left[ F_2(\rho) \right]_{i+\lhalf,j+1}^{n+\gamma}
\left[ \Delta A_2 \right]_{i+\lhalf,j+1}  -
\left[ F_2(\rho) \right]_{i+\lhalf,j}^{n+\gamma}
\left[ \Delta A_2 \right]_{i+\lhalf,j} \right),
\end{eqnarray}
where
$[ F_2(\rho)
]_{i+\lhalf,j}$ represents the mass flux density 
$x_2$ direction across the cell face at
$(i+\lhalf,j)$.  
As in the $x_1$ update the superscripts indicate 
the time level at the beginning and end of the update:
if the
$x_1$ update is carried out before the $x_2$
update we have $\gamma = a$ and $\delta=b$; if the
$x_2$ update is carried out first we have
$\gamma=0$ and $\delta = a$.

Readers who are contrasting our equations with
those of \cite{sn92a} will notice that the ZEUS
algorithm time-centers many of the quantities that
appear in the advection equations. This is
necessitated by the possibility that the grid may
be moving.  Since we are considering only grids
that are fixed in time, the advection prescription
here is more straightforward.  As a consequence of
fixing the grid, the spatial coordinates, the
volume and area elements, as well as the metric
coefficients, are all constant in time.

Simple algebra allows equations
(\ref{eq:mflux_comb_x1}) and
(\ref{eq:mflux_comb_x2}) are readily solved for
the new values of the density in terms of the old
values and the mass fluxes which are computed base on those
old values.

The mass fluxes are constructed as 
\begin{equation}
\left[ F_{1}(\rho) \right]_{i,j+\lhalf} = 
\left[ {\cal I}_1(\rho) \right]_{i,j+\lhalf} 
\left[ \varv_{1} \right]_{i,j+\lhalf}
\end{equation}
and
\begin{equation}
\left[ F_{2}(\rho) \right]_{i+\lhalf,j} = 
\left[ {\cal I}_2(\rho) \right]_{i+\lhalf,j} 
\left[ \varv_{2} \right]_{i+\lhalf,j},
\end{equation}
where the ${\cal I}$'s are van Leer monotonic 
upwind advection functions 
\citep{vanleer77} 
given by
\begin{equation} \label{eq:i1}
\left[ {\cal I}_1(\psi) \right]_{i,j+\lhalf} = 
\begin{cases} \displaystyle{
\left[ \psi \right]_{i-\lhalf,j+\lhalf} + 
\left[ \delta_1(\psi) \right]_{i-\lhalf,j+\lhalf}
\left( 1 - \frac{\left[ \varv_{1}\right]_{i,j+\lhalf} \Delta t }
{\left[x_1\right]_{i} - \left[x_1\right]_{i-1} } \right)
}
 & \text{if} \; \left[ \varv_1 \right]_{i,j+\lhalf} > 0 \vspace{0.1in} \\
\displaystyle{
\left[ \psi \right]_{i+\lhalf,j+\lhalf} - 
\left[ \delta_1(\psi) \right]_{i+\lhalf,j+\lhalf}
\left( 1 + \frac{ \left[ \varv_{1} \right]_{i,j+\lhalf} \Delta t}
{\left[x_1\right]_{i+1} - \left[x_1\right]_{i} }
\right) }
 & \text{if} \; \left[ \varv_1 \right]_{i,j+\lhalf} < 0, \\
\end{cases}
\end{equation}
where
\begin{equation} \label{eq:i1p1}
\left[ \delta_1(\psi) \right]_{i+\lhalf,j+\lhalf} = 
\begin{cases}
\displaystyle{ 
\frac{
\left[ \Delta \psi \right]_{i,j+\lhalf} 
\left[ \Delta \psi \right]_{i+1,j+\lhalf} }
{\left[ \psi \right]_{i+\lthreehalf,j+\lhalf} -
\left[ \psi \right]_{i-\lhalf,j+\lhalf}}
} \vspace{0.1in} \\
\hspace{1.5in}
\text{if} \;
\left[ \Delta \psi \right]_{i,j+\lhalf} 
\left[ \Delta \psi \right]_{i+1,j+\lhalf} 
> 0 \vspace{0.25in} \\
0 \hspace{1.4in} \text{otherwise}, 
\end{cases}
\end{equation}
and
\begin{equation} \label{eq:delsig1}
\left[ \Delta \psi \right]_{i,j+\lhalf} = 
\left[ \psi \right]_{i+\lhalf,j+\lhalf} -
\left[ \psi \right]_{i-\lhalf,j+\lhalf}.
\end{equation}
Here,
and in equations that follow, we have temporarily suspended temporal
superscripts whenever writing expressions at a generic time-level.
Equations for updates will retain the appropriate time superscripts.
For reasons of brevity, we will not write down the analogous 
expressions for the $x_2$ direction.
However, Table~\ref{vanleer} gives a complete prescription for
constructing ${\cal I}_2(\psi)$ from the ${\cal I}_1(\psi)$ definition.

The advection equations for scalar quantities other than the 
density are discretized as
\begin{eqnarray} \label{eq:sflux_comb_x1}
\lefteqn{ \frac{ \left[ \Delta V \right]_{i+\lhalf,j+\lhalf}}{\Delta t} 
\left( \left[ \psi \right]^{n+\beta}_{i+\lhalf,j+\lhalf} - 
\left[ \psi \right]^{n+\alpha}_{i+\lhalf,j+\lhalf} \right) 
= }
\nonumber \\ & & 
- \left( \left[ F_1(\psi) \right]_{i+1,j+\lhalf}^{n+\alpha}
\left[ \Delta A_1 \right]_{i+1,j+\lhalf}  -
\left[ F_1(\psi) \right]_{i,j+\lhalf}^{n+\alpha}
\left[ \Delta A_1 \right]_{i,j+\lhalf} \right),
\end{eqnarray}
for the $x_1$ update and
\begin{eqnarray} \label{eq:sflux_comb_x2}
\lefteqn{ \frac{ \left[ \Delta V \right]_{i+\lhalf,j+\lhalf}}{\Delta t} 
\left( \left[ \psi \right]^{n+\delta}_{i+\lhalf,j+\lhalf} - 
\left[ \psi \right]^{n+\gamma}_{i+\lhalf,j+\lhalf} \right) 
= }
\nonumber \\ & & 
- \left( \left[ F_2(\psi) \right]_{i+\lhalf,j+1}^{n+\gamma}
\left[ \Delta A_2 \right]_{i+\lhalf,j+1}  -
\left[ F_2(\psi) \right]_{i+\lhalf,j}^{n+\gamma}
\left[ \Delta A_2 \right]_{i+\lhalf,j} \right)
\end{eqnarray}
for the $x_2$ update.

One of the hallmarks of the ZEUS advection strategy is the use
of {\em Norman's consistent advection scheme} \citep{nwb80}, which ties the  
fluxes for all advected quantities to the mass flux.  For 
scalar quantities other than the density, the fluxes are written as
\begin{equation}
\left[ F_1(\psi) \right]_{i,j+\lhalf} = 
\left[ {\cal I}_1\left(\frac{\psi}{\rho}\right) \right]_{i,j+\lhalf} 
\left[ F_1(\rho) \right]_{i,j+\lhalf}
\label{eq:nca1}
\end{equation}
and
\begin{equation}
\left[ F_2(\psi) \right]_{i+\lhalf,j} = 
\left[{\cal I}_2\left(\frac{\psi}{\rho} \right) \right]_{i+\lhalf,j} 
\left[ F_2(\rho) \right]_{i+\lhalf,j}.
\label{eq:nca2}
\end{equation}
When equations~(\ref{eq:nca1}) and (\ref{eq:nca2}) are 
substituted into equations~(\ref{eq:sflux_comb_x1}) and
(\ref{eq:sflux_comb_x2}) we can solve for new values of 
all scalar quantities in terms of old values and the fluxes
which are calculate in terms of those old values.

%
%
\begin{figure}[htbp]
\vspace{0.3in}
\begin{center}
\includegraphics[scale=0.50]{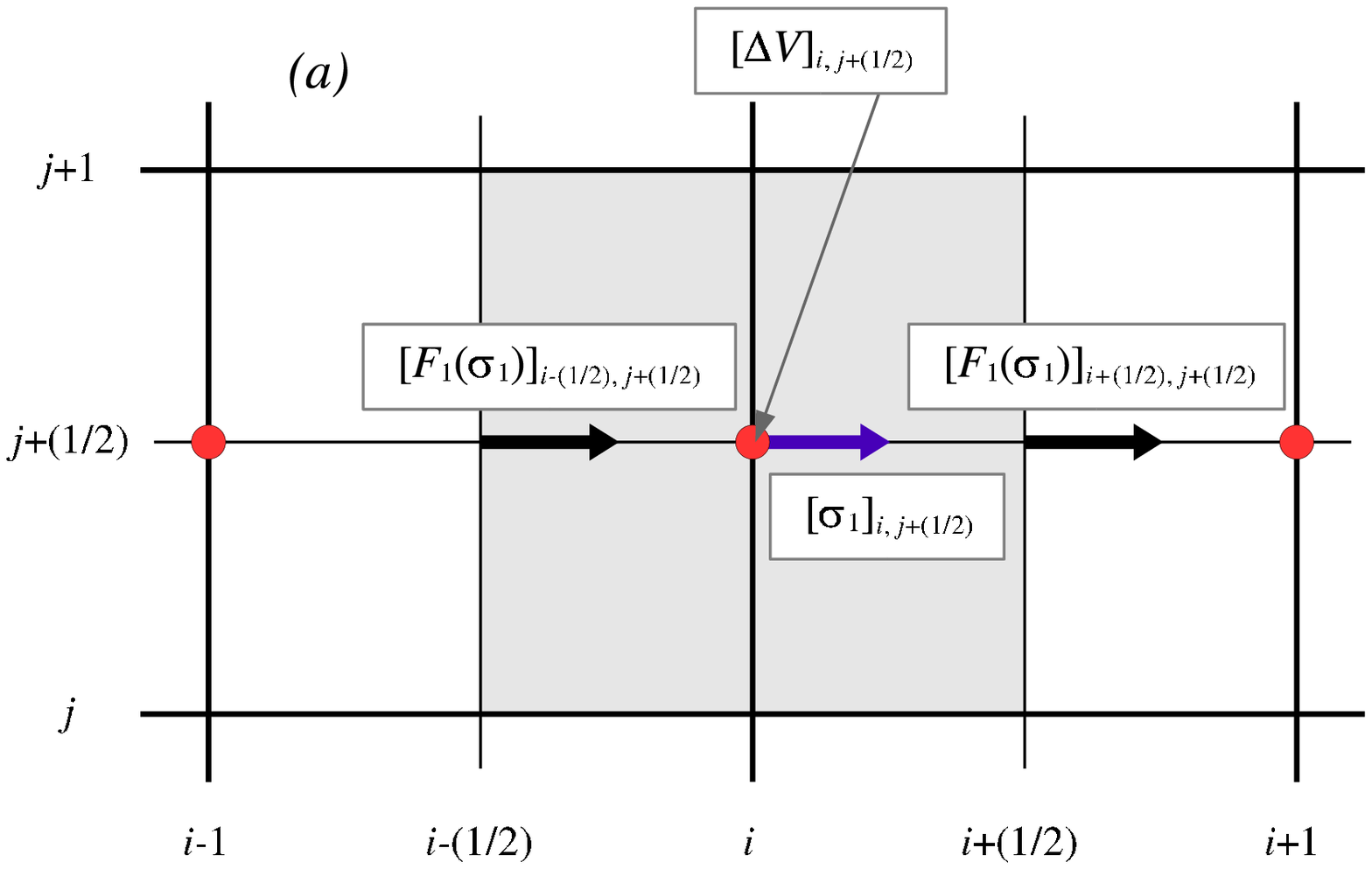}
\end{center}
\vspace{-2.5in}
\begin{center}
\includegraphics[scale=0.50]{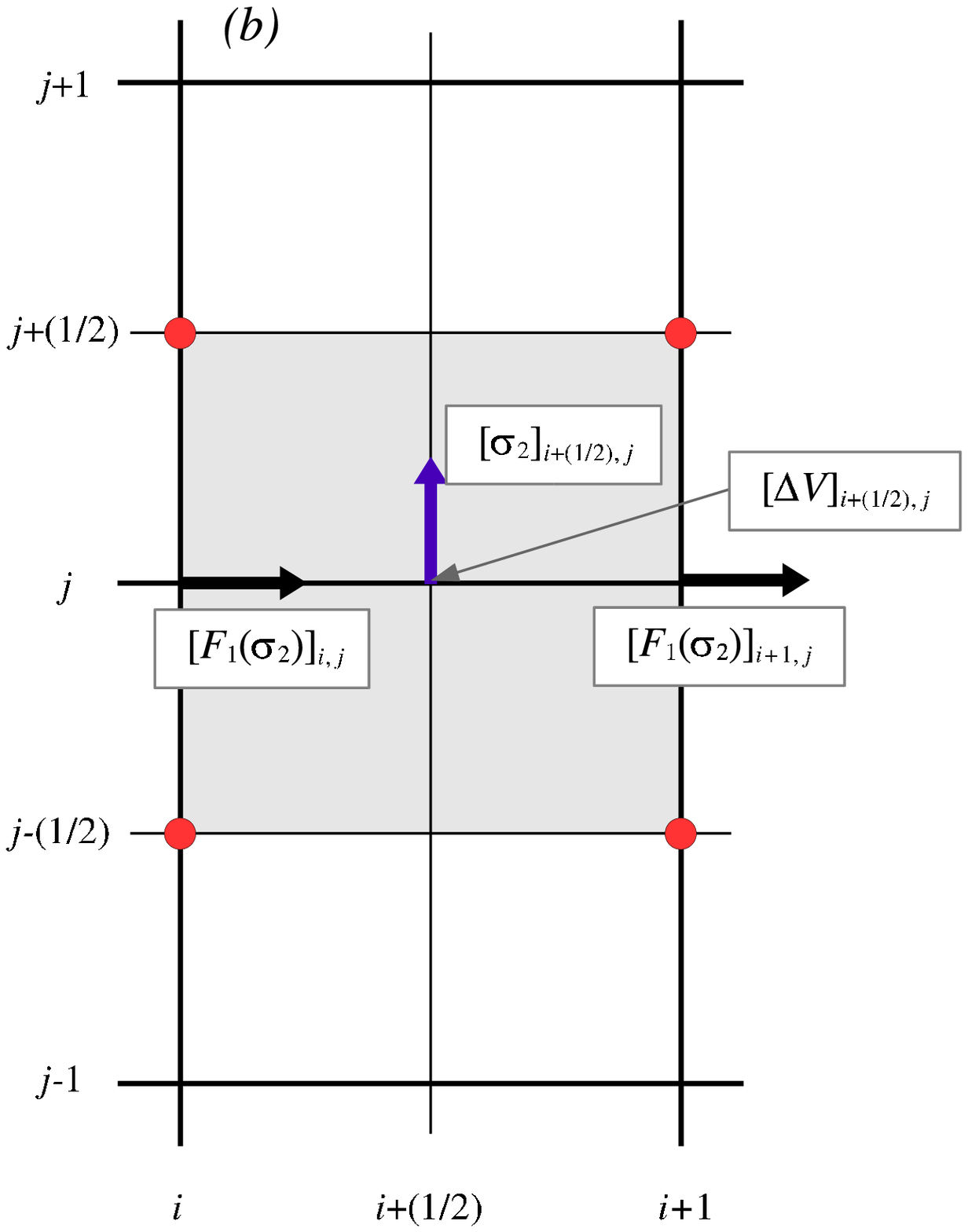}
\vspace{-1.5in}
\end{center}
\caption{\label{fig:vfluxes} Locations of the flux densities needed to
perform an $x_1$ advection sweep of a vector quantity ${\smpmb
\sigma}$. The fluxes used to advect component $\sigma_1$ are shown in
({\em a}), while the fluxes for component $\sigma_2$ are shown in ({\em
b}).  The shaded regions indicates the volume elements used in this
$x_1$ sweep. Volume element, $[\Delta V]_{i,j+\lhalf}$, shown in ({\em
a}) is used in the $\sigma_1$ updates,
while in ({\em b}), $[\Delta V]_{i+\lhalf,j}$ is used in the
$\sigma_2$ updates.  These are different volume elements
than used in the advection of a scalar.  This is because of the
face-centering of vector quantities, whose component locations are
indicated by blue arrows.  Because the volume elements are different,
the locations of the flux densities are also different.  The
interpolations used to obtain them are given by
equations~(\ref{eq:f1sig1}), (\ref{eq:f2sig1}), and their $\sigma_2$
analogues. The face-centered locations that are used to calculate the
interpolations are shown as red circles in the figure. }
\end{figure}
%
%
\begin{figure}[htbp]
\vspace{0.3in}
\begin{center}
\includegraphics[scale=0.50]{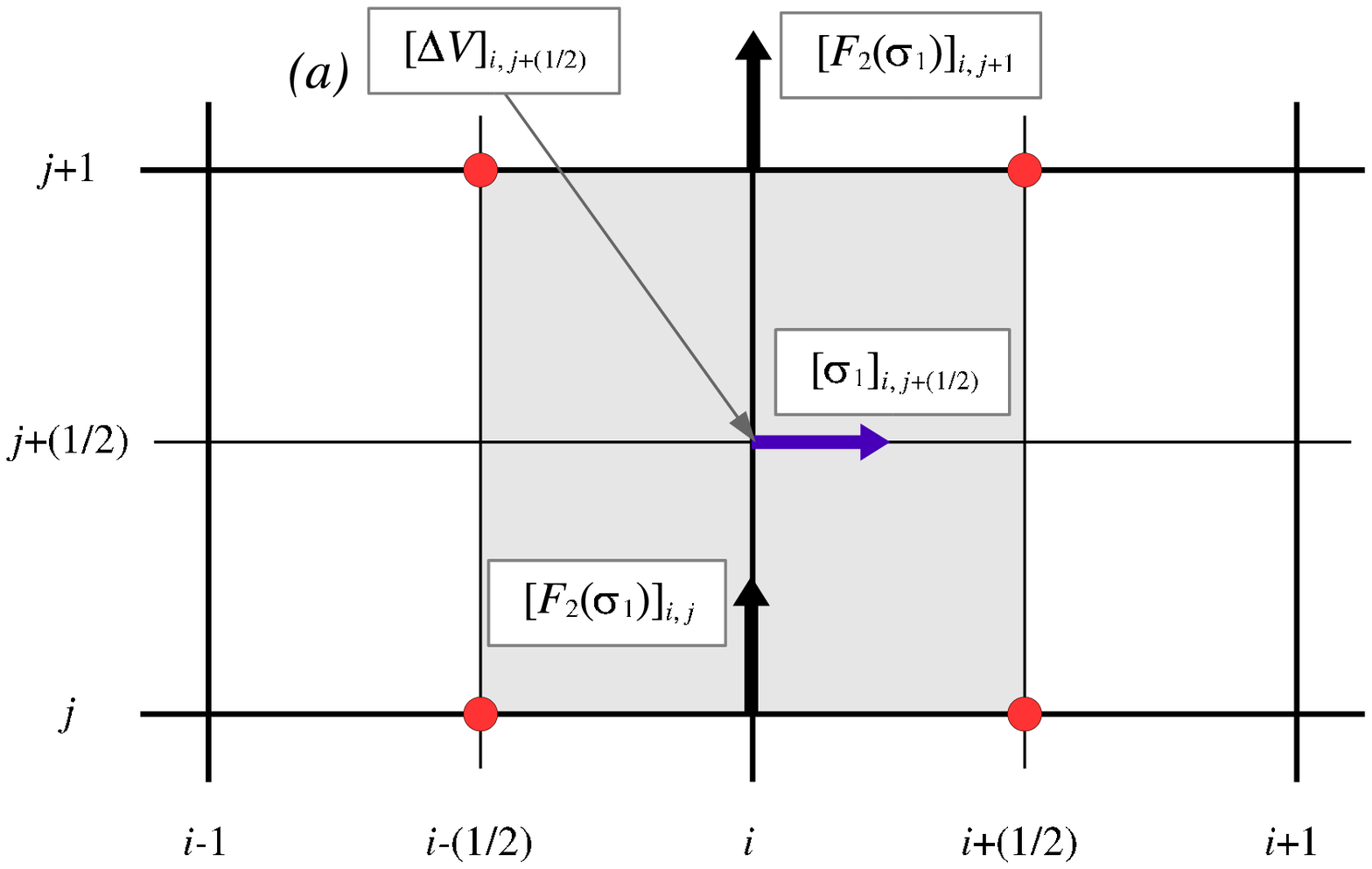}
\end{center}
\vspace{-2.5in}
\begin{center}
\includegraphics[scale=0.50]{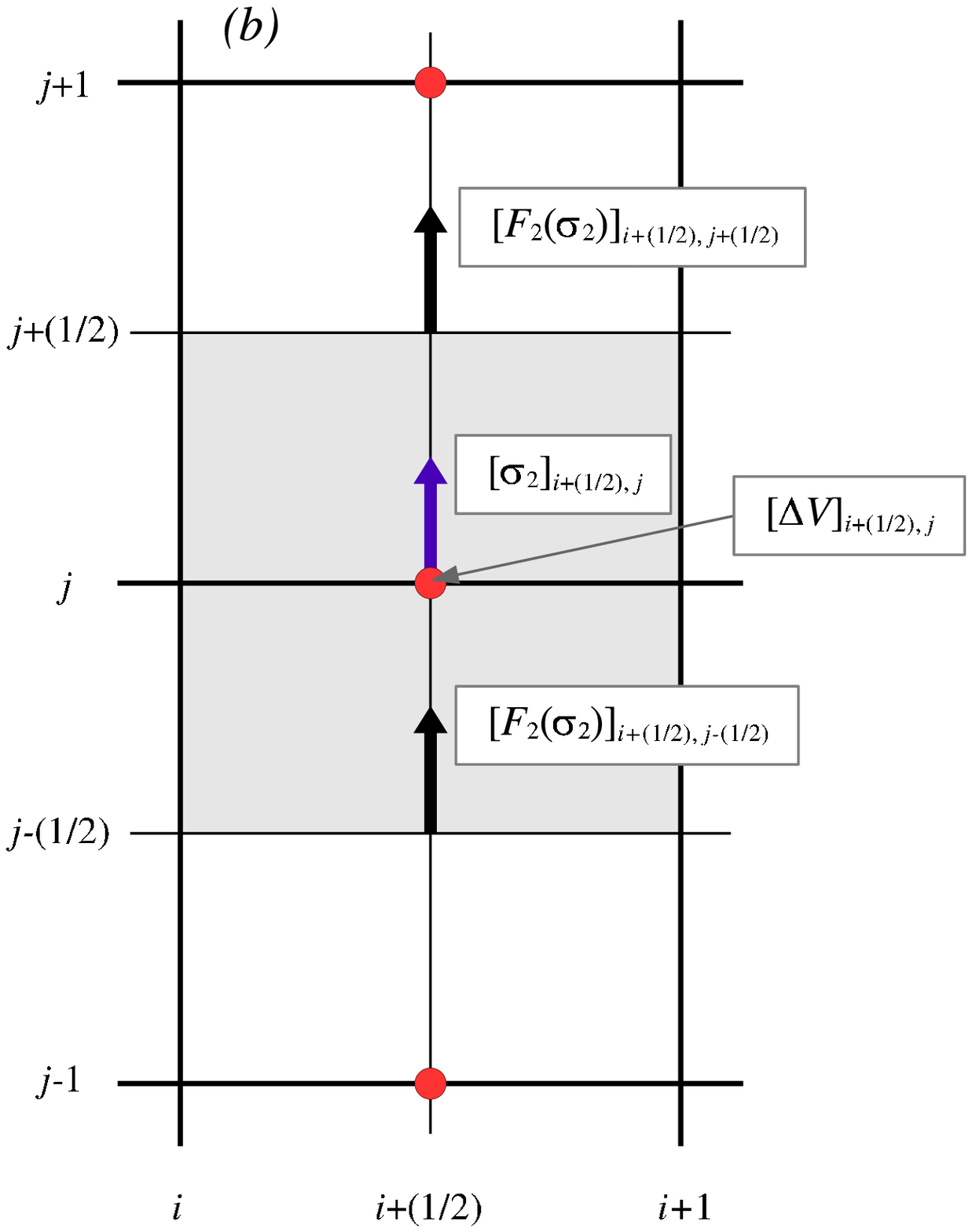}
\vspace{-1.5in}
\end{center}
\caption{\label{fig:vfluxes2} Locations of the flux densities needed to
perform an $x_2$ advection sweep of a vector quantity ${\smpmb
\sigma}$. The fluxes used to advect component $\sigma_1$ are shown in
({\em a}), while the fluxes for component $\sigma_2$ are shown in ({\em
b}).  The comments made in the caption to Figure~\ref{fig:vfluxes}
regarding volume elements, the locations of flux density vectors, and
fluxes used in interpolation formulae also apply here.
}
\end{figure}
%

Advection of vector quantities proceeds slightly differently because
of the location where vector components are defined on the staggered
mesh.  
The analog of equations~(\ref{eq:mflux_comb_x1}) 
and (\ref{eq:mflux_comb_x2}) for the
advection of $\sigma_1$, the $x_1$ component of vector ${\smpmb
\sigma}$, in both the $x_1$ and $x_2$ directions is
\begin{eqnarray} \label{eq:flux_comb_v1x1}
\lefteqn{ \frac{ \left[ \Delta V \right]_{i,j+\lhalf}}{\Delta t} 
\left( \left[ \sigma_1 \right]^{n+\beta}_{i,j+\lhalf} - 
\left[ \sigma_1 \right]^{n+\alpha}_{i,j+\lhalf} \right) 
= }
\nonumber \\ & & 
- \left( \left[ F_1(\sigma_1) 
\right]_{i+\lhalf,j+\lhalf}^{n+\alpha}
\left[ \Delta A_1 \right]_{i+\lhalf,j+\lhalf}  -
\left[ F_1(\sigma_1) \right]_{i-\lhalf,j+\lhalf}^{n+\alpha} 
\left[ \Delta A_1 \right]_{i-\lhalf,j+\lhalf} \right) 
\end{eqnarray}
and
\begin{eqnarray} \label{eq:flux_comb_v1x2}
\lefteqn{ \frac{ \left[ \Delta V \right]_{i,j+\lhalf}}{\Delta t} 
\left( \left[ \sigma_1 \right]^{n+\delta}_{i,j+\lhalf} - 
\left[ \sigma_1 \right]^{n+\gamma}_{i,j+\lhalf} \right) 
= }
\nonumber \\ & & 
\quad \; \:
- \left. \left( \left[ F_2(\sigma_1) \right]_{i,j+1}^{n+\gamma} 
\left[ \Delta A_2 \right]_{i,j+1} -
\left[ F_2(\sigma_1) \right]_{i+1,j}^{n+\gamma} 
\left[ \Delta A_2 \right]_{i,j} \right) 
 \right\},
\end{eqnarray}
where the fluxes are given by
\begin{equation} \label{eq:f1sig1} 
\left[ F_1(\sigma_1) \right]_{i+\lhalf,j+\lhalf} = 
\left[ {\cal J}_1\left(\frac{\sigma_1}{\rho}\right) \right]_{i+\lhalf,j+\lhalf} 
\left[ F_1(\rho) \right]_{i+\lhalf,j+\lhalf} 
\end{equation}
and
\begin{equation} \label{eq:f2sig1} 
\left[ F_2(\sigma_1) \right]_{i,j} = 
\left[ {\cal J}_2\left(\frac{\sigma_1}{\rho}\right) \right]_{i,j} 
\left[ F_2(\rho) \right]_{i,j}.
\end{equation}
We note that fluxes in the $x_1$ direction (the $F_1$'s) are here
needed at cell centers and the fluxes in the $x_2$ direction (the
$F_2$'s) are needed at zone vertices---locations where neither of
these quantities is naturally defined.  (Figures \ref{fig:vfluxes}{\em
(a)} and \ref{fig:vfluxes2}{\em (a)} show the location of fluxes
needed to the advect $\sigma_1$.)  We use the following
averages to supply the needed values.
\begin{equation} \label{eq:f1avg}
\left[ F_1(\rho) \right]_{i+\lhalf,j+\lhalf} = 
\half \left(
\left[ F_1(\rho) \right]_{i,j+\lhalf} +
\left[ F_1(\rho) \right]_{i+1,j+\lhalf}
\right)
\end{equation}
and
\begin{equation} \label{eq:f2avg}
\left[ F_2(\rho) \right]_{i,j} = 
\half \left(
\left[ F_2(\rho) \right]_{i+\lhalf,j} +
\left[ F_2(\rho) \right]_{i-\lhalf,j}
\right).
\end{equation}

The van Leer interpolation function for advection of the $x_1$
component of a vector in the $x_1$ direction, ${\cal J}_1(\sigma_1)$,
is given by 
\begin{equation} \label{eq:j11}
\left[ {\cal J}_1(\sigma_1) \right]_{i+\lhalf,j+\lhalf} = 
\begin{cases} \displaystyle{
\left[ \sigma_1 \right]_{i,j+\lhalf} + 
\left[ \delta_1(\sigma_1) \right]_{i,j+\lhalf}
\left( 1 - \frac{ \left(
\left[ \varv_{1}\right]_{i,j+\lhalf} + \left[ \varv_{1}\right]_{i+1,j+\lhalf} 
\right) \Delta t }
{ 2 \left( 
\left[x_1\right]_{i+\lhalf} - \left[x_1\right]_{i-\lhalf} \right) } \right)
} \vspace{0.1in} \\
\hspace{2.0in}
\text{if} \; 
\left[ \varv_{1}\right]_{i,j+\lhalf} + \left[
\varv_{1}\right]_{i+1,j+\lhalf}   > 0 \vspace{0.5in} \\
\displaystyle{
\left[ \sigma_1 \right]_{i+1,j+\lhalf} - 
\left[ \delta_1(\sigma_1) \right]_{i+1,j+\lhalf}
\left( 1 + \frac{ \left(
\left[ \varv_{1}\right]_{i,j+\lhalf} + \left[\varv_{1}\right]_{i+1,j+\lhalf} 
\right)  \Delta t }
{2 \left( \left[x_1\right]_{i+\lthreehalf} - 
\left[x_1\right]_{i+\lhalf} \right)}
\right) 
} \vspace{0.1in} \\
\hspace{2.0in}
\text{if} \; \left[ \varv_{1}\right]_{i,j+\lhalf} + \left[
\varv_{1}\right]_{i+1,j+\lhalf}  
< 0, \\
\end{cases}
\end{equation}
where
\begin{equation} \label{eq:j11p1}
\left[ \delta_1(\sigma_1) \right]_{i,j+\lhalf} = 
\begin{cases}
\displaystyle{ 
\frac{
\left[ \Delta \sigma_1 \right]_{i-\lhalf,j+\lhalf}
\left[ \Delta \sigma_1 \right]_{i+\lhalf,j+\lhalf}
}
{\left[ \sigma_1 \right]_{i+1,j+\lhalf} -
\left[ \sigma_1 \right]_{i-1,j+\lhalf}}
} \vspace{0.1in} \\
\hspace{1.5in}
\text{if} \;
\left[ \Delta \sigma_1 \right]_{i-\lhalf,j+\lhalf}
\left[ \Delta \sigma_1 \right]_{i+\lhalf,j+\lhalf}
> 0 \vspace{0.25in} \\
0 \hspace{1.4in} \text{otherwise},
\end{cases}
\end{equation}
and
\begin{equation} \label{eq:delsig11}
\left[ \Delta \sigma_1 \right]_{i+\lhalf,j+\lhalf} = 
\left[ \sigma_1 \right]_{i+1,j+\lhalf} -
\left[ \sigma_1 \right]_{i,j+\lhalf}.
\end{equation}
The interpolation function needed for the calculation of the fluxes
in the $x_2$ direction can be obtained by the substitutions 
described in Table~\ref{vanleer}.

The analogs of equations~(\ref{eq:flux_comb_v1x1})
and (\ref{eq:flux_comb_v1x2}) for the $x_2$
components of a vector are straightforward to
obtain:
\begin{eqnarray} \label{eq:flux_comb_v2x1}
\lefteqn{ \frac{ \left[ \Delta V \right]_{i+\lhalf,j}}{\Delta t} 
\left( \left[ \sigma_2 \right]^{n+\delta}_{i+\lhalf,j} - 
\left[ \sigma_2 \right]^{n+\gamma}_{i+\lhalf,j} \right) 
= }
\nonumber \\ & & 
- \left( \left[ F_1(\sigma_2) 
\right]_{i+1,j}^{n+\gamma}
\left[ \Delta A_1 \right]_{i+1,j}  -
\left[ F_1(\sigma_2) \right]_{i,j}^{n+\gamma} 
\left[ \Delta A_1 \right]_{i,j} \right) 
\end{eqnarray}
and
\begin{eqnarray} \label{eq:flux_comb_v2x2}
\lefteqn{ \frac{ \left[ \Delta V \right]_{i+\lhalf,j}}{\Delta t} 
\left( \left[ \sigma_2 \right]^{n+\delta}_{i+\lhalf,j} - 
\left[ \sigma_2 \right]^{n+\gamma}_{i+\lhalf,j} \right) 
= }
\nonumber \\ & & 
- \left( \left[ F_2(\sigma_2) \right]_{i+\lhalf,j+\lhalf}^{n+\gamma} 
\left[ \Delta A_2 \right]_{i+\lhalf,j+\lhalf} -
\left[ F_2(\sigma_2) \right]_{i+\lhalf,j-\lhalf}^{n+\gamma} 
\left[ \Delta A_2 \right]_{i+\lhalf,j-\lhalf} \right).
\end{eqnarray}
For details we direct the reader to \cite{sn92a}.  The
fluxes are given by
\begin{equation} \label{eq:f1avg_2}
\left[ F_1(\sigma_2) \right]_{i,j} = 
\half 
\left[{\cal J}_1\left(\frac{\sigma_2}{\rho}\right)\right]_{i,j}
\left(
\left[ F_1(\rho) \right]_{i,j} +
\left[ F_1(\rho) \right]_{i+1,j}
\right)
\end{equation}
and
\begin{equation} \label{eq:f2avg_2}
\left[ F_2(\sigma_2) \right]_{i+\lhalf,j+\lhalf} = 
\half 
\left[{\cal J}_2\left(\frac{\sigma_2}{\rho}\right)\right]_{i+\lhalf,j+\lhalf}
\left(
\left[ F_2(\rho) \right]_{i+\lhalf,j+1} +
\left[ F_2(\rho) \right]_{i+\lhalf,j}
\right).
\end{equation}

The van Leer
interpolation function needed for the calculation of the fluxes
in the $x_1$ direction, ${\cal J}_1(\sigma_2)$, is
given by
\begin{equation} \label{eq:j12}
\left[ {\cal J}_1(\sigma_2) \right]_{i+\lhalf,j+\lhalf} = 
\begin{cases} \displaystyle{
\left[ \sigma_2 \right]_{i-\lhalf,j} + 
\left[ \delta_1(\sigma_2) \right]_{i-\lhalf,j}
\left( 1 - \frac{ \left(
\left[ \varv_{1}\right]_{i,j-\lhalf} + \left[ \varv_{1}\right]_{i,j+\lhalf} 
\right) \Delta t }
{ 2 \left( 
\left[x_1\right]_{i+\lhalf} - \left[x_1\right]_{i-\lhalf} \right) } \right)
} \vspace{0.1in} \\
\hspace{2.0in}
\text{if} \; 
\left[ \varv_{1}\right]_{i,j-\lhalf} + \left[
\varv_{1}\right]_{i,j+\lhalf}   > 0 \vspace{0.5in} \\
\displaystyle{
\left[ \sigma_2 \right]_{i+\lhalf,j} - 
\left[ \delta_1(\sigma_2) \right]_{i+\lhalf,j}
\left( 1 + \frac{ \left(
\left[ \varv_{1}\right]_{i,j-\lhalf} + \left[\varv_{1}\right]_{i,j+\lhalf} 
\right)  \Delta t }
{2 \left( \left[x_1\right]_{i+\lthreehalf} - 
\left[x_1\right]_{i+\lhalf} \right)}
\right) 
} \vspace{0.1in} \\
\hspace{2.0in}
\text{if} \; \left[ \varv_{1}\right]_{i,j-\lhalf} + \left[
\varv_{1}\right]_{i,j+\lhalf}  
< 0, \\
\end{cases}
\end{equation}
where
\begin{equation} \label{eq:j12p1}
\left[ \delta_1(\sigma_2) \right]_{i,j+\lhalf} = 
\begin{cases}
\displaystyle{ 
\frac{
\left[ \Delta \sigma_2 \right]_{i-1,j}
\left[ \Delta \sigma_2 \right]_{i,j}
}
{\left[ \sigma_2 \right]_{i+\lhalf,j} -
\left[ \sigma_2 \right]_{i-\lthreehalf,j}}
} \vspace{0.1in} \\
\hspace{1.5in}
\text{if} \;
\left[ \Delta \sigma_2 \right]_{i-1,j}
\left[ \Delta \sigma_2 \right]_{i,j}
> 0 \vspace{0.25in} \\
0 \hspace{1.4in} \text{otherwise},
\end{cases}
\end{equation}
and
\begin{equation}
\left[ \Delta \sigma_2 \right]_{i,j} = 
\left[ \sigma_2 \right]_{i+\lhalf,j} -
\left[ \sigma_2 \right]_{i-\lhalf,j}.
\end{equation}
The analogous ${\cal J}_2$ function can be obtained
by the appropriate substitutions as described in Table~\ref{vanleer}.

%
\begin{deluxetable}{llccc}
\tabletypesize{\small}
\tablecaption{{\sc Substitutions for Constructing Scalar and
Vector van Leer Interpolation Functions for Advection in the $x_2$
Direction from the $x_1$-Advection Definitions
}}
\tablewidth{0pt}
\tablehead{
%
%
\colhead{$x_1$ Advection Function \tablenotemark{a}} & 
\colhead{$x_2$ Advection Function \tablenotemark{b}}& 
\colhead{Corresponding Substitutions \tablenotemark{c}}
}
\startdata
$\left[ {\cal I}_1(\psi) \right]_{i,j+\lhalf}$ &
$\left[ {\cal I}_2(\psi) \right]_{i+\lhalf,j}$ &
$\left[ \psi \right]_{i-\lhalf,j+\lhalf} \longrightarrow 
\left[ \psi \right]_{i+\lhalf,j-\lhalf}$  
\\ & &
$\left[ \psi \right]_{i+\lhalf,j+\lhalf} \longrightarrow 
\left[ \psi  \right]_{i+\lhalf,j+\lhalf}$
\\ & &
$\left[ \delta_1(\psi) \right]_{i-\lhalf,j+\lhalf} \longrightarrow 
\left[  \delta_2(\psi) \right]_{i+\lhalf,j-\lhalf}$  
\\ & &
$\left[ \delta_1(\psi) \right]_{i+\lhalf,j+\lhalf} \longrightarrow 
\left[  \delta_2(\psi) \right]_{i+\lhalf,j+\lhalf}$
\\ & &
$\left[ \varv_1 \right]_{i,j+\lhalf} \longrightarrow 
\left[ \varv_2 \right]_{i+\lhalf,j}$ 
\\ & &
$\left[ x_1 \right]_{i} \longrightarrow \left[ x_2 \right]_{j}$
%
\\
%
%
\vspace{-0.1in}\\ 
$\left[ {\cal J}_1(\sigma_1) \right]_{i+\lhalf,j+\lhalf}$ &
$\left[ {\cal J}_2(\sigma_2) \right]_{i+\lhalf,j+\lhalf}$ &
$\left[ \sigma_1  \right]_{i,j+\lhalf} \longrightarrow 
\left[  \sigma_2  \right]_{i+\lhalf,j}$ 
\\ & &
$\left[ \sigma_1  \right]_{i+1,j+\lhalf}
\longrightarrow \left[ \sigma_2  \right]_{i+\lhalf,j+1}$ 
\\ & &
$\left[ \delta_1(\sigma_1) \right]_{i,j+\lhalf} \longrightarrow 
\left[  \delta_2(\sigma_2) \right]_{i+\lhalf,j}$ 
\\ & &
$\left[ \delta_1(\sigma_1) \right]_{i+1,j+\lhalf}
\longrightarrow \left[ \delta_2(\sigma_2)\right]_{i+\lhalf,j+1}$ 
\\ & &
$\left[ \varv_1 \right]_{i,j+\lhalf} \longrightarrow 
\left[ \varv_2 \right]_{i+\lhalf,j}$ 
\\ & &
$\left[ \varv_1 \right]_{i+1,j+\lhalf} \longrightarrow 
\left[ \varv_2 \right]_{i+\lhalf,j+1}$ 
\\ & &
$\left[ x_1 \right]_{i} \longrightarrow \left[ x_2 \right]_{j}$ 
\\
\vspace{-0.1in}\\ 
$\left[ {\cal J}_1(\sigma_1) \right]_{i+\lhalf,j+\lhalf}$ &
$\left[ {\cal J}_2(\sigma_1) \right]_{i,j}$ &
$\left[ \sigma_1 \right]_{i,j+\lhalf} \longrightarrow 
\left[  \sigma_1 \right]_{i,j-\lhalf}$ 
\\ & &
$\left[ \sigma_1  \right]_{i+1,j+\lhalf} \longrightarrow 
\left[  \sigma_1  \right]_{i,j+\lhalf} \quad$ 
\\ & &
$\left[ \delta_1(\sigma_1) \right]_{i,j+\lhalf} \longrightarrow 
\left[  \delta_2(\sigma_1)\right]_{i,j-\lhalf}$ 
\\ & &
$\left[ \delta_1(\sigma_1) \right]_{i+1,j+\lhalf} \longrightarrow 
\left[  \delta_2(\sigma_1) \right]_{i,j+\lhalf} \quad$
\\ & &
$\left[ \varv_1 \right]_{i,j+\lhalf} \longrightarrow 
\left[ \varv_2 \right]_{i-\lhalf,j}$ 
\\ & &
$\left[ \varv_1 \right]_{i+1,j+\lhalf} \longrightarrow 
\left[ \varv_2 \right]_{i+\lhalf,j} \quad$ 
\\ & &
$\; \: \left[ x_1 \right]_{i} \longrightarrow \left[ x_2 \right]_{j}$ 
\vspace{0.1in}
\enddata
\tablenotetext{a}{This column contains the van Leer interpolation
functions for advection in the $x_1$ direction that have been provided
in full in the text .}
\tablenotetext{b}{This column contains the remaining van Leer
interpolation functions not given explicitly in the text. These are
needed to describe advection in the $x_2$ direction.}
\tablenotetext{c}{This column contains the needed substitutions in
order to use the $x_1$ interpolation functions
(eqs.~[\ref{eq:i1}--\ref{eq:delsig1}] and
[\ref{eq:j11}--\ref{eq:delsig11}]) to construct the $x_2$ functions.}
\label{vanleer}
\end{deluxetable}

\subsection{Update Equations for an $x_1$ Advection Sweep}

Using the discretization scheme from the previous subsection
and substituting 
in the appropriate formulae for the cell interface areas,
we arrive at a set of update equations for a $x_1$ sweep.

Density update:
\begin{equation}
\left[ \rho \right]_{i+\lhalf,j+\lhalf}^{n+\beta} = 
\left[ \rho \right]_{i+\lhalf,j+\lhalf}^{n+\alpha}
- \Delta t 
\frac{ \left[ g_2 \right]_{i+1} \left[ g_{31} \right]_{i+1}
\left[ F_1(\rho) \right]^{n+\alpha}_{i+1,j+\lhalf} - 
\left[ g_2 \right]_{i} \left[ g_{31} \right]_{i}
\left[ F_1(\rho) \right]^{n+\alpha}_{i,j+\lhalf}}
{\left[ \Delta \Upsilon_1 \right]_{i+\lhalf}}.
\end{equation}
Electron flux:
\begin{equation}
\left[ F_1(n_e/\rho) \right]^{n+\alpha}_{i,j+\lhalf} = 
\left[ {\cal I}_1 \left(\frac{n_e}{\rho} \right) \right]^{n+\alpha}_{i,j+\lhalf} 
\left[ F_1(\rho) \right]^{n+\alpha}_{i,j+\lhalf}
\end{equation}
Matter-internal-energy-density flux:
\begin{equation}
\left[ F_1(E/\rho) \right]^{n+\alpha}_{i,j+\lhalf} = 
\left[ {\cal I}_1 \left( \frac{E}{\rho}\right) \right]^{n+\alpha}_{i,j+\lhalf} 
\left[ F_1(\rho) \right]^{n+\alpha}_{i,j+\lhalf}
\end{equation}
$x_1$-momentum-density flux:
\begin{equation}
\left[ F_1(s_1/\rho) \right]^{n+\alpha}_{i+\lhalf,j+\lhalf} = 
\left[ {\cal J}_1 \left( 
\varv_{1} \right) \right]]^{n+\alpha}_{i+\lhalf,j+\lhalf}
\left[ F_1(\rho) \right]]^{n+\alpha}_{i+\lhalf,j+\lhalf}
\end{equation}
$x_2$-momentum-density flux:
\begin{equation}
\left[ F_1(s_2/\rho) \right]^{n+\alpha}_{i,j} = 
\left[ {\cal J}_1 \left( 
\varv_{2} \right) \right]^{n+\alpha}_{i,j}
\left[ F_1(\rho) \right]^{n+\alpha}_{i,j}
\end{equation}
Radiation-energy-density flux:
\begin{equation}
\left[ F_1(E_\epsilon/\rho) \right]^{n+\alpha}_{k+\lhalf,i,j+\lhalf} = 
\left[ {\cal I}_1 \left( \frac{E_\epsilon}{\rho} \right) 
\right]^{n+\alpha}_{k+\lhalf,i,j+\lhalf} 
\left[ F_1(\rho) \right]^{n+\alpha}_{i,j+\lhalf}.
\end{equation}

Electron-density update:
\begin{eqnarray}
\left[ n_e \right]_{i+\lhalf,j+\lhalf}^{n+\beta} & = &
\left[ n_e \right]_{i+\lhalf,j+\lhalf}^{n+\alpha}
- \frac{ \Delta t }{\left[ \Delta \Upsilon_1 \right]_{i+\lhalf}}
\nonumber \\ & &
\times
\left(
\left[ g_2 \right]_{i+1} \left[ g_{31} \right]_{i+1}
\left[ F_1(n_e/\rho) \right]^{n+\alpha}_{i+1,j+\lhalf} - 
\left[ g_2 \right]_{i} \left[ g_{31} \right]_{i}
\left[ F_1(n_e/\rho) \right]^{n+\alpha}_{i,j+\lhalf}
\right),
\nonumber \\ 
\end{eqnarray}
Matter-internal-energy-density update:
\begin{eqnarray}
\left[ E \right]_{i+\lhalf,j+\lhalf}^{n+\beta} & = &
\left[ E \right]_{i+\lhalf,j+\lhalf}^{n+\alpha}
- \frac{\Delta t}{\left[ \Delta \Upsilon_1 \right]_{i+\lhalf}} 
\nonumber \\ & &
\times
\left(
\left[ g_2 \right]_{i+1} \left[ g_{31} \right]_{i+1}
\left[ F_1(E) \right]^{n+\alpha}_{i+1,j+\lhalf} - 
\left[ g_2 \right]_{i} \left[ g_{31} \right]_{i}
\left[ F_1(E) \right]^{n+\alpha}_{i,j+\lhalf}
\right),
\end{eqnarray}
$x_1$-momentum-density update:
\begin{eqnarray}
\left[ s_1 \right]_{i,j+\lhalf}^{n+\beta} & = &
\left[ s_1 \right]_{i,j+\lhalf}^{n+\alpha}
- \frac{\Delta t}{\left[ \Delta \Upsilon_1 \right]_{i}}
\left(
\left[ g_2 \right]_{i+\lhalf} \left[ g_{31} \right]_{i+\lhalf}
\left[ F_1(s_1/\rho) \right]^{n+\alpha}_{i+\lhalf,j+\lhalf}
\right.
\nonumber \\ & &
\quad \quad \quad \quad \quad \quad \quad \quad \; \; \;
\left.
- \left[ g_2 \right]_{i-\lhalf} \left[ g_{31} \right]_{i-\lhalf}
\left[ F_1(s_1/\rho) \right]^{n+\alpha}_{i-\lhalf,j+\lhalf}
\right),
\end{eqnarray}
$x_2$-momentum-density update:
\begin{eqnarray}
\left[ s_2 \right]_{i+\lhalf,j}^{n+\beta} & = &
\left[ s_2 \right]_{i+\lhalf,j}^{n+\alpha}
- \frac{\Delta t}{\left[ \Delta \Upsilon_1 \right]_{i+\lhalf}}
\left( \left[ g_2 \right]_{i+1} \left[ g_{31} \right]_{i+1}
\left[ F_1(s_2/\rho) \right]^{n+\alpha}_{i+1,j}
\right.
\nonumber \\ & &
\left.
\quad \quad \quad \quad \quad \quad \quad \quad \quad \quad \quad 
- \left[ g_2 \right]_{i} \left[ g_{31} \right]_{i}
\left[ F_1(s_2/\rho) \right]^{n+\alpha}_{i,j}
\right),
\end{eqnarray}
Radiation-energy-density update:
\begin{eqnarray}
\left[ E_\epsilon \right]_{k+\lhalf,i+\lhalf,j+\lhalf}^{n+\beta} & = &
\left[ E_\epsilon \right]_{k+\lhalf,i+\lhalf,j+\lhalf}^{n+\alpha}
- \frac{\Delta t}{\left[ \Delta \Upsilon_1 \right]_{i+\lhalf}}
\nonumber \\ & &
\times \left( \left[ g_2 \right]_{i+1} \left[ g_{31} \right]_{i+1}
\left[ F_1(E_\epsilon) \right]^{n+\alpha}_{k+\lhalf,i+1,j+\lhalf} 
\right.
\nonumber \\ & &
\quad \; \;
\left.
- \left[ g_2 \right]_{i} \left[ g_{31} \right]_{i}
\left[ F_1(E_\epsilon) \right]^{n+\alpha}_{k+\lhalf,i,j+\lhalf}
\right).
\end{eqnarray}

\subsection{Update Equations for an $x_2$ Advection Sweep}

For the $x_2$ advection sweep we obtain the following set 
of update equations.

Density update:
\begin{eqnarray}
\left[ \rho \right]_{i+\lhalf,j+\lhalf}^{n+\delta} & = &
\left[ \rho \right]_{i+\lhalf,j+\lhalf}^{n+\gamma}
- \Delta t \left[ g_{31} \right]_{i+\lhalf} 
\left( \left[ x_1 \right]_{i+1} - \left[ x_1 \right]_{i} \right)
\nonumber \\ & &
\times
\left(\frac{ \left[ g_{32} \right]_{j+1}
\left[ F_2(\rho) \right]^{n+\gamma}_{i+\lhalf,j+1} - 
\left[ g_{32} \right]_{j}
\left[ F_2(\rho) \right]^{n+\gamma}_{i+\lhalf,j}}
{\left[ \Delta \Upsilon_1 \right]_{i+\lhalf} 
\left[ \Delta \Upsilon_2 \right]_{j+\lhalf}}\right).
\end{eqnarray}
Electron flux:
\begin{equation}
\left[ F_2(n_e/\rho) \right]^{n+\gamma}_{i+\lhalf,j} = 
\left[ {\cal I}_2 \left(\frac{n_e}{\rho}\right) \right]^{n+\gamma}_{i+\lhalf,j} 
\left[ F_2(\rho) \right]^{n+\gamma}_{i+\lhalf,j}
\end{equation}
Matter-internal-energy-density flux:
\begin{equation}
\left[ F_2(E/\rho) \right]^{n+\gamma}_{i+\lhalf,j} = 
\left[ {\cal I}_2 \left( \frac{E}{\rho}\right) \right]^{n+\gamma}_{i+\lhalf,j} 
\left[ F_2(\rho) \right]^{n+\gamma}_{i+\lhalf,j} 
\end{equation}
$x_1$-momentum-density flux:
\begin{equation}
\left[ F_2(s_1/\rho) \right]^{n+\gamma}_{i,j} = 
\left[ {\cal J}_2 (\varv_{1}) \right]^{n+\gamma}_{i,j}
\left[ F_2(\rho) \right]^{n+\gamma}_{i,j} 
\end{equation}
$x_2$-momentum-density flux:
\begin{equation}
\left[ F_2(s_2/\rho) \right]^{n+\gamma}_{i+\lhalf,j+\lhalf} = 
\left[ {\cal J}_2 (\varv_{2}) \right]^{n+\gamma}_{i+\lhalf,j+\lhalf}
\left[ F_2(\rho) \right]^{n+\gamma}_{i+\lhalf,j+\lhalf}
\end{equation}
Radiation-energy-density flux:
\begin{equation}
\left[ F_2(E_\epsilon/\rho) \right]^{n+\gamma}_{k+\lhalf,i+\lhalf,j} = 
\left[ {\cal I}_2 \left( \frac{E_\epsilon}{\rho}\right) 
\right]^{n+\gamma}_{k+\lhalf,i+\lhalf,j} 
\left[ F_2(\rho) \right]^{n+\gamma}_{i+\lhalf,j}.
\end{equation}

Electron-density update:
\begin{eqnarray}
\left[ n_e \right]_{i+\lhalf,j+\lhalf}^{n+\delta} & = &
\left[ n_e \right]_{i+\lhalf,j+\lhalf}^{n+\gamma} -
\frac{\Delta t \left[ g_{31}\right]_{i+\lhalf}
\left( \left[ x_1 \right]_{i+1} - \left[ x_1 \right]_{i} \right)}
{\left[ \Delta \Upsilon_1 \right]_{i+\lhalf}
 \left[ \Delta \Upsilon_2 \right]_{j+\lhalf}} 
\nonumber \\ & & 
\times
\left( \left[ g_{32}\right]_{j+1}
\left[ F_2(n_e/\rho) \right]^{n+\gamma}_{i+\lhalf,j+1} - 
\left[ g_{32}\right]_{j}
\left[ F_2(n_e/\rho) \right]^{n+\gamma}_{i+\lhalf,j}
\right),
\end{eqnarray}
Matter-internal-energy update:
\begin{eqnarray}
\left[ E \right]_{i+\lhalf,j+\lhalf}^{n+\delta} & = &
\left[ E \right]_{i+\lhalf,j+\lhalf}^{n+\gamma} -
\frac{\Delta t \left[ g_{31}\right]_{i+\lhalf}
\left( \left[ x_1 \right]_{i+1} - \left[ x_1 \right]_{i} \right)}
{\left[ \Delta \Upsilon_1 \right]_{i+\lhalf} 
\left[ \Delta \Upsilon_2 \right]_{j+\lhalf}} 
\nonumber \\ & &
\times
\left( \left[ g_{32}\right]_{j+1}
\left[ F_2(E/\rho) \right]^{n+\gamma}_{i+\lhalf,j+1} - 
\left[ g_{32}\right]_{j}
\left[ F_2(E/\rho) \right]^{n+\gamma}_{i+\lhalf,j}
\right),
\end{eqnarray}
$x_1$-momentum-density update:
\begin{eqnarray}
\left[ s_1 \right]_{i,j+\lhalf}^{n+\delta} & = &
\left[ s_1 \right]_{i,j+\lhalf}^{n+\gamma} -
\frac{ \Delta t \left[ g_{31}\right]_{i} 
( \left[ x_1 \right]_{i+\lhalf} - \left[ x_1 \right]_{i-\lhalf} ) }
{\left[ \Delta \Upsilon_1 \right]_{i} 
\left[ \Delta \Upsilon_2 \right]_{j+\lhalf}} 
\nonumber \\ & &
\times
\left(
\left[ g_{32}\right]_{j+1}
\left[ F_2(s_1/\rho) \right]^{n+\gamma}_{i,j+1} - 
\left[ g_{32}\right]_{j}
\left[ F_2(s_1/\rho) \right]^{n+\gamma}_{i,j}
\right),
\end{eqnarray}
$x_2$-momentum-density update:
\begin{eqnarray}
\left[ s_2 \right]_{i+\lhalf,j}^{n+\delta} & = &
\left[ s_2 \right]_{i+\lhalf,j}^{n+\gamma} -
\frac{ \Delta t \left[ g_{31}\right]_{i+\lhalf}
\left( \left[ x_1 \right]_{i+1} - \left[ x_1 \right]_{i} \right) }
{\left[ \Delta \Upsilon_1 \right]_{i+\lhalf} 
\left[ \Delta \Upsilon_2 \right]_{j}} 
\nonumber \\ & &
\times
\left(
\left[ g_{32}\right]_{j+\lhalf}
\left[ F_2(s_2/\rho) \right]^{n+\gamma}_{i+\lhalf,j+\lhalf} - 
\left[ g_{32}\right]_{j-\lhalf}
\left[ F_2(s_2/\rho) \right]^{n+\gamma}_{i+\lhalf,j-\lhalf}
\right)
\end{eqnarray}
Radiation-energy-density update:
\begin{eqnarray}
\left[ E_\epsilon \right]_{k+\lhalf,i+\lhalf,j+\lhalf}^{n+\delta} = 
\left[ E_\epsilon \right]_{k+\lhalf,i+\lhalf,j+\lhalf}^{n+\gamma} -
\frac{ \Delta t \left[ g_{31}\right]_{i+\lhalf}
\left( \left[ x_1 \right]_{i+1} - \left[ x_1 \right]_{i} \right) }
{\left[ \Delta \Upsilon_1 \right]_{i+\lhalf} 
\left[ \Delta \Upsilon_2 \right]_{j+\lhalf}} 
\nonumber \\ 
\times
\left(
\left[ g_{32}\right]_{j+1}
\left[ F_2(E_\epsilon/\rho) \right]^{n+\gamma}_{k+\lhalf,i+\lhalf,j+1} - 
\left[ g_{32}\right]_{j}
\left[ F_2(E_\epsilon/\rho) \right]^{n+\gamma}_{k+\lhalf,i+\lhalf,j}
\right).
\end{eqnarray}

%
\begin{deluxetable}{lccccc}
\tabletypesize{\normalsize}
\tablecaption{{\sc Advection Time-superscript Values}}
\tablewidth{0pt}
\tablehead{
\colhead{Direction of First Sweep} & 
\colhead{Direction of Second Sweep} &
\colhead{$\alpha$} &
\colhead{$\beta$} &
\colhead{$\gamma$} &
\colhead{$\delta$}
}
\startdata
$x_1$           &   $x_2$   &   $0$   &   $a$    & $a$     &   $b$   \\
$x_2$           &   $x_1$   &   $a$   &   $b$    & $0$     &   $a$   \\
\enddata
\label{tab:timelev}
\end{deluxetable}
%
%

\subsection{Post-Sweep Update Equations}

After the advected quantities have been updated during an advection sweep,
there are several related updates that must
be performed to render certain related quantities, such as the momentum
densities and velocities, consistent with one another.  
The electron fraction $Y_e$ should be updated to make it consistent
with the electron number density $n_e$.
The two quantities are related by $Y_e = n_e m_b/\rho$, where
$m_b$ is the baryon mass.  In discretized form, this
becomes
\begin{equation}
\left[ Y_e \right]^\omega_{i+\lhalf, j+\lhalf} =
m_b \frac{ \left[ n_e \right]^\omega_{i+\lhalf, j+\lhalf} }
{ \left[ \rho \right]^\omega_{i+\lhalf, j+\lhalf} }
\end{equation}
where $\omega=a$ after the first advection sweep and 
$\omega=b$ after the second advection sweep.
We must also update the velocity field that has
changed as a result of momentum being advected.  This is 
accomplished as follows
\begin{equation}
\left[ \varv_1 \right]^\omega_{i,j+\lhalf} = 
\frac{2 \left[ s_1 \right]^\omega_{i,j+\lhalf}}
{\left[ \rho \right]^\omega_{i+\lhalf,j+\lhalf} + 
\left[ \rho \right]^\omega_{i-\lhalf,j+\lhalf}}
\end{equation}
and
\begin{equation}
\left[ \varv_2 \right]^\omega_{i+\lhalf,j} = 
\frac{2 \left[ s_2 \right]^\omega_{i+\lhalf,j}}
{{\left[ g_2 \right]^\omega_{i+\lhalf}}
\left(
\left[ \rho \right]^\omega_{i+\lhalf,j+\lhalf} + 
\left[ \rho \right]^\omega_{i+\lhalf,j-\lhalf}
\right)}
.
\end{equation}
Finally, the temperature and pressure should be related to 
render them consistent with the updated density, electron fraction,
and matter internal energy density.
The procedure for doing this update is the same
as that described in appendix \ref{app:collision}, and we refer the 
reader there for detail of this procedure.  Since the 
temperature and pressure play no role in the advection equations,
we perform this update only after the completion of the 
second advection sweep.  The equation that must be iteratively
solved for the new temperature $[T]^{n+b}$ is:
\begin{equation} \label{eq:eos_inv_h_2}
E\left(
\left[T \right]^{n+b}_{i+(1/2),j+(1/2)},
\left[\rho \right]^{n+b}_{i+(1/2),j+(1/2)},
\left[Ye \right]^{n+b}_{i+(1/2),j+(1/2)}
\right)
-
\left[E \right]^{n+b}_{i+(1/2),j+(1/2)}
= 0.
\end{equation}

\subsection{The Complete Advective Update}

With the complete prescription for the two advection sweeps now derived
and differenced, we are finally prepared to assemble the complete
sequence of operations of the advection algorithm.  This sequence of
steps is shown in Figure~\ref{fig:advection} and corresponds to 
box $(a),(b)$ of Figure~\ref{fig:timestep}.  The values for the 
time superscripts $\alpha$, $\beta$, $\gamma$, and $\delta$,
which depend on the order of the advection sweeps, are
given in Table~\ref{tab:timelev}.
%
%
\begin{figure}[htbp]
\begin{center}
\vspace{0.5in}
\includegraphics[scale=0.70]{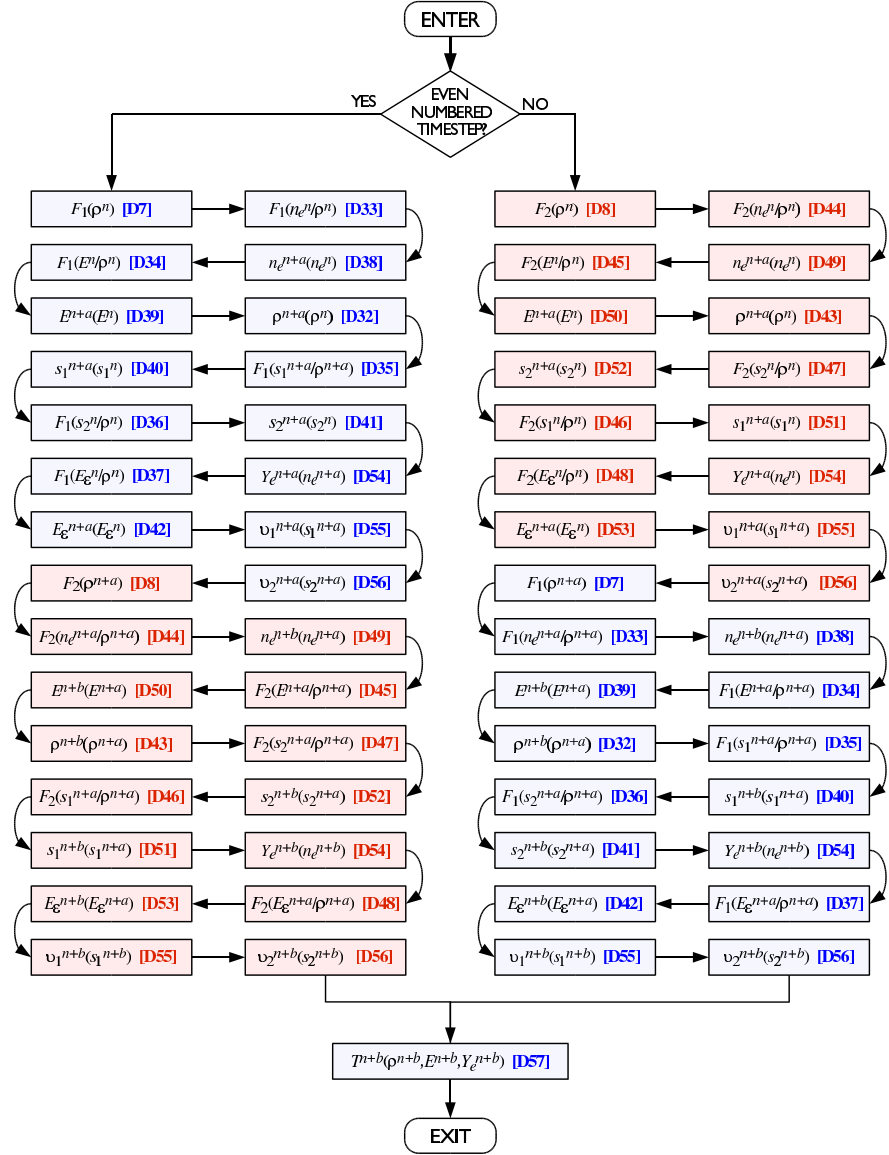}
\end{center}
\vspace{-1.0in}
\caption{\label{fig:advection} A flowchart of the advection sweep
algorithm.  The blue-shaded sections of the figure show the
calculations necessary for an advection sweep in the $x_1$
direction. The red-shaded regions show the analogous steps for an
$x_2$ sweep.  The equation numbers corresponding to each step of the
algorithm are indicated.  The initial decision at the top of the
algorithm sets up the directional sweeps so that they alternate order
with timestep.  }
\end{figure}
%
%

\section{Discretization of Viscous Dissipation Equations\label{app:artvisc}}


To insure the accurate treatment of shocks, the ZEUS algorithm \citep{sn92a}
includes viscous dissipation in the form of a 
term that mimics the viscous stresses actually
present at a real shock front.  
Were such a term omitted from our
model, the hydrodynamic equations would cause oscillations to develop
in the solution.  Such oscillations would result from the lack of any
means to generate entropy in the finite-difference scheme and satisfy
the Rankine-Hugoniot jump conditions across the shock. (For more
detail on the choice and implementation of artificial viscosity
schemes, see \citet{bw91}.)

In this work we follow \cite{sn92a}, but consider only pseudo-tensorial 
artificial viscosities.  It has
been widely known for sometime \citep{tw79} that tensor artificial
viscosities offer superior qualities to scalar schemes.  In future
work, we will consider such prescriptions, but for now we restrict
ourselves to a scalar approach.  
Readers interested in tensorial forms for the viscous dissipation
should consult \cite{sn92a}.
The ZEUS scheme employs a form of 
the von Neumann-Richtmyer scheme \citep{noh87,rm67},
where the viscous stress in the $i$th direction is given by
\begin{equation}
Q_i = \left\{
\begin{array}{ll} 
\displaystyle{
w^2 \rho \left( \frac{\partial \varv_i}{\partial x_i} \right)^2} & 
\mbox{if $\partial \varv_i/\partial x_i < 0$} \\
0 & \mbox{if $\partial \varv_i/\partial x_i > 0$,}
\end{array}
\right.
\end{equation}
where $w$ is a characteristic length.  With the substitution of 
$l_q = w/\Delta x_i$, where $\Delta x_i$ is the width of the
computational zone in the $i$th direction, we have, upon
differencing, 
\begin{equation}
\left[ Q_1 \right]_{i+\lhalf, j+\lhalf} = 
l_q^2 \left[ \rho \right]_{i+\lhalf, j+\lhalf}^{n+1}
\left\{
\min \left( \left[ \varv_1 \right]_{i+1,j}^{n+1} - 
\left[ \varv_1 \right]_{i,j}^{n+1}, 0 \right)
\right\}^2
\end{equation}
\begin{equation}
\left[ Q_2 \right]_{i+\lhalf, j+\lhalf} = 
l_q^2 \left[ \rho \right]_{i+\lhalf, j+\lhalf}^{n+1}
\left\{
\min \left( \left[ \varv_2 \right]_{i+1,j}^{n+1} - 
\left[ \varv_2 \right]_{i,j}^{n+1}, 0 \right)
\right\}^2.
\end{equation}
By doing this, we have rewritten the characteristic length $w$ in
terms of a characteristic zone count.  For all calculations, 
we set $l_q^2=2$.  The quantity $l_q$ corresponds to the
approximate number of zones over which the shock is spread.

These forms enter into the discretized analogs of equations 
(\ref{eq:e-visc}) and (\ref{eq:s-visc}) exactly as described by 
equations~(35)--(37) of \cite{sn92a} and we refer the interested
reader there for more detail.
We difference equation~(\ref{eq:e-visc}) as
\begin{eqnarray} 
\left[ E \right]^{n+1}_{i+\lhalf,j+\lhalf} & = &
\left[ E \right]^{n}_{i+\lhalf,j+\lhalf} 
\nonumber \\ & &
- \Delta t \left\{ \left[ Q_1 \right]_{i+\lhalf,j+\lhalf}
\left( 
\frac{\left[ \varv_{1} \right]_{i+1,j+\lhalf} - 
\left[ \varv_{1} \right]_{i,j+\lhalf}} 
{ \left[ x_1 \right]_{i+1} - \left[ x_1 \right]_{i}}
\right) \right.
\nonumber \\ & & \nonumber \\ & & 
\left.
\quad\:\:\:\:\:
+ \frac{\left[ Q_2 \right]_{i+\lhalf,j+\lhalf}}
{ \left[ g_2 \right]_{i+\lhalf}} 
\left( 
\frac{\left[ \varv_{2} \right]_{i+\lhalf,j+1} - 
\left[ \varv_{2} \right]_{i+\lhalf,j}}
{\left[ x_2 \right]_{j+1} - \left[ x_2 \right]_{j}}
\right) \right\}.
\end{eqnarray} 
The components of equation~(\ref{eq:s-visc}) are differenced as
\begin{eqnarray} \label{eq:visc_x1}
\left[ \varv_{1} \right]^{n+i}_{i,j+\lhalf}  =
\left[ \varv_{1} \right]^{n+b}_{i,j+\lhalf} -
\frac{2 \Delta t}
{\left[ x_1 \right]_{i+\lhalf} - \left[ x_1 \right]_{i-\lhalf}}
\left(
\frac{ \left[ Q_1 \right]^{n+h}_{i+\lhalf,j+\lhalf} - 
\left[ Q_1 \right]^{n+h}_{i-\lhalf,j+\lhalf}}
{ \left[ \rho \right]^{n+1}_{i+\lhalf,j+\lhalf} + 
\left[ \rho \right]^{n+1}_{i-\lhalf,j+\lhalf}}
\right)
\end{eqnarray}
and
\begin{eqnarray} \label{eq:visc_x2}
\left[ \varv_{2} \right]^{n+i}_{i+\lhalf,j} & = &
\left[ \varv_{1} \right]^{n+b}_{i+\lhalf,j}  -  
\frac{2 \Delta t}{ \left[ g_2 \right]_{i+\lhalf}
\left( 
\left[ x_2 \right]_{j+\lhalf} - \left[ x_2 \right]_{j-\lhalf}
\right) }  \nonumber \\ \nonumber \\
& & \times \left(
\frac{ \left[ Q_2 \right]^{n+h}_{i+\lhalf,j+\lhalf} - 
\left[ Q_2 \right]^{n+h}_{i+\lhalf,j-\lhalf}}
{ \left[ \rho \right]^{n+1}_{i+\lhalf,j+\lhalf} + 
\left[ \rho \right]^{n+1}_{i+\lhalf,j-\lhalf}}
\right).
\end{eqnarray}

\section{Discretization of Gas-Momentum Source Equation  \label{app:gmomsrc}}


The solution of equation~(\ref{eq:s-source}), in substep $i$, accounts
for accelerations due to fluid pressure gradients and gravitational
forces.
Since $\rho$ remains unchanged during this step of the algorithm, it
can emerge from the time derivative in equation
(\ref{eq:s-source}).  Thus, we rewrite equation~(\ref{eq:s-source}) as
\begin{equation}
\left\ldbrack \frac{ \partial \left( \varv_1 \: {\hat{\bf x}}_1 + 
\varv_2 \: {\hat{\bf x}}_2\right) }
{\partial t}\right\rdbrack_{\rm matter}  = 
\frac{1}{\rho}{\bfnabla} P + {\bfnabla} \Phi,
\end{equation}
where ${\hat{\bf x}}_1$ and ${\hat{\bf x}}_2$ are unit vectors in the
$x_1$ and $x_2$ directions.  Upon differencing, we obtain expressions for
updated velocities.  In the $x_1$ direction, we have
\begin{eqnarray} \label{eq:mom_x1}
\lefteqn {
\left[ \varv_{1} \right]^{n+i}_{i,j+\lhalf}  =  }
\nonumber \\ & &
\left[ \varv_{1} \right]^{n+b}_{i,j+\lhalf} -
\frac{2 \Delta t}
{ \left[ x_1 \right]_{i+\lhalf} - \left[ x_1 \right]_{i-\lhalf}}
\left(
\frac{ \left[ P \right]^{n+h}_{i+\lhalf,j+\lhalf} - 
\left[ P \right]^{n+h}_{i-\lhalf,j+\lhalf}}
{ \left[ \rho \right]^{n+1}_{i+\lhalf,j+\lhalf} + 
\left[ \rho \right]^{n+1}_{i-\lhalf,j+\lhalf}}
\right)
\nonumber \\ & & 
- \frac{2 \Delta t}{ \left[ x_1 \right]_{i+\lhalf} - 
\left[ x_1 \right]_{i-\lhalf}}
\left( \left[ \Phi \right]^n_{i+\lhalf,j+\lhalf} - 
\left[ \Phi \right]^n_{i-\lhalf,j+\lhalf} \right)
\nonumber \\ & & 
+ \frac{\Delta t}{ \left[ g_2 \right]_{i}}
\left[ \frac{\partial g_2}{\partial x_1} \right]_{i}
\left\{ \frac{1}{4} 
\left( \left[ \varv_2 \right]^{n+b}_{i+\lhalf,j} + 
\left[ \varv_2 \right]^{n+b}_{i+\lhalf,j+1} +
\left[ \varv_2 \right]^{n+b}_{i-\lhalf,j} + 
\left[ \varv_2 \right]^{n+b}_{i-\lhalf,j+1} \right) \right\}^2,
\end{eqnarray}
where the last term is the contribution of the time derivative of
${\hat{\bf x}}_2$.  (The time derivative of ${\hat{\bf x}}_1$ yields
zero contribution to this term.)  The averaging of the nearest four
face-centered values of $\varv_2$ in the $x_1$--$x_3$ plane gives the
value at the need location for calculation of an $\varv_1$ velocity,
{\em viz.}, at $i,j+(1/2)$.

Similarly, in the $x_2$ direction, we have
\begin{eqnarray} \label{eq:mom_x2}
\left[ \varv_{2} \right]^{n+i}_{i+\lhalf,j} & = & 
\left[ \varv_{1} \right]^{n+b}_{i+\lhalf,j}
- \frac{2 \Delta t}{ \left[ g_2 \right]_{i+\lhalf}
\left( 
\left[ x_2 \right]_{j+\lhalf} - \left[ x_2 \right]_{j-\lhalf}
\right) }
\nonumber \\ & &
\times \left(
\frac{ \left[ P \right]^{n+h}_{i+\lhalf,j+\lhalf} - 
\left[ P \right]^{n+h}_{i+\lhalf,j-\lhalf}}
{ \left[ \rho \right]^{n+1}_{i+\lhalf,j+\lhalf} + 
\left[ \rho \right]^{n+1}_{i+\lhalf,j-\lhalf}}
\right).
\end{eqnarray}
Here, we note that for spherical self gravity, the $x_2$ component of
${\bf \nabla} \Phi$ vanishes.  In the case of a general gravitational
potential, arising from a non-symmetric solution to the Poisson Equation,
a non-zero $\partial \Phi / \partial x_2$ would be present.  In the
$x_2$ direction, the contribution of the time derivative of ${\bf
x}_1$ is absorbed into the definition of the velocity.  This follows
from our definition of $\varv_i$, which {\em always} has the dimension
of length over time, even if the dimension of $x_i$, itself, is one of
angle (see \S\ref{sec-covariant}).

In this algorithm, we assume a 1-D approximation to 
self-gravity by assuming the mass
distribution behaves as if it were spherically symmetric.
Although not exact, the highly condensed nature of a collapsed stellar
core makes this approximation satisfactory for supernova models.  With
this assumption of a spherically symmetric mass distribution, we obtain
a spherically symmetric gravitational potential of the form
\begin{equation}
\Phi(r,t) = -G \frac{M(r,t)}{r},
\end{equation}
where $M(r,t)$ is the gravitational mass contained within the radius
$r$ at time $t$.  Note that we we have now specified spherical-polar
coordinates for this calculation.  Although, this is the most
convenient form for problems in spherical geometry, another coordinate
system could be used, with the appropriate transformation.

We evaluate $M(r,t)$ as follows.  In general, an element of mass is
given by
\begin{equation}
dM({\bf x},t) = \rho({\bf x},t) d\Upsilon_1 d\Upsilon_2 \Upsilon_3,
\end{equation}
which in spherical geometry is
\begin{equation}
dM(r,\theta,t) = 2 \pi \rho(r,\theta,t) r^2 dr \sin \theta d\theta.
\end{equation}
The mass, $M(r)$, is evaluated by
performing a radial integral out to the point of interest.  When
combined with an angular integration, we obtain an angularly averaged
mass distribution,
\begin{equation}
M(r,t) = \frac{\pi}{\theta_2 - \theta_1}
\int_0^r \int^{\theta_2}_{\theta_1} 2 \pi \rho(r,\theta) r^2 dr 
\sin \theta d\theta, 
\end{equation}
where the averaging pre-factor in front of the integral allows the
computational grid, which runs from $\theta_1$ to $\theta_2$, to be of
arbitrary extent---not necessarily reaching either pole.

Finally, the differenced version of these expressions can be written.
Note, that although the overall timestep advancement is from times
$[t]^n$ to $[t]^{n+1}$, since a all quantities that go into the
gravitational-potential calculation are evaluated at time $t^n$, the
result is also given a time subscript of $n$:
\begin{equation} \label{eq:gmass}
\left[ M \right]_{i+1}^{n} = 
\frac{2 \pi^2}
{ \left[ \theta \right]_{j_{\rm max}} - 
\left[ \theta \right]_{j_{\rm min}} }
\sum_{i=i_{\rm min}}^{i}
\sum_{j=j_{\rm min}}^{j_{\rm max}}
\left[ \rho \right]^{n}_{i+\lhalf, j+\lhalf} 
\left[ \Delta \Upsilon_1 \right]_{i+\lhalf}
\left[ \Delta \Upsilon_2 \right]_{j+\lhalf},
\end{equation}
with
\begin{equation} 
\left[ \Phi \right]_i^{n} = 
-G \frac{ \left[ M \right]^{n}_{i} }{\left[ r \right]_{i}}.
\label{eq:potential_sph}
\end{equation}
We use $i_{\rm min}, i_{\rm max}, j_{\rm min}$, and $j_{\rm max}$ to
indicate lower and upper bounds of $x_1$ and $x_2$ dimensions,
respectively (see Appendix~\ref{app:bconds}). Expressions in
equations~(\ref{eq:gmass}) and (\ref{eq:potential_sph}) are
evaluated once at the beginning of a timestep and kept constant until
the next timestep.

\section{Discretization of Lagrangean Gas-Energy Equation \label{app:gas-lagrange}}


Equation~(\ref{eq:e-source}) accounts for thermodynamic work done on the
fluid internal energy by compression and/or expansion.   Our discretization 
is a generalization of scheme of \cite{sn92a}, which allows an arbitrary
equation of state in which the internal energy and pressure are 
expressed as a function of temperature and density.  This solution method
makes no assumptions about convexity of the EOS.

We implicitly difference equation~(\ref{eq:e-source}) as
\begin{eqnarray} 
& & 
E\left(
\left[T \right]^{n+1}_{i+(1/2),j+(1/2)}, 
\left[\rho \right]^{n+1}_{i+(1/2),j+(1/2)}, 
\left[Y_e \right]^{n+1}_{i+(1/2),j+(1/2)} 
\right) - 
\left[ E \right]^{n}_{i+\lhalf,j+\lhalf}
\nonumber \\
& & + \frac{\Delta t }{2}
\left[ {\bfnabla}\cdot{\bf v} \right]^{n+1}_{i+\lhalf,j+\lhalf}
\nonumber \\ & &
\times \left\{ 
P\left(
\left[T \right]^{n+1}_{i+(1/2),j+(1/2)}, 
\left[\rho \right]^{n+1}_{i+(1/2),j+(1/2)}, 
\left[Y_e \right]^{n+1}_{i+(1/2),j+(1/2)} 
\right)
+
\left[ P \right]^{n}_{i+\lhalf,j+\lhalf} \right\} = 0,
\label{eq:e-source-id}
\end{eqnarray} 
where the first and third terms of equation~(\ref{eq:e-source})
contain the EOS in the form of equations~(\ref{eq:eos1}) and
(\ref{eq:eos2}), respectively, and where
\begin{eqnarray} 
\left[ {\bfnabla}\cdot{\bf v} \right]^{n+1}_{i+\lhalf,j+\lhalf} & = &
\frac{1}{\left[ \Delta \Upsilon_1 \right]_{i+\lhalf}} 
\left( \left[ g_2 \right]_{i+1} \left[ g_{31} \right]_{i+1}
\left[ \varv_{1} \right]_{i+1,j+\lhalf} -
\left[ g_2 \right]_{i} \left[ g_{31} \right]_{i} 
\left[ \varv_{1} \right]_{i,j+\lhalf} \right) 
\nonumber \\ & &
+ \frac{1}{ \left[ g_2 \right]_{i+\lhalf}
\left[ \Delta \Upsilon_2 \right]_{j+\lhalf}}
\left( \left[ g_{32} \right]_{j+1} 
\left[ \varv_{2} \right]_{i+\lhalf,j+1} -
\left[ g_{32} \right]_{j} 
\left[ \varv_{2} \right]_{i+\lhalf,j} \right).
\end{eqnarray} 
For a realistic equation of state, this equation cannot be solved
explicitly, since there is no explicit way of expressing matter
pressure as a simple function of internal energy.  
This equation must be solved iteratively for the new temperature
$[ T]^\npo_{i+\lhalf,j+\lhalf}$, which is unknown and
enters parametrically into the new internal energy and pressure
through the equation of state as described in equations
(\ref{eq:eos1}) and (\ref{eq:eos2}).
Since the solution of equation~(\ref{eq:e-source}) is the last step
in the algorithm, the new value of the electron fraction and density
are known prior to the solution of equation~(\ref{eq:e-source-id}).
In practice, we employ a combination of Newton-Raphson and
bisection algorithms to accomplish the iterative solution
of equation~(\ref{eq:e-source-id}).  The presence of first-order
phase transitions 
in complex equations of state, such as those describing 
matter in core-collapse supernovae \citep{ls91}, 
can create convergence problems for the solution 
of equation~(\ref{eq:e-source-id}) Newton-Raphson iteration.
These same phase transitions also exhibit non-convex behavior, 
which can also pose problems for Riemann solver based methods.
When non-convergence occurs during the iterative solution of 
equation~(\ref{eq:e-source-id}), we make use of the 
slower converging, but more robust, bisection method for
single-variable nonlinear equations.

\section{Discretization of Radiation Transport Diffusive/Collision Equation \label{app:rad-trans}}


\subsection{Pair-Coupling of Equations}

In this appendix, we discuss the implicit solution
of the sets of diffusion equations for the
radiation represented by equation
(\ref{eq:nu-diff}).  These equations are solved as
pair-coupled sets described by boxes ($c$), ($e$), and
($g$) of Figure~\ref{fig:timestep}.  By the use of
the terminology ``pair-coupled,'' we means that
discretized version of equation~(\ref{eq:nu-diff})
and its antineutrino analog are solved
simultaneously for all energy groups for a given
type of neutrino (electron, muon, or tauon).  For
example, in the step represented by box ({\em c}) of
Figure~\ref{fig:timestep}, the nonlinear equations
\begin{equation}
\label{eq:dc-e}
\left\ldbrack \frac{\partial (^eE_{\epsilon})}{\partial t} 
\right\rdbrack_{\rm radiation} -
{\bfnabla} \cdot \left(^eD_\epsilon ^eE_{\epsilon} \right) -
\epsilon \frac{\partial}{\partial \epsilon} 
\left( ^e{\mathsf P}_{\epsilon}:
{\bfnabla} {\bf v} \right) - ^e{\mathbb S}_{\epsilon} = 0
\end{equation}
and
\begin{equation}
\label{eq:dc-eb}
\left\ldbrack \frac{\partial (^e\bar{E}_{\epsilon})}{\partial t} 
\right\rdbrack_{\rm radiation} -
{\bfnabla} \cdot \left(^e\bar{D}_\epsilon ^e\bar{E}_{\epsilon} \right) -
\epsilon \frac{\partial}{\partial \epsilon} 
\left( ^e\bar{{\mathsf P}}_{\epsilon}:
{\bfnabla} {\bf v} \right) - ^e\bar{\mathbb S}_{\epsilon} = 0
\end{equation}
are solved simultaneously over all groups for the
electron neutrino energy density $^eE_\epsilon$
and the electron antineutrino energy density
$^e\bar{E}_\epsilon$.  There is one such pair of
equations for each of the $N_g$ energy groups that span
the radiation spectrum.
Note that the diffusion
coefficients and the source-terms in equations
(\ref{eq:dc-e}) and (\ref{eq:dc-eb}) are functions
of the neutrino energy and differ between the
neutrinos and antineutrinos.  The analogous
pair-coupled sets of equations are solved in the
step represented by box ({\em e}) of Figure~\ref{fig:timestep} for the muon
neutrino-antineutrino case and in the step
represented by box ($g$) of Figure~\ref{fig:timestep} for the tauon
neutrino-antineutrino case.

For the sake of compactness of notation in the
remainder of this appendix, we use the generic
notation $E_\epsilon$, without the leading
superscripts $e$, $\mu$, or $\tau$ to refer to the
neutrino energy density and $\bar{E}_\epsilon$,
without the leading superscripts $e$, $\mu$, or
$\tau$ to refer to the antineutrino density.  The
finite-difference equations for each neutrino
species can then be obtained by straightforwardly
substituting $^eE_\epsilon$ for $E_\epsilon$,
$^eD_\epsilon$ for $D_\epsilon$, {\em etc.}

\subsection{Implicit Finite Differencing of Diffusion Equation}

We discretize equations~(\ref{eq:dc-e}) and
(\ref{eq:dc-eb}) using the spatial differencing
scheme developed by \cite{Crank} for diffusion
equations and using the standard implicit
backward-Euler approach for the time evolution
scheme.  The backward-Euler time discretization
scheme has the advantage that method is L-stable
for all timesteps \citep{hv03}.  

Equation~(\ref{eq:dc-e}) is implicitly discretized as
\begin{eqnarray} \label{eq:nu_tr_fd}
\frac{ \left[ E_\epsilon \right]^\npo_{k+\lhalf,i+\lhalf,j+\lhalf} - 
\left[E_\epsilon\right]^n_{k+\lhalf,i+\lhalf,j+\lhalf}}{\Delta t}-
\left[{\bfnabla} \cdot D_\epsilon \nabla E_{\epsilon}
\right]^\npo_{k+\lhalf,i+\lhalf,j+\lhalf} \nonumber \\
-\left[ \epsilon \frac{\partial \left({\mathsf P}_{\epsilon}:
{\bfnabla} {\bf v}\right)}{\partial \epsilon} 
\right]^\npo_{k+\lhalf,i+\lhalf,j+\lhalf}
-\left[{\mathbb S}_{\epsilon} \right]^\npo_{k+(1/2),i+(1/2),j+)1/2)}
= 0 
\end{eqnarray}
where
\begin{eqnarray} \label{eq:divfdiff}
\lefteqn{
\left[{\bfnabla} \cdot D_\epsilon \nabla E_{\epsilon}
\right]^\npo_{k+\lhalf,i+\lhalf,j+\lhalf} \equiv }
\nonumber \\ & & 
\frac{1}{ \left[ g_2 \right]_{i+\lhalf} 
\left[ g_{31} \right]_{i+\lhalf} 
\left[ g_{32} \right]_{j+\lhalf}}
\Biggl\{
\frac{1}{ \left[ x_1 \right]_{i+\lthreehalf} - 
\left[ x_1 \right]_{i+\lhalf}} 
\Biggr.
\nonumber \\ & & 
\quad \quad
\Biggl.
\times 
\left( \left[ g_2 \right]_{i+1}
\left[ g_{31} \right]_{i+1} 
\left[ g_{32} \right]_{j+\lhalf} 
\left [D_\epsilon(x_1) \right]^{n+t}_{k+\lhalf,i+1,j+\lhalf}
\Biggr. \right.
\nonumber \\ & & 
\quad \quad \; \; \;
\Biggl. \left.
\times
\frac{ \left[ E_\epsilon \right]^\npo_{k+\lhalf,i+\lthreehalf,j+\lhalf} - 
\left[ E_\epsilon \right]^\npo_{k+\lhalf,i+\lhalf,j+\lhalf}}
{ \left[ x_1 \right]_{i+\lthreehalf} - 
\left[ x_1 \right]_{i+\lhalf}}
\Biggr. \right.
\nonumber \\ & & 
\quad \quad \; \; \;
- \Biggl. \left. \left[ g_2 \right]_{i}
\left[ g_{31} \right]_{i} 
\left[ g_{32} \right]_{j+\lhalf}
\left[ D_\epsilon(x_1) \right]^{n+t}_{k+\lhalf,i,j+\lhalf} 
\Biggr. \right.
\nonumber \\ & & 
\quad \quad \; \; \;
\Biggl. \left.
\times
\frac{ \left[ E_\epsilon \right]^\npo_{k+\lhalf,i+\lhalf,j+\lhalf} - 
\left[ E_\epsilon \right]^\npo_{k+\lhalf,i-\lhalf,j+\lhalf}}
{ \left[ x_1 \right]_{i+\lhalf} - 
\left[ x_1 \right]_{i-\lhalf}}  \right) 
\Biggr.
\nonumber \\ & & 
\quad \quad \quad \quad \quad \quad \quad \quad \quad \quad \quad \quad \quad \quad 
\Biggl.
+ \frac{1}{ \left[ x_2 \right]_{j+\lthreehalf} - 
\left[ x_2 \right]_{j+\lhalf}}
\Biggr.
\nonumber \\ & & 
\quad \quad
\Biggl.
\times
\left( \frac{ \left[ g_{31} \right]_{i+\lhalf} 
\left[ g_{32} \right]_{j+1}}
{ \left[ g_2 \right]_{i+\lhalf}} 
\left[ D_\epsilon(x_2) \right]^{n+t}_{k+\lhalf,i+\lhalf,j+1}
\Biggr. \right.
\nonumber \\ & & 
\quad \quad \; \; \;
\Biggl. \left.
\times
\frac{ \left[ E_\epsilon \right]^\npo_{k+\lhalf,i+\lhalf,j+\lthreehalf} -
\left[ E_\epsilon \right]^\npo_{k+\lhalf,i+\lhalf,j+\lhalf}}
{\left[ x_2 \right]_{j+\lthreehalf} - 
\left[ x_2 \right]_{j+\lhalf}} 
\Biggr. \right.
\nonumber \\ & & 
\quad \quad \; \; \;
\Biggl. \left.
- \frac{ \left[ g_{31} \right]_{i+\lhalf} 
\left[ g_{32} \right]_{j}}
{ \left[ g_2 \right]_{i+\lhalf}} 
\left[ D_\epsilon(x_2) \right]^{n+t}_{k+\lhalf,i+\lhalf,j}
\Biggr. \right.
\nonumber \\ & & 
\quad \quad \; \; \;
\Biggl. \left.
\times
\frac{ \left[ E_\epsilon \right]^\npo_{k+\lhalf,i+\lhalf,j+\lhalf} -
\left[ E_\epsilon \right]^\npo_{k+\lhalf,i+\lhalf,j-\lhalf}}
{ \left[ x_2 \right]_{j+\lhalf} - 
\left[ x_2 \right]_{j-\lhalf}}
\right) 
\Biggr\},
\end{eqnarray}
\begin{eqnarray} \label{eq:pv_diff}
\lefteqn{
\left[ \epsilon \frac{\partial \left({\mathsf P}_{\epsilon}:
{\bfnabla} {\bf v}\right)}{\partial \epsilon} 
\right]^\npo_{k+\lhalf,i+\lhalf,j+\lhalf}
\equiv
\frac{\left[ \epsilon \right]_{k+\lhalf}}
{\left[ \epsilon \right]_{k+1} - \left[ \epsilon \right]_k} 
 }
\nonumber \\
& &
\times \biggl( \left[ \mathsf X_{\epsilon}:{\bfnabla} {\bf v} 
\right]^{n+t}_{k+(3/2),i+(1/2),j+(1/2)} 
\left[ E_\epsilon \right]^\npo_{k+(3/2),i+(1/2),j+(1/2)} 
\nonumber \\
& &  - \left[ \mathsf X_{\epsilon}:{\bfnabla} {\bf v} 
\right]^{n+t}_{k-(1/2),i+(1/2),j+(1/2)} 
\left[ E_\epsilon \right]^\npo_{k-(1/2),i+(1/2),j+(1/2)} 
\biggl),
\end{eqnarray}
and
\begin{eqnarray} \label{eq:nu_tr_src_trm}
\lefteqn{
\left[{\mathbb S}_{\epsilon} \right]^\npo_{k+(1/2),i+(1/2),j+(1/2)}
 \equiv 
}
 \nonumber \\
& &                    
 -\left[ S_\epsilon \right]^{n+t}_{k+(1/2),i+(1/2),j+(1/2)}
\left( 1 + \frac{\eta\alpha}{\left( \left[\epsilon \right]_{k+(1/2)}\right)^3}
\left[E_\epsilon\right]^\npo_{k+(1/2),i+(1/2),j+(1/2)} \right)
\nonumber \\
& &                    
+ c \left[\kappa^a_\epsilon\right]^{n+t}_{k+(1/2),i+(1/2),j+(1/2)}
\left[E_\epsilon\right]^\npo_{k+(1/2),i+(1/2),j+(1/2)}
\nonumber \\
& &                    
-
\left( 1 + \frac{\eta\alpha}{\left( \left[\epsilon \right]_{k+(1/2)}\right)^3}
\left[E_\epsilon\right]^\npo_{k+(1/2),i+(1/2),j+(1/2)} \right)
\left[ \epsilon \right]_{k+(1/2)}
\nonumber \\
& &
\times
\sum_{\ell=0}^{N_g-1} \left[\Delta\epsilon\right]_{\ell+(1/2)}
\left[G\right]^{n+t}_{k+(1/2),\ell+(1/2),i+(1/2),j+(1/2)}
\left( 1 + \frac{\eta\alpha}{\left( \left[\epsilon \right]_{\ell+(1/2)}\right)^3}
\left[\bar{E}_\epsilon\right]^\npo_{\ell+(1/2),i+(1/2),j+(1/2)} \right)
\nonumber \\
& &                    
-
c\left( 1 + \frac{\eta\alpha}{\left( \left[\epsilon \right]_{k+(1/2)}\right)^3}
\left[E_\epsilon\right]^\npo_{k+(1/2),i+(1/2),j+(1/2)} \right)
\nonumber \\
& & 
\times
\sum_{\ell=0}^{N_g-1} \left[\Delta\epsilon\right]_{\ell+(1/2)}
\left[\kappa^s\right]^{n+t}_{k+(1/2),\ell+(1/2),i+(1/2),j+(1/2)}
\left[E_\epsilon\right]^\npo_{\ell+(1/2),i+(1/2),j+(1/2)}
\nonumber \\
& &                    
+ 
c\left[E_\epsilon\right]^\npo_{k+(1/2),i+(1/2),j+(1/2)}
\nonumber \\
& & 
\times
\sum_{\ell=0}^{N_g-1} \left[\Delta\epsilon\right]_{\ell+(1/2)}
\left[\kappa^s\right]^{n+t}_{\ell+(1/2),k+(1/2),i+(1/2),j+(1/2)}
\left( 1 + \frac{\eta\alpha}{\left( \left[\epsilon \right]_{\ell+(1/2)}\right)^3}
\left[E_\epsilon\right]^\npo_{\ell+(1/2),i+(1/2),j+(1/2)} \right)
.
\nonumber \\
\end{eqnarray}
In equations~(\ref{eq:divfdiff}),
(\ref{eq:pv_diff}), and (\ref{eq:nu_tr_src_trm})
the superscript $n+t$ is used to indicate
evaluation of the superscripted quantity at times
$n+b$, $n+d$ and $n+f$ according to whether the
equations for electron, muon, or tauon type
neutrinos, respectively, are being solved (see
Figure~\ref{fig:timestep} for the order of
updates).
For readability, we have omitted displaying all the explicit
functional dependences of the microphysics factors ($S_\epsilon$,
$\kappa^a_\epsilon$, $G$, $\kappa^{\rm in}$, $\kappa^{\rm
out}$) in the finite-differenced equations. 
These microphysics factors are also functions of the physical
conditions of the matter at the spatial point in question
(temperature, density, chemical composition, {\em etc.}).
In addition, they are functions of the particle energy or energies
in the radiation field.  
We denote this energy dependence in the discretized quantity through 
the use of additional subscripts:
\begin{equation}
\left[ \kappa^a \right]^{n+t}_{k+(1/2),i+(1/2),j+(1/2)}
\equiv
\left[ \kappa^a_{\left[\epsilon\right]_{k+(1/2)}} \right]^{n+t}_{i+(1/2),j+(1/2)},
\end{equation}
\begin{equation}
\left[ \kappa^s \right]^{n+t}_{\ell+(1/2),k+(1/2),i+(1/2),j+(1/2)}
\equiv
\left[ \kappa^s
  (\left[\epsilon \right]_{\ell+(1/2)},
   \left[\epsilon\right]_{k+(1/2)}
  ) 
\right]^{n+t}_{i+(1/2),j+(1/2)},
\end{equation}
and
\begin{equation}
\left[ G \right]^{n+t}_{\ell+(1/2),k+(1/2),i+(1/2),j+(1/2)}
\equiv
\left[ G
  (\left[\epsilon \right]_{\ell+(1/2)},
   \left[\epsilon\right]_{k+(1/2)}
  ) 
\right]^{n+t}_{i+(1/2),j+(1/2)}.
\end{equation}

\sloppypar{ Equation~(\ref{eq:nu_tr_fd}) is fully implicit in
terms of the unknowns 
$[ E_\epsilon]^\npo_{k+(1/2),i+(1/2),j+(1/2)}$ 
and
$[\bar{E}_\epsilon]^\npo_{k+(1/2),i+(1/2),j+(1/2)}$ 
but some
of the coefficients arising from the microphysics
are evaluated at the beginning of the substep.
Following the practice of \cite{ts01}, we evaluate
the diffusion coefficient and Eddington tensor based on information at
the beginning of the substep.  The reason for this
is that the absolute value function that appears
in the evaluation of the Knudsen number would
introduce a discontinuity in the Jacobian of the
nonlinear system if evaluated implicitly.  Such
discontinuities can often produce failures in
Newton-type iterative procedures for nonlinear
systems.  By evaluating the diffusion coefficient
at the beginning of the substep, we avoid this
difficulty altogether.  This approximation is
accurate as long as the change in the radiation
energy density is small over a timestep.  For most
astrophysical heating/cooling problems, this is
indeed the case.  In equation~(\ref{eq:nu_tr_fd}),
all quantities that are temperature-, density-, or
$Y_e$-dependent are also evaluated at the
beginning of the substep.  Doing so avoids the
difficult problem of simultaneously solving the
radiative diffusion equations while also solving
the Lagrangean portion of the gas-energy equation.
All quantities, such as the opacities,
emissivities, {\em etc.,} are evaluated by using the
values of the temperature, density, and electron
fraction that are known at the beginning of the
substep.  If the exchange of energy, or lepton
number, that takes place during a timestep is small,
this approximation is accurate.   The validity of 
these approximations are tested in the verification
problems that we consider in \S\ref{sec:validation}.
} 

The antineutrino counterpart of equation~(\ref{eq:nu_tr_fd}) 
is given by
\begin{eqnarray} \label{eq:anu_tr_fd}
\frac{ \left[ \bar{E}_\epsilon \right]^\npo_{k+\lhalf,i+\lhalf,j+\lhalf} - 
\left[\bar{E}_\epsilon\right]^n_{k+\lhalf,i+\lhalf,j+\lhalf}}{\Delta t}-
\left[{\bfnabla} \cdot \bar{D}_\epsilon \nabla \bar{E}_{\epsilon}
\right]^\npo_{k+\lhalf,i+\lhalf,j+\lhalf} \nonumber \\
-\left[ \epsilon \frac{\partial \left(\bar{{\mathsf P}}_{\epsilon}:
{\bfnabla} {\bf v}\right)}{\partial \epsilon} 
\right]^\npo_{k+\lhalf,i+\lhalf,j+\lhalf}
-\left[\bar{\mathbb S}_{\epsilon} \right]^\npo_{k+(1/2),i+(1/2),j+)1/2)}
= 0 
\end{eqnarray}
where
\begin{eqnarray} \label{eq:adivfdiff}
\lefteqn{
\left[{\bfnabla} \cdot \bar{D}_\epsilon \nabla \bar{E}_{\epsilon}
\right]^{n+1}_{k+\lhalf,i+\lhalf,j+\lhalf} \equiv }
\nonumber \\ & & 
\frac{1}{ \left[ g_2 \right]_{i+\lhalf} 
\left[ g_{31} \right]_{i+\lhalf} 
\left[ g_{32} \right]_{j+\lhalf}}
\Biggl\{
\frac{1}{ \left[ x_1 \right]_{i+\lthreehalf} - 
\left[ x_1 \right]_{i+\lhalf}} 
\Biggr.
\nonumber \\ & & 
\quad \quad
\times
\Biggl.
\left( \left[ g_2 \right]_{i+1}
\left[ g_{31} \right]_{i+1} 
\left[ g_{32} \right]_{j+\lhalf} 
\left[\bar{D}_\epsilon(x_1) \right]^{n+t}_{k+\lhalf,i+1,j+\lhalf}
\Biggr. \right.
\nonumber \\ & & 
\quad \quad \; \; \;
\times
\Biggl. \left.
\frac{ \left[ \bar{E}_\epsilon \right]^\npo_{k+\lhalf,i+\lthreehalf,j+\lhalf} - 
\left[ \bar{E}_\epsilon \right]^\npo_{k+\lhalf,i+\lhalf,j+\lhalf}}
{ \left[ x_1 \right]_{i+\lthreehalf} - 
\left[ x_1 \right]_{i+\lhalf}}
\Biggr. \right.
\nonumber \\ & & 
\quad \quad \; \; \;
- \Biggl. \left. \left[ g_2 \right]_{i}
\left[ g_{31} \right]_{i} 
\left[ g_{32} \right]_{j+\lhalf}
\left[ \bar{D}_\epsilon(x_1) \right]^{n+t}_{k+\lhalf,i,j+\lhalf} \times
\Biggr. \right.
\nonumber \\ & & 
\quad \quad \; \; \;
\Biggl. \left.
\frac{ \left[ \bar{E}_\epsilon \right]^\npo_{k+\lhalf,i+\lhalf,j+\lhalf} - 
\left[ \bar{E}_\epsilon \right]^\npo_{k+\lhalf,i-\lhalf,j+\lhalf}}
{ \left[ x_1 \right]_{i+\lhalf} - 
\left[ x_1 \right]_{i-\lhalf}}  \right) 
\Biggr.
\nonumber \\ & & 
\quad \quad \quad \quad \quad \quad \quad \quad \quad \quad \quad \quad \quad \quad 
\Biggl.
+ \frac{1}{ \left[ x_2 \right]_{j+\lthreehalf} - 
\left[ x_2 \right]_{j+\lhalf}} 
\Biggr.
\nonumber \\ & & 
\quad \quad
\Biggl.
\times
\left( \frac{ \left[ g_{31} \right]_{i+\lhalf} 
\left[ g_{32} \right]_{j+1}}
{ \left[ g_2 \right]_{i+\lhalf}} 
\left[ \bar{D}_\epsilon(x_2) \right]^{n+t}_{k+\lhalf,i+\lhalf,j+1}
\Biggr. \right.
\nonumber \\ & & 
\quad \quad \; \; \;
\Biggl. \left.
\times
\frac{ \left[ \bar{E}_\epsilon \right]^\npo_{k+\lhalf,i+\lhalf,j+\lthreehalf} -
\left[ \bar{E}_\epsilon \right]^\npo_{k+\lhalf,i+\lhalf,j+\lhalf}}
{\left[ x_2 \right]_{j+\lthreehalf} - 
\left[ x_2 \right]_{j+\lhalf}} 
\Biggr. \right.
\nonumber \\ & & 
\quad \quad \; \; \;
\Biggl. \left.
- \frac{ \left[ g_{31} \right]_{i+\lhalf} 
\left[ g_{32} \right]_{j}}
{ \left[ g_2 \right]_{i+\lhalf}} 
\left[ \bar{D}_\epsilon(x_2) \right]^{n+t}_{k+\lhalf,i+\lhalf,j}
\Biggr. \right.
\nonumber \\ & & 
\quad \quad \; \; \;
\Biggl. \left.
\times
\frac{ \left[ \bar{E}_\epsilon \right]^\npo_{k+\lhalf,i+\lhalf,j+\lhalf} -
\left[ \bar{E}_\epsilon \right]^\npo_{k+\lhalf,i+\lhalf,j-\lhalf}}
{ \left[ x_2 \right]_{j+\lhalf} - 
\left[ x_2 \right]_{j-\lhalf}}
\right) 
\Biggr\},
\end{eqnarray}
\begin{eqnarray} \label{eq:apv_diff}
\lefteqn{
\left[ \epsilon \frac{\partial \left(\bar{\mathsf P}_{\epsilon}:
{\bfnabla} {\bf v}\right)}{\partial \epsilon} 
\right]^{n+1}_{k+\lhalf,i+\lhalf,j+\lhalf}
\equiv
\frac{\left[ \epsilon \right]_{k+\lhalf}}
{\left[ \epsilon \right]_{k+1} - \left[ \epsilon \right]_k} 
 }
\nonumber \\
& &
\times \biggl(
\left[ \bar{\mathsf X}_{\epsilon}:{\bfnabla} {\bf v} 
\right]^{n+t}_{k+(3/2),i+(1/2),j+(1/2)} 
\left[ \bar{E}_\epsilon \right]^\npo_{k+(3/2),i+(1/2),j+(1/2)} 
\nonumber \\
& &  
\;\:\:\:
- \left[ \bar{\mathsf X}_{\epsilon}:{\bfnabla} {\bf v} 
\right]^{n+t}_{k-(1/2),i+(1/2),j+(1/2)} 
\left[ \bar{E}_\epsilon \right]^\npo_{k-(1/2),i+(1/2),j+(1/2)} 
\biggl),
\end{eqnarray}
and
\begin{eqnarray} \label{eq:anu_tr_src_trm}
\lefteqn{
\left[\bar{\mathbb S}_{\epsilon} \right]^\npo_{k+(1/2),i+(1/2),j+(1/2)}
 \equiv 
}
 \nonumber \\
& &                    
 -\left[ \bar{S}_\epsilon \right]^{n+t}_{k+(1/2),i+(1/2),j+(1/2)}
\left( 1 + \frac{\eta\alpha}{\left( \left[\epsilon \right]_{k+(1/2)}\right)^3}
\left[\bar{E}_\epsilon\right]^\npo_{k+(1/2),i+(1/2),j+(1/2)} \right)
\nonumber \\
& &                    
+ c \left[\bar{\kappa}^a_\epsilon\right]^{n+t}_{k+(1/2),i+(1/2),j+(1/2)}
\left[\bar{E}_\epsilon\right]^\npo_{k+(1/2),i+(1/2),j+(1/2)}
\nonumber \\
& &                    
-
\left( 1 + \frac{\eta\alpha}{\left( \left[\epsilon \right]_{k+(1/2)}\right)^3}
\left[\bar{E}_\epsilon\right]^\npo_{k+(1/2),i+(1/2),j+(1/2)} \right)
\left[ \epsilon \right]_{k+(1/2)} 
\nonumber \\
& &
\times
\sum_{\ell=0}^{N_g-1} \left[\Delta\epsilon\right]_{\ell+(1/2)}
\left[G\right]^{n+t}_{\ell+(1/2),k+(1/2),i+(1/2),j+(1/2)}
\left( 1 + \frac{\eta\alpha}{\left( \left[\epsilon \right]_{\ell+(1/2)}\right)^3}
\left[{E}_\epsilon\right]^\npo_{\ell+(1/2),i+(1/2),j+(1/2)} \right)
\nonumber \\
& &                    
-
c\left( 1 + \frac{\eta\alpha}{\left( \left[\epsilon \right]_{k+(1/2)}\right)^3}
\left[\bar{E}_\epsilon\right]^\npo_{k+(1/2),i+(1/2),j+(1/2)} \right)
\nonumber \\
& & 
\times
\sum_{\ell=0}^{N_g-1} \left[\Delta\epsilon\right]_{\ell+(1/2)}
\left[\bar{\kappa}^s\right]^{n+t}_{k+(1/2),\ell+(1/2),i+(1/2),j+(1/2)}
\left[\bar{E}_\epsilon\right]^\npo_{\ell+(1/2),i+(1/2),j+(1/2)}
\nonumber \\
& &                    
+ 
c\left[\bar{E}_\epsilon\right]^\npo_{k+(1/2),i+(1/2),j+(1/2)}
\nonumber \\
& & 
\times
\sum_{\ell=0}^{N_g-1} \left[\Delta\epsilon\right]_{\ell+(1/2)}
\left[\bar{\kappa}^s\right]^{n+t}_{\ell+(1/2),k+(1/2),i+(1/2),j+(1/2)}
\left( 1 + \frac{\eta\alpha}{\left( \left[\epsilon \right]_{\ell+(1/2)}\right)^3}
\left[\bar{E}_\epsilon\right]^\npo_{\ell+(1/2),i+(1/2),j+(1/2)} \right)
.
\nonumber \\
\end{eqnarray}
The finite-differenced antineutrino
equations~(\ref{eq:anu_tr_fd})--(\ref{eq:anu_tr_src_trm})  
are almost analogous to the
finite-differenced neutrino equations
(\ref{eq:nu_tr_fd})--(\ref{eq:nu_tr_src_trm}).  The one exception to this
one-to-one correspondence is in the pair-production terms where the 
same discretized function 
$[ G ]^{n+t}_{\ell+(1/2),k+(1/2),i+(1/2),j+(1/2)}$ appears
in both equation~(\ref{eq:nu_tr_src_trm}) and
equation~(\ref{eq:anu_tr_src_trm}).  In the latter, the summation is
over the  
first index of the function, while in the former, the summation is over
the second index.   Because both $E_\epsilon$ and $\bar{E}_\epsilon$
appear as factors in this set of terms, the process of pair-production 
is responsible for coupling the 
neutrino diffusion equations to the antineutrino equations.   This requires
that the two sets of equations be solved simultaneously for each of the 
three neutrino flavors: electron, muon, and tauon.  We discuss the method
of solution of these implicitly differenced equations 
in \S\ref{sec:iter_soln}.

\subsection{Implementation of Boundary Conditions}

Boundary conditions are specified for equations
(\ref{eq:dc-e}) and (\ref{eq:dc-eb}) by altering the 
differencing of the ${\bfnabla} \cdot {\bf F}_\epsilon$  and 
${\mathsf P}_\epsilon : {\bfnabla}{\bf v}$ terms.
In this subsection, we only present boundary conditions corresponding to
equation~(\ref{eq:dc-e}), with the understanding that similar boundary
conditions for equation  (\ref{eq:dc-eb}) can be trivially obtained by 
substitution of $\bar{E}_\epsilon$ for ${E}_\epsilon$ etc.

We consider three types of boundary conditions in the $x_1$ direction:
Dirichlet boundary conditions ($E_\epsilon$ is specified), 
zero-flux boundary conditions,  and free-streaming boundary conditions.  
The first of these conditions is handled by equation~(\ref{eq:divfdiff}) 
with no modifications necessary.   Zero-flux boundary conditions 
are usually applied at the inner or left-edge of the grid.  To implement
zero-flux boundary conditions on a given edge, the 
corresponding $D{\bfnabla} E$ terms
in equation~(\ref{eq:divfdiff}) for zones at that edge are set to zero. 
In the case of free-streaming boundary conditions at the outer (or right)
edge of the grid, the corresponding $D{\bfnabla} E$ is replaced 
by $cE_\epsilon$ in an upwinded sense.  Equation  (\ref{eq:divfdiff})
then replaced in the outermost zone by
\begin{eqnarray} \label{eq:divfdiff_fs}
\lefteqn{
\left[{\bfnabla} \cdot D_\epsilon \nabla E_{\epsilon}
\right]^\npo_{k+\lhalf,i+\lhalf,j+\lhalf} \equiv }
\nonumber \\ & & 
\frac{1}{ \left[ g_2 \right]_{i+\lhalf} 
\left[ g_{31} \right]_{i+\lhalf} 
\left[ g_{32} \right]_{j+\lhalf}}
\Biggl\{
\frac{1}{ \left[ x_1 \right]_{i+\lthreehalf} - 
\left[ x_1 \right]_{i+\lhalf}} 
\Biggr.
\nonumber \\ & & 
\quad \quad
\Biggl.
\times
\left( \left[ g_2 \right]_{i+1}
\left[ g_{31} \right]_{i+1} 
\left[ g_{32} \right]_{j+\lhalf} 
c
\left[ E_\epsilon \right]^\npo_{k+\lhalf,i+\lhalf,j+\lhalf}
\Biggr. \right.
\nonumber \\ & & 
\quad \quad \; \; \;
- \Biggl. \left. \left[ g_2 \right]_{i}
\left[ g_{31} \right]_{i} 
\left[ g_{32} \right]_{j+\lhalf}
\left[ D_\epsilon(x_1) \right]^{n+t}_{k+\lhalf,i,j+\lhalf} 
\Biggr. \right.
\nonumber \\ & & 
\quad \quad \; \; \;
\Biggl. \left.
\times
\frac{ \left[ E_\epsilon \right]^\npo_{k+\lhalf,i+\lhalf,j+\lhalf} - 
\left[ E_\epsilon \right]^\npo_{k+\lhalf,i-\lhalf,j+\lhalf}}
{ \left[ x_1 \right]_{i+\lhalf} - 
\left[ x_1 \right]_{i-\lhalf}}  \right)
\Biggr.
\nonumber \\ & & 
\quad \quad \quad \quad \quad \quad \quad \quad \quad \quad \quad \quad \quad \quad 
\Biggl.
+ \frac{1}{ \left[ x_2 \right]_{j+\lthreehalf} - 
\left[ x_2 \right]_{j+\lhalf}} 
\Biggr.
\nonumber \\ & & 
\quad \quad
\Biggl.
\times
\left( \frac{ \left[ g_{31} \right]_{i+\lhalf} 
\left[ g_{32} \right]_{j+1}}
{ \left[ g_2 \right]_{i+\lhalf}} 
\left[ D_\epsilon(x_2) \right]^{n+t}_{k+\lhalf,i+\lhalf,j+1} 
\Biggr. \right.
\nonumber \\ & & 
\quad \quad \; \; \;
\Biggl. \left.
\times
\frac{ \left[ E_\epsilon \right]^\npo_{k+\lhalf,i+\lhalf,j+\lthreehalf} -
\left[ E_\epsilon \right]^\npo_{k+\lhalf,i+\lhalf,j+\lhalf}}
{\left[ x_2 \right]_{j+\lthreehalf} - 
\left[ x_2 \right]_{j+\lhalf}} 
\Biggr. \right.
\nonumber \\ & & 
\quad \quad \; \; \;
\Biggl. \left.
- \frac{ \left[ g_{31} \right]_{i+\lhalf} 
\left[ g_{32} \right]_{j}}
{ \left[ g_2 \right]_{i+\lhalf}} 
\left[ D_\epsilon(x_2) \right]^{n+t}_{k+\lhalf,i+\lhalf,j}
\Biggr. \right.
\nonumber \\ & & 
\quad \quad \; \; \;
\Biggl. \left.
\times
\frac{ \left[ E_\epsilon \right]^\npo_{k+\lhalf,i+\lhalf,j+\lhalf} -
\left[ E_\epsilon \right]^\npo_{k+\lhalf,i+\lhalf,j-\lhalf}}
{ \left[ x_2 \right]_{j+\lhalf} - 
\left[ x_2 \right]_{j-\lhalf}}
\right) 
\Biggr\},
\end{eqnarray}
In this equation, we have made the assumption that
radiation is flowing off the grid at the outer
edge in order to determine the upwind direction.
This boundary condition must be used with care and
only in cases where one is sure of the accuracy of this
assumption.

In the $x_2$ direction, we consider three types of
boundary conditions: Dirichlet, zero-flux, and
periodic.  The first of these is again trivial and
requires no modification of equation
(\ref{eq:divfdiff}).  The second, zero-flux
conditions, is treated as we previously
described, by setting the appropriate $D{\smpmb
\nabla}E$ terms to zero in equation
(\ref{eq:divfdiff}) for the zones at the
corresponding edge of the mesh.  The third case,
periodic boundary conditions, is treated by
using the ability of MPI to establish periodic
process topologies.  Using MPI send and receive
subroutines to exchange ghost zones, together with
a process topology that is periodic in the $x_2$
direction, permits equation~(\ref{eq:divfdiff})
to be used to apply periodic boundary conditions
without modification.

Spectral boundary conditions for the 
${\mathsf P}_\epsilon:{\bfnabla} {\bf v}$
term in the $\epsilon$ direction are handled similarly to the 
previously described zero-flux spatial boundary conditions.
The appropriate terms of 
${\mathsf P}_\epsilon:{\bfnabla} {\bf v}$
in equation~(\ref{eq:pv_diff}) are set to zero at the lower and
upper edges of the energy grid. This ensures that no energy flows
out of the spectrum at either end.

\subsection{Evaluation of Diffusion Coefficient and Eddington-Tensor Terms}

In the verification tests described in this paper,
we rely on the \cite{lp81} prescription for
flux-limiting, which relates the diffusion
coefficient and the Eddington tensor to the
Knudsen number for the radiation flow.  However,
the modification of our scheme to utilize other
flux-limiters is trivial so long as the diffusion
coefficient and the Eddington tensor can be cast
into a form that relies on the Knudsen number.

In what follows, we will generically describe the
evaluation of Knudsen numbers, diffusion
coefficients, and Eddington tensors for
neutrinos.  The analogous equations for 
antineutrinos can be obtained by straightforward
substitution of $\bar{E_\epsilon}$ for
$E_\epsilon$, $\bar{\kappa}^a$ for $\kappa^a$,
{\em etc.,} and is not presented here for reasons of
brevity.

In our calculation of the diffusion coefficients
and Eddington tensor, we emulate \cite{ts01}
and make use of separate Knudsen numbers $\kn_1$
and $\kn_2$ to describe the radiation flow in each
of the two orthogonal coordinate directions.
Since Knudsen numbers are based on
components of the gradient of radiation energy
density, which is defined at cell-centers,
the gradients, and hence Knudsen numbers, are
naturally evaluated at cell faces---precisely where
they are needed.  First, for $\kn_1$, we have
\begin{equation}
\left[\kn_1 \right]^{n+t}_{k+\lhalf,i,j+\lhalf}  \equiv 
\left[
\frac{\left|{\bfnabla} E_\epsilon \cdot {\bf \hat{x}_1}\right|}
{\kappa^T_\epsilon E_\epsilon}\right]^{n+t}_{k+\lhalf,i,j+\lhalf} = 
\left[
\frac{1}{\kappa^T_\epsilon E_\epsilon}
\left|
\frac{\partial E_\epsilon}{\partial x_1} 
\right|\right]^{n+t}_{k+\lhalf,i,j+\lhalf},
\end{equation}
expanding the right-hand-side we can write $\kn_1$ as
\begin{eqnarray}
\left[\kn_1 \right]^{n+t}_{k+\lhalf,i,j+\lhalf} & = &
\left(
\frac{2}{\left[ \kappa^T_\epsilon \right]^{n+t}_{k+\lhalf,i+\lhalf,j+\lhalf}+
\left[ \kappa^T_\epsilon \right]^{n+t}_{k+\lhalf,i-\lhalf,j+\lhalf}}
\right) 
\nonumber \\ & &
\nonumber \\ & &
\; \; \: \:
\times
\left(
\frac{2}{ \left[ E_\epsilon \right]^{n+b}_{k+\lhalf,i+\lhalf,j+\lhalf} + 
\left[ E_\epsilon \right]^{n+b}_{k+\lhalf,i-\lhalf,j+\lhalf}} 
\right) 
\nonumber \\ & &
\nonumber \\ & &
\; \; \: \:
\times 
\left( \frac{ \left[ E_\epsilon \right]^{n+b}_{k+\lhalf+i+\lhalf,j+\lhalf} - 
\left[ E_\epsilon \right]^{n+b}_{k+\lhalf,i-\lhalf,j+\lhalf}}
{ \left[ x_1 \right]_{i+\lhalf} - 
\left[ x_1 \right]_{i-\lhalf}} \right).
\label{eq:fdr1}
\end{eqnarray}
In equation~(\ref{eq:fdr1}), we have arithmetically averaged the opacities and
radiation-energy densities in adjacent cells to obtain values at the
zone interface in the $x_1$ direction.   The most recent values of the
radiation-energy densities, obtained at the end of the advection substep,
are used in equation~(\ref{eq:fdr1}).
The values for the opacities are obtained at the 
beginning of the substep at point $n+t$, where $t$ takes on the values
$b$, $d$, or $f$ for electron, muon, and tauon type neutrinos respectively. 
In the $x_2$ direction, an analogous expression can be written
\begin{eqnarray}
\left[\kn_2 \right]^{n+t}_{k+\lhalf,i+\lhalf,j} & = &
\left(
\frac{2}{ \left[ \kappa^T_\epsilon \right]^{n+t}_{k+\lhalf,i+\lhalf,j+\lhalf} +
\left[ \kappa^T_\epsilon \right]^{n+t}_{k+\lhalf,i+\lhalf,j-\lhalf}}
\right) 
\nonumber \\ & &
\nonumber \\ & &
\; \; \: \:
\times 
\left(
\frac{2}{ \left[ E_\epsilon \right]^{n+b}_{k+\lhalf, i+\lhalf,j+\lhalf} +
\left[ E_\epsilon \right]^{n+b}_{k+\lhalf, i+\lhalf,j-\lhalf}} \right)
\nonumber \\ & &
\nonumber \\ & &
\; \; \: \:
\times 
\left( \frac{ \left[ E_\epsilon \right]^{n+b}_{k+\lhalf,i+\lhalf,j+\lhalf} -
\left[ E_\epsilon \right]^{n+b}_{k+\lhalf,i+\lhalf,j-\lhalf}}
{\left[ g_2 \right]_{i+\lhalf}
(\left[ x_2 \right]_{j+\lhalf} -
\left[ x_2 \right]_{j-\lhalf})} \right).
\end{eqnarray}

To obtain the diffusion coefficients in each direction,
$D_\epsilon(x_1)$ and $D_\epsilon(x_2)$, we note
that $D_\epsilon$, in the Levermore-Pomraning
formalism, is given by equation~(\ref{eq:lpd}) as
\begin{equation}
D_\epsilon = \frac{c \lambda_\epsilon(\kn)}
{\kappa^T_\epsilon},
\end{equation}
where $\lambda_\epsilon$ is as given in
equation~(\ref{eq:lpfl}) and ${\kappa^T_\epsilon}$
is as given in equation~(\ref{eq:ktot}). Many
other flux-limiting prescriptions can be cast into
this form and could easily be accommodated in this
algorithm by replacing equation~(\ref{eq:lpfl})
with some other function (see \cite{janka91} for a
description of how numerous flux-limiters fit this
form).  Thus, we have:
\begin{eqnarray}
\left[D_\epsilon(x_1)\right]^{n+t}_{k+\lhalf,i,j+\lhalf} & = &
\left[ \frac{c \lambda(\kn_1^{n+t})}
{\kappa^T_\epsilon} \right]^{n+t}_{k+\lhalf,i,j+\lhalf}
\\ \nonumber \\ & = &
\frac{2c 
\lambda (\left[\kn_1\right]^{n+t}_{k+\lhalf,i,j+\lhalf})} 
{ \left[ \kappa^T_\epsilon \right]^{n+t}_{k+\lhalf,i+\lhalf,j+\lhalf} + 
\left[ \kappa^T_\epsilon \right]^{n+t}_{k+\lhalf,i-\lhalf,j+\lhalf}} 
\end{eqnarray}
and
\begin{eqnarray}
\left[D_\epsilon(x_2)\right]^{n+t}_{k+\lhalf,i+\lhalf,j} & = &
\left[ \frac{c \lambda(\kn_2)}
{\kappa^T_\epsilon} \right]^{n+t}_{k+\lhalf,i+\lhalf,j}
\\ \nonumber \\ & = &
\frac{2c 
\lambda (\left[\kn_2\right]^{n+t}_{k+\lhalf,i+\lhalf,j})} 
{ \left[ \kappa^T_\epsilon \right]^{n+t}_{k+\lhalf,i+\lhalf,j+\lhalf} +
\left[ \kappa^T_\epsilon \right]^{n+t}_{k+\lhalf,i+\lhalf,j-\lhalf}}.
\end{eqnarray}

  The evaluation of the Eddington tensor components in terms of the 
Knudsen number is similarly straightforward.  Equations 
(\ref{eq:chismdef}) and (\ref{eq:chi_tensor}) describe the 
Levermore-Pomraning prescription for the Eddintgton tensor.  
The algorithm we present is trivially modifiable to handle 
other closures replacing these equations with other alternative functions 
describing the tensor components in terms of the Knudsen number. 

We now focus on the components of the Eddington tensor needed to evaluate
the tensorial double-contraction terms of equation~(\ref{eq:pv_diff}).  This
double-contraction can be expressed as
\begin{eqnarray} \label{eq:chidotdot}
{\mathsf X}_{\epsilon}:
{\bfnabla} {\bf v} 
& = &
\left\{ {\mathsf X_{\epsilon}} \right\}_{11}
\left\{ {\bfnabla} {\bf v} \right\}_{11} +
\left\{ {\mathsf X_{\epsilon}} \right\}_{12}
\left( \left\{ {\bfnabla} {\bf v} \right\}_{12} + 
\left\{ {\bfnabla} {\bf v} \right\}_{21} \right) 
\nonumber \\ & &
+ \left\{ {\mathsf X_{\epsilon}} \right\}_{22}
\left\{ {\bfnabla} {\bf v} \right\}_{22} +
\left\{ {\mathsf X_{\epsilon}} \right\}_{33}
\left\{ {\bfnabla} {\bf v} \right\}_{33}.
\end{eqnarray}
We have dropped terms
that are identically zero in a two-dimensional formulation, namely,
those proportional to 
$\left\{ {\mathsf X}_\epsilon \right\}_{23} 
\left\{  {\smpmb\nabla} {\bf v}\right\}_{23}$ 
and 
$\left\{ {\mathsf X}_\epsilon\right\}_{32} 
\left\{  {\bfnabla} {\bf v}\right\}_{32}$.

The components of ${\bfnabla} {\bf v}$ are evaluated using the
expressions derived in Appendix~\ref{app:velgrad}.  The required
components of ${\mathsf
X_{\epsilon}}$ are evaluated as follows. Using
equation~(\ref{eq:chidef}), we have 
\begin{eqnarray} \label{eq:chi11diff}
\left[ \left\{{\mathsf X}_\epsilon\right\}_{11} 
\right]^{n+t}_{k+\lhalf,i+\lhalf,j+\lhalf} & = &
\half
\left[
\left( 1 - \chi_\epsilon \right) +
\left( 3\chi_\epsilon - 1 \right) 
\frac{ (\partial E_\epsilon / \partial x_1)^2 }
{\left| {\bfnabla} E_\epsilon \right|^2 }
\right]^{n+t}_{k+\lhalf,i+\lhalf,j+\lhalf},
\end{eqnarray}
\begin{eqnarray}  \label{eq:chi22diff}
\left[ \left\{ {\mathsf X}_\epsilon 
\right\}_{22} \right]^{n+t}_{k+\lhalf,i+\lhalf,j+\lhalf} & = &
\half
\left[
\left( 1 - \chi_\epsilon \right) +
\left( 3\chi_\epsilon - 1 \right)
\frac{ (\partial E_\epsilon / \partial x_2)^2 }
{\left| {\bfnabla} E_\epsilon \right|^2 }
\right]^{n+t}_{k+\lhalf,i+\lhalf,j+\lhalf}, 
\end{eqnarray}
\begin{eqnarray}  \label{eq:chi33diff}
\left[ \left\{ {\mathsf X}_\epsilon 
\right\}_{33} \right]^{n+t}_{k+\lhalf,i+\lhalf,j+\lhalf} & = &
\left[
\onehalf\left( 1 - \chi_\epsilon \right) 
\right]^{n+t}_{k+\lhalf,i+\lhalf,j+\lhalf},
\end{eqnarray}
and
\begin{eqnarray}  \label{eq:chi12diff}
\left[ \left\{ {\mathsf X}_\epsilon 
\right\}_{12} \right]^{n+t}_{k+\lhalf,i+\lhalf,j+\lhalf} & = &
\left[\onehalf
\left(3 \chi_\epsilon - 1 \right) \frac{ (\partial E_\epsilon / \partial x_1) 
(\partial E_\epsilon / \partial x_2) }
{\left| {\bfnabla} E_\epsilon \right|^2 }
\right]^{n+t}_{k+\lhalf,i+\lhalf,j+\lhalf}.
\end{eqnarray}
The scalar Eddington factor $\chi_\epsilon$ is evaluated using
equation~(\ref{eq:chismdef}), which requires Knudsen numbers, which we
have evaluated at cell faces.  To
perform the evaluation of $\chi_\epsilon$, which is required at cell
centers, we average over the surrounding (directionally decomposed)
face-centered values, {\em viz.},
\begin{eqnarray}
\left[ \chi_\epsilon \right]^{n+t}_{k+\lhalf,i+\lhalf,j+\lhalf} 
& = &
\frac{1}{4} \biggl( \left[
\lambda_\epsilon \left(1 + \lambda_\epsilon (\kn_1)^2 \right)
\right]^{n+t}_{k+\lhalf,i+1,j+\lhalf}  
\biggr.
\nonumber \\ & &
\quad \; \; 
\biggl.
+ \left[
\lambda_\epsilon \left(1 + \lambda_\epsilon (\kn_1)^2 \right)
\right]^{n+t}_{k+\lhalf,i,j+\lhalf}  
\biggr.
\nonumber \\ & &
\quad \; \; 
\biggl.  
+ \left[
\lambda_\epsilon \left(1 + \lambda_\epsilon (\kn_2)^2 \right)
\right]^{n+t}_{k+\lhalf,i+\lhalf,j+1}  
\biggr.
\nonumber \\ & &
\quad \; \; 
\biggl.   
+ \left[
\lambda_\epsilon \left(1 + \lambda_\epsilon (\kn_2)^2 \right)
\right]^{n+t}_{k+\lhalf,i+\lhalf,j}
\biggr),
\end{eqnarray}
where we use equation~(\ref{eq:lpfl}) in our evaluation of the
$\lambda_\epsilon$'s. 

The factors in equations~(\ref{eq:chi11diff}--\ref{eq:chi12diff})
containing ${\bfnabla} E_\epsilon$ and $\partial E_\epsilon/\partial x_i$
arise from the definition of the radiation flux direction ${\bf n}$.
They are evaluated at cell centers by centered differences:
\begin{eqnarray}
\left[ \frac{\partial E_\epsilon}{\partial x_1} 
\right]^{n+b}_{k+\lhalf,i+\lhalf,j+\lhalf} & = &
\frac{ \left[E_\epsilon \right]^{n+b}_{k,i+\lthreehalf,j+\lhalf} - 
\left[ E_\epsilon \right]^{n+b}_{k,i-\lhalf,j+\lhalf}}
{ \left[ x_1 \right]_{i+\lthreehalf} - \left[ x_1 \right]_{i-\lhalf}}
\end{eqnarray}
and
\begin{eqnarray}
\left[ \frac{\partial E_\epsilon}{\partial x_2} 
\right]^{n+b}_{k+\lhalf,i+\lhalf,j+\lhalf} & = &
\frac{ \left[ E_\epsilon \right]^{n+b}_{k,i+\lhalf,j+\lthreehalf} - 
\left[ E_\epsilon \right]^{n+b}_{k,i+\lhalf,j-\lhalf}}
{ \left[ g_2 \right]_{i+(1/2)} \left(
\left[ x_2 \right]_{j+\lthreehalf} - \left[ x_2 \right]_{j-\lhalf}
\right) },
\end{eqnarray}
with
\begin{eqnarray}
\left[ {\left| {\bfnabla} E_\epsilon \right|^2 } 
\right]^{n+b}_{k+\lhalf,i+\lhalf,j+\lhalf} & = &
\left(\left[ \frac{\partial E_\epsilon}{\partial x_1} 
\right]^{n+b}_{k+\lhalf,i+\lhalf,j+\lhalf}\right)^2 
\nonumber \\
& & 
+ \left( \left[ \frac{\partial E_\epsilon}{\partial x_2} 
\right]^{n+b}_{k+\lhalf,i+\lhalf,j+\lhalf}\right)^2.
\end{eqnarray}

With the preceding derivations, we now have complete information
regarding the differencing of the velocity-dependent terms.

\subsection{Velocity Gradients\label{app:velgrad}}


The evaluation of equation~(\ref{eq:chidotdot}), described in the
previous subsection,
requires the gradient of the velocity field,
${\bfnabla} {\bf v}$. Since ${\bfnabla} {\bf v}$ is a
second-rank tensor, we can use equation~(126) in Appendix~A of
\citet{sn92a} to help with the evaluation.  Upon inspecting that
equation, it is easy to extract the set of non-zero elements of ${\smpmb
\nabla}{\bf v}$ and obtain
\begin{equation}
\left\{ {\bfnabla} {\bf v} \right\}_{11} = 
\frac{\partial \varv_{1}}{\partial x_1},
\end{equation}
\begin{equation}
\left\{ {\bfnabla} {\bf v} \right\}_{22} = 
\frac{1}{g_2} \frac{\partial \varv_{2}}{\partial x_2} +
\frac{\varv_{1}}{g_2} \frac{\partial g_2}{\partial x_1},
\end{equation}
\begin{equation}
\left\{ {\bfnabla} {\bf v} \right\}_{33} = 
\frac{\varv_{1}}{g_{31}} \frac{\partial g_{31}}{\partial x_1} +
\frac{\varv_{2}}{g_{2} g_{32}} \frac{\partial g_{32}}{\partial x_2},
\end{equation}
\begin{equation} \label{eq:v12}
\left\{ {\bfnabla} {\bf v} \right\}_{12} = 
\frac{\partial \varv_{2}}{\partial x_1},
\end{equation}
and
\begin{equation} \label{eq:v21}
\left\{ {\bfnabla} {\bf v} \right\}_{21} = 
\frac{1}{g_2} \frac{\partial \varv_{1}}{\partial x_2} -
\frac{\varv_{2}}{g_2} \frac{\partial g_2}{\partial x_1}.
\end{equation}
When differenced, these equations yield the following expressions for
the three diagonal elements,
\begin{eqnarray}
\left[ \left\{ {\bfnabla} {\bf v} \right\}_{11}\right]_{i+\lhalf,j+\lhalf} = 
\frac{ \left[\varv_{1}\right]_{i+1,j+\lhalf} - 
\left[\varv_{1}\right]_{i,j+\lhalf}}
{\left[x_1\right]_{i+1} - \left[x_1\right]_i},
\end{eqnarray}
\begin{eqnarray}
\left[ \left\{ {\bfnabla} {\bf v} \right\}_{22}\right]_{i+\lhalf,j+\lhalf} = 
\frac{ \left[\varv_{2}\right]_{i+\lhalf,j+1} - 
\left[ \varv_{2} \right]_{i+\lhalf,j}} 
{\left[g_2 \right]_{i+\lhalf} 
\left(\left[x_2\right]_{j+1} - \left[x_2\right]_j \right)} +
\frac{ \left[\varv_{1}\right]_{i+1,j+\lhalf} +
\left[ \varv_{1} \right]_{i,j+\lhalf}}{2 \left[g_2\right]_{i+\lhalf}}
\left[ \frac{\partial g_2}{\partial x_1} \right]_{i+\lhalf},
\end{eqnarray}
and
\begin{eqnarray}
\left[ \left\{ {\bfnabla} {\bf v} \right\}_{33}\right]_{i+\lhalf,j+\lhalf} 
& = &
\frac{ \left[ \varv_{1} \right]_{i+1,j+\lhalf} + 
\left[ \varv_{1} \right]_{i,j+\lhalf}}
{2 \left[ g_{31} \right]_{i+\lhalf}}
\left[ \frac{\partial g_{31}}{\partial x_1}\right]_{i+\lhalf}  
\nonumber \\ & &
+ \frac{ \left[ \varv_{2} \right]_{i+\lhalf,j} + 
\left[ \varv_{2} \right]_{i+\lhalf,j+1}}
{2 \left[g_{2}\right]_{i+\lhalf} \left[ g_{32} \right]_{j+\lhalf}}
\left[ \frac{\partial g_{32}}{\partial x_2} \right]_{j+\lhalf}.
\end{eqnarray}
The evaluation of equation~(\ref{eq:chidotdot})
requires that
components of ${\bfnabla} {\bf v}$ be evaluated at cell centers.
Although this is a natural location for evaluation of the diagonal
components of ${\bfnabla} {\bf v}$, this evaluation location does not
fall naturally for $\{{\bfnabla} {\bf v} \}_{12}$ and
$\{ {\bfnabla} {\bf v} \}_{21}$. Because of cross
derivatives, these off-diagonal terms are more naturally evaluated at
the vertices of the integer mesh.  To obtain the values at the needed
locations, we evaluate these expressions using a four-way arithmetic
average of the surrounding vertex values, {\em i.e.},
\begin{equation}
\left[ \left\{ {\bfnabla} {\bf v} \right\}_{12} \right]_{i+\lhalf,j+\lhalf} =
\frac{1}{4} \left(
\left[ \left\{ {\bfnabla} {\bf v} \right\}_{12} \right]_{i,j} +
\left[ \left\{ {\bfnabla} {\bf v} \right\}_{12} \right]_{i,j+1} +
\left[ \left\{ {\bfnabla} {\bf v} \right\}_{12} \right]_{i+1,j} +
\left[ \left\{ {\bfnabla} {\bf v} \right\}_{12} \right]_{i+1,j+1}
\right),
\end{equation}
which, upon use of equation~(\ref{eq:v12}), gives
\begin{eqnarray} \label{eq:v12avg}
\left[ \left\{ {\bfnabla} {\bf v} \right\}_{12} 
\right]_{i+\lhalf,j+\lhalf} & = & 
\frac{1}{4} \left(
\frac{ \left[ \varv_{2} \right]_{i+\lhalf,j} - 
\left[ \varv_{2} \right]_{i-\lhalf,j}}
{ \left[ x_1 \right]_{i+\lhalf} - \left[ x_1 \right]_{i-\lhalf}} +
\frac{ \left[ \varv_{2} \right]_{i+\lhalf,j+1} - 
\left[ \varv_{2} \right]_{i-\lhalf,j+1}}
{ \left[ x_1 \right]_{i+\lhalf} - \left[ x_1 \right]_{i-\lhalf}}
\right. 
\nonumber \\ & &
\quad \; \;
\left.
+ \frac{ \left[ \varv_{2} \right]_{i+\lthreehalf,j} -
\left[ \varv_{2} \right]_{i+\lhalf,j}}
{\left[ x_1 \right]_{i+\lthreehalf} - \left[ x_1 \right]_{i+\lhalf}} + 
\frac{ \left[ \varv_{2} \right]_{i+\lthreehalf,j+1} - 
\left[ \varv_{2} \right]_{i+\lhalf,j+1}}
{\left[ x_1 \right]_{i+\lthreehalf} - \left[ x_1 \right]_{i+\lhalf}} 
\right).
\end{eqnarray}
Following the same procedure for $\left\{ {\bfnabla} {\bf v}
\right\}_{21}$, and using equation~(\ref{eq:v21}), we arrive at
\begin{eqnarray} \label{eq:v21avg}
\left[ \left\{ {\bfnabla} {\bf v} 
\right\}_{21} \right]_{i+\lhalf,j+\lhalf} & = &
\frac{1}{4} \Biggl\{ \Biggl(
\frac{ \left[ \varv_{1} \right]_{i,j+\lhalf} - 
\left[ \varv_{1} \right]_{i,j-\lhalf} }
{\left[ g_2 \right]_i 
\left( \left[ x_2 \right]_{j+\lhalf} - 
\left[ x_2 \right]_{j-\lhalf} \right)}
\vspace{-0.1in}
\nonumber \\ & &
\quad \; \; \;
- \frac{ \left[ \varv_{2} \right]_{i+\lhalf,j} + 
\left[ \varv_{2} \right]_{i-\lhalf,j} }
{ \left[ g_2 \right]_i}
\left[ \frac{\partial g_{2}}{\partial x_1} \right]_{i+\lhalf} \Biggr)
\vspace{-0.1in}
\nonumber \\ & &
\quad \; \; \;
+ \Biggl( \frac{ \left[ \varv_{1} \right]_{i,j+\lthreehalf} - 
\left[ \varv_{1} \right]_{i,j+\lhalf} }
{\left[ g_2 \right]_i 
\left( \left[ x_2 \right]_{j+\lthreehalf} - 
\left[ x_2 \right]_{j+\lhalf} \right)} 
\vspace{-0.1in}
\nonumber \\ & &
\quad \; \; \;
- \frac{ \left[ \varv_{2} \right]_{i+\lhalf,j+\lhalf} + 
\left[ \varv_{2} \right]_{i-\lhalf,j+\lhalf} }
{ \left[ g_2 \right]_i }
\left[ \frac{\partial g_{2}}{\partial x_1} \right]_{i+\lhalf} \Biggr)
\vspace{-0.1in}
\nonumber \\ & &
\quad \; \; \;
+ \Biggr( \frac{ \left[ \varv_{1} \right]_{i+1,j+\lhalf} - 
\left[ \varv_{1} \right]_{i+1,j-\lhalf} }
{\left[ g_2 \right]_{i+1} 
\left( \left[ x_2 \right]_{j+\lhalf} - \left[ x_2 \right]_{j-\lhalf} \right)}
\vspace{-0.1in}
\nonumber \\ & &
\quad \; \; \;
- \frac{ \left[ \varv_{2} \right]_{i+\lthreehalf,j} + 
\left[ \varv_{2} \right]_{i+\lhalf,j} }
{ \left[ g_2 \right]_{i+1}}
\left[ \frac{\partial g_{2}}{\partial x_1} \right]_{i+\lthreehalf} \Biggr)
\vspace{-0.1in}
\nonumber \\ & &
\quad \; \; \;
+ \Biggl( \frac{ \left[ \varv_{1} \right]_{i+1,j+\lthreehalf} - 
\left[ \varv_{1} \right]_{i+1,j+\lhalf} }
{ \left[ g_2 \right]_{i+1} 
\left( \left[ x_2 \right]_{j+\lthreehalf} - 
\left[ x_2 \right]_{j+\lhalf}\right)}
\vspace{-0.1in}
\nonumber \\ & &
\quad \; \; \;
- \frac{ \left[ \varv_{2} \right]_{i+\lthreehalf,j+1} + 
\left[ \varv_{2} \right]_{i+\lhalf,j+1} }
{ \left[ g_2 \right]_{i+1}} 
\left[ \frac{\partial g_{2}}{\partial x_1} \right]_{i+\lthreehalf} 
\Biggr) \Biggr\}.
\end{eqnarray}

\subsection{Iterative Solution of the Implicit Finite-Difference Equations
\label{sec:iter_soln}}

The implicit finite-differencing scheme for the numerical 
solutions of equations~(\ref{eq:dc-e}) and (\ref{eq:dc-eb})
is fully described in the previous three subsections of this appendix.
This finite-differencing results in a coupled set of nonlinear
algebraic equations that must be solved for the complete set of 
radiation energy densities for each flavor of radiation.
In this section, we briefly describe the iterative method employed 
to find the solution of this set of nonlinear equations.

The complete set of implicitly discretized equations that must be solved 
for each radiation flavor is given by equations~(\ref{eq:nu_tr_fd}) and 
(\ref{eq:anu_tr_fd}).  The unknowns in this set of equations are
$[E_\epsilon]^\npo_{k+(1/2),i+(1/2),j+(1/2)}$ 
and
$[\bar{E}_\epsilon]^\npo_{k+(1/2),i+(1/2),j+(1/2)}$ 
for $k=1,\ldots,N_g$; $i=i_{\rm min},\ldots,i_{\rm max}$;
$j=j_{\rm min},\ldots,j_{\rm max}$.  
We use $i_{\rm min}, i_{\rm max}, j_{\rm min}$, and $j_{\rm max}$ to
indicate lower and upper bounds of $x_1$ and $x_2$ dimensions,
respectively (see Appendix~\ref{app:bconds}).

Nonlinearities in this set of equations arise in two forms.
First, the pair-production terms in equations
(\ref{eq:nu_tr_src_trm}) and (\ref{eq:anu_tr_src_trm}) give rise to
a bilinear coupling between neutrinos and antineutrinos if the 
pair-production kernel $G$ is non-zero.  This is what necessitates the
simultaneous solution of equations~(\ref{eq:nu_tr_fd}) and 
(\ref{eq:anu_tr_fd}).
The second set of nonlinearities are quadratic in form and arise 
from the Pauli-blocking factors that appear in the scattering 
integrals in  equations~(\ref{eq:nu_tr_src_trm}) and 
(\ref{eq:anu_tr_src_trm}).   In the case where $\alpha=0$ the 
nonlinearities disappear and the set of equations decouples 
into two separate sets of linear equations: one set describing 
the energy densities of the radiation particles 
and one set describing the energy densities of the radiation
antiparticles.  

In the nonlinear case where $\alpha$ and $G$ are non-zero, this 
set of equations could be solved by various means.  Because of
the sparseness of the nonlinear system, we have used a Newton-BiCGSTAB
variant of Newton-Krylov iteration \citep{kelley95}.  
This iterative algorithm consists of an outer Newton iteration 
\citep{kelley03} loop surrounding a Krylov subspace iteration
\citep{saad03} to solve the linearized Jacobian system.  The 
overall flow of the algorithm is depicted in Figure~\ref{fig:nk_algorithm}.
\begin{figure}[htbp]
\vspace{0.5in}
\begin{center}
\includegraphics[scale=0.7]{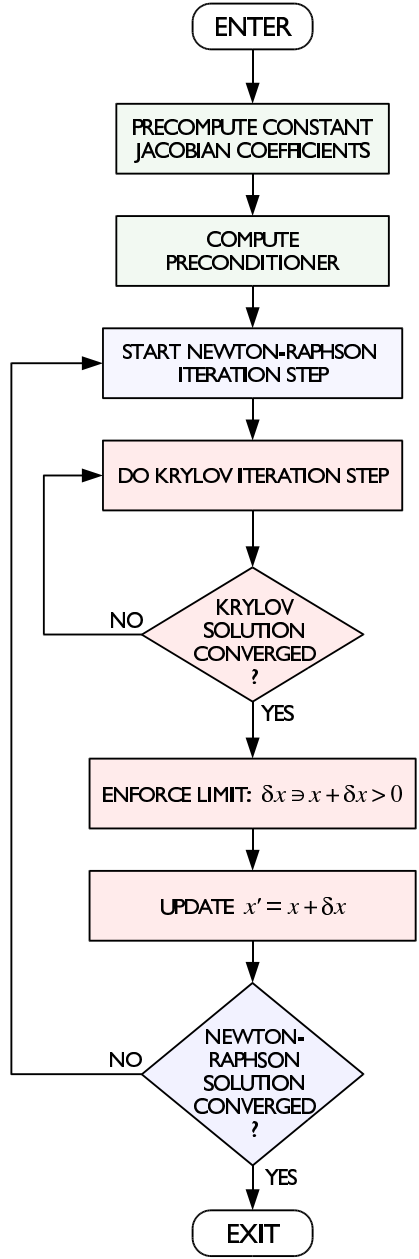}
\end{center}
\caption{\label{fig:nk_algorithm} A flowchart of the Newton-Krylov
iteration algorithm used to solve the implicitly
discretized nonlinear radiation diffusion
equations. In this flow chart, for simplicity, we
have omitted error conditions, such as
non-convergence of either the Newton or the Krylov
loops.  Such conditions are always considered as
fatal, and the calculation terminates.}
\end{figure}
The BiCGSTAB iteration \citep{kelley95,saad03} is implemented exactly as 
described in \cite{Barrett94}.  The iterative convergence of this method is 
hastened, or enabled, through the use of preconditioners \citep{chen05}.
The particular preconditioning strategy that we employ is a 2-D extension of
the sparse approximate inverse approach used for 1-D multigroup flux-limited
diffusion problems \citep{sss04}.

This iterative Newton-BiCGSTAB approach approach has several distinct 
advantages.  First, the matrix corresponding
to the Jacobian need not be computed and stored; instead, the Jacobian
can be applied in operator form.  Second, the knowledge of 
effective preconditioners from the linear flux-limited diffusion 
equation can be exploited to accelerate iterative convergence.
Third, the method, including preconditioning, is easily parallelizable.

One minor deviation from standard Newton-Krylov iteration that we 
employ is the enforcement of positivity during the nonlinear
iteration step.  In Figure~\ref{fig:nk_algorithm}, we indicate
this by the red box prior to the solution update.  During this step,
the Newton step for a particular variable is limited if it would
result in a negative value for the energy density.

The stopping criterion for the Newton iteration is
\begin{equation}
\max\left\{
\frac{ {{\bf X }^{(\ell+1)}_j}-{{\bf X }^{(\ell)}_j} }
{{\bf X }^{(\ell)}_j} 
\right\} < \epsilon_{\rm Newt},
\label{eq:stop}
\end{equation}
where the solution tolerance, $\epsilon_{\rm Newt}$, is chosen to be some value
between $10^{-6}$ and $10^{-8}$.
In equation~(\ref{eq:stop}), ${\bf X}^\ell$ is the $\ell$th estimate of the
unknown vector is given by
\begin{equation}
{\bf X }^T \equiv \left(
E^\npo_1,\ldots,E^\npo_{N_g},
\bar{E}^\npo_1,\ldots,\bar{E}^\npo_{N_g},
\right).
\end{equation}
The stopping criterion for the Krylov iteration (BiCGSTAB) is based on the
norm of the linear system residual ${\bf r}$, and is given by
\begin{equation}
|{\bf r}| < \epsilon_{\rm Krylov} |{\bf b}|,
\label{eq:stop2}
\end{equation}
where ${\bf b}$ is the right-hand-side of the linearized system
of equations on each Newton iteration.  We typically choose 
$\epsilon_{\rm Krylov} = 10^{-2} \epsilon_{\rm Newt}$, based
on extensive numerical trials, to yield an inexact-Newton method.

\section{Discretization of Energy and Lepton Exchange Equations
\label{app:collision}} 


In the radiation-transport equations in
Appendix~\ref{app:rad-trans}, collisional processes that change lepton 
number and/or energy of the components of the radiation field are 
included in the
terms ${\mathbb S_\epsilon}$ and ${\bar{\mathbb S}_\epsilon}$
in equations~(\ref{eq:dc-e}) and (\ref{eq:dc-eb}) respectively.
Since these terms depend on the radiation energy densities
$E_\epsilon$ and $\bar{E}_\epsilon$, once
equations~(\ref{eq:dc-e}) and (\ref{eq:dc-eb})
have been solved, the amount of energy and lepton (if any) exchange
is thus determined.  Once the values of 
${\mathbb S_\epsilon}$ and ${\bar{\mathbb S}_\epsilon}$ are known,
equations~(\ref{eq:ye-source}) and (\ref{eq:e-coll}) can then be
solved to accomplish steps $d$, $f$, and $h$ of Figure~\ref{fig:timestep}.

The right-hand side of equation~(\ref{eq:ye-source}) gives the total
amount of lepton number exchange with a given flavor of radiation in terms
of an integral over the spectrum:
\begin{equation} \label{eq:n_xch}
{\mathbb N} = - \int 
\left( \frac{ ^e{\mathbb S}_\epsilon -
^e\bar{{\mathbb S}}_\epsilon}{\epsilon}\right) d\epsilon.
\end{equation}
The leading superscript $e$ on quantities in equation~(\ref{eq:n_xch})
is used to indicate radiation of the electron neutrino flavor
(muon and tauon type neutrinos do not contribute to lepton number
exchange since they are only produced in particle-antiparticle pairs).
The first term in the 
integrand of equation  (\ref{eq:n_xch}) gives the lepton number exchange
between matter and radiation particles while the second term gives
the lepton number exchange with antiparticles (if they are present).
The minus sign in the second term reflects the negative quantum number
assigned to the antiparticles.

In our discretization of equation~(\ref{eq:ye-source}), we replace the
integral with a midpoint-rule summation and the time derivative
with a forward difference to get
\begin{equation} \label{eq:n_xch_fd}
\left[n_e \right]^{n+1}_{i+(1/2),j+(1/2)} = 
\left[n_e \right]^{n+b}_{i+(1/2),j+(1/2)}
- \Delta t \sum_{\ell=0}^{N_g-1} \left[\Delta\epsilon\right]_{\ell+(1/2)} 
\left( 
\frac{ \left[^e{\mathbb S}_\epsilon\right]^{n+b}_{i+(1/2),j+(1/2)} 
-
\left[^e\bar{{\mathbb S}}_\epsilon\right]^{n+b}_{i+(1/2),j+(1/2)}}
{\left[\epsilon\right]_{\ell+(1/2)}}
\right),
\end{equation}
which is the discretization of equation~(\ref{eq:ye-source}).
Once the new value of the electron number density is known, the new
electron fraction can be computed trivially as
\begin{equation} \label{eq:new_ye}
\left[Y_e \right]^{n+1}_{i+(1/2),j+(1/2)} = 
\frac{\left[n_e \right]^{n+1}_{i+(1/2),j+(1/2)}}
{\left[\rho \right]^{n+1}_{i+(1/2),j+(1/2)}}.
\end{equation}

The solution of the energy exchange equation (eq.~[\ref{eq:e-coll}])
is similar to the solution of equation~(\ref{eq:ye-source}), but 
involves several additional steps.  Although lepton number 
exchange only takes place for neutrinos of the electron flavor, 
the exchange of energy takes place for all flavors.  The right-hand-side
of equation~(\ref{eq:e-coll}) is given by
\begin{equation} \label{eq:e_xch}
{\mathbb S} = 
- \sum_\ell \int 
\left({ ^\ell{\mathbb S}_\epsilon +
^\ell\bar{{\mathbb S}}_\epsilon}\right) d\epsilon,
\end{equation}
where the leading superscript $\ell$ is used to indicate the 
flavor of the radiation-species (in the case of neutrinos
$\ell$ equals $e$, $\mu$ or $\tau$).
Equation~(\ref{eq:e-coll}) thus becomes
\begin{equation}
\left\ldbrack \frac{\partial E}{\partial t} \right\rdbrack_{\rm collision} = 
- \sum_\ell \int 
\left({ ^\ell{\mathbb S}_\epsilon +
^\ell\bar{{\mathbb S}}_\epsilon}\right) d\epsilon.
\label{eq:e-coll-sum}
\end{equation}
The concept of operator splitting can be applied to equation
(\ref{eq:e-coll-sum}), so we can solve separate equations of the form
\begin{equation}
\left\ldbrack \left(\frac{\partial E}{\partial t} \right)\right\rdbrack_{\rm collision-\ell} = 
- \int 
\left({ ^\ell{\mathbb S}_\epsilon +
^\ell\bar{{\mathbb S}}_\epsilon}\right) d\epsilon
\label{eq:e-coll-F}
\end{equation}
for each radiation species.  These operator split sub-equations are 
each solved immediately after the solution of the radiation diffusion
equation for the corresponding radiation flavor.  In the case of electron
type neutrinos in substep $d$ (box $d$ of Figure~\ref{fig:timestep}), 
equation~(\ref{eq:e-coll-F}) is solved after equation
(\ref{eq:n_xch_fd}) is solved.  Equation~(\ref{eq:e-coll-F}) is solved
for muon neutrinos in substep $f$ and for tauon neutrinos in substep $h$.

  The discretization of equation~(\ref{eq:e-coll-F}) proceeds 
almost identically to the discretization of equation~(\ref{eq:n_xch}).
In substep $d$ the discretized equation solved to account for energy
exchange with electron neutrinos is
\begin{equation} \label{eq:e_xch_fde}
\left[E \right]^{n+d}_{i+(1/2),j+(1/2)} = 
\left[E \right]^{n+b}_{i+(1/2),j+(1/2)}
- \Delta t \sum_{\ell=0}^{N_g-1} \left[\Delta\epsilon\right]_{\ell+(1/2)} 
\left( 
{ \left[^e{\mathbb S}_\epsilon\right]^{n+c}_{i+(1/2),j+(1/2)} 
-
\left[^e\bar{{\mathbb S}}_\epsilon\right]^{n+c}_{i+(1/2),j+(1/2)}}
\right).
\end{equation}
In substep $f$ the discretized equation solved to account for energy
exchange with muon neutrinos is
\begin{equation} \label{eq:e_xch_fdm}
\left[E \right]^{n+f}_{i+(1/2),j+(1/2)} = 
\left[E \right]^{n+d}_{i+(1/2),j+(1/2)}
- \Delta t \sum_{\ell=0}^{N_g-1} \left[\Delta\epsilon\right]_{\ell+(1/2)} 
\left( 
{ \left[^\mu{\mathbb S}_\epsilon\right]^{n+e}_{i+(1/2),j+(1/2)} 
-
\left[^\mu\bar{{\mathbb S}}_\epsilon\right]^{n+e}_{i+(1/2),j+(1/2)}}
\right).
\end{equation}
In substep $h$ the discretized equation solved to account for energy
exchange with tauon neutrinos is
\begin{equation} \label{eq:e_xch_fdt}
\left[E \right]^{n+h}_{i+(1/2),j+(1/2)} = 
\left[E \right]^{n+f}_{i+(1/2),j+(1/2)}
- \Delta t \sum_{\ell=0}^{N_g-1} \left[\Delta\epsilon\right]_{\ell+(1/2)} 
\left( 
{ \left[^\tau{\mathbb S}_\epsilon\right]^{n+g}_{i+(1/2),j+(1/2)} 
-
\left[^\tau\bar{{\mathbb S}}_\epsilon\right]^{n+g}_{i+(1/2),j+(1/2)}}
\right).
\end{equation}

  In each of substeps $d$, $f$, and $h$, it is necessary, after the
new value of $E$ is obtained, to use the equation of state to obtain 
new values of the temperature and the pressure corresponding to the 
updated internal energy density and, in the case of substep $d$, the new
updated electron fraction.  This is performed by inverting the EOS to
find the temperature corresponding to the new energy density
via a combination of Newton-Raphson and bisection methods for
nonlinear equations.

In substep $d$ the new temperature is found by iteratively solving
\begin{equation} \label{eq:eos_inv_d}
E\left(
\left[T \right]^{n+d}_{i+(1/2),j+(1/2)}, 
\left[\rho \right]^{n+1}_{i+(1/2),j+(1/2)}, 
\left[Ye \right]^{n+1}_{i+(1/2),j+(1/2)} 
\right) 
-
\left[E \right]^{n+d}_{i+(1/2),j+(1/2)} 
= 0
\end{equation}
for the new temperature, based on the updated values of $E$ and $Y_e$.
In the case of substep $f$ the equation solved for the new temperature is
\begin{equation} \label{eq:eos_inv_f}
E\left(
\left[T \right]^{n+f}_{i+(1/2),j+(1/2)}, 
\left[\rho \right]^{n+1}_{i+(1/2),j+(1/2)}, 
\left[Ye \right]^{n+1}_{i+(1/2),j+(1/2)} 
\right) 
-
\left[E \right]^{n+f}_{i+(1/2),j+(1/2)} 
= 0,
\end{equation}
while in substep $h$ the equation solved is
\begin{equation} \label{eq:eos_inv_h}
E\left(
\left[T \right]^{n+h}_{i+(1/2),j+(1/2)}, 
\left[\rho \right]^{n+1}_{i+(1/2),j+(1/2)}, 
\left[Ye \right]^{n+1}_{i+(1/2),j+(1/2)} 
\right) 
-
\left[E \right]^{n+h}_{i+(1/2),j+(1/2)} 
= 0.
\end{equation}
In each case, once the new temperature is known, the pressure is
determined by the equation of state.  The computational cost of the
equation of state computations needed to solve iteratively
equations~(\ref{eq:eos_inv_d})-(\ref{eq:eos_inv_h}) is negligible
compared to the computational effort involved in solving the
nonlinear systems corresponding to the radiation diffusion equations.

\section{Discretization of the Gas-Momentum Radiation Equation 
\label{app:gmom-rad}}


In this appendix we describe the solution of equation
(\ref{eq:s-rad}).
Following the update to gas momentum from the matter contribution, as
described in Appendix~\ref{app:gmomsrc}, we next update the gas
momentum to account for radiation-matter interaction.  This update
corresponds to the solution of the $\ldbrack\partial (\rho {\bf
v})/\partial t\rdbrack_{\rm radiation}$ term in
equation~(\ref{eq:s-split}),
\begin{equation} \label{eq:rad_acc}
\left\ldbrack \frac{ \partial \left( \varv_1 \: {\hat{\bf x}}_1 +
\varv_2 \: {\hat{\bf x}}_2\right) }
{\partial t}\right\rdbrack_{\rm radiation}
= - {\bfnabla} \cdot  {\mathsf P}_{{\rm rad}}.
\end{equation}
For radiation transport algorithms that have a non-zero
radiation-matter momentum exchange, this step of the algorithm can
also be used to solve for the $\ldbrack\partial (\rho {\bf
v})/\partial t\rdbrack_{\rm collision}$ term in
equation~(\ref{eq:s-split}).  In such a case, the right-hand side of
equation~(\ref{eq:rad_acc}) will have an additional term ${\mathbb
P}$.  As previously stated in \S \ref{sec-equations}, we assume
${\mathbb P} = 0$.  (However, see \S\ref{sec-coup-eq} for a
discussion of radiation-transport algorithms and
non-zero value of ${\mathbb P}$.)

To evaluate ${\bfnabla} \cdot  {\mathsf P}_{{\rm rad}}$, we use
\citet{sn92a}, equations~(130)--(132).  Note that several terms
of these equations in \cite{sn92a} have sign errors.
Specifically,
the $\partial h_1/\partial x_1$ term in  their equation~(130) and 
the $\partial h_2/\partial x_2$ term in  their equation~(131) both
have the incorrect signs. 
Fortunately, these errors are irrelevant
since the choice of the $h_1$ and $h_2$ for the coordinate systems
under consideration makes these terms vanish.  
In writing the following
expressions, we have omitted terms in the divergence that are always
zero in our implementation. These include all terms in
$\partial/\partial x_3$, terms proportional to zero-valued tensor
elements ${\mathsf P}_{13}$, ${\mathsf P}_{23}$, terms
proportional to derivatives of the constant function $h_1$, and terms
containing $\partial g_{31}/\partial x_2$ and $\partial
g_{32}/\partial x_1$, which are zero by definition.  The
remaining non-zero terms give
\begin{eqnarray} 
\left\{{\bfnabla} \cdot  {\mathsf P}_{{\rm rad}}\right\}_{(1)} & = & 
\frac{1}{g_2 g_{31} g_{32}} \left\{ \frac{\partial}{\partial x_1}
\left( g_2 g_{31} g_{32} \left\{{\mathsf P}_{rad}\right\}_{11} \right) +
\frac{\partial}{\partial x_2} \left( g_{31} g_{32} 
\left\{{\mathsf P}_{rad}\right\}_{12} \right)
\right\}
\nonumber \\ \nonumber \\
& & - \frac{\left\{{\mathsf P}_{rad}\right\}_{22}}{g_2} \frac{\partial g_2}{\partial x_1}
- \frac{\left\{{\mathsf P}_{rad}\right\}_{33}}{g_{31} g_{32}} 
\frac{\partial ( g_{31} g_{32} )}{\partial x_1}
\end{eqnarray} 
\begin{eqnarray} 
\left\{{\bfnabla} \cdot  {\mathsf P}_{{\rm rad}}\right\}_{(2)} & = &
\frac{1}{g_{31} g_{32}} 
\left\{ \frac{\partial}{\partial x_1}
\left( g_{31} g_{32} \left\{{\mathsf P}_{rad}\right\}_{21} \right) 
+
\frac{\partial}{\partial x_2} \left( \frac{g_{31} g_{32}}{g_2} 
\left\{{\mathsf P}_{rad}\right\}_{22} \right)  \right\}
\nonumber \\ & & \nonumber \\ & &
- \frac{\left\{{\mathsf P}_{rad}\right\}_{33}}{g_2 g_{31} g_{32}} 
\frac{\partial ( g_{31} g_{32} )}{\partial x_2}
+\frac{(\left\{{\mathsf P}_{rad}\right\}_{12} 
+ \left\{{\mathsf P}_{rad}\right\}_{21})}{g_2}
\frac{\partial g_2}{\partial x_1}
\end{eqnarray} 
\begin{equation} 
\left\{{\bfnabla} \cdot  {\mathsf P}_{{\rm rad}}\right\}_{(3)} = 0.
\end{equation}

The quantity ${\mathsf P}_{{\rm rad}}$ represents
the radiation pressure tensor containing
contributions from the complete spectrum of
radiation.  Hence,
\begin{equation}
{\mathsf P}_{{\rm rad}} = \sum_\ell \int_0^\infty 
{^\ell {\mathsf P}}_\epsilon 
\left( + ^\ell\bar{{\mathsf P}}_\epsilon \right)
d\epsilon,
\end{equation}
where the sum over $l$ represents contributions from all species of
radiation. The added term in parentheses indicates the possible
presence of a distinct antiparticle that is also evolved.  Remembering
the definition of ${\mathsf P}_{{\rm rad}}$ (see
eq.~[\ref{eq:edddef}]), this expands in terms of the Eddington tensor
to give
\begin{equation}
{\mathsf P}_{{\rm rad}} = \sum_\ell \int_0^\infty 
{^\ell{\mathsf X}}_\epsilon {^\ell E}_\epsilon 
\left(+ ^\ell\bar{{\mathsf X}}_\epsilon ^\ell\bar{E}_\epsilon \right)
d\epsilon,
\end{equation}
where, once again, the possible presence of an antiparticle is
indicated in parentheses.

Expressing equation~(\ref{eq:rad_acc}) in differenced form
(remembering that the density $\rho$ is constant over this step in the
operator splitting and emerges from the time derivative), we have the
following, where here, the possible antiparticle has been suppressed to
make the equations readable, and the radiation-energy spectra have
been discretized using index $k$. This update corresponds to
substep~$j$ ~(see~\S\ref{sec:oos}):
\begin{eqnarray} \label{eq:radmom_x1}
\left[ \varv_{1} \right]^{n+j}_{i,j+\lhalf} & = &
\left[ \varv_{1} \right]^{n+i}_{i,j+\lhalf} - 
\frac{2 \Delta t}
{ \left[ \rho \right]^\npo_{i+\lhalf,j+\lhalf} + 
\left[ \rho \right]^\npo_{i-\lhalf,j+\lhalf} }
\nonumber \\ & &
\times \sum_\ell \sum_k
\Biggl( \frac{1}{ 
\left\{ g_2 \right]_i \left[ g_{31} \right]_i
\left(\left[ x_1 \right]_{i+\lhalf} - \left[ x_1 \right]_{i-\lhalf}\right)} 
\Biggr.
\nonumber \\ & &
\times \Biggl\{ 
\left[ g_2 \right]_{i+\lhalf} \left[ g_{31} \right]_{i+\lhalf}
\left( 
\left[ ^\ell\left\{ {\mathsf X}_\epsilon \right\}_{11} 
\right]_{k+\lhalf,i+\lhalf,j+\lhalf}
{^\ell\left[ E_\epsilon \right]}_{k+\lhalf,i+\lhalf,j+\lhalf} \right)
\nonumber \\ & &
\quad\:\:\:
- \left[ g_2 \right]_{i-\lhalf} \left[ g_{31} \right]_{i-\lhalf}
\left( 
\left[ ^\ell\left\{ {\mathsf X}_\epsilon \right\}_{11} 
\right]_{k+\lhalf,i-\lhalf,j+\lhalf} 
{^\ell\left[ E_\epsilon \right]}_{k+\lhalf,i-\lhalf,j+\lhalf} \right) \Biggr\} 
\nonumber \\ & &
+ \frac{1}{ 4 \left[ g_2 \right]_{i} \left[ g_{31} \right]_{j+\lhalf}
 \left( \left[ x_2 \right]_{j+1} - 
\left[ x_2 \right]_{j} \right)}
\nonumber \\ & &
\times \Biggl\{ 
\left[ g_{32} \right]_{j+1}
\left( \left[ ^\ell\left\{ {\mathsf X}_\epsilon \right\}_{12} 
\right]_{k+\lhalf,i+\lhalf,j+\lthreehalf} 
{^\ell\left[ E_\epsilon \right]}_{k+\lhalf,i+\lhalf,j+\lthreehalf}
\right.
\nonumber \\ & &
\left.
\quad\quad\quad\quad\:\:\:
+ \left[ ^\ell\left\{ {\mathsf X}_\epsilon \right\}_{12} 
\right]_{k+\lhalf,i-\lhalf,j+\lthreehalf} 
{^\ell\left[ E_\epsilon \right]}_{k+\lhalf,i-\lhalf,j+\lthreehalf}
\right. 
\nonumber \\ & &
\nonumber \\ & &
\left.
\quad\quad\quad\quad\:\:\:
+ \left[ ^\ell\left\{ {\mathsf X}_\epsilon \right\}_{12} 
\right]_{k+\lhalf,i+\lhalf,j+\lhalf} 
{^\ell\left[ E_\epsilon \right]}_{k+\lhalf,i+\lhalf,j+\lhalf}
\right.
\nonumber \\ & &
\left. 
\quad\quad\quad\quad\:\:\:
+ \left[ ^\ell\left\{ {\mathsf X}_\epsilon \right\}_{12} 
\right]_{k+\lhalf,i-\lhalf,j+\lhalf}
{^\ell\left[ E_\epsilon \right]}_{k+\lhalf,i-\lhalf,j+\lhalf} \right) 
\Biggr. 
\nonumber \\ & &
\Biggl. \quad
- \left[ g_{32} \right]_{j}
\left( 
\left[ ^\ell\left\{ {\mathsf X}_\epsilon \right\}_{12}
\right]_{k+\lhalf,i+\lhalf,j+\lhalf}
{^\ell\left[ E_\epsilon \right]}_{k+\lhalf,i+\lhalf,j+\lhalf}  
\right.
\nonumber \\ & &
\left.
\quad\quad\quad\;\;\;\;
+ \left[ ^\ell\left\{ {\mathsf X}_\epsilon \right\}_{12}
\right]_{k+\lhalf,i-\lhalf,j+\lhalf}
{^\ell\left[ E_\epsilon \right]}_{k+\lhalf,i-\lhalf,j+\lhalf}  
\right.
\nonumber \\ & &
\nonumber \\ & &
\left.
\quad\quad\quad\;\;\;\;
+ \left[ ^\ell\left\{ {\mathsf X}_\epsilon \right\}_{12}
\right]_{k+\lhalf,i+\lhalf,j-\lhalf}
{^\ell\left[ E_\epsilon \right]}_{k+\lhalf,i+\lhalf,j-\lhalf}  
\right.
\nonumber \\ & &
\left. 
\quad\quad\quad\;\;\;\;
+ \left[ ^\ell\left\{ {\mathsf X}_\epsilon \right\}_{12}
\right]_{k+\lhalf,i-\lhalf,j-\lhalf}
{^\ell\left[ E_\epsilon \right]}_{k+\lhalf,i-\lhalf,j-\lhalf} \right) 
\Biggr\}
\nonumber \\ & & 
- \frac{1}{ 2 \left[ g_2 \right]_{i}} 
\left[ \frac{\partial g_2}{\partial x_1} \right]_{i}
\Biggl\{
\left[ ^\ell\left\{ {\mathsf X}_\epsilon \right\}_{22}
\right]_{k+\lhalf,i+\lhalf,j+\lhalf}
{^\ell\left[ E_\epsilon \right]}_{k+\lhalf,i+\lhalf,j+\lhalf}
\nonumber \\ & & 
\quad\quad\quad\quad\quad\quad\:\:\:\:
+ \left[ ^\ell\left\{ {\mathsf X}_\epsilon \right\}_{22}
\right]_{k+\lhalf,i-\lhalf,j+\lhalf} 
{^\ell\left[ E_\epsilon \right]}_{k+\lhalf,i-\lhalf,j+\lhalf} 
\Biggr\}
\nonumber \\ & &
- \Biggl. 
\frac{1}{ 2 \left[ g_{31} \right]_{i}} 
\left[ \frac{\partial g_{31}}{\partial x_1} \right]_{i}
\Biggl\{
\left[ ^\ell\left\{ {\mathsf X}_\epsilon \right\}_{33}
\right]_{k+\lhalf,i+\lhalf,j+\lhalf}
{^\ell\left[ E_\epsilon \right]}_{k+\lhalf,i+\lhalf,j+\lhalf} 
\nonumber \\ & &
\quad\quad\quad\quad\quad\quad\quad\:\:
+ \left[ ^\ell\left\{ {\mathsf X}_\epsilon \right\}_{33}
\right]_{k+\lhalf,i-\lhalf,j+\lhalf}
{^\ell\left[ E_\epsilon \right]}_{k+\lhalf,i-\lhalf,j+\lhalf} 
\Biggr\} 
\Biggr)
\end{eqnarray}
and
\begin{eqnarray} \label{eq:radmom_x2}
\left[ \varv_{2} \right]^{n+j}_{i+\lhalf,j} & = &
\left[ \varv_{2} \right]^{n+i}_{i+\lhalf,j} - 
\frac{2 \Delta t}
{\left[ \rho \right]^\npo_{i+\lhalf,j+\lhalf} + 
\left[ \rho \right]^\npo_{i+\lhalf,j-\lhalf}}
\nonumber \\ & &
\times \sum_\ell \sum_k
\Biggl( \frac{1}
{4 \left[ g_{31} \right]_{i+\lhalf}
\left(\left[ x_1 \right]_{i+1} - \left[ x_1 \right]_{i}\right)} \Biggr.
\nonumber \\ & &
\times \Biggl\{ \left[ g_{31} \right]_{i+1}
\left( \left[ ^\ell\left\{ {\mathsf X}_\epsilon \right\}_{21} 
\right]_{k+\lhalf,i+\lthreehalf,j+\lhalf}
{^\ell\left[ E_\epsilon \right]}_{k+\lhalf,i+\lthreehalf,j+\lhalf}
\right.
\nonumber \\ & &
\left.
\quad\quad\quad\quad\:\:
+ \left[ ^\ell\left\{ {\mathsf X}_\epsilon \right\}_{21} 
\right]_{k+\lhalf,i+\lthreehalf,j-\lhalf}
{^\ell\left[ E_\epsilon \right]}_{k+\lhalf,i+\lthreehalf,j-\lhalf}
\right.
\nonumber \\ & &
\nonumber \\ & &
\left.
\quad\quad\quad\quad\:\:
+ \left[ ^\ell\left\{ {\mathsf X}_\epsilon \right\}_{21} 
\right]_{k+\lhalf,i+\lhalf,j+\lhalf}
{^\ell\left[ E_\epsilon \right]}_{k+\lhalf,i+\lhalf,j+\lhalf}
\right.
\nonumber \\ & &
\left. 
\quad\quad\quad\quad\:\:
+ \left[ ^\ell\left\{ {\mathsf X}_\epsilon \right\}_{21}
\right]_{k+\lhalf,i+\lhalf,j-\lhalf} 
{^\ell\left[ E_\epsilon \right]}_{k+\lhalf,i+\lhalf,j-\lhalf} \right) 
\Biggr.
\nonumber \\ & &
\Biggl.
- \left[ g_{31} \right]_{i}
\left( 
\left[ ^\ell\left\{ {\mathsf X}_\epsilon \right\}_{21} 
\right]_{k+\lhalf,i+\lhalf,j+\lhalf}
{^\ell\left[ E_\epsilon \right]}_{k+\lhalf,i+\lhalf,j+\lhalf}  
\right.
\nonumber \\ & &
\left.
\quad\quad\quad
+ \left[ ^\ell\left\{ {\mathsf X}_\epsilon \right\}_{21} 
\right]_{k+\lhalf,i+\lhalf,j-\lhalf}
{^\ell\left[ E_\epsilon \right]}_{k+\lhalf,i+\lhalf,j-\lhalf}
\right.
\nonumber \\ & &
\nonumber \\ & &
\left.
\quad\quad\quad
+ \left[ ^\ell\left\{ {\mathsf X}_\epsilon \right\}_{21} 
\right]_{k+\lhalf,i-\lhalf,j+\lhalf}
{^\ell\left[ E_\epsilon \right]}_{k+\lhalf,i-\lhalf,j+\lhalf}
\right.
\nonumber \\ & &
\Biggl. \left. 
\quad\quad\quad
+ \left[ ^\ell\left\{ {\mathsf X}_\epsilon \right\}_{21}
\right]_{k+\lhalf,i-\lhalf,j-\lhalf}
{^\ell\left[ E_\epsilon \right]}_{k+\lhalf,i-\lhalf,j-\lhalf}
\right)
\Biggr\}
\nonumber \\ & &
+ \frac{1}{2 \left[ g_2 \right]_{i+\lhalf} \left[ g_{32} \right]_{j}
\left( \left[ x_2 \right]_{j+\lhalf} - 
\left[ x_2 \right]_{j-\lhalf} \right)}
\nonumber \\ & &
\times
\Biggl\{ \left[ g_{32} \right]_{j+\lhalf} \left(
\left[ ^\ell\left\{ {\mathsf X}_\epsilon \right\}_{22}
\right]_{k+\lhalf,i+\lhalf,j+\lhalf}
{^\ell\left[ E_\epsilon \right]}_{k+\lhalf,i+\lhalf,j+\lhalf}
\right)
\nonumber \\ & &
\;\:\:
- \left[ g_{32} \right]_{j-\lhalf} \left(
\left[ ^\ell\left\{ {\mathsf X}_\epsilon \right\}_{22}
\right]_{k+\lhalf,i+\lhalf,j-\lhalf}
{^\ell\left[ E_\epsilon \right]}_{k+\lhalf,i+\lhalf,j-\lhalf} \right)
\Biggr\}
\nonumber \\ & &
+ \frac{1}{\left[ g_2 \right]_{i+\lhalf}} 
\left[ \frac{ \partial g_2}{\partial x_1} \right]_{i+\lhalf}
\Biggl\{ \left[ ^\ell\left\{ {\mathsf X}_\epsilon \right\}_{21}
\right]_{k+\lhalf,i+\lhalf,j+\lhalf}
{^\ell\left[ E_\epsilon \right]}_{k+\lhalf,i+\lhalf,j+\lhalf} 
\nonumber \\ & &
\quad\quad\quad\quad\quad\quad\quad\quad\quad\quad\:\:\:
+ \left[ ^\ell\left\{ {\mathsf X}_\epsilon \right\}_{21}
\right]_{k+\lhalf,i+\lhalf,j-\lhalf}
{^\ell\left[ E_\epsilon \right]}_{k+\lhalf,i+\lhalf,j-\lhalf}
\Biggr\}
\nonumber \\ & &
+ \frac{1}{2 \left[ g_2 \right]_{i+\lhalf}\left[ g_{32} \right]_{j}} 
\left[ \frac{ \partial g_{32}}{\partial x_2} \right]_{j} 
\nonumber \\ & &
\times \Biggl\{\left(
\left[ ^\ell\left\{ {\mathsf X}_\epsilon \right\}_{33}
\right]_{k+\lhalf,i+\lhalf,j+\lhalf}
{^\ell\left[ E_\epsilon \right]}_{k+\lhalf,i+\lhalf,j+\lhalf}
\right.
\nonumber \\ & &
\quad\:\:
\left.
+ \left[ ^\ell\left\{ {\mathsf X}_\epsilon \right\}_{33}
\right]_{k+\lhalf,i+\lhalf,j-\lhalf}
{^\ell\left[ E_\epsilon \right]}_{k+\lhalf,i+\lhalf,j-\lhalf} \right) 
\Biggr\}
\Biggr).
\end{eqnarray}
In this two-dimensional formulation, $\varv_3 = 0$ at all times.
The summations run over energy index $k$ and species index $\ell$.  These
expressions are evaluated over the entire spatial domain, {\em i.e}.,
for each index $i$ and $j$.  To include the effects of antiparticles 
in
equations~(\ref{eq:radmom_x1}) and (\ref{eq:radmom_x2}), the following
substitution should be made in equations 
(\ref{eq:radmom_x1}) and (\ref{eq:radmom_x2}) 
for each instance of the expression to the left
of the arrow:
\begin{eqnarray}
\lefteqn{
\left[ ^\ell\left\{ {\mathsf X}_\epsilon \right\}_{mn}
\right]_{k+\lhalf,i+\lhalf,j+\lhalf}
{^\ell\left[ E_\epsilon \right]}_{k+\lhalf,i+\lhalf,j+\lhalf} \longrightarrow  }
\nonumber \\  & & 
\quad
\left[ ^\ell\left\{ {\mathsf X}_\epsilon \right\}_{mn}
\right]_{k+\lhalf,i+\lhalf,j+\lhalf}
{^\ell\left[ E_\epsilon \right]}_{k+\lhalf,i+\lhalf,j+\lhalf} 
\nonumber \\ & &
\:
+ \left[ ^\ell\left\{ \bar{{\mathsf X}}_\epsilon \right\}_{mn}
\right]_{k+\lhalf,i+\lhalf,j+\lhalf}
^\ell\left[ \bar{E}_\epsilon \right]_{k+\lhalf,i+\lhalf,j+\lhalf}.
\end{eqnarray}

\section{Implementation of Explicit Boundary 
Conditions\label{app:bconds}\label{sec-bc}}



In this appendix we discuss the formulation and implementation
of boundary conditions for the equations that are solved explicitly.
Boundary conditions for the implicitly solved radiation-diffusion
equations are discussed in Appendix~\ref{app:rad-trans}.

To permit solution of a wide range of problems, we have implemented
multiple options for boundary conditions into V2D.  In this appendix,
we describe four such options: (i) flat, (ii) periodic, (iii)
reflecting, and (iv) flux conserving.  V2D makes no requirement that
the same option be chosen in each direction or at each of the two
boundaries in a given direction---in many problems, it is common for
them to be different.

As shown in Figures~\ref{fig:bc_1} and \ref{fig:bc_2}, V2D is
structured to use boundaries that are centered on the integer mesh,
{\em i.e.}, boundary interfaces lie on cell faces, not cell centers.
In this appendix, for each option listed above, we present the
necessary boundary-value assignments for both scalar and vector
quantities.  Since our differencing scheme uses values no farther from
the point of interest than that of next-to-nearest neighbor, we need
carry no more than two boundary cells.

In what follows, we describe the application of boundary conditions
only in the $x_1$ direction.  Application of the same sets of
conditions in the $x_2$ direction is straightforward, as long as one
remembers that components of vectors normal and tangential to the
boundary interface are swapped.

%
\begin{figure}[htbp]
\begin{center}
\vspace{0.5in}
\includegraphics[scale=0.60]{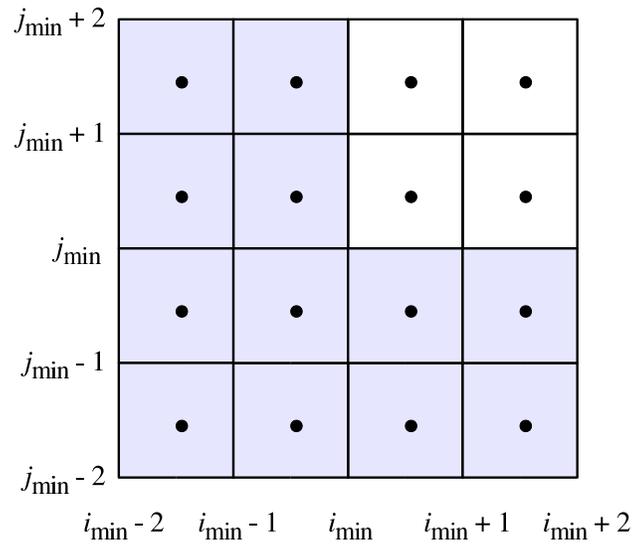}
\vspace{-2.0in}
\end{center}
\caption{\label{fig:bc_1} The bottom left-hand corner of the staggered
mesh. The shaded area indicates boundary cells to which boundary
conditions need to be supplied.  The unshaded area, where both $i \geq
i_{\rm min}$ and $j \geq j_{\rm min}$, is a portion of physical domain
where a solution is sought. The dots represent the positions of
cell-centered variables at half-integer locations.
\vspace{0.5in}}
\end{figure}
%
%
\begin{figure}[htbp]
\begin{center}
\includegraphics[scale=0.6]{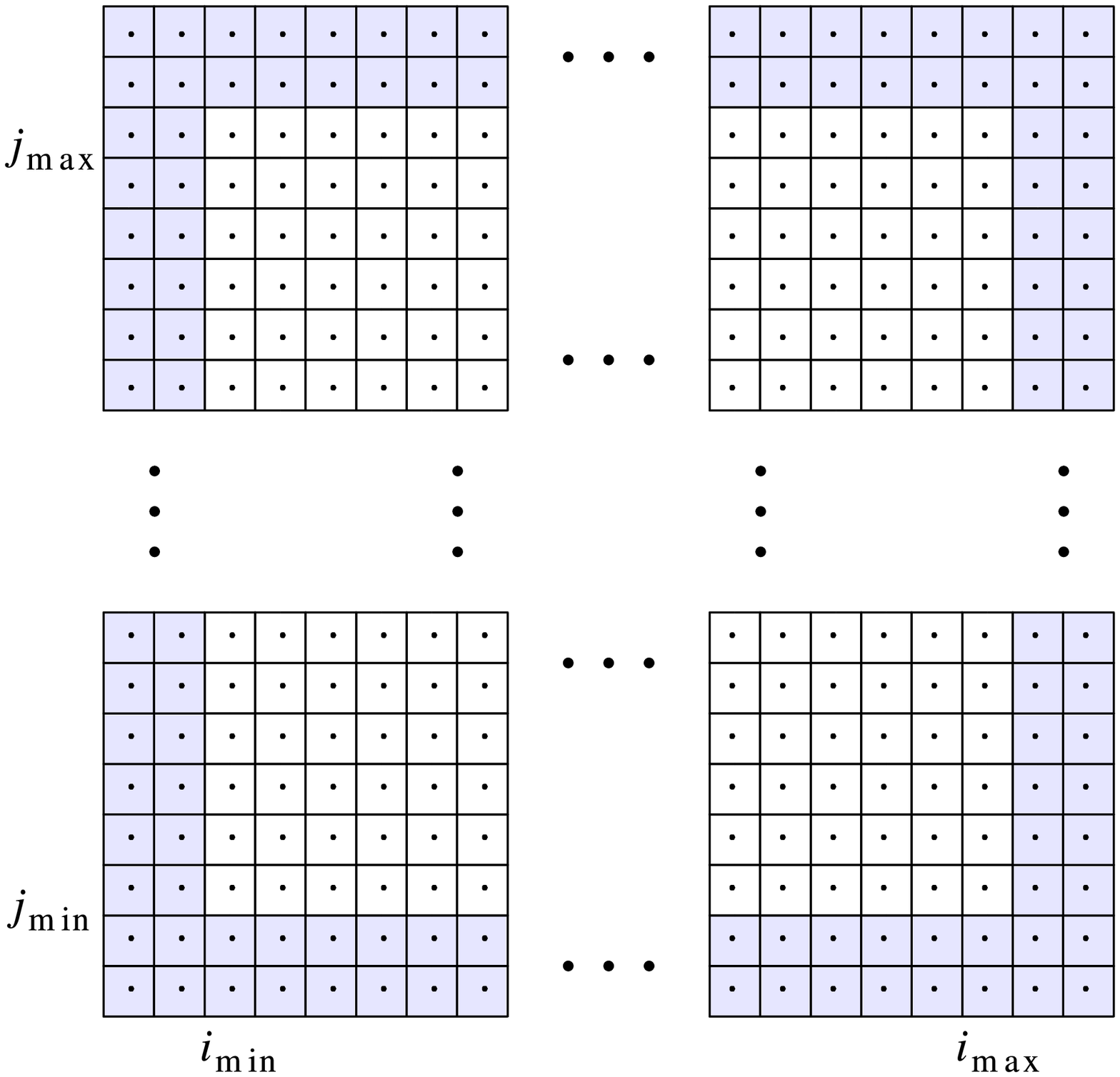}
\end{center}
\caption{\label{fig:bc_2} The four corners of the staggered mesh. As
in Figure~\ref{fig:bc_1}, shaded areas indicate boundary zones for
which boundary conditions need to be supplied.  Unshaded areas
represent portions of the physical domain, where both $i_{\rm min} \leq i
\leq i_{\rm max+1}$ and $j_{\rm min} \leq j \leq j_{\rm max+1}$.  The dots
represent the locations of cell-centered variables at half-integer
locations.  
}
\end{figure}
%

1. {\em Flat Boundary Conditions.} To implement flat boundary
conditions in the $x_1$ direction, we apply the following procedure:
First, for any scalar quantity $\psi$, at the left-hand boundary, we
set the values at the two left-hand boundary zones to the value at the
first physical zone:
\begin{equation}
\left[ \psi \right]_{i_{\rm min}-\lthreehalf,j+\lhalf} =
\left[ \psi \right]_{i_{\rm min}-\lhalf,j+\lhalf} =
\left[ \psi \right]_{i_{\rm min}+\lhalf,j+\lhalf}.
\end{equation}
This applies to all values of the $x_2$ index $j$, which is also the
case for all such equations in this appendix.  Similarly, to
implement flat conditions at the right-hand boundary, we set the scalar
values of the two right-hand boundary zones to have the value of the last
physical zone:
\begin{equation}
\left[ \psi \right]_{i_{\rm max}+(5/2),j+\lhalf} =
\left[ \psi \right]_{i_{\rm max}+\lthreehalf,j+\lhalf} =
\left[ \psi \right]_{i_{\rm max}+\lhalf,j+\lhalf}.
\end{equation}
For any vector quantity ${\smpmb \sigma}$, the procedure is
analogous. In the $x_1$ direction, for the $x_1$ component of vector
$\sigma$, we use at left-hand boundary,
\begin{equation}
\left[ \sigma_1 \right]_{i_{\rm min}-1,j+\lhalf} =
\left[ \sigma_1 \right]_{i_{\rm min},j+\lhalf} =
\left[ \sigma_1 \right]_{i_{\rm min}+1,j+\lhalf},
\end{equation}
where we note that $[ \sigma_1 ]_{i_{\rm min}-2,j}$ is
undefined.  At the right-hand boundary we have
\begin{equation}
\left[ \sigma_1 \right]_{i_{\rm max}+2,j+\lhalf} =
\left[ \sigma_1 \right]_{i_{\rm max}+1,j+\lhalf} =
\left[ \sigma_1 \right]_{i_{\rm max},j+\lhalf}.
\end{equation}
We also need to specify how $\sigma_2$ behaves at the both right- and
left-hand boundaries. At the left-hand boundary,
\begin{equation}
\left[ \sigma_2 \right]_{i_{\rm min}-\lthreehalf,j} =
\left[ \sigma_2 \right]_{i_{\rm min}-\lhalf,j} =
\left[ \sigma_2 \right]_{i_{\rm min}+\lhalf,j},
\end{equation}
and, at the right,
\begin{equation}
\left[ \sigma_2 \right]_{i_{\rm max}+(5/2),j} =
\left[ \sigma_2 \right]_{i_{\rm max}+\lthreehalf,j} =
\left[ \sigma_2 \right]_{i_{\rm max}+\lhalf,j}.
\end{equation}

2. {\em Periodic Boundary Conditions.} Next, we apply periodic
boundary conditions in the $x_1$ direction.  For a scalar, we have at
the left-hand boundary,
\begin{equation}
\left[ \psi \right]_{i_{\rm min}-\lthreehalf,j+\lhalf} =
\left[ \psi \right]_{i_{\rm max}-\lhalf,j+\lhalf}
\end{equation}
\begin{equation}
\left[ \psi \right]_{i_{\rm min}-\lhalf,j+\lhalf} =
\left[ \psi \right]_{i_{\rm max}+\lhalf,j+\lhalf},
\end{equation}
while at the right,
\begin{equation}
\left[ \psi \right]_{i_{\rm max}+\lthreehalf,j+\lhalf} =
\left[ \psi \right]_{i_{\rm min}+\lhalf,j+\lhalf}
\end{equation}
\begin{equation}
\left[ \psi \right]_{i_{\rm max}+(5/2),j+\lhalf} =
\left[ \psi \right]_{i_{\rm min}+\lthreehalf,j+\lhalf}.
\end{equation}
For vectors, the procedure is again analogous.  In the $x_1$
direction, for $\sigma_1$ at the left-hand boundary,
\begin{equation}
\left[ \sigma_1 \right]_{i_{\rm min}-1,j+\lhalf} =
\left[ \sigma_1 \right]_{i_{\rm max}-1,j+\lhalf},
\end{equation}
\begin{equation}
\left[ \sigma_1 \right]_{i_{\rm min},j+\lhalf} =
\left[ \sigma_1 \right]_{i_{\rm max},j+\lhalf},
\end{equation}
and at the right,
\begin{equation}
\left[ \sigma_1 \right]_{i_{\rm max}+1,j+\lhalf} =
\left[ \sigma_1 \right]_{i_{\rm min}+1,j+\lhalf},
\end{equation}
\begin{equation}
\left[ \sigma_1 \right]_{i_{\rm max}+2,j+\lhalf} =
\left[ \sigma_1 \right]_{i_{\rm min}+2,j+\lhalf}.
\end{equation}
For $\sigma_2$ at the left-hand boundary,
\begin{equation}
\left[ \sigma_2 \right]_{i_{\rm min}-\lthreehalf,j} =
\left[ \sigma_2 \right]_{i_{\rm max}-\lhalf,j},
\end{equation}
\begin{equation}
\left[ \sigma_2 \right]_{i_{\rm min}-\lhalf,j} =
\left[ \sigma_2 \right]_{i_{\rm max}+\lhalf,j},
\end{equation}
and at the right,
\begin{equation}
\left[ \sigma_2 \right]_{i_{\rm max}+\lthreehalf,j} =
\left[ \sigma_2 \right]_{i_{\rm min}+\lhalf,j},
\end{equation}
\begin{equation}
\left[ \sigma_2 \right]_{i_{\rm max}+(5/2),j} =
\left[ \sigma_2 \right]_{i_{\rm min}+\lthreehalf,j}.
\end{equation}

3. {\em Reflecting Boundary Conditions.} For scalars, applying reflecting
boundary conditions in the $x_1$ direction, at the left boundary gives
\begin{equation}
\left[ \psi \right]_{i_{\rm min}-\lhalf,j} =
\left[ \psi \right]_{i_{\rm min}+\lhalf,j},
\end{equation}
\begin{equation}
\left[ \psi \right]_{i_{\rm min}-\lthreehalf,j} =
\left[ \psi \right]_{i_{\rm min}+\lthreehalf,j},
\end{equation}
and at the right-hand boundary,
\begin{equation}
\left[ \psi \right]_{i_{\rm max}+\lthreehalf,j} =
\left[ \psi \right]_{i_{\rm max}+\lhalf,j},
\end{equation}
\begin{equation}
\left[ \psi \right]_{i_{\rm max}+(5/2),j} =
\left[ \psi \right]_{i_{\rm max}-\lhalf,j}.
\end{equation}
For vector quantities, reflection symmetry requires that the normal
component of a vector vanish at planes of symmetry, Thus, for
reflecting boundary conditions in the $x_1$ direction, we have the
following at the left-hand boundary for normal component, $\sigma_1$:
\begin{equation}
\left[ \sigma_1 \right]_{i_{\rm min},j+\lhalf} = 0,
\end{equation}
\begin{equation}
\left[ \sigma_1 \right]_{i_{\rm min}-1,j+\lhalf} =
-\left[ \sigma_1 \right]_{i_{\rm min}+1,j+\lhalf}.
\end{equation}
Similarly, at the right-hand boundary, we have
\begin{equation}
\left[ \sigma_1 \right]_{i_{\rm max}+1,j+\lhalf} = 0,
\end{equation}
\begin{equation}
\left[ \sigma_1 \right]_{i_{\rm max}+2,j+\lhalf} =
-\left[ \sigma_1 \right]_{i_{\rm max},j+\lhalf}.
\end{equation}
Since the tangential vector component, $\sigma_2$, is not defined at
the $x_1$ planes of symmetry, we do not have to apply a zero-value
requirement at boundaries.  Hence, boundary values are reflected in
the same way as scalars.  At the left-hand boundary,
\begin{equation}
\left[ \sigma_2 \right]_{i_{\rm min}-\lhalf,j} = 
\left[ \sigma_2 \right]_{i_{\rm min}+\lhalf,j},
\end{equation}
\begin{equation}
\left[ \sigma_2 \right]_{i_{\rm min}-\lthreehalf,j} = 
\left[ \sigma_2 \right]_{i_{\rm min}+\lthreehalf,j},
\end{equation}
while at the right,
\begin{equation}
\left[ \sigma_2 \right]_{i_{\rm max}+\lthreehalf,j} =
\left[ \sigma_2 \right]_{i_{\rm max}+\lhalf,j},
\end{equation}
\begin{equation}
\left[ \sigma_2 \right]_{i_{\rm max}+(5/2),j} = 
\left[ \sigma_2 \right]_{i_{\rm max}-\lhalf,j}.
\end{equation}

4. {\em Flux-Conserving Boundary Conditions.} This option is
especially useful for problems with spherical geometry, when applied
to the radial coordinate.  Rather than keeping a physical quantity
(such as density) constant across the boundary, this option allows
flux of a physical quantity (such as mass flux) to be kept constant.
At the outer boundary, a constant flux of $\psi$ is maintained when
\begin{equation}
\left[ \psi \right]_{i_{\rm max}+\lthreehalf,j} =
\left[ \psi \right]_{i_{\rm max}+\lhalf,j} 
\left( \frac{\left[ r \right]_{i_{\rm max}+\lhalf,j}}
{\left[ r \right]_{i_{\rm max}+\lthreehalf,j}} \right)^2,
\end{equation}
\begin{equation}
\left[ \psi \right]_{i_{\rm max}+(5/2),j} =
\left[ \psi \right]_{i_{\rm max}+\lhalf,j} 
\left( \frac{\left[ r \right]_{i_{\rm max}+\lhalf,j}}
{\left[ r \right]_{i_{\rm max}+(5/2),j}} \right)^2
\end{equation}
If the same condition is desired at an inner boundary, for a problem
where the computational grid does not extend all the way to $r=0$, we
have
\begin{equation}
\left[ \psi \right]_{i_{\rm min}-\lhalf,j} =
\left[ \psi \right]_{i_{\rm min}+\lhalf,j} 
\left( \frac{\left[ r \right]_{i_{\rm min}-\lhalf,j}}
{\left[ r \right]_{i_{\rm min}+\lhalf,j}} \right)^2,
\end{equation}
\begin{equation}
\left[ \psi \right]_{i_{\rm min}-\lthreehalf,j} =
\left[ \psi \right]_{i_{\rm min}-\lhalf,j} 
\left( \frac{\left[ r \right]_{i_{\rm min}-\lthreehalf,j}}
{\left[ r \right]_{i_{\rm min}-\lhalf,j}} \right)^2.
\end{equation}

Finally, it should be noted that the spatial boundary conditions we
have just outlined are also applied to each spectral component of
radiation quantities.  Additionally, with spectral quantities, there
is an another dimension for which boundary conditions must be
applied---the energy dimension.  In V2D, we always apply energy-space
boundary conditions that prevent both inflow below the lowest group
(the lower boundary is typically at zero anyway) and outflow beyond
the highest group.  In the case of the latter, we always carry a
sufficient number of groups such that radiation occupancy in the
higher groups is always small. Thus, it is largely irrelevant how we
treat this upper boundary.


\section{Enforcing the Pauli Exclusion Principal for Neutrinos
\label{app:pauli}}

As we discuss in \S\ref{sec-coll}, it is necessary to enforce 
the Pauli exclusion principle for neutrinos.  Neither 
of the operator split portions of equation~(\ref{eq:bte0}), 
the radiation diffusion
equation~(\ref{eq:dc-e}) and  the neutrino advection
equation~(\ref{eq:nu-advect}), guarantee that the neutrino occupancy
of a given energy state will remain less than unity.
This reflects the fact that equation~(\ref{eq:bte0}) is semiclassical
in the sense that only the collision integral is treated in a 
quantum-mechanical fashion.  Therefore, we must enforce the Pauli
exclusion principle as a separate step in our algorithm.  Other
authors \citep{bruenn85,mb93b} have adopted similar strategies.

The Pauli exclusion principle requires that the constraint 
(\ref{eq:pauli}) be satisfied.  In general, when a numerical solution
of equations~(\ref{eq:dc-e}) or  (\ref{eq:nu-advect}) are obtained,
the newly calculated value of the neutrino radiation energy density
$E_\epsilon$ may not satisfy this constraint.  As neutrinos are produced
during the collapse of a stellar core, the lower energy states in the center
of the core become fully populated.   The continued hydrodynamic
compression of neutrinos within a spatial zone, as 
described by equation~(\ref{eq:dc-e}) or  
(\ref{eq:nu-advect}), can result in occupation numbers greater than
unity in a given group.
In our finite difference notation, the Pauli constraint 
corresponds to 
\begin{equation}
\frac{\alpha \left[E_\epsilon\right]_{k+(1/2),i+(1/2),j+(1/2)}} 
{\left(\left[\epsilon\right]_{k+(1/2)}\right)^3}
\leq 1,
\end{equation}
where $\alpha = (hc)^3/4\pi g = $ 9.4523 ${\rm MeV}^4$ ${\rm cm}^3$
${\rm erg}^{-1}$ for both photons and neutrinos (for which $g=1$).

In order to counter this unphysical effect, we examine the radiation
energy densities each time equation~(\ref{eq:dc-e}) or (\ref{eq:nu-advect})
is solved.  If the values of the radiation energy density in a specific
group exceeds $\epsilon^3/\alpha$, the excess neutrinos are removed from that
group and placed in the next highest energy group or groups where phase 
space is available.  After this process is completed for all groups,
the matter internal energy is corrected to account for any change
in the total (integrated over the spectrum) 
neutrino energy density. A new temperature
and pressure are then computed for the zone.

 This enforcement algorithm is applied as the final operation in all 
substeps corresponding to the red boxes in Figure~\ref{fig:timestep}.

\clearpage
\end{document}